\documentclass[12pt,preprint]{aastex}

\usepackage{hyperref}

\usepackage{color}
\usepackage{ulem}
\usepackage{natbib}

\usepackage{lscape}
\usepackage{rotating}
\usepackage{float}
\usepackage{graphicx}
\usepackage{epstopdf}

\usepackage{xspace}

\newcommand{\Fermi}{{\textit{Fermi} }}

\newcommand{\pointlike}{\ensuremath{\mathtt{pointlike}}\xspace}
\newcommand{\gtlike}{\ensuremath{\mathtt{gtlike}}\xspace}

\newcommand{\ts}{\ensuremath{TS}\xspace}
\newcommand{\tspoint}{\ts\xspace}
\newcommand{\tsext}{\ensuremath{\ts_\mathrm{ext}}\xspace}

\newcommand{\tscutoff}{\ensuremath{\ts_\mathrm{cutoff}}\xspace}
\newcommand{\Ecutoff}{\ensuremath{E_\mathrm{cutoff}}\xspace}

\newcommand{\ph}{\rm{ph}\xspace}
\newcommand{\erg}{\rm{erg}\xspace}
\newcommand{\cm}{\rm{cm}\xspace}
\newcommand{\s}{\rm{s}\xspace}
\newcommand{\mev}{MeV\xspace}
\newcommand{\gev}{GeV\xspace}

\newcommand{\fluxunits}{\ensuremath{\ph\;\cm^{-2}\s^{-1}}}
\newcommand{\efluxunits}{\ensuremath{\erg\;\cm^{-2}\s^{-1}}}
\newcommand{\prefunits}{\ensuremath{\ph\;\cm^{-2}\s^{-1}\mev^{-1}}}

\newcommand{\fitsformat}[1]{\texttt{#1}}

\newcommand{\dnde}{\frac{dN}{dE}}

\newcommand{\figref}[1]{Figure~\ref{fig:#1}}
\newcommand{\eqnref}[1]{Eq.~\ref{eqn:#1}}

\newcommand{\subseclabel}[1]{\label{subsec:#1}}

\newcommand{\figlabel}[1]{\label{fig:#1}}
\newcommand{\eqnlabel}[1]{\label{eqn:#1}}

\newcommand{\PLSuperExpCutoff}{PLSuperExpCutoff}

\newcommand{\degree}{\ensuremath{^\circ}\xspace}

\slugcomment{Submitted to ApJ Suppl \\ {\bf
    Revision: 9828:
Last update: 2013-07-10 15:08:01 +0200

    }}

\shorttitle{$2^{nd}$ \textit{Fermi} LAT Pulsar Catalog}
\shortauthors{Fermi-LAT Collaboration}

\begin{document}

\title{The Second $Fermi$ Large Area Telescope Catalog of Gamma-ray Pulsars}
\author{
A.~A.~Abdo\altaffilmark{1}, 
M.~Ajello\altaffilmark{2}, 
A.~Allafort\altaffilmark{3}, 
L.~Baldini\altaffilmark{4}, 
J.~Ballet\altaffilmark{5}, 
G.~Barbiellini\altaffilmark{6,7}, 
M.~G.~Baring\altaffilmark{8}, 
D.~Bastieri\altaffilmark{9,10}, 
A.~Belfiore\altaffilmark{11,12,13}, 
R.~Bellazzini\altaffilmark{14}
% ...
}
\altaffiltext{1}{Center for Earth Observing and Space Research, College of Science, George Mason University, Fairfax, VA 22030, resident at Naval Research Laboratory, Washington, DC 20375, USA}
\altaffiltext{2}{Space Sciences Laboratory, 7 Gauss Way, University of California, Berkeley, CA 94720-7450, USA}
\altaffiltext{3}{W. W. Hansen Experimental Physics Laboratory, Kavli Institute for Particle Astrophysics and Cosmology, Department of Physics and SLAC National Accelerator Laboratory, Stanford University, Stanford, CA 94305, USA}
\altaffiltext{4}{Universit\`a  di Pisa and Istituto Nazionale di Fisica Nucleare, Sezione di Pisa I-56127 Pisa, Italy}
\altaffiltext{5}{Laboratoire AIM, CEA-IRFU/CNRS/Universit\'e Paris Diderot, Service d'Astrophysique, CEA Saclay, 91191 Gif sur Yvette, France}
\altaffiltext{6}{Istituto Nazionale di Fisica Nucleare, Sezione di Trieste, I-34127 Trieste, Italy}
\altaffiltext{7}{Dipartimento di Fisica, Universit\`a di Trieste, I-34127 Trieste, Italy}
\altaffiltext{8}{Rice University, Department of Physics and Astronomy, MS-108, P. O. Box 1892, Houston, TX 77251, USA}
\altaffiltext{9}{Istituto Nazionale di Fisica Nucleare, Sezione di Padova, I-35131 Padova, Italy}
\altaffiltext{10}{Dipartimento di Fisica e Astronomia ``G. Galilei'', Universit\`a di Padova, I-35131 Padova, Italy}
\altaffiltext{11}{Santa Cruz Institute for Particle Physics, Department of Physics and Department of Astronomy and Astrophysics, University of California at Santa Cruz, Santa Cruz, CA 95064, USA}
\altaffiltext{12}{Universit\`a degli Studi di Pavia, 27100 Pavia, Italy}
\altaffiltext{13}{INAF-Istituto di Astrofisica Spaziale e Fisica Cosmica, I-20133 Milano, Italy}
\altaffiltext{14}{Istituto Nazionale di Fisica Nucleare, Sezione di Pisa, I-56127 Pisa, Italy}
% ...

\begin{abstract}

%

This catalog summarizes 117 high-confidence $\geq0.1$ GeV gamma-ray pulsar detections using three years of data acquired by the Large Area Telescope
(LAT) on the \textit{Fermi} satellite.  Half are neutron stars discovered using LAT data, through  
periodicity searches in gamma-ray and radio data around LAT unassociated source positions.
The 117 pulsars are evenly divided into three groups: millisecond pulsars, young radio-loud pulsars, and young radio-quiet pulsars.
We characterize the pulse profiles and energy spectra and derive luminosities when distance information exists. 
Spectral analysis of the off-peak phase intervals indicates probable pulsar wind nebula emission for four pulsars, 
and off-peak magnetospheric emission for several young and millisecond pulsars. 
We compare the gamma-ray properties with those in the radio, optical, and X-ray bands.
We provide flux limits for pulsars with no observed gamma-ray emission, 
highlighting a small number of gamma-faint, radio-loud pulsars. 
The large, varied gamma-ray pulsar sample constrains emission models. 
\textit{Fermi}'s selection biases complement those of radio surveys, enhancing comparisons with predicted population distributions.

\end{abstract}


\tableofcontents

\section{Introduction}

%

Pulsars have featured prominently in the gamma-ray sky since the birth of gamma-ray astronomy. 
The Crab and Vela pulsars were the first two sources identified in the 1970's by SAS-2 \citep{SAS75} and COS-B \citep{COSB81}. 
In the 1990's the \textit{Compton Gamma-Ray Observatory} brought the pulsar grand total to at least seven, along with three other strong candidates \citep{Thompson08}.  
One of these early gamma-ray pulsars, Geminga, was undetected at radio wavelengths \citep{BnGgeminga96}.
Despite the meager number, neutron stars 
were estimated to represent a sizeable fraction of the EGRET unassociated low-latitude gamma-ray sources \citep{Romani1995}.  
The Large Area Telescope (LAT) on the \textit{Fermi Gamma-ray Space Telescope} did not just confirm the expectation: 
by discovering dozens of radio-quiet gamma-ray pulsars and  millisecond pulsars \citep[MSPs, thought to be old pulsars spun up to rapid periods via accretion from a companion,][]{Alpar1982},
the LAT established pulsars as the dominant GeV gamma-ray source class in the Milky Way \citep[][The First {\it Fermi} Large Area Telescope Catalog of Gamma-ray Pulsars, hereafter 1PC]{1PC}.

A pulsar is a rapidly-rotating, highly-magnetized neutron star, surrounded by a plasma-filled magnetosphere. 
Modeling its emission drives ever-more sophisticated electrodynamic 
calculations \citep[e.g.][]{WangHirotani2011,2012ApJ...746...60L,Kalapotharakos2012b,Petri2012}. 
Throughout this paper, we will call pulsars in the main population of the spin period ($P$) and period derivative ($\dot P$) plane `young' 
to distinguish them from the much older `recycled' pulsars, including MSPs. 
All known gamma-ray pulsars, and the most promising candidates to date, are rotation-powered pulsars (RPPs). 
The LAT has yet to detect significant gamma-ray pulsations from any accretion-powered pulsar or from the magnetars, anomalous X-ray pulsars, 
and soft gamma repeaters 
for which the dominant energy source is not electromagnetic braking, but magnetic field decay \citep{FermiMagnetars}.

Here we present 117 gamma-ray pulsars unveiled in three years of on-orbit observations with {\it Fermi}.
Extensive radio observations by the ``Pulsar Timing Consortium'' \citep{TimingForFermi} greatly enhanced the gamma-ray data analysis.
Our analysis of the gamma-ray pulsars is as uniform as is feasible given the  widely varying pulsar characteristics.
In addition to 1PC, this catalog builds on the 2$^\mathrm{nd}$ {\it Fermi} LAT source catalog \citep[][hereafter 2FGL]{2FGL},
which reported pulsations for 83 of the 2FGL sources, included here. An additional 27 pulsars were found to be spatially associated with 2FGL sources and pulsations have since been established for 12 of these, included here. 
The remaining 22 new pulsars with strong pulsations were either unassociated in 2FGL (pointing to subsequent pulsar discoveries, see Section \ref{pulsation_discovery}) 
or were below the 2FGL detection threshold and seen to pulse after 2FGL was completed. 
We provide our results in FITS files\footnote{\url{http://fits.gsfc.nasa.gov/}}
available in the journal electronic supplement as well as on the {\it Fermi} Science Support Center (FSSC) servers at \url{http://fermi.gsfc.nasa.gov/ssc/data/access/lat/2nd_PSR_catalog/}.

\section{Observations}
\label{obsvSection}
%

%
\textit{Fermi} was launched on 2008 June 11, carrying two gamma-ray
instruments: the LAT and the Gamma-ray Burst Monitor \citep{GBMinstrument}; the latter was not used to prepare this catalog.
\citet{LATinstrument} describe {\it Fermi}'s main instrument,
the LAT, and on-orbit performance of the LAT is reported by \citet{OnOrbit} and \citet{P7Paper}. The LAT is a pair-production
telescope composed of a $4\times4$ grid of towers. Each tower consists of a stack of tungsten foil converters interleaved with silicon-strip particle tracking detectors, mated with a hodoscopic cesium-iodide calorimeter. A segmented plastic scintillator anti-coincidence detector covers the grid to help discriminate charged particle backgrounds from gamma-ray photons. 
The LAT field of view is $\sim$2.4 sr. 
The primary operational mode is a sky survey where the satellite rocks between a pointing above the orbital plane and one below the plane after each orbit.
The entire sky is imaged every two orbits ($\sim$3 hours) 
and any given point on the sky is observed $\sim1/6^{\rm th}$ of the time. 
Each event classified as a gamma ray in the ground data processing has
its incident direction, incident energy ($E$), and time of arrival recorded in
the science data stream. 

The LAT is sensitive to gamma rays with energies $E$ from 20 MeV to over 300 GeV, 
with an on-axis effective area of $\sim8000$ cm$^{2}$ above 1
GeV. Multiple Coulomb scattering of the electron-positron pairs
created by converted gamma rays degrades the per-photon angular resolution with
decreasing energy as $\theta^2_{68}(E) = (3\fdg3)^{2}(100 \,{\rm MeV} /E)^{1.56} + (0\fdg1)^{2}$,
averaged over the acceptance for events converting in the front section of the LAT, 
where $\theta_{68}$ is the 68\% containment radius. 
The energy- and direction-dependent effective area and PSF are part of the Instrument Response Functions (IRFs).
The analysis in this paper used the \texttt{Pass7} V6 IRFs selecting events in the ``Source'' class \citep{P7Paper}.

The data used here to search for gamma-ray pulsars span 2008 August 4 to 2011 August 4.  Events were selected with reconstructed
energies from 0.1 to 100 GeV and directions within 2$^{\circ}$ of each pulsar position for pulsation searches (Section \ref{pulsation_discovery})
and 15$^{\circ}$ for spectral analyses (Section \ref{spectralSection}).
We excluded gamma rays collected when the LAT
was not in nominal science operations mode or the spacecraft rocking angle exceeded $52^{\circ}$, as well as those with measured
zenith angles $>100^{\circ}$,
to greatly reduce the residual gamma rays from the bright limb of the Earth.  
For PSRs J0205+6449, J1838$-$0537, and J2215+5135 we did not have timing solutions that were coherent over the full three years.  
For these pulsars the data sets for pulsation searches and light curve generation only include events within the validity range of the corresponding timing solutions.
For the first two pulsars, the data loss is $<7$\% but for PSR J2215+5135 it is 60\%.




\section{Pulsation Discovery}
\label{pulsation_discovery}
Events recorded by the LAT have timestamps derived from GPS
clocks integrated into the satellite's Guidance, Navigation, and Control (GNC) subsystem,
accurate to $<1$ $\mu$s relative to UTC \citep{OnOrbit}. The GNC subsystem provides the instantaneous spacecraft position with corresponding accuracy.
We compute pulsar rotational phases $\phi_i$ using \textsc{Tempo2} \citep{Hobbs2006} with the \texttt{fermi} plug-in \citep{Ray2011}. 
The \texttt{fermi} plug-in uses the recorded times and spacecraft positions combined with a pulsar timing ephemeris (specified
in a \textsc{Tempo2} parameter, or `par', file). 
The timing chain from the instrument clocks through the barycentering and epoch folding software is accurate to
better than a few $\mu$s for binary orbits, and significantly better for isolated pulsars \citep{TimingForFermi}.
The accuracy of the phase computation is thus determined by the ephemeris. 
The par file is created using radio or gamma-ray data, or a mix, depending on the LAT pulsar discovery method, as described in the following three subsections.
 
We required a $\geq5\sigma$ confidence level detection of modulation in the phase histogram for a pulsar to be included in this catalog, as described below.
Gamma-ray pulsar data are extremely sparse, often with fewer than one 
photon detected in tens of thousands (or in the case of MSPs, millions!) of pulsar rotations. 
In these circumstances, the favored techniques are unbinned tests for periodic signals. 
%
%
We use the H-test \citep{DeJager1989,deJager2010}, a
statistical test for discarding the null hypothesis that a set of
photon phases is uniformly distributed. For $N_{\gamma}$ gamma-rays, the H-test statistic is $H \equiv \max(Z^2_m -
4\times(m-1), 1 \leq m \leq 20)$, with $Z^2_m\equiv \frac{2}{N_{\gamma}}\sum_{k=1}^{m}
\alpha_k^2+\beta_k^2$, and $\alpha_k$ and $\beta_k$ the empirical
trigonometric coefficients $\alpha_k\equiv \sum_{i=1}^{N_{\gamma}}
\sin(2\pi k\phi_i)$ and $\beta_k\equiv \sum_{i=1}^{N_{\gamma}}
\cos(2\pi k\phi_i)$.  By including a search over a range of harmonics, the
H-test maintains sensitivity to light curves with a large range of
morphologies (e.g., sharp vs.~broad peaks).  
The sharpness of the peaks in the gamma-ray profile has a large impact on the detectability of the pulsar; 
in particular, pulsars with narrow, sharp peaks are easier to detect than pulsars with broad peaks covering more of the pulse phase.

Early in the mission most pulsation searches \citep[for example,][]{BSP} selected events with arrival directions within a fixed angular distance of the pulsar (the region of interest, or ROI) and a minimum energy cut ($E_\mathrm{min}$). 
Because of the range of pulsar spectra, fluxes, and levels of diffuse gamma-ray background, combined with the strongly energy-dependent PSF of the LAT, a number of trials must be done over a 
range of ROI sizes and $E_\mathrm{min}$ to optimize the detection significance for each candidate pulsar.

Using the probability that each event originates from the pulsar, computed from a spectral model of the region and the LAT IRFs, the H-test can be extended, using these probabilities as weights \citep{KerrWeighted}. 
This both improves the sensitivity of the H-test and removes the need for trials over event selection criteria.
The weights, $w_i$, representing the probability that the $i^{\rm th}$ event originates from the pulsar are
\begin{equation}
w_i = \frac{dN/dE_{\mathrm{psr}}(E_i,\vec{x}_i)} {\sum_j dN/dE_j(E_i,\vec{x}_i)},
\label{weightsEq}
\end{equation}
where $E_i$ and $\vec{x}_i$ are the observed energy and position on
the sky of the $i^{\rm th}$ event and $dN/dE_j$ is the phase-averaged spectra for the $j^{\rm th}$ source in the ROI (See Section \ref{spectralSection}).  Incorporating the weights yields the weighted H-test, 
$H \equiv \mathrm{max}(Z^2_{mw} - 4\times(m-1); 1\leq m \leq 20)$, with
\begin{equation}
Z^2_{mw} \equiv
\frac{2}{N_{\gamma}}\,\left(\frac{1}{N_{\gamma}}\sum\limits_{i=1}^{N_{\gamma}}{w_i^2}\right)^{-1}
\sum_{k=1}^{m} \alpha_{kw}^2 + \beta_{kw}^2
\end{equation}
where
$\alpha_{kw} = \sum_{i=1}^{N_{\gamma}} w_i\,\cos(2\pi k\phi_i) $ and $
\beta_{kw} = \sum_{i=1}^{N_{\gamma}} w_i\,\sin(2\pi k\phi_i)$.
\citet{KerrWeighted} provides the probability that a detection is a statistical fluctuation for a given $H$ value, approximated by $e^{-0.4H}$. 
$H=36$ (15) corresponds to a $5\sigma$ ($3\sigma$) detection.

\begin{figure}[!ht]
\centering
\includegraphics[width=0.9\textwidth]{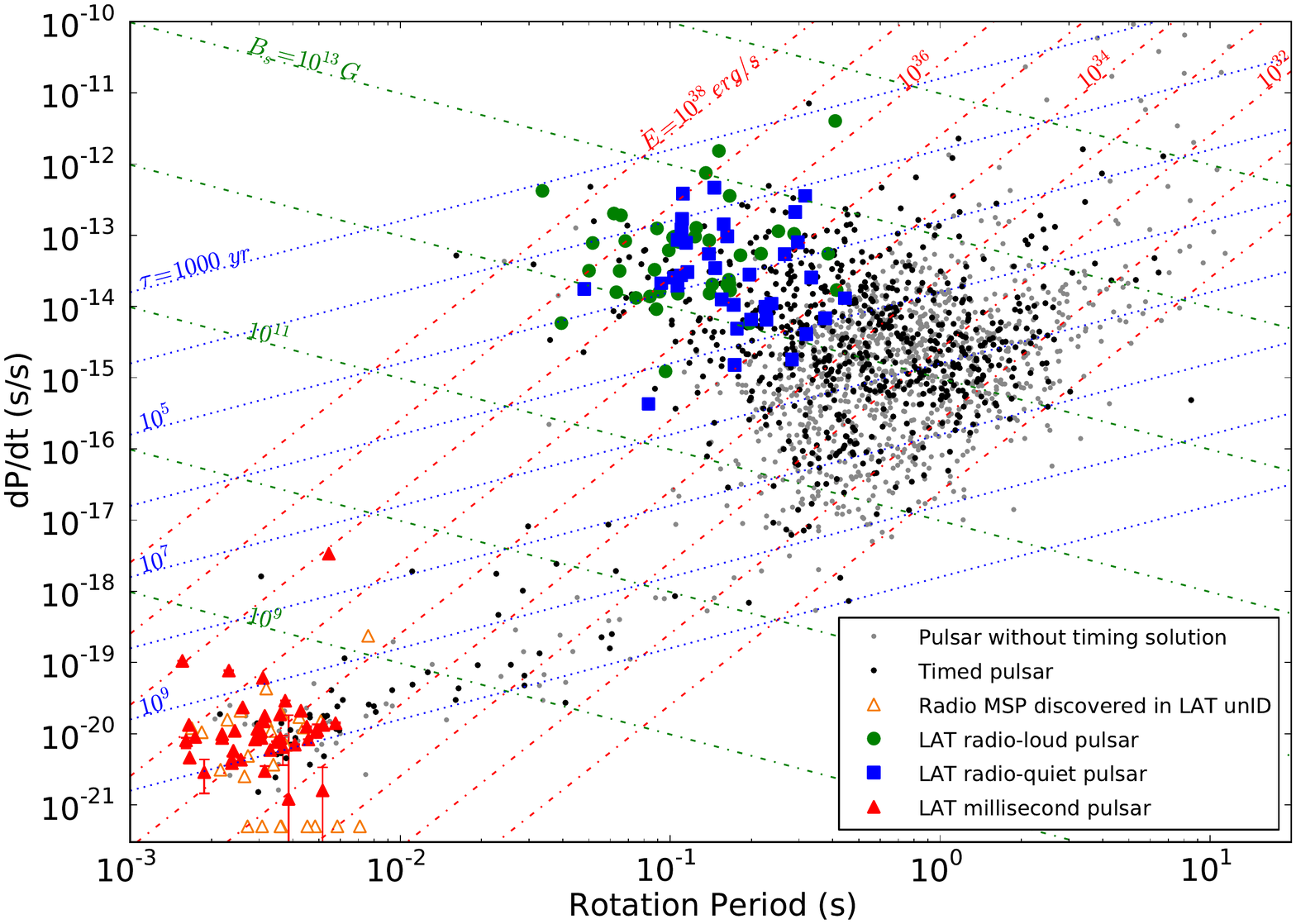}
\caption{ 
Pulsar spindown rate, $\dot P$, versus the rotation period $P$. 
Green dots indicate the 42 young, radio-loud gamma-ray pulsars and 
blue squares show the 35 young, `radio-quiet' pulsars, defined as
$S_{1400}< 30\, \mu$Jy, where $S_{1400}$ is the radio flux density at 1400 MHz.
Red triangles are the 40 millisecond gamma-ray pulsars. 
The 710 black dots indicate pulsars phase-folded in gamma rays using rotation models
provided by the ``Pulsar Timing consortium'' for which no significant pulsations were observed.
Phase-folding was not performed for the 1337 pulsars outside of globular clusters indicated by gray dots. 
Orange open triangles indicate radio MSPs discovered at the positions of previously unassociated LAT sources 
for which we have not yet seen gamma pulsations.
We plot them at $\dot P \equiv 5\times 10^{-22}$ when $\dot P$ is unavailable.
Shklovskii corrections to $\dot P$ have been applied to the pulsars
with proper motion measurements (see Section \ref{doppler}). 
For clarity, error bars are shown only for the gamma-detected pulsars.
\label{PPdotplot}}
\end{figure}

\subsection{Using Known Rotation Ephemerides}
\label{radioselected}

\begin{figure}[!ht]
\centering
\includegraphics[width=0.7\textwidth]{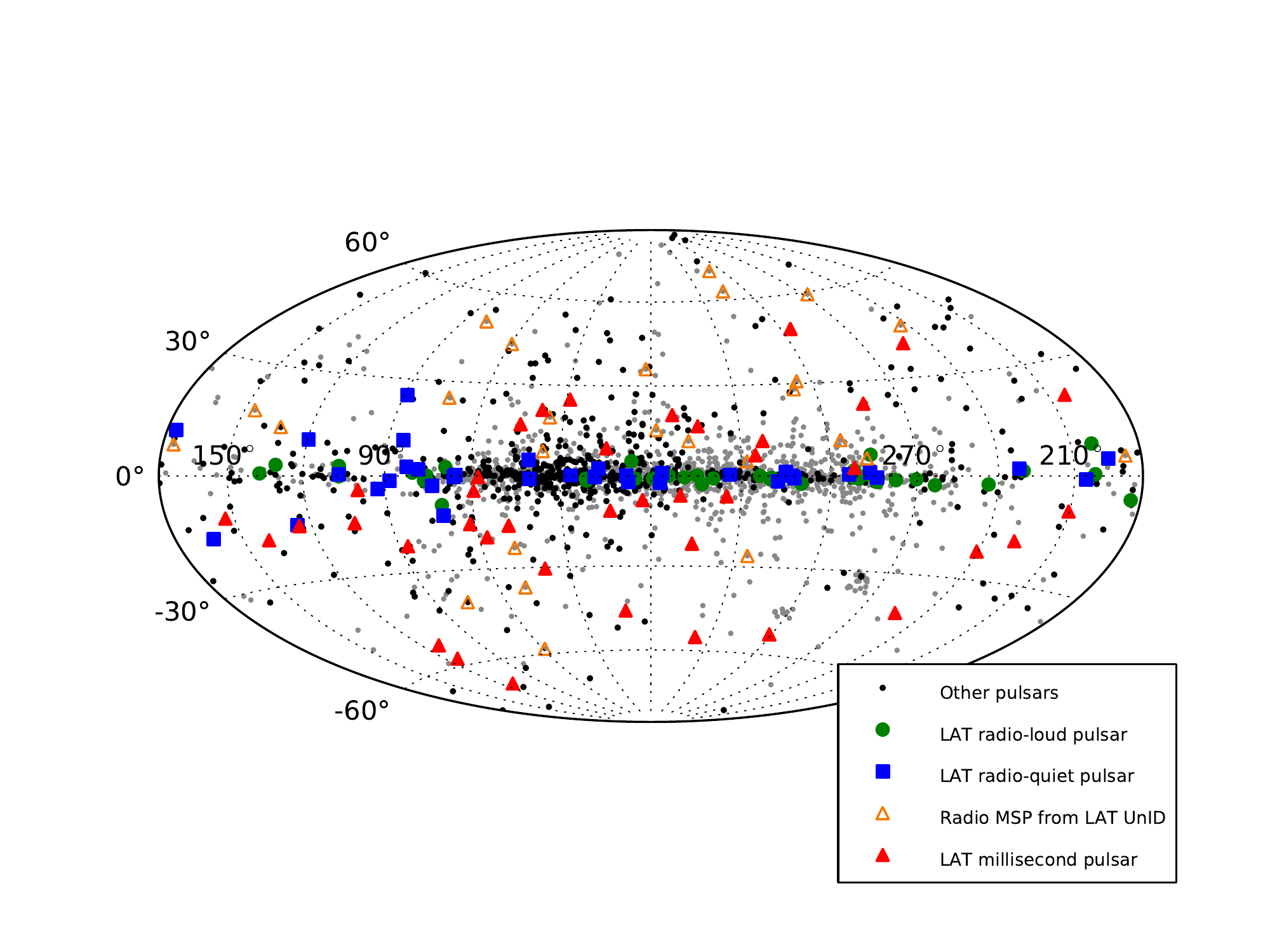}
\caption{Pulsar sky map in Galactic coordinates.  The markers are the same as in Figure \ref{PPdotplot}.
\label{Aitoff}}
\end{figure}

The first gamma-ray pulsar discovery method, described above, found 61 of the gamma-ray pulsars in this catalog.
It uses known rotational ephemerides from radio or X-ray observatories.
The $2286$ known rotation-powered pulsars 
(mostly from the ATNF Pulsar Catalog\footnote{\url{http://www.atnf.csiro.au/research/pulsar/psrcat}}  \citep{ATNFcatalog}, see Table \ref{tab:tallies}) are all candidate gamma-ray pulsars. 
Nearly all of these were discovered in radio searches, with a handful coming from X-ray observations. 
Phase-folding with a radio or X-ray ephemeris is the most sensitive way to find gamma-ray pulsations, 
since no penalities are incurred for trials in position, $P, \dot P$, or other search parameters.   
Having a current ephemeris for as many known pulsars as possible is of critical importance to LAT science and is the key goal of the Pulsar Timing Consortium \citep{TimingForFermi}. 
EGRET results \citep{Thompson08} as well as theoretical expectations indicated that young pulsars with 
large spindown power\footnote{$\dot E = 4\pi^2 I_{0} \dot P / P^3$, for which we use $I_{0} = 10^{45}$ g cm$^2$ as the neutron star moment of inertia.}, 
$\dot E > 1\times 10^{34}$ erg s$^{-1}$, are the most likely gamma-ray pulsar candidates.  
Because of their intrinsic instabilities, such as timing noise and glitches, 
these pulsars are also the most resource intensive to maintain ephemerides of sufficient accuracy. To allow for unexpected discoveries, 
the Timing Consortium also provides ephemerides for essentially all known pulsars that are regularly timed, 
spanning the $P\dot{P}$ space of known pulsars (Figure \ref{PPdotplot}).
In addition to $\dot E$, the $P\dot{P}$ diagram shows two other physical parameters derived from the timing information:
the magnetic field at the neutron star surface, $B_{\rm S} = (1.5 I_{0} c^{3} P \dot{P})^{1/2}/2\pi R_{\rm NS}^{3} $, 
assuming an orthogonal rotator with neutron star radius $R_{\rm NS} = 10$ km and the speed of light in a vacuum, $c$; and 
the characteristic age $\tau_c = P/2\dot{P}$, assuming magnetic dipole braking as the only energy-loss mechanism and an initial spin period much less than the current period. 
The black dots in Figure \ref{PPdotplot} show 710 pulsars that we have phase-folded without detecting gamma pulsations, 
in addition to the 117 gamma-ray pulsars.  The locations of all 117 gamma-ray pulsars on the sky are shown in Figure \ref{Aitoff}. 

For known pulsars we use years of radio and/or X-ray time-of-arrival measurements (``TOAs'') to fit the timing model parameters using the standard
pulsar timing codes \textsc{Tempo} \citep{TaylorWeisberg1989} or \textsc{Tempo2} \citep{Hobbs2006}.  In addition to providing a model for folding the gamma-ray
data, the radio observations also provide the information needed to measure the absolute phase alignment (after correcting for interstellar dispersion)
between the radio or X-ray and gamma-ray pulses, providing key information about the relative geometry of the different emission regions.

\subsection{Blind Periodicity Searches}
\label{blindsearch}

The second method of discovering gamma-ray pulsars, which produced 36 (approximately one-third) of the gamma-ray pulsars in this catalog, involves detecting the rotational period in the LAT data.
Both these searches and the radio searches described in the next subsection begin with a target list of candidate pulsars. Some targets are 
sources known at other wavelengths that are suspected of harboring pulsars.  These include supernova remnants (SNRs), pulsar wind nebulae (PWNe), compact central
objects (CCOs), unidentified TeV sources, and other high-energy sources, mostly along the Galactic plane. 
Generally, these sources had already been subjected to deep radio searches independent of \textit{Fermi}.

In addition, as the LAT surveys the sky, an increasing number of gamma-ray sources are discovered and characterized that are not associated with previously known objects.
Several methods have been used to rank these according to their probabilities of being yet-undiscovered pulsars. Most of these rely
on the tendency of gamma-ray pulsars to be non-variable and have spectra that can be fit with exponential cutoffs in the few GeV band
\citep{UNASSOC_SOURCE_PAPER,KJLee}. 

Blind searches for pulsars in gamma rays are challenging, due to the wide pulsar parameter ranges that must be searched and due to the sparseness of the data 
(a few photons per hour for the brightest sources). This results in very long integration times (months to years) making standard Fast Fourier Transform search techniques computationally prohibitive.  
New semi-coherent search techniques \citep{Atwood2006,Pletsch_2012} have been extremely successful at discovering gamma-ray pulsars with modest computational requirements.


LAT blind search sensitivity depends on a number of parameters: the rotation frequency, energy spectrum, pulsed fraction, level of diffuse gamma-ray background, 
event extraction choices (e.g. ROI and $E_\mathrm{min}$), and the accuracy of the position used to barycenter the data. 
The one-year sensitivity was evaluated using a Monte Carlo study by \citet{Dormody_2011}. 
Newer searches \citep{Pletsch_2012,Pletsch_2012b} have mitigated dependence on event selection criteria and source localization by weighting events 
and searching over a grid of positions. 

In all, well over one hundred LAT sources have been subjected to blind period searches. Pulsars might have been missed for a few reasons: (1) low pulsed
fraction or very high backgrounds, (2) broad pulse profiles (our algorithms detect sharp pulses more easily), (3) high levels of timing noise or glitches, 
(4) being in an unknown binary system. Most MSPs are in binary systems, where the Doppler shifts from the orbital motion smear the signal. 
In some cases, multiwavelength observations constrain the orbit and position to make the search more like that of an isolated MSP.
Optical studies \citep{Romani_2011,Kong_2012,Romani2012} led to the first discovery of a millisecond pulsar, PSR J1311$-$3430, 
in a blind search of LAT data \citep{Pletsch_2012c}. Detection of radio pulsations followed shortly \citep{RayJ1311}.
Even isolated MSP searches require massive computation with fine frequency and position gridding. 
The Einstein@home\footnote{\url{http://einstein.phys.uwm.edu}} project applies the power of global volunteer computing to this problem.

For the LAT pulsars undetected in the radio (see Section \ref{S1400}), or too faint for regular radio timing, 
we must determine the pulsar timing ephemeris directly from the LAT data.  
Techniques for TOA determination optimized for sparse photon data have been developed and applied to generate the timing models required for the profile analysis \citep{Ray2011}. 
This timing provides much more precise pulsar positions than can be determined from the LAT event directions, which is important for multiwavelength counterpart searches. 
It also allows study of timing noise and glitch behavior (Dormody et al., in prep).



\subsection{Radio Pulsar Discoveries Leading to Gamma-ray Pulsations}
\label{psc}
In the third discovery method that we applied, which yielded 20 of this catalog's MSPs, 
unassociated LAT source positions are searched for radio pulsations.
When found, the resulting ephemeris enables gamma-ray phase-folding, as in Section \ref{radioselected}.
A key feature of radio pulsar searches is that they are sensitive to binary systems with the application
of techniques to correct for the orbital acceleration in short data sets \citep[with durations much less than the binary period,][]{RansomLongOrbit2002}. This allows for the discovery of binary MSPs, which are largely
inaccessible to gamma-ray blind searches, as described above.

Radio searches of several hundred LAT sources by the \textit{Fermi} Pulsar Search Consortium (PSC), an international collaboration of radio observers with access to large radio telescopes, have
resulted in the discovery of 47 pulsars, including 43 MSPs and four young or middle-aged pulsars \citep{Ray2012psc}.
As the LAT Collaboration generates internal source lists and preliminary 
catalogs of gamma-ray sources from the accumulating sky-survey data, these target localizations are provided 
to the PSC for searching, with rankings of how strongly their characteristics resemble those of gamma-ray pulsars, as described in Section \ref{blindsearch}. 
This technique was employed during the EGRET era as well, but with 
modest success, in part due to the relatively poor source localizations.  With the LAT, there are many more gamma-ray 
sources detected and each one is localized to an accuracy that is comparable to, or smaller than, the beam 
width of the radio telescopes being used.  This enables deep searches by removing the need to 
mosaic a large region. It also facilitates repeated searches of the same source, which is important because 
discoveries can be missed as a result of scintillation or eclipses in binary systems \citep[e.g., PSR J0101$-$6422, see][]{Kerr2012}.

Guided by these ranked lists of pulsar-like gamma-ray sources, the 43 radio MSPs were
discovered in a tiny fraction of the radio telescope time that would have been required to find them in undirected radio pulsar surveys.  
In particular, because the MSP population out to the LAT's detection limit ($\sim 2$ kpc) is distributed nearly uniformly across the sky, 
full sky surveys are required, whereas most young pulsar searches have concentrated on the Galactic plane.
For comparison, after analyzing thousands of pointings carried out since 2007, 
the High Time Resolution Universe surveys \citep[][and references therein]{Keith2012_IAU,Ng2012_IAU} found 29 new radio MSPs.

Interestingly, the success rate for radio searches of LAT sources in the plane has been much poorer.  
Only four young pulsars have been discovered, and only one of those turned 
out to be a gamma-ray pulsar \citep[PSR J2030+3641,][]{Camilo2012}, the others being chance associations. 
This is probably due to a combination of young pulsars having smaller radio beaming fractions than MSPs (as evidenced by the large number of young, radio-quiet pulsars discovered) 
and the fact that the Galactic plane has been well surveyed for radio pulsars.
The great success of the blind gamma searches in the plane is because young pulsars mainly reside there.
Their smaller radio beaming fractions leave a large number of radio-quiet pulsars that can only be discovered in high-energy data.


Once a radio pulsar has been discovered positionally coincident with a LAT source, it must be observed for a 
substantial period (typically six months to a year or more) to determine a timing model that allows a 
periodicity search in the LAT data, as described in Section \ref{radioselected}.  In several cases, an initial radio model has allowed 
discovery of the gamma-ray pulsations, then the LAT data themselves have been used to extend the validity of 
the timing model back through the launch of \textit{Fermi}, a few years before the radio discovery. 
This radio follow up has resulted in the confirmed detection of LAT pulsations from 20 of these MSPs.
Five more were detected using data beyond the set described in Section \ref{obsvSection}.  
Of the remainder, most will have LAT pulsations detected once their radio timing 
models are well determined, but a few (e.g., PSR J1103$-$5403, see \cite{Keith_2011}) are likely to be just chance coincidences with the 
target LAT source.


\section{The Gamma-ray Pulsars}\label{cat}
The discovery strategies discussed in Section \ref{pulsation_discovery} yielded 117 gamma-ray pulsars in three years of
data. Of the gamma-ray pulsars in this catalog, roughly half (41 young and 20 MSPs) were
known in radio and/or X-rays prior to the launch of \textit{Fermi}. The remaining pulsars were
discovered by or with the aid of the LAT, with 36 being young pulsars found in blind
searches of LAT data and the remaining being MSPs found in radio searches of unassociated
LAT sources. \textit{Fermi} has not only significantly increased the number of known energetic
young and millisecond pulsars, but has done so with selection biases complementary to those of
previous surveys. The LAT all-sky survey has its greatest sensitivity in regions of the sky away from the Galactic plane (see Section \ref{sensitivity}), 
increasing the diversity and the uniformity of the sampled neutron star
population. As an example, Figure \ref{Aitoff} shows the broad range of Galactic latitude of the \textit{Fermi} pulsars.

\tabletypesize{\scriptsize}
\begin{deluxetable}{lrrr}
\tablecaption{Pulsar varieties\label{tab:tallies}}
\tablewidth{0pt}

\tablehead{
\colhead{Category}      & 
\colhead{Count}   & 
\colhead{Sub-count}      &
\colhead{Fraction}}
\startdata
Known rotation-powered pulsars (RPPs)\tablenotemark{a} &  2286&  & \\
RPPs with measured $\dot P > 0$       & & 1944 & \\
RPPs with measured $\dot E > 3\times 10^{33}$ erg s$^{-1}$       & & 552 & \\
 & & & \\
Millisecond pulsars (MSPs, $P < 16$ ms) & 292&  & \\
Field MSPs  & & 169 & \\
MSPs in globular clusters  & & 123 & \\
Field MSPs with measured $\dot E > 3\times 10^{33}$ erg s$^{-1}$ & & 96 & \\
Globular cluster MSPs with measured $\dot E > 3\times 10^{33}$ erg s$^{-1}$ &  & 25 & \\
 & & & \\
Gamma-ray pulsars in this catalog & 117 & & \\
Young or middle-aged &  & 77  & \\
Radio-loud gamma-ray\tablenotemark{b} &  & 42 & 36\% \\
Radio-quiet gamma-ray &  & 35 & 30\% \\
Gamma-ray MSPs (isolated + binary) & & (10+30) = 40 & 34\% \\
 & & & \\
Radio MSPs discovered in LAT sources &  46 & & \\
with gamma-ray pulsations\tablenotemark{c} &  & 34 & \\ 
\enddata
\tablenotetext{a}{Includes the 2193 pulsars, which are all RPPs, in the ATNF Pulsar Catalog \citep[v1.46, ][]{ATNFcatalog}, see \url{http://www.atnf.csiro.au/research/pulsar/psrcat},
as well as more recent discoveries. D. Lorimer maintains a list of known field MSPs at \url{http://astro.phys.wvu.edu/GalacticMSPs/}.}
\tablenotetext{b}{$S_{1400}> 30\, \mu$Jy, where $S_{1400}$ is the radio flux density at 1400 MHz.}
\tablenotetext{c}{Only 20 of the new radio MSPs showed gamma-ray pulsations when the dataset for this catalog was frozen.}



\end{deluxetable}

Table \ref{tab:tallies} summarizes the census of known pulsars, independent of the method by which the pulsars were discovered.
Tables \ref{tbl-charPSR} and \ref{tbl-charMSP} list the characteristics of the 117 gamma-ray pulsars, divided into young and millisecond gamma-ray pulsars, respectively. 
All have large spindown powers, $\dot E > 3\times 10^{33}$ erg s$^{-1}$, apparent in Figure \ref{PPdotplot}.
The large uncertainties on the two seeming exceptions, PSRs J0610$-$2100 and J1024$-$0719, are discussed in Section \ref{lumin}.

Pulsar discoveries
continue as increased statistics bring light curves above our $5\sigma$ 
detection threshold, improved methods for event selection and blind searches
allow increased sensitivity, and multiwavelength studies either detect radio pulsations
or constrain the blind-search space for likely pulsar candidates. Table \ref{Beyond117} lists a number
of LAT pulsars announced since the sample was frozen for the uniform analysis of the
present paper.

\tabletypesize{\scriptsize}
\begin{deluxetable}{lrrrrrrrr}
\tablewidth{0pt}
\tablecaption{Some parameters of young LAT-detected pulsars
\label{tbl-charPSR}}

\tablehead{
\colhead{PSR} &   \colhead{History} &    \colhead{$l$} & \colhead{$b$} & \colhead{$P$} & \colhead{$\dot P$}  & \colhead{$\dot E$}& \colhead{$S_{1400}$}  \\
&      &  \colhead{($^\circ$)} & \colhead{($^\circ$)} & \colhead{(ms)} & \colhead{($10^{-15}$)}  & \colhead{($10^{34}$ erg s$^{-1}$)} & \colhead{(mJy)} 
}

\startdata
 J0007+7303 & g & 119.66 & 10.46 &  315.9 & 357. & 44.8 & $<$0.005$^1$ \\ 
 J0106+4855 & gu & 125.47 & $-$13.87 &  83.2 & 0.43 & 2.9 & 0.008 \\ 
 J0205+6449 & x & 130.72 & 3.08 &  65.7 & 190. & 2644. & 0.045 \\ 
 J0248+6021 & r & 136.90 & 0.70 &  217.1 & 55.0 & 21.2 & 13.7 \\ 
 J0357+3205 & gu & 162.76 & $-$16.01 &  444.1 & 13.1 & 0.6 & $<$0.004$^1$ \\ 
 J0534+2200 & re & 184.56 & $-$5.78 &  33.6 & 420. & 43606. & 14.0 \\ 
 J0622+3749 & gu & 175.88 & 10.96 &  333.2 & 25.4 & 2.7 & $<$0.012$^2$ \\ 
 J0631+1036 & r & 201.22 & 0.45 &  287.8 & 104. & 17.3 & 0.8 \\ 
 J0633+0632 & gu & 205.09 & $-$0.93 &  297.4 & 79.6 & 11.9 & $<$0.003$^1$ \\ 
 J0633+1746 & xe & 195.13 & 4.27 &  237.1 & 11.0 & 3.3 & $<$0.507$^3$ \\ 
 J0659+1414 & r & 201.11 & 8.26 &  384.9 & 55.0 & 3.8 & 3.7 \\ 
 J0729$-$1448 & r & 230.39 & 1.42 &  251.7 & 114. & 28.2 & 0.7 \\ 
 J0734$-$1559 & gu & 232.06 & 2.02 &  155.1 & 12.5 & 13.2 & $<$0.005$^4$ \\ 
 J0742$-$2822 & r & 243.77 & $-$2.44 &  166.8 & 16.8 & 14.3 & 15.0 \\ 
 J0835$-$4510 & re & 263.55 & $-$2.79 &  89.4 & 125. & 690. & 1100. \\ 
 J0908$-$4913 & r & 270.27 & $-$1.02 &  106.8 & 15.1 & 49.0 & 10.0 \\ 
 J0940$-$5428 & r & 277.51 & $-$1.29 &  87.6 & 32.8 & 193. & 0.66 \\ 
 J1016$-$5857 & r & 284.08 & $-$1.88 &  107.4 & 80.6 & 257. & 0.46 \\ 
 J1019$-$5749 & r & 283.84 & $-$0.68 &  162.5 & 20.1 & 18.4 & 0.8 \\ 
 J1023$-$5746 & gu & 284.17 & $-$0.41 &  111.5 & 382. & 1089. & $<$0.030$^5$ \\ 
 J1028$-$5819 & r & 285.06 & $-$0.50 &  91.4 & 16.1 & 83.3 & 0.36 \\ 
 J1044$-$5737 & gu & 286.57 & 1.16 &  139.0 & 54.6 & 80.2 & $<$0.020$^5$ \\ 
 J1048$-$5832 & r & 287.42 & 0.58 &  123.7 & 95.7 & 200. & 6.5 \\ 
 J1057$-$5226 & re & 285.98 & 6.65 &  197.1 & 5.8 & 3.0 & 9.5 \\ 
 J1105$-$6107 & r & 290.49 & $-$0.85 &  63.2 & 15.8 & 248. & 0.75 \\ 
 J1112$-$6103 & r & 291.22 & $-$0.46 &  65.0 & 31.5 & 454. & 1.4 \\ 
 J1119$-$6127 & r & 292.15 & $-$0.54 &  408.7 & 4028. & 233. & 0.8 \\ 
 J1124$-$5916 & r & 292.04 & 1.75 &  135.5 & 750. & 1190. & 0.08 \\ 
 J1135$-$6055 & gu & 293.79 & 0.58 &  114.5 & 78.4 & 206. & $<$0.030$^4$ \\ 
 J1357$-$6429 & r & 309.92 & $-$2.51 &  166.2 & 357. & 307. & 0.44 \\ 
 J1410$-$6132 & r & 312.20 & $-$0.09 &  50.1 & 31.8 & 1000. & 6.56$^6$ \\ 
 J1413$-$6205 & gu & 312.37 & $-$0.74 &  109.7 & 27.4 & 81.8 & $<$0.024$^5$ \\ 
 J1418$-$6058 & gu & 313.32 & 0.13 &  110.6 & 169. & 494. & $<$0.029$^1$ \\ 
 J1420$-$6048 & r & 313.54 & 0.23 &  68.2 & 82.9 & 1032. & 0.9 \\ 
 J1429$-$5911 & gu & 315.26 & 1.30 &  115.8 & 30.5 & 77.4 & $<$0.021$^5$ \\ 
 J1459$-$6053 & gu & 317.89 & $-$1.79 &  103.2 & 25.3 & 90.9 & $<$0.037$^1$ \\ 
 J1509$-$5850 & r & 319.97 & $-$0.62 &  88.9 & 9.2 & 51.5 & 0.15 \\ 
 J1513$-$5908 & xe & 320.32 & $-$1.16 &  151.5 & 1529. & 1735. & 0.94 \\ 
 J1531$-$5610 & r & 323.90 & 0.03 &  84.2 & 13.8 & 91.2 & 0.6 \\ 
 J1620$-$4927 & gu & 333.89 & 0.41 &  171.9 & 10.5 & 8.1 & $<$0.040$^2$ \\ 
 J1648$-$4611 & r & 339.44 & $-$0.79 &  165.0 & 23.7 & 20.9 & 0.58 \\ 
 J1702$-$4128 & r & 344.74 & 0.12 &  182.2 & 52.3 & 34.2 & 1.1 \\ 
 J1709$-$4429 & re & 343.10 & $-$2.69 &  102.5 & 92.8 & 340. & 7.3 \\ 
 J1718$-$3825 & r & 348.95 & $-$0.43 &  74.7 & 13.2 & 125. & 1.3 \\ 
 J1730$-$3350 & r & 354.13 & 0.09 &  139.5 & 84.8 & 123. & 3.2 \\ 
 J1732$-$3131 & gu & 356.31 & 1.01 &  196.5 & 28.0 & 14.6 & $<$0.015$^1$ \\ 
 J1741$-$2054 & gu & 6.43 & 4.91 &  413.7 & 17.0 & 0.9 & 0.16 \\ 
 J1746$-$3239 & gu & 356.96 & $-$2.18 &  199.5 & 6.6 & 3.3 & $<$0.034$^2$ \\ 
 J1747$-$2958 & r & 359.31 & $-$0.84 &  98.8 & 61.3 & 251. & 0.25 \\ 
 J1801$-$2451 & r & 5.25 & $-$0.88 &  125.0 & 127. & 257. & 0.85 \\ 
 J1803$-$2149 & gu & 8.14 & 0.19 &  106.3 & 19.5 & 64.1 & $<$0.024$^2$ \\ 
 J1809$-$2332 & g & 7.39 & $-$1.99 &  146.8 & 34.4 & 43.0 & $<$0.025$^1$ \\ 
 J1813$-$1246 & gu & 17.24 & 2.44 &  48.1 & 17.6 & 624. & $<$0.017$^1$ \\ 
 J1826$-$1256 & g & 18.56 & $-$0.38 &  110.2 & 121. & 358. & $<$0.013$^1$ \\ 
 J1833$-$1034 & r & 21.50 & $-$0.89 &  61.9 & 202. & 3364. & 0.071 \\ 
 J1835$-$1106 & r & 21.22 & $-$1.51 &  165.9 & 20.6 & 17.8 & 2.2 \\ 
 J1836+5925 & g & 88.88 & 25.00 &  173.3 & 1.5 & 1.1 & $<$0.004$^1$ \\ 
 J1838$-$0537 & gu & 26.51 & 0.21 &  145.7 & 465. & 593. & $<$0.017$^7$ \\ 
 J1846+0919 & gu & 40.69 & 5.34 &  225.6 & 9.9 & 3.4 & $<$0.005$^5$ \\ 
 J1907+0602 & g & 40.18 & $-$0.89 &  106.6 & 86.7 & 282. & 0.0034 \\ 
 J1952+3252 & re & 68.77 & 2.82 &  39.5 & 5.8 & 372. & 1.0 \\ 
 J1954+2836 & gu & 65.24 & 0.38 &  92.7 & 21.2 & 105. & $<$0.005$^5$ \\ 
 J1957+5033 & gu & 84.60 & 11.00 &  374.8 & 6.8 & 0.5 & $<$0.010$^5$ \\ 
 J1958+2846 & gu & 65.88 & $-$0.35 &  290.4 & 212. & 34.2 & $<$0.006$^1$ \\ 
 J2021+3651 & r & 75.22 & 0.11 &  103.7 & 95.6 & 338. & 0.1 \\ 
 J2021+4026 & g & 78.23 & 2.09 &  265.3 & 54.2 & 11.4 & $<$0.020$^1$ \\ 
 J2028+3332 & gu & 73.36 & $-$3.01 &  176.7 & 4.9 & 3.5 & $<$0.005$^2$ \\ 
 J2030+3641 & ru & 76.12 & -1.44 &  200.1 & 6.5 & 3.2 & 0.15 \\ 
 J2030+4415 & gu & 82.34 & 2.89 &  227.1 & 6.5 & 2.2 & $<$0.008$^2$ \\ 
 J2032+4127 & gu & 80.22 & 1.03 &  143.2 & 20.4 & 27.3 & 0.23 \\ 
 J2043+2740 & r & 70.61 & -9.15 &  96.1 & 1.2 & 5.5 & 9.35 \\ 
 J2055+2539 & gu & 70.69 & -12.52 &  319.6 & 4.1 & 0.5 & $<$0.007$^5$ \\ 
 J2111+4606 & gu & 88.31 & -1.45 &  157.8 & 143. & 144. & $<$0.013$^2$ \\ 
 J2139+4716 & gu & 92.63 & -4.02 &  282.8 & 1.8 & 0.3 & $<$0.014$^2$ \\ 
 J2229+6114 & r & 106.65 & 2.95 &  51.6 & 77.9 & 2231. & 0.25 \\ 
 J2238+5903 & gu & 106.56 & 0.48 &  162.7 & 97.0 & 88.8 & $<$0.011$^1$ \\ 
 J2240+5832 & r & 106.57 & -0.11 &  139.9 & 15.2 & 21.9 & 2.7 \\ 

\enddata

\tablecomments{
Column 2 gives a discovery/detection code: g=gamma-ray blind search, r=radio, u=candidate location was that of an unassociated LAT source, x=X-ray, e=seen by \textit{EGRET}.
Columns 3 and 4 give Galactic coordinates for each pulsar. 
Columns 5 and 6 list the period ($P$) and its first derivative ($\dot{P}$), and  
Column 7 gives the spindown luminosity $\dot E$ . 
The Shklovskii correction to $\dot{P}$ and $\dot E$ is negligible for these young pulsars (see Section \ref{doppler}).
Column 8 gives the radio flux density (or upper limit) at 1400 MHz ($S_{1400}$, see Section \ref{S1400}), 
taken from the ATNF database except for the noted entries where: (1) \cite{Ray2011}; (2) \cite{Pletsch_2012}; (3) Geminga: \cite{Spoelstra1984}; 
(4) GBT (this paper); (5) \cite{Saz_Parkinson_2010}; (6) \cite{OBrien_2008}; (7) \cite{Pletsch_2012b}. 
PSR J1509$-$5850 should not be confused with PSR B1509$-$58 (= J1513$-$5908) observed by instruments on the \textit{Compton Gamma-Ray Observatory}.}



\end{deluxetable}

\normalsize

\tabletypesize{\scriptsize}
\begin{deluxetable}{lrrrrrrrr}
\tablewidth{0pt}
\tablecaption{Some parameters of LAT-detected millisecond pulsars
\label{tbl-charMSP}}

\tablehead{
\colhead{PSR} &   \colhead{Type,} &    \colhead{$l$} & \colhead{$b$} & \colhead{$P$} & \colhead{$\dot P$}  & \colhead{$\dot E$}& \colhead{$S_{1400}$}  \\
&  \colhead{history.}  &  \colhead{($^\circ$)} &  \colhead{($^\circ$)} & \colhead{(ms)} & \colhead{($10^{-20}$)}  & \colhead{($10^{-33}$ erg s$^{-1}$)} & \colhead{(mJy)} 
}

\startdata
 J0023+0923 & bwru & 111.15 & $-$53.22 &  3.05 & 1.08 & 15.1 &  0.19$^1$ \\ 
 J0030+0451 & r & 113.14 & $-$57.61 &  4.87 & 1.02 & 3.49 &  0.60 \\ 
 J0034$-$0534 & br & 111.49 & $-$68.07 &  1.88 & 0.50 & 29.7 &  0.61 \\ 
 J0101$-$6422 & bru & 301.19 & $-$52.72 &  2.57 & 0.48 & 12.0 &  0.28 \\ 
 J0102+4839 & bru & 124.93 & $-$14.83 &  2.96 & 1.17 & 17.5 &  0.22$^2$ \\ 
 J0218+4232 & br & 139.51 & $-$17.53 &  2.32 & 7.74 & 243. &  0.90 \\ 
 J0340+4130 & ru & 154.04 & $-$11.47 &  3.30 & 0.59 & 7.9 &  0.17$^2$ \\ 
 J0437$-$4715 & br & 253.39 & $-$41.96 &  5.76 & 5.73 & 11.8 &  149 \\ 
 J0610$-$2100 & bwr & 227.75 & $-$18.18 &  3.86 & 1.23 & 8.5 &  0.40 \\ 
 J0613$-$0200 & br & 210.41 & $-$9.30 &  3.06 & 0.96 & 13.2 &  2.3 \\ 
 J0614$-$3329 & bru & 240.50 & $-$21.83 &  3.15 & 1.78 & 22.0 &  0.60$^3$ \\ 
 J0751+1807 & br & 202.73 & 21.09 &  3.48 & 0.78 & 7.30 &  3.2 \\ 
 J1024$-$0719 & r & 251.70 & 40.52 &  5.16 & 1.85 & 5.30 &  1.5 \\ 
 J1124$-$3653 & bwru & 283.74 & 23.59 &  2.41 & 0.58 & 17.1 &  0.04$^4$ \\ 
 J1125$-$5825 & br & 291.89 & 2.60 &  3.10 & 6.09 & 80.5 &  0.44 \\ 
 J1231$-$1411 & bru & 295.53 & 48.39 &  3.68 & 2.12 & 17.9 &  0.16$^3$ \\ 
 J1446$-$4701 & br & 322.50 & 11.43 &  2.19 & 0.98 & 36.8 &  0.37 \\ 
 J1514$-$4946 & bru & 325.22 & 6.84 &  3.58 & 1.87 & 16.0 &   \nodata \\ 
 J1600$-$3053 & br & 344.09 & 16.45 &  3.60 & 0.95 & 8.05 &  2.5 \\ 
 J1614$-$2230 & br & 352.64 & 20.19 &  3.15 & 0.96 & 12.1 &  1.2$^5$ \\ 
 J1658$-$5324 & ru & 334.87 & $-$6.63 &  2.43 & 1.10 & 30.2 &  0.50$^6$ \\ 
 J1713+0747 & br & 28.75 & 25.22 &  4.57 & 0.85 & 3.53 &  10.2 \\ 
 J1741+1351 & br & 37.90 & 21.62 &  3.75 & 3.02 & 22.7 &  0.93 \\ 
 J1744$-$1134 & r & 14.79 & 9.18 &  4.07 & 0.89 & 5.20 &  3.1 \\ 
 J1747$-$4036 & ru & 350.19 & $-$6.35 &  1.64 & 1.33 & 116. &  1.22$^6$ \\ 
 J1810+1744 & bwru & 43.87 & 16.64 &  1.66 & 0.46 & 39.7 &  1.89$^1$ \\ 
 J1823$-$3021A & r & 2.79 & $-$7.91 &  5.44 & 338. & 828. &  0.72 \\ 
 J1858$-$2216 & bru & 13.55 & $-$11.45 &  2.38 & 0.39 & 11.3 &   \nodata \\ 
 J1902$-$5105 & bru & 345.59 & $-$22.40 &  1.74 & 0.90 & 68.6 &  0.90$^6$ \\ 
 J1939+2134 & r & 57.51 & $-$0.29 &  1.56 & 10.5 & 1097. &  13.9 \\ 
 J1959+2048 & bwr & 59.20 & $-$4.70 &  1.61 & 1.68 & 160. &  0.40 \\ 
 J2017+0603 & bru & 48.62 & $-$16.03 &  2.90 & 0.83 & 13.0 &  0.50 \\ 
 J2043+1711 & bru & 61.92 & $-$15.31 &  2.38 & 0.57 & 15.3 &  0.17$^7$ \\ 
 J2047+1053 & bru & 57.06 & $-$19.67 &  4.29 & 2.10 & 10.5 &   \nodata \\ 
 J2051$-$0827 & bwr & 39.19 & $-$30.41 &  4.51 & 1.28 & 5.49 &  2.8 \\ 
 J2124$-$3358 & r & 10.93 & $-$45.44 &  4.93 & 2.06 & 6.77 &  3.6 \\ 
 J2214+3000 & bwru & 86.86 & $-$21.67 &  3.12 & 1.50 & 19.2 &  0.85$^3$ \\ 
 J2215+5135 & bkru & 99.46 & $-$4.60 &  2.61 & 2.34 & 51.9 &  0.47$^1$ \\ 
 J2241$-$5236 & bwru & 337.46 & $-$54.93 &  2.19 & 0.87 & 26.0 &  4.1 \\ 
 J2302+4442 & bru & 103.40 & $-$14.00 &  5.20 & 1.33 & 3.82 &  1.2 \\ 

\enddata

\tablecomments{ 
Column 2: b=binary, r=radio detected, u=seed position was that of an unassociated LAT source, w=white dwarf companion, k=``redback''. 
Columns 3 and 4 give the Galactic coordinates, with the rotation period $P$ in column 5. 
The first period time derivative $ \dot{P}$ and the spindown luminosity $\dot E$ in Columns 6 and 7 are \textit{uncorrected} for the Shklovskii effect, in this Table. 
The corrected values are used throughout the rest of the paper, and are listed in Table \ref{tbl-doppler} in Section \ref{doppler}. 
Column 9 gives the radio flux density (or upper limit) at 1400 MHz (Section \ref{S1400}), taken from the ATNF database except for the noted entries: 
(1) \citet{Hessels_2011}; (2) Bangale et al. (in prep.); (3) \citet{rrc+11}; (4) This paper; (5) \citet{dpr+10}; (6) \citet{Kerr5MSPs}; (7) \citet{gfc+12}. 
The three MSPs with no $S_{1400}$ listed scintillate too much to obtain a good flux measurement (PSR J1514$-$4946), 
or the radio flux has not yet been measured (PSRs J1858$-$2216 and J2047+1053).}

%
\end{deluxetable}

\normalsize

%
%
%
\tabletypesize{\scriptsize}
\begin{deluxetable}{lrcrl}
\tablecaption{Gamma-ray pulsars not in this catalog
\label{Beyond117}}
\tablewidth{0pt}

\tablehead{
\colhead{PSRJ}      & 
\colhead{$P$}  & 
\colhead{$\dot E$}  & 
\colhead{Codes}      &
\colhead{References}      
\\
\colhead{ }      & 
\colhead{(ms)}  & 
\colhead{($10^{34}$ erg s$^{-1}$) }  & 
\colhead{}      &
\colhead{}      
}
\startdata
 J0307+7443    &   3.16 & 2.2 & mbr & \citet{Ray2012psc}  \\
 J0737$-$3039A &  22.7  & 0.59 & r & \citet{GuillemotJ0737}  \\
 J1055$-$6028  &  99.7  & 120  & r    & \citet{Hou2013}   \\
 J1311$-$3430  &   2.56 & 4.9  & mbgu & \citet{Pletsch_2012c} \\
 J1544+4937	&   2.16 & 1.2  & mbru & \citet{Bhattacharyya}\\
 J1640+2224    &   3.16 & 0.35 & mbr & \citet{Hou2013}  \\
 J1705$-$1906  & 299.0  & 0.61 & r    & \citet{Hou2013}   \\
 J1732$-$5049  &   5.31  & 0.37 & mrb  &  \citet{Hou2013}    \\ 
 J1745+1017    &   2.65 & 0.53 & mbru & \citet{BarrJ1745}   \\
 J1816+4510    &   3.19 & 5.2 & mbru & \citet{Kaplan2012}   \\
 J1824$-$2452A &   3.05 & 220 & mr & \citet{J1824m2452A}; Johnson et al. (in prep.)  \\
 J1843$-$1113  &   1.85 & 6.0  & mr   & \citet{Hou2013}   \\
 J1913+0904    & 163.2  & 16   & r    & \citet{Hou2013}   \\
 J2256$-$1024  &   2.29 & 5.2 & mbru & \citet{Boyles2011}  \\
 J2339$-$0533  &   2.88 & 2.3 & mbru & Ray et al. (in prep.)  
\enddata

\tablecomments{Beyond the 117 pulsars. 
The above 15 pulsars were discovered in gamma rays as this catalog neared completion.
An additional 13 gamma-ray pulsars discovered by the LAT collaboration or by other groups using public LAT data
have publications in preparation, for a total of 145 as we go to submission (8 May 2013).
We maintain a list at \url{https://confluence.slac.stanford.edu/display/GLAMCOG/Public+List+of+LAT-Detected+Gamma-Ray+Pulsars}.
The codes are: u=discovered in a LAT unassociated source, 
g=discovered in a gamma-ray blind period search, r=radio detection, m=MSP, b=binary system.
}
\end{deluxetable}


\subsection{Radio Intensities}
\label{S1400}

\begin{figure}[bp]
\centering
\includegraphics[width=0.9\textwidth]{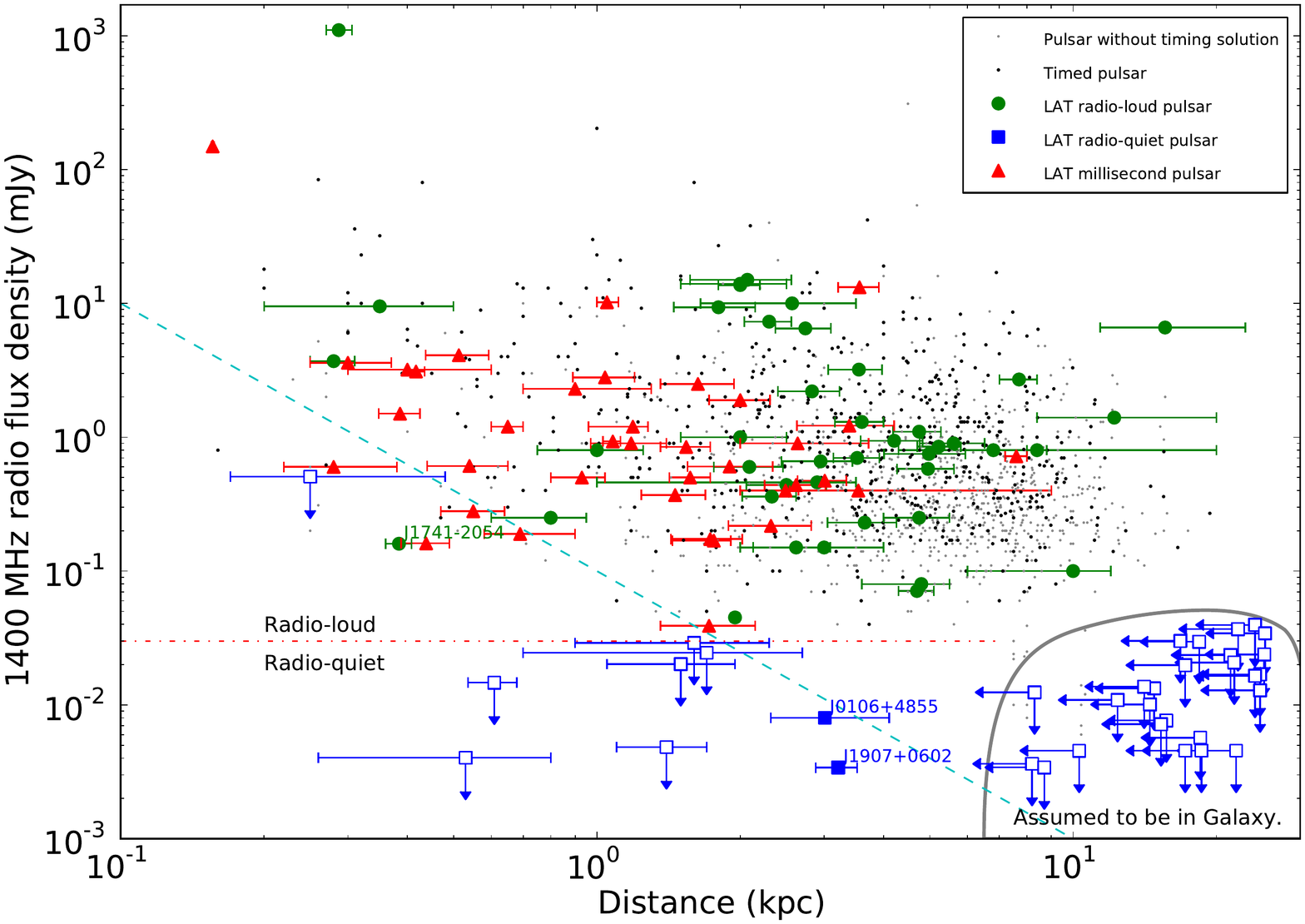}
\caption{Radio flux density at $1400$ MHz versus pulsar distance. 
Markers are as in Figure \ref{PPdotplot}, except that blue open squares show pulsars discovered in gamma-ray blind period searches for which
no radio signal has been detected.
The horizontal line at 30 $\mu$Jy is our convention for distinguishing radio ``loud'' from radio ``quiet'' pulsars.
The diagonal line shows a threshold in pseudo-luminosity of 100 $\mu$Jy - kpc$^2$.
Four gamma-discovered pulsars have been detected at radio frequencies: two are radio-quiet and are labeled.
Of the two that are radio loud, one is labeled while PSR J2032+4127 is in the cloud of points.
The pulsars at lower-right are assigned distance limits along the Milky Way's rim in Figure \ref{galaxie}.
\label{S1400Dist}}
\end{figure}

The 1400\,MHz flux densities, $S_{1400}$, of the young LAT-detected pulsars are listed in Table~\ref{tbl-charPSR},
and in Table \ref{tbl-charMSP} for the MSPs.
Figure \ref{S1400Dist} shows how they compare with the overall pulsar population.
Whenever possible, we report $S_{1400}$ as given in the ATNF Pulsar Catalog. 
For radio-loud pulsars with no published value at 1400\,MHz, we extrapolate to $S_{1400}$ from measurements at other frequencies, 
assuming $S_{\nu} \propto \nu^{\alpha}$, where $\alpha$ is the spectral index.
For most pulsars $\alpha$ has not been measured, and we use an average value $\langle\alpha\rangle = -1.7$, a middle
ground between $-1.6$ from \citet{Lorimer_1995} and $-1.8$ from \citet{Maron_2000}.
For those pulsars with measured spectral indices, we use the the published value of $\alpha$ for the extrapolation.
In the Table notes, we list those pulsars for which we have extrapolated $S_{1400}$ from another frequency and/or used a
value of $\alpha$ other than $-1.7$ for the extrapolation.

Table~\ref{tbl-charPSR} also reports upper limits on $S_{1400}$ for blind search pulsars that have been observed, 
but not detected, at radio frequencies.  
We define these upper limits as the sensitivity of the observation given by the pulsar radiometer equation \citep[Eq.~7.10 on page 174 of][]{PulsarHandbook} assuming a 
minimum signal-to-noise ratio of 5 for a detection and a pulse duty cycle of 10\%.
We mention here the unconfirmed radio detections of Geminga and PSR J1732$-$3131 at low radio frequencies, 
consistent with their non-detection above 300 MHz \citep{radioGemingaNature,Maanetal}. 

All pulsars discovered in blind searches have been searched deeply for radio pulsations \citep{Saz_Parkinson_2010,Ray2011,Ray2012psc},
and four of the 36 have been detected \citep{Camilo2009,LATPSR1907,Pletsch_2012}. 
In 1PC we labeled the young pulsars by how they were discovered (radio-selected vs.~gamma-ray selected), 
whereas we now define a pulsar as `radio-loud' if $S_{1400} > 30\, \mu$Jy, 
and `radio-quiet' if the measured flux density is lower, 
as for the two pulsars with detections of very faint radio pulsations, or if no radio detection has been achieved. 
The horizontal line in Figure \ref{S1400Dist} shows the threshold.
This definition favors observational characteristics instead of discovery history.  
Of the four radio-detected blind-search pulsars, two remain radio-quiet whereas the other two could in principle have been discovered in a sensitive radio survey.
The diagonal line in Figure \ref{S1400Dist} shows a possible alternate threshold at pseudo-luminosity 100 $\mu$Jy-kpc$^2$, for reference. 
Three of the four have pseudo-luminosities lower than for any previously known young pulsar, and comparable to only a small number of MSPs.

\subsection{Distances}
	\label{Distances}
%
%

\begin{figure}[!ht]
\centering
\includegraphics[width=0.7\textwidth]{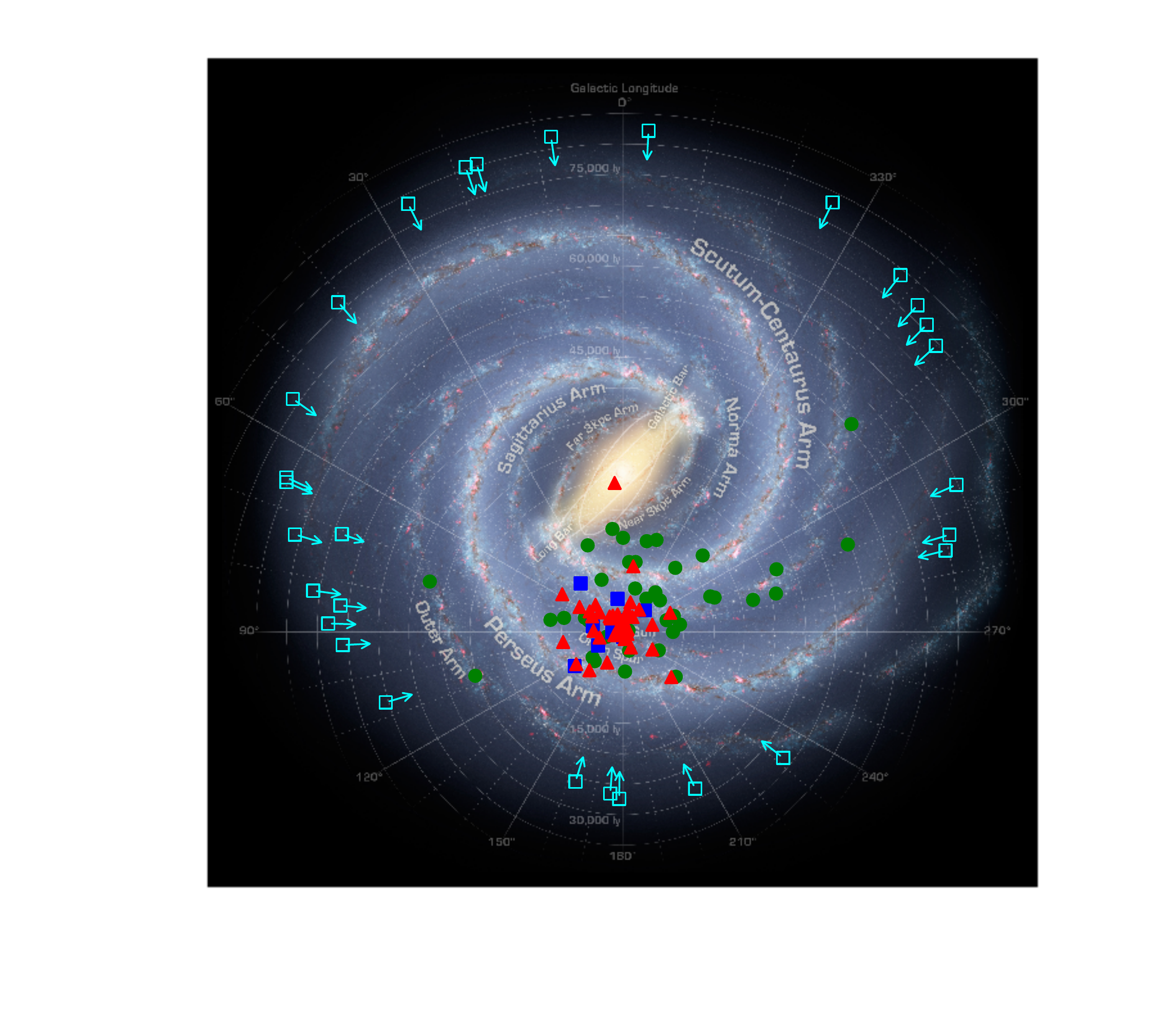}
\caption{Gamma-ray pulsar positions projected onto the Milky Way model of \citet{AnatomicallyCorrect}.
The pulsar that appears to be coincident with the Galactic center, PSR J1823$-$3021A in the globular
cluster NGC 6624, lies well above the Galactic plane. Distance uncertainties are not shown, for clarity, however they can be quite large especially
for the more distant objects. The open squares with arrows indicate the lines of sight toward pulsars for which no
distance estimates exist, placed at the distances where 95\% of the electron column density has been integrated in the NE2001 model.
The markers are the same as in Figure \ref{PPdotplot}. 
\label{galaxie}}
\end{figure}

Converting measured pulsar fluxes to emitted luminosities $L_\gamma$ (detailed in Section \ref{lumin}) requires the distances to the sources. 
Knowing the distances also allows mapping neutron star distributions relative to the Galaxy's spiral arms, as in Figure \ref{galaxie}, 
or evaluating their scale height above the plane.
Several methods can be used to estimate pulsar distances; however, the methods vastly differ in reliability. Deciding which method to use can be subjective.
Tables \ref{tbl-PSRdist} and \ref{tbl-doppler} list the distance estimates that we adopt, the methods with which these estimates were acquired, 
and the appropriate references. 

The most accurate distance estimator is the annual trigonometric parallax. 
Unfortunately, parallax can only be measured for relatively nearby pulsars,  
using X-ray or optical images, radio interferometric imaging, or accurate timing. 
For 14 \textit{Fermi} pulsars a parallax has been measured. 
We rejected two with low-significance ($< 2 \sigma$).
For the remaining 12 pulsars we consider this the best distance estimate.
One caveat when converting parallax measurements to distances is the Lutz-Kelker effect, 
an overestimate of parallax values (and hence underestimate of distances) that must be corrected for the larger volume of space traced by smaller parallax values \citep{lk73}. 
We use the Lutz-Kelker corrected distance estimates determined by \citet{Verbiest12}. 

The dispersion measure (DM) is by far the most commonly used pulsar distance estimator. DM is the column
density of free electrons along the path from Earth to the pulsar, in units of pc cm$^{-3}$.
The electrons delay the radio pulse arrival by $\Delta t = {\rm DM} (p\nu^2)^{-1}$ where $\nu$ is the observation frequency in MHz and
$p = 2.410 \times 10^{-4}$ MHz$^{-2}$ pc cm$^{-3}$ s$^{-1}$. Given a model for the electron density $n_e$ in the various structures of our Galaxy,
integrating DM = $\int_0^d n_e dl$ along the line of sight $dl$ yields the distance $d$ for which DM matches the radio measurement.
In this work we use the NE2001 model \citep{Cordes2002}, available as off-line code. 
To estimate the distance errors we re-run NE2001 twice, using DM $\pm$ 20\%, as the authors recommend.
The measured DM uncertainty is much lower than this, but this accommodates unmodelled electron-rich or poor regions. 
This yields distance uncertainties less than 30\% for many pulsars. 
Nevertheless, significant discrepancies with the true pulsar distances along some lines of sight still occur. 
As examples, the DM distances for PSR J2021+3651 \citep{PSR2021LAT} and PSR J0248+6021 \citep{Theureau2011} may be more than three times greater than the true distances.

For some pulsars, an absorbing hydrogen column density $N_{\rm H}$ below 1 keV has been obtained (see Section \ref{Xrays}). 
Comparing $N_{\rm H}$ with the total hydrogen column density for that line of sight obtained from 21 cm radio surveys yields a rough distance estimate.
The Doppler shift of neutral hydrogen (H~{\sc I}) absorption or emission lines measured from clouds on the line of sight, 
together with a Galactic rotation model as described in Section \ref{doppler}, can give ``kinematic'' distances to the clouds.
The pulsar distance is then constrained if there is evidence that the pulsar is in one of the clouds, or between some of them. 
Associations can be uncertain and these distance estimates can be controversial.

With the growing number of gamma-ray pulsars not detected at radio wavelengths, and thus without a DM, and the difficulties of the other methods,
we face an ever-growing pulsar distance problem. We have 26 objects with no distance estimates, compared to nine in 1PC. 
To mitigate this, we determine a maximum distance by \textit{assuming} that the pulsar is within the Galaxy. 
We define the Galaxy edge as the distance for a given line of sight where the NE2001 DM reaches its maximum value
(``DMM'' in Table \ref{tbl-PSRdist}, illustrated in Figure \ref{galaxie}).

\tabletypesize{\scriptsize}
\begin{deluxetable}{lccc}
\tablecaption{Distance estimates for young LAT-detected pulsars
\label{tbl-PSRdist}}
\tablewidth{0pt}

\tablehead{
\colhead{Pulsar Name}      & 
\colhead{Distance (kpc)}   & 
\colhead{Method }      &
\colhead{Reference} }
\startdata
 J0007+7303 &   $1.4\pm {0.3}$  &  K & \citet{Pineault1993}\\ 
 J0106+4855 &   $3.0^{+1.1}_{-0.7}$  &  DM & \citet{Pletsch_2012}\\ 
 J0205+6449 &   $1.95\pm {0.04}$  &  KP & \citet{Xu_2006}\\ 
 J0248+6021 &   $2.0\pm {0.2}$  &  K & \citet{Theureau2011}\\ 
 J0357+3205 &   $<8.2$ & DMM &   \nodata   \\ 
 J0534+2200 &   $2.0\pm {0.5}$  &  O & \citet{Trimble1973}\\ 
 J0622+3749 &   $<8.3$ & DMM &   \nodata  \\ 
 J0631+1036 &   $1.0\pm {0.2}$  &  O & \citet{Zepka1996}\\ 
 J0633+0632 &   $<8.7$ & DMM &   \nodata  \\ 
 J0633+1746 &   $0.25^{+0.23}_{-0.08}$  &  P & \citet{Verbiest12}\\ 
 J0659+1414 &   $0.28\pm {0.03}$  &  P & \citet{Verbiest12}\\ 
 J0729$-$1448 &   $3.5\pm {0.4}$  &  DM & \citet{Morris02_parkes}\\ 
 J0734$-$1559 &   $<10.3$ & DMM &   \nodata  \\ 
 J0742$-$2822 &   $2.1\pm {0.5}$  &  DM & \citet{Janssen_2006}\\ 
 J0835$-$4510 &   $0.29^{+0.02}_{-0.02}$  &  P & \citet{Dodson2003}\\ 
 J0908$-$4913 &   $2.6\pm {0.9}$  &  DM & \citet{Hobbs_pmbs}\\ 
 J0940$-$5428 &   $3.0\pm {0.5}$  &  DM & \citet{Manchester2001}\\ 
 J1016$-$5857 &   $2.9^{+0.6}_{-1.9}$  &  K & \citet{Ruiz86}\\ 
 J1019$-$5749 &   $6.8^{+13.2}_{-2.5}$  &  DM & \citet{Kramer03}\\ 
 J1023$-$5746 &   $<16.8$ & DMM &   \nodata  \\ 
 J1028$-$5819 &   $2.3\pm {0.3}$  &  DM & \citet{Keith2008}\\ 
 J1044$-$5737 &   $<17.2$ & DMM &   \nodata  \\ 
 J1048$-$5832 &   $2.7\pm {0.4}$  &  DM & \citet{Johnston_45}\\ 
 J1057$-$5226 &   $0.3\pm {0.2}$  &  O & \citet{Mignani2010a}\\ 
 J1105$-$6107 &   $5.0\pm {1.0}$  &  DM & \citet{Kaspi_97}\\ 
 J1112$-$6103 &   $12.2^{+7.8}_{-3.8}$  &  DM & \citet{Manchester2001}\\ 
 J1119$-$6127 &   $8.4\pm {0.4}$  &  K & \citet{Caswell_2004}\\ 
 J1124$-$5916 &   $4.8^{+0.7}_{-1.2}$  &  X & \citet{Gonzalez2003}\\ 
 J1135$-$6055 &   $<18.4$ & DMM &   \nodata  \\ 
 J1357$-$6429 &   $2.5^{+0.5}_{-0.4}$  &  DM & \citet{lfl+06}\\ 
 J1410$-$6132 &   $15.6^{+7.4}_{-4.2}$  &  DM & \citet{OBrien_2008}\\ 
 J1413$-$6205 &   $<21.4$ & DMM &   \nodata  \\ 
 J1418$-$6058 &   $1.6\pm {0.7}$  &  O & \citet{Yadigaroglu1997}\\ 
 J1420$-$6048 &   $5.6\pm {0.9}$  &  DM & \citet{LAT6pulsars}\\ 
 J1429$-$5911 &   $<21.8$ & DMM &   \nodata  \\ 
 J1459$-$6053 &   $<22.2$ & DMM &   \nodata  \\ 
 J1509$-$5850 &   $2.6\pm {0.5}$  &  DM & \citet{LAT6pulsars}\\ 
 J1513$-$5908 &   $4.2\pm {0.6}$  &  DM & \citet{Hobbs_pmbs}\\ 
 J1531$-$5610 &   $2.1^{+0.4}_{-0.3}$  &  DM & \citet{Kramer03}\\ 
 J1620$-$4927 &   $<24.1$ & DMM &   \nodata  \\ 
 J1648$-$4611 &   $5.0\pm {0.7}$  &  DM & \citet{Kramer03}\\ 
 J1702$-$4128 &   $4.8\pm {0.6}$  &  DM & \citet{Kramer03}\\ 
 J1709$-$4429 &   $2.3\pm {0.3}$  &  DM & \citet{Johnston_45}\\ 
 J1718$-$3825 &   $3.6\pm {0.4}$  &  DM & \citet{Manchester2001}\\ 
 J1730$-$3350 &   $3.5^{+0.4}_{-0.5}$  &  DM & \citet{Jodrell}\\ 
 J1732$-$3131 &   $0.6\pm {0.1}$  &  DM & \citet{Maanetal}\\ 
 J1741$-$2054 &   $0.38\pm {0.02}$  &  DM & \citet{Camilo2009}\\ 
 J1746$-$3239 &   $<25.3$ & DMM &   \nodata  \\ 
 J1747$-$2958 &   $4.8\pm {0.8}$  &  X & \citet{Gaensler2004}\\ 
 J1801$-$2451 &   $5.2^{+0.6}_{-0.5}$  &  DM & \citet{Jodrell}\\ 
 J1803$-$2149 &   $<25.2$ & DMM &   \nodata  \\ 
 J1809$-$2332 &   $1.7\pm {1.0}$  &  K & \citet{Oka1999}\\ 
 J1813$-$1246 &   $<24.7$ & DMM &   \nodata  \\ 
 J1826$-$1256 &   $<24.7$ & DMM &   \nodata  \\ 
 J1833$-$1034 &   $4.7\pm {0.4}$  &  K & \citet{J1833discovery,Camilo2006}\\ 
 J1835$-$1106 &   $2.8\pm {0.4}$  &  DM & \citet{DAmico_98}\\ 
 J1836+5925 &   $0.5\pm {0.3}$  &  X & \citet{Halpern2002}\\ 
 J1838$-$0537 &   $<24.1$ & DMM &   \nodata  \\ 
 J1846+0919 &   $<22.0$ & DMM &   \nodata  \\ 
 J1907+0602 &   $3.2\pm {0.3}$  &  DM & \citet{LATPSR1907}\\ 
 J1952+3252 &   $2.0\pm {0.5}$  &  K & \citet{Greidanus1990}\\ 
 J1954+2836 &   $<18.6$ & DMM &   \nodata  \\ 
 J1957+5033 &   $<14.5$ & DMM &   \nodata  \\ 
 J1958+2846 &   $<18.5$ & DMM &   \nodata  \\ 
 J2021+3651 &   $10.0^{+2.0}_{-4.0}$  &  O & \citet{Hessels_2004}\\ 
 J2021+4026 &   $1.5\pm {0.4}$  &  K & \citet{Landecker1980}\\ 
 J2028+3332 &   $<17.2$ & DMM &   \nodata  \\ 
 J2030+3641 &   $3.0\pm {1.0}$  &  O & \citet{Camilo2012}\\ 
 J2030+4415 &   $<15.7$ & DMM &   \nodata  \\ 
 J2032+4127 &   $3.7\pm {0.6}$  &  DM & \citet{Camilo2009}\\ 
 J2043+2740 &   $1.8\pm {0.3}$  &  DM & \citet{Ray1996}\\ 
 J2055+2539 &   $<15.3$ & DMM &   \nodata  \\ 
 J2111+4606 &   $<14.8$ & DMM &   \nodata  \\ 
 J2139+4716 &   $<14.1$ & DMM &   \nodata  \\ 
 J2229+6114 &   $0.80^{+0.15}_{-0.20}$  &  K & \citet{Kothes2001}\\ 
 J2238+5903 &   $<12.4$ & DMM &   \nodata  \\ 
 J2240+5832 &   $7.7\pm {0.7}$  &  O & \citet{Theureau2011}\\ 

\enddata


\tablecomments{The best known distances of the 77 young pulsars detected by {\it Fermi}. 
The methods are: K -- kinematic method; P -- parallax; 
DM -- from dispersion measure using the \citet{Cordes2002} NE2001 model;  X -- from X-ray measurements ; O -- other methods. 
For DM, the reference gives the DM measurement.
For the 26 pulsars with no distance estimate, DMM is the distance to the Galaxy's edge, taken as an upper limit, determined from the maximum NE2001 DM value for that line of sight.}



\end{deluxetable}
\normalsize


\begin{landscape}
\tabletypesize{\scriptsize}
\begin{deluxetable}{lcccccccccc}

\tablecaption{Millisecond pulsar distances and spindown Doppler corrections
\label{tbl-doppler}}
\tablewidth{0pt}

\tablehead{
\colhead{PSR}                               & 
\colhead{$ d $}     & \colhead{Method}   & \colhead{Ref\tablenotemark{b}}   &
\colhead{$ \mu$} & \colhead{ Ref\tablenotemark{c}}   & 
\colhead{$\dot P^{\rm \, int}$}  & \colhead{$\dot P^{\rm \, shk}$}  & \colhead{$\dot P^{\rm \, gal}$}  &
\colhead{$\dot E^{\rm \, int}$}  & \colhead{$\xi$} \\ 
\colhead{}  & \colhead{(pc)} & \colhead{} & \colhead{} & \colhead{($\rm mas\; yr^{-1}$)} & \colhead{} &  
\colhead{($\rm 10^{-21}$)} & \colhead{($\rm 10^{-21}$)} & \colhead{($\rm 10^{-21}$)} & \colhead{($\rm 10^{33} \,erg \, s^{-1}$)} & \colhead{(\%)} 
}

\startdata
 J0023+0923 &  $690^{+210}_{-110}$ &    DM & (1) &	&	&	&	&	&	&     \\ 
 J0030+0451 &  $280^{+100}_{-60}$ &    P & (2) & $5.7 \pm 1.1 $ 	&(1)  & $10.7\pm {0.1}$	& $0.11$	& $-0.60$ &  $3.64\pm {0.02}$	& $-5$ \\ 
 J0034$-$0534 &  $540\pm 100$ &    DM & (3) & $31.0 \pm 9.0 $ 	&(2)  & $2.9\pm {1.4}$	& $2.37$	& $-0.31$ &  $17.3\pm {8.6}$	& $41$ \\ 
 J0101$-$6422 &  $550^{+90}_{-80}$ &    DM & (4) & $15.6 \pm 1.7 $ 	&(3)  & $4.4\pm {0.2}$	& $0.84$	& $-0.39$ &  $10.1\pm {0.5}$	& $9$ \\ 
 J0102+4839 &  $2320^{+500}_{-430}$ &    DM & (1) &	&	&	&	&	&	&     \\ 
 J0218+4232 &  $2640^{+1080}_{-640}$ &    DM & (5) & $5.0 \pm 6.0 $ 	&(2)  & $76.9\pm {0.9}$	& $0.37$	& $0.09$ &  $243.2\pm {2.8}$	& $0.6$ \\ 
 J0340+4130 &  $1730\pm 300$ &    DM & (1) &	&	&	&	&	&	&     \\ 
 J0437$-$4715 &  $156\pm 1$ &    P & (2) & $141.3 \pm 0.1 $ 	&(4)  & $14.1\pm {0.3}$	& $43.59$	& $-0.40$ &  $2.9\pm {0.1}$	& $75$ \\ 
 J0610$-$2100 &  $3540^{+5460}_{-1000}$ &    DM & (6) & $18.2 \pm 0.2 $ 	&(5)  & $1.2^{+17.0}_{-1.1}$	& $11.00$	& $0.10$ &   $0.8^{+11.7}_{-0.8}$	& $90$ \\ 
 J0613$-$0200 &  $900^{+400}_{-200}$ &    P & (2) & $10.8 \pm 0.2 $ 	&(6)  & $8.7^{+0.3}_{-0.2}$	& $0.77$	& $0.08$ &   $12.0^{+0.5}_{-0.2}$	& $9$ \\ 
 J0614$-$3329 &  $1900^{+440}_{-350}$ &    DM & (7) &	&	&	&	&	&	&     \\ 
 J0751+1807 &  $400^{+200}_{-100}$ &    P & (2) & $6.0 \pm 2.0 $ 	&(7)  & $7.7\pm {0.1}$	& $0.12$	& $-0.02$ &  $7.2\pm {0.1}$	& $1$ \\ 
 J1024$-$0719 &  $390\pm 40$ &    DM\tablenotemark{a} & (8) & $59.9 \pm 0.2 $ 	&(5)  & $1.6^{+1.8}_{-1.4}$	& $17.37$	& $-0.44$ &   $0.4^{+0.5}_{-0.4}$	& $92$ \\ 
 J1124$-$3653 &  $1720^{+430}_{-360}$ &    DM & (1) &	&	&	&	&	&	&     \\ 
 J1125$-$5825 &  $2620\pm 370$ &    DM & (9) &	&	&	&	&	&	&     \\ 
 J1231$-$1411 &  $440\pm 50$ &    DM & (7) & $62.2 \pm 4.7 $ 	&(8)\tablenotemark{a}  & $6.5\pm {2.9}$	& $15.15$	& $-0.41$ &  $5.1\pm {2.3}$	& $70$ \\ 
 J1446$-$4701 &  $1460\pm 220$ &    DM & (10) &	&	&	&	&	&	&     \\ 
 J1514$-$4946 &  $940\pm 120$ &    DM & (4) &	&	&	&	&	&	&     \\ 
 J1600$-$3053 &  $1630^{+310}_{-270}$ &    DM & (11) & $7.2 \pm 0.3 $ 	&(6)  & $8.6^{+0.2}_{-0.1}$	& $0.74$	& $0.15$ &   $7.3\pm {0.1}$	& $9$ \\ 
 J1614$-$2230 &  $650\pm 50$ &    P & (12) & $36.5 \pm 0.2 $ 	&(9)  & $3.0\pm {0.5}$	& $6.65$	& $-0.003$ &  $3.8\pm {0.6}$	& $69$ \\ 
 J1658$-$5324 &  $930^{+110}_{-130}$ &    DM & (4) &	&	&	&	&	&	&     \\ 
 J1713+0747 &  $1050^{+60}_{-50}$ &    P & (2) & $6.30 \pm 0.01 $ 	&(10)  & $8.28^{+0.03}_{-0.02}$	& $0.46$	& $-0.21$ &   $3.42\pm {0.01}$	& $3$ \\ 
 J1741+1351 &  $1080^{+40}_{-50}$ &    P & (13) & $11.71 \pm 0.01 $ 	&(11)  & $29.1\pm 0.1$	& $1.35$	& $-0.20$ &   $21.76^{+0.04}_{-0.05}$	& $4$ \\ 
 J1744$-$1134 &  $417\pm 17$ &    P & (11) & $21.02 \pm 0.03 $ 	&(6)  & $7.0\pm {0.1}$	& $1.82$	& $0.08$ &  $4.11\pm {0.04}$	& $21$ \\ 
 J1747$-$4036 &  $3390\pm 760$ &    DM & (4) &	&	&	&	&	&	&     \\ 
 J1810+1744 &  $2000^{+310}_{-280}$ &    DM & (1) &	&	&	&	&	&	&     \\ 
 J1823$-$3021A &  $7600\pm 400$ &    O & (14) &	&	&	&	&	&	&     \\ 
 J1858$-$2216 &  $940^{+200}_{-130}$ &    DM & (15) &	&	&	&	&	&	&     \\ 
 J1902$-$5105 &  $1180\pm 210$ &    DM & (4) &	&	&	&	&	&	&     \\ 
 J1939+2134 &  $3560\pm 350$ &    DM & (16) & $0.80 \pm 0.02 $ 	&(12)  & $105.5\pm {0.1}$	& $0.01$	& $-0.36$ &  $1096.6\pm {0.5}$	& $-0.3$ \\ 
 J1959+2048 &  $2490^{+160}_{-490}$ &    DM & (17) & $30.4 \pm 0.6 $ 	&(13)  & $8.1^{+0.7}_{-1.8}$	& $9.00$	& $-0.25$ &   $76.3^{+6.4}_{-17.1}$	& $52$ \\ 
 J2017+0603 &  $1570\pm 150$ &    DM & (18) &	&	&	&	&	&	&     \\ 
 J2043+1711 &  $1760^{+150}_{-320}$ &    DM & (1) & $13.0 \pm 2.0 $ 	&(14)  & $4.3\pm {0.6}$	& $1.72$	& $-0.35$ &  $12.7^{+1.6}_{-1.8}$	& $24$ \\ 
 J2047+1053 &  $2050^{+320}_{-290}$ &    DM & (15) &	&	&	&	&	&	&     \\ 
 J2051$-$0827 &  $1040\pm 150$ &    DM & (19) & $7.3 \pm 0.4 $ 	&(15)  & $12.6\pm {0.1}$	& $0.61$	& $-0.47$ &  $5.43\pm {0.05}$	& $1$ \\ 
 J2124$-$3358 &  $300^{+70}_{-50}$ &    P & (2) & $52.3 \pm 0.3 $ 	&(6)  & $11.2^{+2.3}_{-1.6}$	& $9.83$	& $-0.46$ &   $3.7^{+0.8}_{-0.5}$	& $46$ \\ 
 J2214+3000 &  $1540\pm 180$ &    DM & (7) &	&	&	&	&	&	&     \\ 
 J2215+5135 &  $3010^{+330}_{-370}$ &    DM & (1) &	&	&	&	&	&	&     \\ 
 J2241$-$5236 &  $510\pm 80$ &    DM & (20) &	&	&	&	&	&	&     \\ 
 J2302+4442 &  $1190^{+90}_{-230}$ &    DM & (18) &	&	&	&	&	&	&     \\

\enddata

\tablecomments{Columns 2 and 3 give the distances for the 40 MSPs detected by {\it Fermi}, and the method used to find them:
P -- parallax; 
DM -- from dispersion measure using the \citet{Cordes2002} NE2001 model; 
O -- other methods.
For DM, the references in Column 4 give the DM measurement.
For the 20 MSPs with a proper motion measurement, it is listed in Column 5, obtained from the reference in Column 6. 
Columns 7 to 10 list the intrinsic $\dot P^{\rm \, int}$, the contributions of the Shklovskii effect $\dot P^{\rm \, shk}$
and of acceleration due to the Galactic rotation, $\dot P^{\rm \, gal}$, and the intrinsic spindown power $\dot E^{\rm \, int}$. 
The relative correction $\xi$ is defined from $\dot E^{\rm \, int} = \dot E \,(1-\xi)$.
}


\tablenotetext{a}{Proper motion for J1231$-$1411 from this work: $\mu_{\rm \alpha}\cos\delta = -60.5(44)\,\rm mas\; yr^{-1}$,  $\mu_{\rm \delta} = -14.3(82)\,\rm mas\; yr^{-1}$. 
For J1024$-$0719, the timing parallax distance measurement gives negative spindown and we use instead the DM distance following \citet{egc+12}. }

\tablenotetext{b}{Distance and DM references:
(1)  \citet{Hessels_2011};
(2)  \citet{Verbiest12};
(3)  \citet{Bailes_1994};
(4)  \citet{Kerr2012};
(5)  \citet{Jodrell};
(6)  \citet{bjd+06};
(7)  \citet{rrc+11};
(8)  \citet{hbo06};
(9)  \citet{Bates_2011};
(10)  \citet{Keith_2012};
(11)  \citet{Verbiest09};
(12)  \citet{antoine+13};
(13)  Freire et al. (in prep.);
(14)  \citet{Kuulkers_2003};
(15)  \citet{Ray2012psc};
(16)  \citet{cbl+95};
(17)  \citet{aft94};
(18)  \citet{Cognard2011};
(19)  \citet{Doroshenko_2001};
(20)  \citet{Keith_2011}.
}

\tablenotetext{c}{Proper motion references: 
(1)  \citet{aaa+09e};
(2)  \citet{hllk05};
(3)  \citet{Kerr2012};
(4)  \citet{dtbr09};
(5)  \citet{egc+12};
(6)  \citet{Verbiest09};
(7)  \citet{Nice2005};
(8)  This work;
(9)  \citet{antoine+13};
(10)  \citet{sns+05};
(11)  Freire et al. (in prep.);
(12)  \citet{cbl+95};
(13)  \citet{aft94};
(14)  \citet{gfc+12};
(15)  \citet{lvt+11}.
}




\end{deluxetable}
\normalsize
\end{landscape}



\subsection{Doppler corrections}
\label{doppler}
Many pulsar characteristics, including some listed in Tables \ref{tbl-charPSR} and \ref{tbl-charMSP},
depend on the \textit{intrinsic} spin period $P^{\rm \,int}$ and spindown rate $\dot P^{\rm \,int}$.
The Doppler shift of the \textit{observed} period is $P = (1 + v_{\rm \,R}/c)P^{\rm \,int}$,
where $v_{\rm \,R}$  is the pulsar's radial velocity along the unit vector $\textbf{n}_{10}$ from the solar system. 
The Doppler correction to $\dot P$ is obtained by differentiating the equation and 
separating the effects of the system's proper motion \citep{Shklovskii} from the acceleration due to
Galactic rotation:
\begin{equation}
\dot P^{\rm \,int} = \dot P -\dot P^{\rm \, shk}-\dot P^{\rm \, gal}
\label{DoppEq}
\end{equation}
with
\begin{equation}
\dot P^{\rm \,shk} = \frac{1}{c}\,\mu^2 \,d\,P = k\,({\mu \over {\rm mas\, yr^{-1}}})^2 \,({d \over {\rm kpc}})\,({P \over {\rm s}})
\end{equation}
and
\begin{equation}
\dot P^{\rm \,gal} = \frac{1}{c}\, \textbf{n}_{10} \cdot (\textbf{a}_{1}-\textbf{a}_{0})P
\end{equation}
where $k=2.43\times 10^{-21}$ for pulsar distance $d$ and proper motion transverse to the line of sight $\mu$.
The Galactic potential model of \cite{Carlberg_Innanen1987} and \cite{Kuijken_Gilmore1989}
provides the accelerations $\textbf{a}_{1}$ of the pulsar and $\textbf{a}_{0}$ of the Sun. 
Since the constant $k$ is small, the corrections are negligible for the young gamma-ray pulsars, which all have $\dot P > 10^{-16}$.
However, for MSPs $\mu^2 \,d$ can be large enough that $\dot P^{\rm \,int}$ differs noticeably from the \textit{observed} $\dot P$ values
and quantities derived from $\dot P$ will also be affected. 

From the literature we compiled proper motion measurements for 243 pulsars, all but one of which also have
distance estimates and $\dot P$ measurements, and we calculated $\dot P^{\rm \,int}$ and its uncertainties for those 242 pulsars. 
Of these, 69 have $P < 30$ ms, and 20 are gamma-ray MSPs, listed in Table \ref{tbl-doppler}. 
The magnitude of the correction is $\xi = (\dot P - \dot P^{\rm \, int} ) /\dot P$, or, equivalently,  
$\dot E^{\rm \, int} = \dot E \,(1-\xi)$ since $\dot E \propto \dot P$. 
For $\left| \xi \right|$ greater than a few percent, $\dot P^{\rm \,shk} > | \dot P^{\rm \,gal}| $ and corrected $\dot E^{\rm \,int}$
is \textit{less} than the observed value. Hence flagging gamma-ray pulsar candidates that have large $\dot E$ selects
some with lower $\dot E^{\rm \,int}$, but we would not miss candidates by neglecting proper motion.
For large Doppler corrections, the Galactic term is negligible,
$\dot P^{\rm \,shk} \gg \dot P^{\rm \,gal}$, and we calculate only the uncertainty due to $\dot P^{\rm \,shk}$.
Unless otherwise noted, throughout this paper we use $\dot P^{\rm \, int}$ from Table \ref{tbl-doppler} to 
replace $\dot P$  and the derived quantities ($\dot E$, $\tau$, et cetera) in the Figures and Tables.
In Section \ref{lumin} we discuss the Doppler correction's effect on the gamma-ray luminosity for a few cases.


\section{Profile Characterization}
\label{profiles}
\subsection{Gamma-ray and Radio Light Curves}
Appendix \ref{App-Samples} contains a small sample of gamma-ray pulse
profiles (Figures \ref{J0034-0534lc} to \ref{J2240+5832lc}), overlaid
with the radio profiles when available, and showing the fits described
in Section \ref{lightCurveFitting}, below.  All pulse profiles are
provided in the online material.  We display gamma-ray light curves by
computing a weighted histogram of gamma-ray rotational phases $\phi$. 
The error on the $i^{\rm th}$ histogram bin containing $N_i$ photons is estimated as $\sigma_i^2 = 1 + \sum_{j=0}^{N_i} w_j^2$, 
using the weights $w_j$ defined in Eq. \ref{weightsEq}. 
The ``1'' term mitigates a bias towards low histogram levels and $\sigma_i$ values caused by background-dominated bins. 
For these bins, the typical photon weight is very low, but the large weights of
rare background photons near the pulsar position (within the point spread function) can substantially increase the bin level.  
The additional term compensates for the spread due to the presence or absence of a single such photon.
We choose the number of bins in the histogram
according to the weighted H-test: 25 bins (H $<$ 100); 50 bins (100
$<$ H $<$ 1000); and 100 bins (H $>$ 1000).


We estimate the background contribution from diffuse sources and
neighboring point sources by computing the expectation value of $w$
under the hypothesis that the photon does not originate from the pulsar:
$b\equiv\int_0^1\,dw\,w\times[1-f(w)]\approx
\sum_{j=0}^{N_{\gamma}}w_j-\sum_{j=0}^{N_{\gamma}}w_j^2$, where $f(w)$
is the probability distribution of the photon weights and we have used
a Monte Carlo approximation to evaluate the integral.  The
corresponding background level for a weighted histogram, shown as a
horizontal dashed line in the gamma-ray light curves, is $b/N_{bins}$.
The dominant error in this quantity arises from systematic errors in
the normalization of the diffuse background.  We estimate this
contribution by increasing/reducing the overall normalization of the
background by 6\% (see Section \ref{spectralMethodSection} for
discussion.)



For pulsars with known radio profiles, we also display by preference the flux density
at 1400 MHz.  When 1400 MHz data are unavailable or highly scattered by
the ISM, we include lower- or higher-frequency profiles, noting the
frequency and provenance of the profile on the figures (Appendix
\ref{App-Samples}, and the online material).

The propagation of radio pulses through the dispersive ISM is delayed
by $\Delta t\propto \mathrm{DM}\,\nu^{-2}$.  This delay, as well as
Roemer-type delays associated with the configuration of the telescope
relative to the solar system barycenter, are accounted for by
\textsc{Tempo2}, allowing precise alignment of radio and gamma-ray
light curves.  The absolute time at which $\phi\equiv0$ is indicated
in the pulsar par files, provided with the online material, by the
parameters TZRMJD (giving the time of arrival of the zero phase),
TZRFRQ (giving the frequency for which this time is correct), and
TZRSITE (encoding the radio telescope/site of arrival).

The zero of phase---the fiducial phase---is ideally the rotational
phase when the magnetic axis, the spin axis, and the line of sight lie
in the same plane.  For some pulsars, this phase can be identified by
fitting the rotating vector model \citep[][]{RVM} to radio polarization
position angle versus phase. However, radio pulses often cover too narrow a phase interval
to constrain fits using the rotating vector model.

If radio emission from the polar cap is symmetric, the peak intensity
can also be used as a proxy for the fiducial phase, and we adopt this
approach here for radio-loud pulsars.    In detail, this approach can
fail if there is appreciable asymmetry in the radio beam or profile
evolution with frequency.  Additionally, for pulsars with radio interpulses and for some
MSPs, we observe emission from the field lines from both poles and must choose with
which to associate the fiducial phase.  In these cases, we generally
choose the hemisphere furthest separated in phase from the gamma rays,
consistent with an interpretation of gamma-ray emission arising from the
outer magnetosphere.  Finally, some pulsars (e.g.,  J0034$-$0534 shown
in Figure \ref{J0034-0534lc}) exhibit a clear double symmetry in the
radio light curve.  In these cases, we choose a fiducial point near
the point of symmetry.  We note which prescription we have followed in
Tables \ref{tbl-pulsePSR} and \ref{tbl-pulseMSP} with a `p' (fiducial
point at peak intensity), `h' (fiducial point from hemisphere opposite
to peak intensity), `s' (fiducial point placed at point of symmetry
rather than peak).
For radio-quiet pulsars, we put the first gamma-ray peak (identified by
looking for sharp rises and bridge emission) at $\phi=0.1$ for display
purposes.

\begin{figure}[!ht]
\centering
\includegraphics[width=0.7\textwidth]{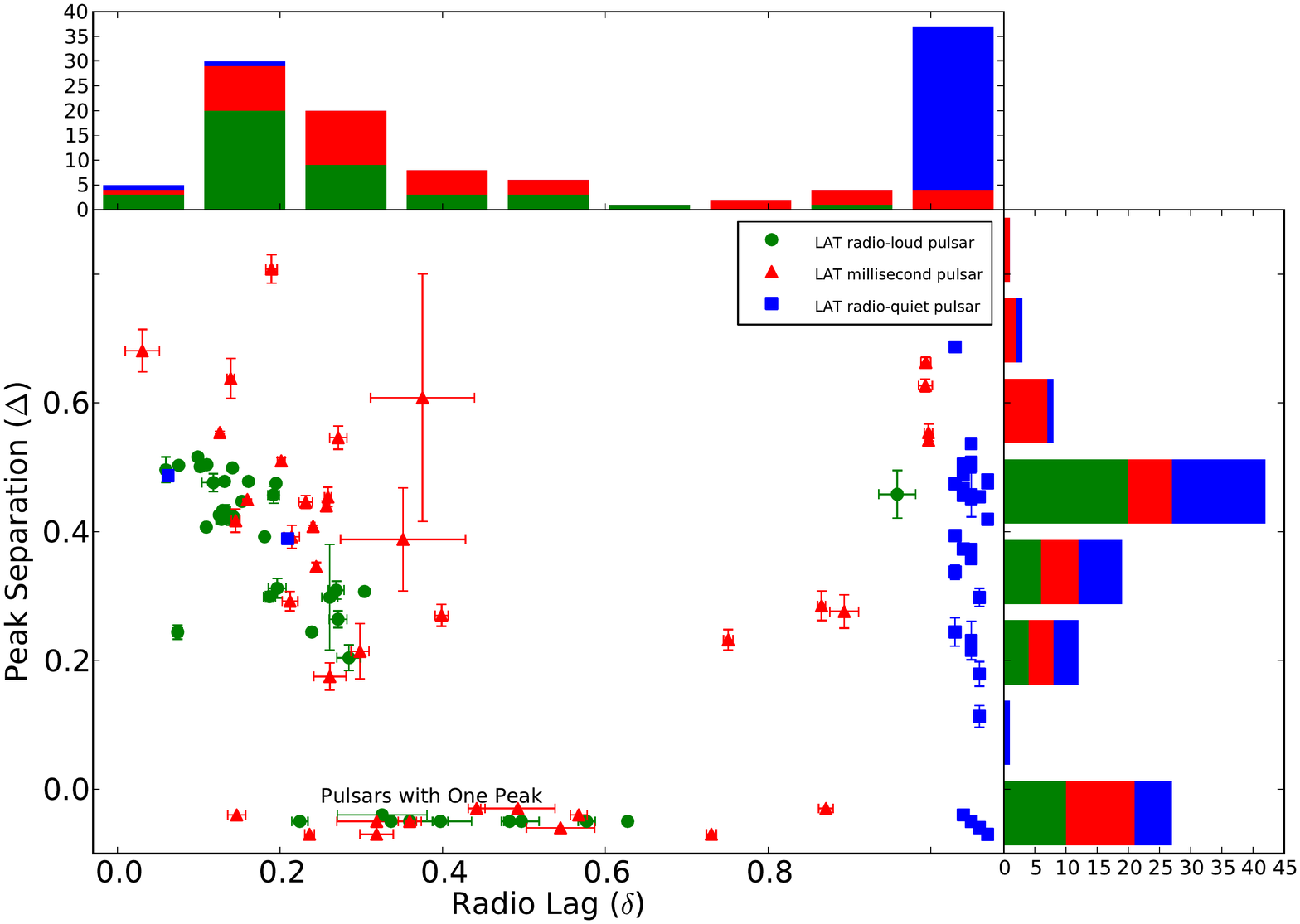}
\caption{Phase lag $\delta$ of the gamma peak relative to the fiducial phase versus the phase separation $\Delta$ between the gamma-ray peaks. 
The artificial staggering of the points along the horizontal axis (single-peaked pulsars)
and the right-hand vertical axis (pulsars with no radio detection) is to enhance clarity. 
The markers are the same as in Figure \ref{PPdotplot}, as is the color code of the histograms projected onto the axes. 
\label{deltaDelta}}
\end{figure}

\subsection{Gamma-ray Light Curve Fitting}
\label{lightCurveFitting}

Generally, the gamma-ray light curve of a pulsar can be
represented by a wrapped probability density function (pdf) of
$\phi\in[0,1)$.  A compact
approximation of the true pdf for photon phases can be constructed
as a linear combination of $N$ unimodal, possibly skew distributions:
\begin{equation}
f(\phi) = \sum_{i=1}^{N} n_i\,g_i(\phi) + 
\left(1-\sum_{i=1}^{N} n_i\right),
\end{equation}
where each of the $g_i$ is an individually normalized distribution
and $1-\sum_{i=1}^{N} n_i \leq 1$ is a uniform distribution representing
an unpulsed component.  To explicitly enforce normalization, we use
spherical polar coordinates lying within the unit sphere as internal
parameters.  With three components, e.g., the normalizations $n_i$ are
given in terms of the internal parameters $\chi_i$ as $n_1 =
\sin\chi_1\cos\chi_2$; $n_2 = \sin\chi_1\sin\chi_2\cos\chi_3$; $n_3 =
\sin\chi_1\sin\chi_2\sin\chi_3$.

Although generalized wrapped distributions exist, analytic forms for
such distributions are typically unavailable.  Instead we adopt
well-known pdfs, viz. the Gaussian (normal) and Lorentzian (Cauchy)
distributions, by wrapping them onto a circle:
\begin{equation}
g(\phi) = \sum_{i=-\infty}^{+\infty}g'(\phi+i),
\end{equation}
where $g'$ is defined on the real line and $g$ on the circle.
For practical reasons this sum must be truncated, and we define a truncated,
wrapped distribution as 
\begin{equation}
g_T(\phi,N) = \sum_{i=-N}^{+N}g'(\phi+i) +
\left(1-\int_{-N}^{+N}g'(x)dx\right),
\end{equation}
i.e., we approximate the tails as a uniform distribution.  In the fits
discussed below, $N\equiv10$.

Because the peaks of gamma-ray light curves may have a caustic
origin, asymmetric distributions are needed to model their fast rise
and slow fall.  We generalize the symmetric Gaussian and Lorentzian
distributions by matching two distributions with differing width
parameters, $\sigma_1$ and $\sigma_2$, at the maximum, $x_0$.  Defining
$z\equiv(x-x_0)/\sigma$ with $\sigma=\sigma_1$ if $z\leq0$ and
$\sigma=\sigma_2$ otherwise,  the functional forms of the resulting
distributions on the real line are
\begin{equation}
g'(x) = \frac{2}{\pi(\sigma_1+\sigma_2)(1+z^2)}\ \mathrm{ (Lorentzian),}
\end{equation}
\begin{equation}
g'(x) =
\sqrt{\frac{2}{\pi}}\frac{\exp(\frac{-z^2}{2})}{\sigma_1+\sigma_2}\ \mathrm{
(Gaussian),}\\
\end{equation}

We employ maximum likelihood to determine the best-fit parameters of the
mixture distribution.  If $w_i$ is a probability that a photon
originates from the pulsar, then the logarithm of the likelihood is
\begin{equation}
\log\mathcal{L} = \sum_{j=1}^{N_{\gamma}} \log \left[ w_i\, f(\phi_i) +
(1-w_i)\right].
\end{equation}
\noindent{}For large data sets ($N_{\gamma}>10^4$) we speed up the computation
by binning $f(\phi)$ to 512 values.

\begin{figure}[!ht]
\centering
\includegraphics[width=0.7\textwidth]{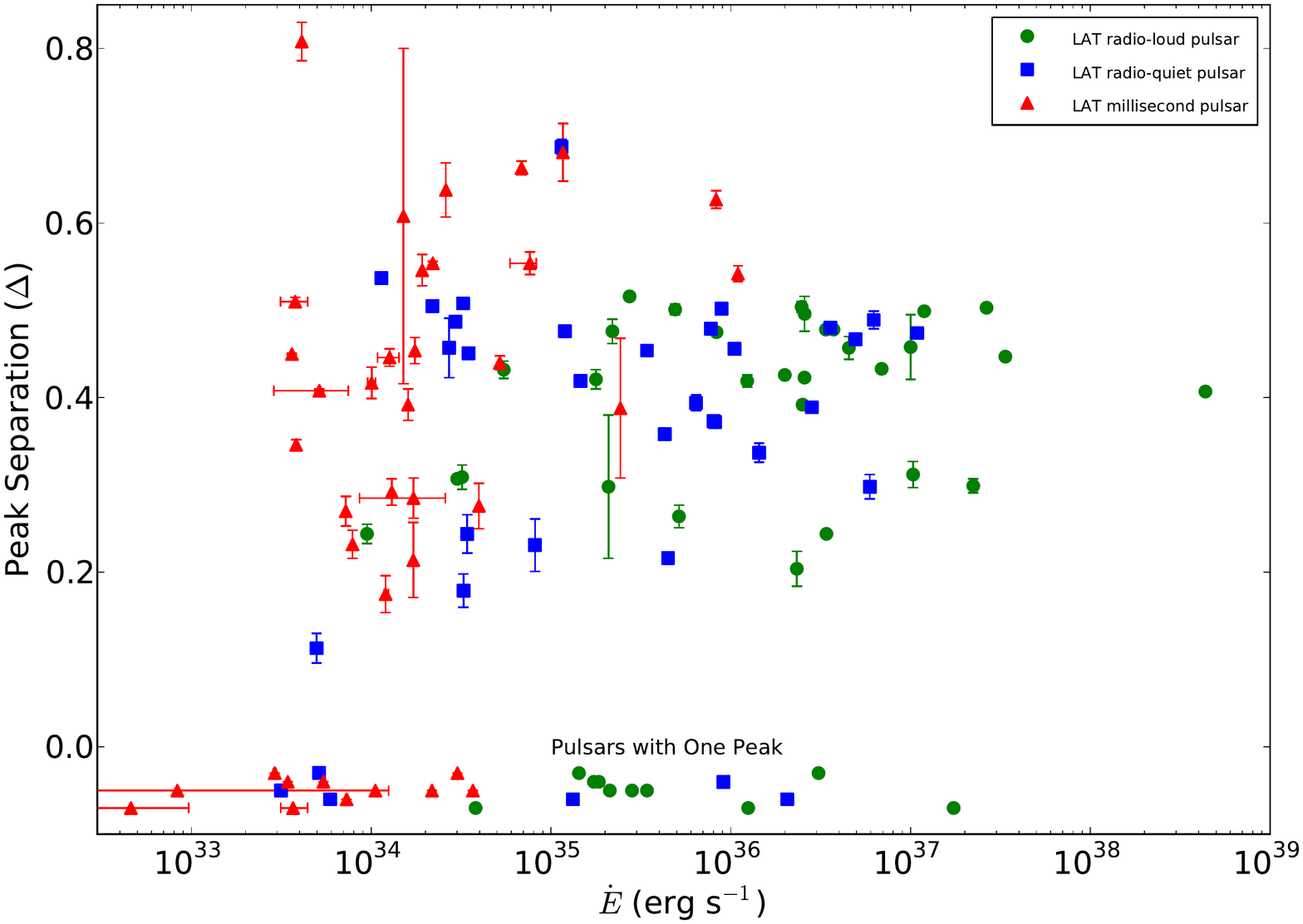}
\caption{
The phase separation $\Delta$ between gamma-ray peaks versus the Shlovkskii-corrected spindown power. 
The markers are the same as in Figure \ref{PPdotplot}. 
\label{DeltaEdot}}
\end{figure}

Most LAT light curves can be modelled with good fidelity by one or
two two-sided narrow peak distributions and a single broad
bridge component.  From these fits, we determine the following
quantities: $N_{\mathrm{peaks}}$, the number of non-bridge components;
$\delta$, the offset of the mode of the leading peak from the fiducial
phase (see above); and $\Delta$, the difference between the modes of
the leading and trailing peak (for $N_{\mathrm{peaks}}>1$).  
These parameters appear in Tables \ref{tbl-pulsePSR} and \ref{tbl-pulseMSP}.
The strong correlation between $\Delta$ and $\delta$ in 
Figure \ref{deltaDelta}, as well as the dependence on spindown power
(Figure \ref{DeltaEdot}) are discussed in Section \ref{discuss}. 
The peak widths are included in the online material.

The uncertainty on $\delta$ is estimated by combining in quadrature
the statistical uncertainty on the position of the relevant
gamma-ray peak with that incurred from uncertainty in the DM.  The
statistical uncertainty naturally includes a contribution from
uncertainty in the timing solution (the \textsc{Tempo} TRES quantity) which
serves to smear out light curve features by a characteristic width
$\delta\phi\approx\mathrm{TRES}/P$. The statistical uncertainty on
$\Delta$ is determined by the sum in quadrature of the position
uncertainty of the relevant gamma-ray peaks.

The representation of the light curve (Gaussian vs.~Lorentzian,
presence or absence of additional components) affects the results for
both $\delta$ and $\Delta$.  In addition to the uncertainties above,
we estimate a ``model'' uncertainty of $\sim$0.01 in phase. 

\clearpage

\tabletypesize{\scriptsize}
\begin{deluxetable}{lccccc}
\tablewidth{0pt}
\tablecaption{Pulse shape parameters of young LAT-detected pulsars
\label{tbl-pulsePSR}}

\tablehead{
\colhead{PSR\tablenotemark{a}} & \colhead{Peaks} & \colhead{Radio lag} & \colhead{Shift} & \colhead{gamma-peak separation} & \colhead{Off-peak definition} \\
\colhead{}        & \colhead{}      & \colhead{$\delta$}  & \colhead{Method}& \colhead{$\Delta$}                     & \colhead{$\phi$} }
 
\startdata
 J0007+7303 &  2 &  \nodata  &     &   $0.216 \pm 0.005$ &   $0.50 - 0.86$  \\ 
   J0106+4855 &  2 &  $0.062 \pm 0.002 $ &    p &   $0.487 \pm 0.003$ &   $0.14 - 0.50,\, 0.75 - 0.01$  \\ 
   J0205+6449 &  2 &  $0.075 \pm 0.004 $ &    p &   $0.503 \pm 0.004$ &   $0.64 - 0.99$  \\ 
   J0248+6021 &  1 &  $0.336 \pm 0.017 $ &    p &   \nodata  &   $0.61 - 0.18$  \\ 
   J0357+3205 &  1 &  \nodata  &     &   \nodata  &   $0.39 - 0.90$  \\ 
   J0534+2200 &  2 &  $0.109 \pm 0.001 $ &    o &   $0.407 \pm 0.001$ &   $0.71 - 0.99$  \\ 
   J0622+3749 &  2 &  \nodata  &     &   $0.457 \pm 0.034$ &   $0.37 - 0.52,\, 0.69 - 0.89$  \\ 
   J0631+1036 &  1 &  $0.497 \pm 0.022 $ &    s &   \nodata  &   $0.63 - 0.17$  \\ 
   J0633+0632 &  2 &  \nodata  &     &   $0.476 \pm 0.003$ &   $0.24 - 0.52,\, 0.67 - 1.00$  \\ 
   J0633+1746 &  2 &  \nodata  &     &   $0.508 \pm 0.001$ &   $0.85 - 0.95$  \\ 
   J0659+1414 &  1 &  $0.224 \pm 0.010 $ &    p &   \nodata  &   $0.40 - 0.04$  \\ 
   J0729$-$1448 &  1 &  $0.577 \pm 0.010 $ &    p &   \nodata  &   $0.70 - 0.47$  \\ 
   J0734$-$1559 &  1 &  \nodata  &     &   \nodata  &   $0.28 - 0.84$  \\ 
   J0742$-$2822 &  1 &  $0.627 \pm 0.005 $ &    p &   \nodata  &   $0.74 - 0.01,\, 0.01 - 0.48$  \\ 
   J0835$-$4510 &  3 &  $0.129 \pm 0.001 $ &    p &   $0.433 \pm 0.001$ &   $0.81 - 0.03$  \\ 
   J0908$-$4913 &  2 &  $0.102 \pm 0.005 $ &    p &   $0.501 \pm 0.006$ &   $0.66 - 0.04,\, 0.17 - 0.54$  \\ 
   J0940$-$5428\dag &  1 &  $0.451 \pm 0.035 $ &    s &   \nodata  &   $0.59 - 0.19$  \\ 
   J1016$-$5857 &  2 &  $0.143 \pm 0.003 $ &    s &   $0.423 \pm 0.004$ &   $0.64 - 0.04$  \\ 
   J1019$-$5749 &  1 &  $0.482 \pm 0.010 $ &    p &   \nodata  &   $0.74 - 0.36$  \\ 
   J1023$-$5746 &  2 &  \nodata  &     &   $0.474 \pm 0.002$ &   $0.76 - 0.02$  \\ 
   J1028$-$5819 &  2 &  $0.195 \pm 0.001 $ &    p &   $0.475 \pm 0.001$ &   $0.75 - 0.08$  \\ 
   J1044$-$5737 &  2 &  \nodata  &     &   $0.373 \pm 0.004$ &   $0.56 - 0.98$  \\ 
   J1048$-$5832 &  2 &  $0.125 \pm 0.001 $ &    s &   $0.426 \pm 0.001$ &   $0.65 - 0.02$  \\ 
   J1057$-$5226 &  3 &  $0.304 \pm 0.003 $ &    sh &   $0.307 \pm 0.004$ &   $0.72 - 0.14$  \\ 
   J1105$-$6107 &  2 &  $0.110 \pm 0.001 $ &    s &   $0.504 \pm 0.006$ &   $0.74 - 0.04,\, 0.22 - 0.47$  \\ 
   J1112$-$6103 &  2 &  $0.192 \pm 0.007 $ &    p &   $0.457 \pm 0.013$ &   $0.79 - 0.04,\, 0.31 - 0.56$  \\ 
   J1119$-$6127 &  2 &  $0.285 \pm 0.015 $ &    p &   $0.204 \pm 0.020$ &   $0.59 - 0.18$  \\ 
   J1124$-$5916 &  2 &  $0.141 \pm 0.003 $ &    p &   $0.499 \pm 0.004$ &   $0.69 - 0.05$  \\ 
   J1135$-$6055 &  1 &  \nodata  &     &   \nodata  &   $0.49 - 0.92$  \\ 
   J1357$-$6429 &  1 &  $0.359 \pm 0.028 $ &    p &   \nodata  &   $0.64 - 0.13$  \\ 
   J1410$-$6132 &  2 &  $0.959 \pm 0.023 $ &    p &   $0.458 \pm 0.037$ &   $0.04 - 0.24,\, 0.55 - 0.85$  \\ 
   J1413$-$6205 &  2 &  \nodata  &     &   $0.372 \pm 0.003$ &   $0.57 - 0.01$  \\ 
   J1418$-$6058$\star$ &  2 &  \nodata  &     &   $0.467 \pm 0.003$ &   $0.65 - 0.93$  \\ 
   J1420$-$6048 &  2 &  $0.196 \pm 0.011 $ &    s &   $0.312 \pm 0.015$ &   $0.63 - 0.12$  \\ 
   J1429$-$5911 &  2 &  \nodata  &     &   $0.479 \pm 0.004$ &   $0.28 - 0.40$  \\ 
   J1459$-$6053$\star$ &  1 &  \nodata  &     &   \nodata  &   $0.59 - 0.91$  \\ 
   J1509$-$5850 &  2 &  $0.271 \pm 0.011 $ &    p &   $0.264 \pm 0.013$ &   $0.64 - 0.14$  \\ 
   J1513$-$5908 &  1 &  $0.325 \pm 0.055 $ &    p &   \nodata  &   $0.53 - 0.15$  \\ 
   J1531$-$5610\dag &  1 &  $0.413 \pm 0.035 $ &    s &   \nodata  &   $0.57 - 0.23$  \\ 
   J1620$-$4927 &  2 &  \nodata  &     &   $0.231 \pm 0.030$ &   $0.51 - 0.95$  \\ 
   J1648$-$4611 &  2 &  $0.261 \pm 0.010 $ &    p &   $0.298 \pm 0.082$ &   $0.64 - 0.17$  \\ 
   J1702$-$4128 &  1 &  $0.397 \pm 0.038 $ &    p &   \nodata  &   $0.57 - 0.17$  \\ 
   J1709$-$4429 &  2 &  $0.239 \pm 0.001 $ &    p &   $0.244 \pm 0.002$ &   $0.72 - 0.07$  \\ 
   J1718$-$3825 &  1 &  $0.397 \pm 0.009 $ &    s &   \nodata  &   $0.64 - 0.09$  \\ 
   J1730$-$3350 &  2 &  $0.128 \pm 0.007 $ &    p &   $0.419 \pm 0.007$ &   $0.63 - 0.04,\, 0.20 - 0.38$  \\ 
   J1732$-$3131 &  2 &  \nodata  &     &   $0.419 \pm 0.002$ &   $0.59 - 0.95$  \\ 
   J1741$-$2054 &  2 &  $0.074 \pm 0.006 $ &    s &   $0.244 \pm 0.011$ &   $0.44 - 0.96$  \\ 
   J1746$-$3239 &  2 &  \nodata  &     &   $0.179 \pm 0.019$ &   $0.44 - 0.02$  \\ 
   J1747$-$2958 &  2 &  $0.181 \pm 0.003 $ &    p &   $0.392 \pm 0.005$ &   $0.64 - 0.08$  \\ 
   J1801$-$2451 &  2 &  $0.060 \pm 0.005 $ &    p &   $0.496 \pm 0.020$ &   $0.13 - 0.48,\, 0.61 - 0.97$  \\ 
   J1803$-$2149 &  2 &  \nodata  &     &   $0.394 \pm 0.009$ &   $0.59 - 0.03$  \\ 
   J1809$-$2332 &  2 &  \nodata  &     &   $0.358 \pm 0.002$ &   $0.55 - 0.93$  \\ 
   J1813$-$1246 &  2 &  \nodata  &     &   $0.489 \pm 0.010$ &   $0.74 - 0.98$  \\ 
   J1826$-$1256 &  2 &  \nodata  &     &   $0.480 \pm 0.001$ &   $0.67 - 0.98$  \\ 
   J1833$-$1034 &  2 &  $0.153 \pm 0.002 $ &    p &   $0.447 \pm 0.004$ &   $0.69 - 0.10$  \\ 
   J1835$-$1106 &  2 &  $0.139 \pm 0.006 $ &    p &   $0.421 \pm 0.011$ &   $0.64 - 0.04$  \\ 
   J1836+5925 &  2 &  \nodata  &     &   $0.537 \pm 0.006$ &   $0.76 - 0.92$  \\ 
   J1838$-$0537$\star$ &  2 &  \nodata  &     &   $0.298 \pm 0.014$ &   $0.51 - 0.01$  \\ 
   J1846+0919 &  2 &  \nodata  &     &   $0.244 \pm 0.022$ &   $0.43 - 0.95$  \\ 
   J1907+0602 &  2 &  $0.209 \pm 0.003 $ &    p &   $0.389 \pm 0.004$ &   $0.69 - 0.12$  \\ 
   J1952+3252 &  3 &  $0.161 \pm 0.002 $ &    p &   $0.478 \pm 0.003$ &   $0.71 - 0.06$  \\ 
   J1954+2836 &  2 &  \nodata  &     &   $0.456 \pm 0.004$ &   $0.67 - 0.02$  \\ 
   J1957+5033 &  1 &  \nodata  &     &   \nodata  &   $0.46 - 0.93$  \\ 
   J1958+2846 &  2 &  \nodata  &     &   $0.454 \pm 0.004$ &   $0.65 - 0.04$  \\ 
   J2021+3651 &  2 &  $0.132 \pm 0.001 $ &    p &   $0.478 \pm 0.001$ &   $0.73 - 0.99$  \\ 
   J2021+4026 &  2 &  \nodata  &     &   $0.687 \pm 0.009$ &   $0.24 - 0.40$  \\ 
   J2028+3332 &  2 &  \nodata  &     &   $0.451 \pm 0.003$ &   $0.57 - 0.97$  \\ 
   J2030+3641 &  2 &  $0.269 \pm 0.010 $ &    p &   $0.309 \pm 0.014$ &   $0.67 - 0.18$  \\ 
   J2030+4415 &  2 &  \nodata  &     &   $0.505 \pm 0.007$ &   $0.65 - 0.01$  \\ 
   J2032+4127 &  2 &  $0.099 \pm 0.001 $ &    p &   $0.516 \pm 0.001$ &   $0.22 - 0.55,\, 0.68 - 1.00$  \\ 
   J2043+2740 &  2 &  $0.132 \pm 0.007 $ &    p &   $0.432 \pm 0.010$ &   $0.63 - 0.05$  \\ 
   J2055+2539$\star$ &  2 &  \nodata  &     &   $0.113 \pm 0.017$ &   $0.39 - 0.89$  \\ 
   J2111+4606$\star$ &  2 &  \nodata  &     &   $0.337 \pm 0.011$ &   $0.53 - 0.01$  \\ 
   J2139+4716 &  1 &  \nodata  &     &   \nodata  &   $0.25 - 0.90$  \\ 
   J2229+6114 &  2 &  $0.187 \pm 0.007 $ &    p &   $0.299 \pm 0.008$ &   $0.68 - 0.10$  \\ 
   J2238+5903 &  2 &  \nodata  &     &   $0.502 \pm 0.002$ &   $0.68 - 0.02$  \\ 
   J2240+5832 &  2 &  $0.118 \pm 0.014 $ &    p &   $0.476 \pm 0.014$ &   $0.70 - 0.05,\, 0.16 - 0.46$  \\

\enddata




\tablenotetext{a}{A dagger (\dag) means the pulse profile fit is unreliable. 
A star ($\star$) means that systematic offset uncertainties from the radio timing residuals (TRES variable in TEMPO) are between 10 and 19 milliperiods.}

\tablecomments{
Column 2 gives the gamma-ray peak multiplicity. 
Columns 3 and 5 give the gamma-radio phase lag $\delta$ and separation $\Delta$ between the gamma-ray peaks.  
Column 4 gives the method used to define the radio fiducial phase: 
`p' -- peak radio intensity;
`s' -- point of symmetry in the radio profile ;
`h' -- opposite hemisphere ($0.5$ phase shift from `p' or `s' point).
PSR J0534+2200 (the Crab pulsar) is exceptional, and we align its profile with the low frequency component observed by \citet{crab_lfc} and denote this method `o' for `other'.
Column 6 gives the off-peak interval definition from Section \ref{unpulsed}.  
}

\end{deluxetable}
\normalsize

\clearpage
\tabletypesize{\scriptsize}
\begin{deluxetable}{lccccc}
\tablewidth{0pt}
\tablecaption{Pulse shape parameters of LAT-detected millisecond pulsars
\label{tbl-pulseMSP}}

\tablehead{
\colhead{PSR\tablenotemark{a}} & \colhead{Peaks} & \colhead{Radio lag} & \colhead{Shift} & \colhead{gamma-peak separation} & \colhead{Off-peak definition} \\
\colhead{}        & \colhead{}      & \colhead{$\delta$}  & \colhead{Method}& \colhead{$\Delta$}                     & \colhead{$\phi$} }
 
\startdata
 J0023+0923 &  2 &  $0.375 \pm 0.064 $ &    p &   $0.608 \pm 0.192$ &   $0.60 - 0.80$  \\ 
   J0030+0451 &  2 &  $0.160 \pm 0.001 $ &    p &   $0.450 \pm 0.001$ &   $0.72 - 0.08$  \\ 
   J0034$-$0534 &  2 &  $0.866 \pm 0.005 $ &    s &   $0.285 \pm 0.023$ &   $0.28 - 0.75$  \\ 
   J0101$-$6422 &  2 &  $0.145 \pm 0.005 $ &    sh &   $0.417 \pm 0.018$ &   $0.85 - 0.09$  \\ 
   J0102+4839 &  2 &  $0.259 \pm 0.004 $ &    p &   $0.454 \pm 0.015$ &   $0.83 - 0.17,\, 0.32 - 0.59$  \\ 
   J0218+4232 &  2 &  $0.351 \pm 0.077 $ &    s &   $0.388 \pm 0.080$ &   $0.84 - 0.02,\, 0.02 - 0.23$  \\ 
   J0340+4130 &  2 &  $0.751 \pm 0.006 $ &    p &   $0.232 \pm 0.016$ &   $0.11 - 0.62$  \\ 
   J0437$-$4715 &  1 &  $0.442 \pm 0.010 $ &    p &   \nodata  &   $0.60 - 0.13$  \\ 
   J0610$-$2100 &  1 &  $0.236 \pm 0.006 $ &    h &   \nodata  &   $0.98 - 0.20$  \\ 
   J0613$-$0200 &  2 &  $0.261 \pm 0.020 $ &    p &   $0.175 \pm 0.021$ &   $0.60 - 0.10$  \\ 
   J0614$-$3329 &  2 &  $0.126 \pm 0.002 $ &    s &   $0.554 \pm 0.002$ &   $0.40 - 0.52$  \\ 
   J0751+1807 &  3 &  $0.398 \pm 0.008 $ &    s &   $0.270 \pm 0.017$ &   $0.77 - 0.03,\, 0.03 - 0.32$  \\ 
   J1024$-$0719 &  1 &  $0.492 \pm 0.046 $ &    p &   \nodata  &   $0.88 - 0.37$  \\ 
   J1124$-$3653 &  2 &  $0.298 \pm 0.011 $ &    s &   $0.214 \pm 0.043$ &   $0.88 - 0.19$  \\ 
   J1125$-$5825\dag &  1 &  $0.645 \pm 0.002 $ &    p &   \nodata  &   $0.74 - 0.48$  \\ 
   J1231$-$1411 &  3 &  $0.241 \pm 0.002 $ &    p &   $0.408 \pm 0.002$ &   $0.72 - 0.12$  \\ 
   J1446$-$4701 &  1 &  $0.319 \pm 0.021 $ &    p &   \nodata  &   $0.65 - 0.21$  \\ 
   J1514$-$4946 &  2 &  $0.214 \pm 0.009 $ &    s &   $0.392 \pm 0.018$ &   $0.69 - 0.15$  \\ 
   J1600$-$3053 &  1 &  $0.147 \pm 0.011 $ &    p &   \nodata  &   $0.51 - 0.09$  \\ 
   J1614$-$2230 &  2 &  $0.201 \pm 0.005 $ &    p &   $0.510 \pm 0.005$ &   $0.78 - 0.14$  \\ 
   J1658$-$5324 &  1 &  $0.359 \pm 0.014 $ &    s &   \nodata  &   $0.64 - 0.26$  \\ 
   J1713+0747 &  1 &  $0.319 \pm 0.049 $ &    p &   \nodata  &   $0.67 - 0.00,\, 0.00 - 0.19$  \\ 
   J1741+1351 &  1 &  $0.730 \pm 0.006 $ &    p &   \nodata  &   $0.91 - 0.59$  \\ 
   J1744$-$1134 &  2 &  $0.189 \pm 0.007 $ &    sh &   $0.808 \pm 0.022$ &   $0.05 - 0.10$  \\ 
   J1747$-$4036 &  2 &  $0.031 \pm 0.021 $ &    p &   $0.681 \pm 0.033$ &   $0.24 - 0.44,\, 0.73 - 0.88$  \\ 
   J1810+1744 &  2 &  $0.894 \pm 0.018 $ &    p &   $0.276 \pm 0.026$ &   $0.23 - 0.68$  \\ 
   J1823$-$3021A &  2 &  $0.993 \pm 0.009 $ &    p &   $0.627 \pm 0.010$ &   $0.08 - 0.55$  \\ 
   J1858$-$2216\dag &  1 &  $0.727 \pm 0.011 $ &    p &   \nodata  &   $0.16 - 0.66$  \\ 
   J1902$-$5105 &  2 &  $0.994 \pm 0.006 $ &    p &   $0.663 \pm 0.008$ &   $0.70 - 0.93$  \\ 
   J1939+2134 &  2 &  $0.997 \pm 0.004 $ &    p &   $0.542 \pm 0.009$ &   $0.08 - 0.47,\, 0.59 - 0.95$  \\ 
   J1959+2048 &  2 &  $0.997 \pm 0.005 $ &    p &   $0.554 \pm 0.013$ &   $0.19 - 0.38$  \\ 
   J2017+0603 &  2 &  $0.212 \pm 0.010 $ &    p &   $0.292 \pm 0.015$ &   $0.60 - 0.12$  \\ 
   J2043+1711 &  2 &  $0.231 \pm 0.008 $ &    p &   $0.446 \pm 0.010$ &   $0.73 - 0.10$  \\ 
   J2047+1053 &  1 &  $0.567 \pm 0.010 $ &    p &   \nodata  &   $0.16 - 0.49$  \\ 
   J2051$-$0827 &  1 &  $0.545 \pm 0.042 $ &    p &   \nodata  &   $0.80 - 0.28$  \\ 
   J2124$-$3358 &  1 &  $0.871 \pm 0.009 $ &    p &   \nodata  &   $0.01 - 0.51$  \\ 
   J2214+3000 &  2 &  $0.271 \pm 0.010 $ &    p &   $0.546 \pm 0.018$ &   $0.09 - 0.24,\, 0.63 - 0.73$  \\ 
   J2215+5135 &  2 &  $0.257 \pm 0.004 $ &    p &   $0.440 \pm 0.008$ &   \nodata  \\ 
 J2241$-$5236 &  3 &  $0.139 \pm 0.004 $ &    p &   $0.638 \pm 0.031$ &   $0.60 - 0.67$  \\ 
   J2302+4442 &  2 &  $0.244 \pm 0.003 $ &    s &   $0.346 \pm 0.006$ &   $0.70 - 0.18$  \\ 
  
\enddata



\tablenotetext{a}{A dagger (\dag) means the pulse profile fit is unreliable.}

\tablecomments{
Column 2 gives the gamma-ray peak multiplicity.  
Columns 3 and 5 give the phase lag $\delta$ and peak separation $\Delta$.  
Column 4 gives the method used to define the radio fiducial phase: 
`p' -- peak radio intensity;
`s' -- point of symmetry in the radio profile ;
`h' -- opposite hemisphere ($0.5$ phase shift from `p' or `s' point).
The off-peak phase interval (Section \ref{unpulsed}) is in column 6.  
}

\end{deluxetable}
\normalsize


\section{Spectral Analyses}
\label{spectralSection}
Most models of pulsar gamma-ray emission predict that the spectrum in the LAT energy range should be dominated by curvature radiation, in the radiation-reaction limited regime \citep[e.g.,][]{Muslimov04}.  
This mechanism predicts that pulsar spectra should be exponentially cut off near energies of a few GeV.  
The detection of pulsed emission above 100 GeV from the Crab (see Sections \ref{TeVPSRs} and \ref{discuss}) suggests that, for some pulsars, 
either an additional component becomes dominant above the cutoff or a different mechanism, e.g., inverse Compton scattering, may be responsible for the LAT emission.  
For the purposes of this catalog, we assume that all gamma-ray pulsars have a cutoff spectrum but test for any deviations from this model and report the pulsars that exhibit a different spectral shape.  

GeV emission from pulsars is largely modulated at the rotational period, with the emission concentrated in one or more narrow peaks.  Depending on the viewing geometry and emission model, a pulsar can
have a 100\% duty cycle and significant magnetospheric emission can also exist away from the peaks.  Young pulsars power PWNe, some of which are detected with
the LAT, spatially overlapping their associated pulsars. PWN-like emission could be confused with the magnetospheric signal but
would not be modulated at the rotational period. Significant contamination from a PWN-like background source needs to be accounted for to properly study the pulsar emission. 

With this goal, we first analyzed the off-peak phase intervals as described in Section~\ref{unpulsed} and characterized significantly detected off-peak emission, to
identify constant magnetospheric emission or PWN-like background emission. In this section, we use the data with all rotation phases to 
characterize the magnetospheric emission, removing background PWN-like contamination, and report the phase-averaged spectra.  

\subsection{Spectral Method}
\label{spectralMethodSection}

For spectral analysis we used the data set described in Section \ref{obsvSection} for phase-averaged fits.  We evaluated the point-source detection
significance and the spectral parameters for each LAT pulsar in this catalog with an analysis similar to that performed for 2FGL using the standard LAT Science Tools package\footnote{\url{http://fermi.gsfc.nasa.gov/ssc/data/analysis/scitools/overview.html}}. The spectrum of each LAT pulsar was modeled as a power law with an exponential cutoff,
\begin{equation}\label{expcutoff}
\frac{{\rm d} N}{{\rm d} E} = K \Big(\frac{E}{E_{0}}\Big)^{-\Gamma} \exp \left(- \frac{E}{E_{\rm cut}} \right)^{b}.
\end{equation}
\noindent{}The four parameters are the normalization factor ($K$), the photon index at low energy ($\Gamma$), the cutoff energy ($E_{\rm cut}$), and a parameter representing the sharpness of the cutoff ($b$) which is fixed to 1 in the default fit.
The energy $E_{0}$ at which $K$ is defined is arbitrary; thus, we used the 2FGL pivot energy when it exists and 1 GeV otherwise.  In the likelihood analysis we constrained $\Gamma$ to be between 0 and 5 and $E_{\rm cut}$ to be between 0.1 and 100 GeV.

We constructed models including all 2FGL sources within $20^{\circ}$ of each pulsar but only the spectral parameters of point sources within 8$^{\circ}$ were left free.  
For sources known to be significantly extended we used the same spatial templates as 2FGL and fixed the spectral parameters to the 2FGL values.
For pulsars with no 2FGL counterpart and an unassociated source within $0\fdg1$ of the pulsar, we moved the source to the timing position; 
otherwise, we added a new point source to the model at the timing position. 
The study of the off-peak phase interval in Section~\ref{unpulsed} yielded four pulsars with wind-like emission (`W'-type), 
and four pulsars where the origin of the off-peak emission is unidentified (`U'-type) but the emission appears spatially extended without evidence of a spectral cutoff,
likely due to poorly modeled diffuse emission. We added a new point source to the model for each of these eight pulsars.
We used the same models for diffuse gamma-ray emission as 2FGL 
to account for the Galactic, isotropic, and Earth limb emission:  
{\tt gal\_2yearp7v6\_v0.fits}, {\tt iso\_p7v6source.txt}, {\tt limb\_2year\_P76\_source\_v0\_smooth.txt}, and {\tt limb\_smooth.fits}.  
These are available from the {\it Fermi} Science Support Center\footnote{\url{http://fermi.gsfc.nasa.gov/ssc/data/access/lat/BackgroundModels.html}}.

For each pulsar, we selected a $20^{\circ}\times20^{\circ}$ square region, centered on the timing position, 
for a binned maximum likelihood \gtlike analysis using the {\it pyLikelihood} python module included with the {\it Fermi} Science Tools.
The best-fit parameters are obtained by maximizing the log-likelihood surface that represents the input model using the MINUIT2 fitting
engine\footnote{\url{http://project-mathlibs.web.cern.ch/project-mathlibs/sw/5_15_04/Minuit2/html}}.
The statistical uncertainties on the parameters were estimated from the quadratic development of the log-likelihood around the best fit.
We first performed a fit with a weak convergence criterion and evaluated the point-source detection significance \citep[the ``Test Statistic'' $TS$,][]{Mattox96} for each point source that had free parameters, except the pulsar of interest.  
We removed all point sources with $TS<2$ and re-optimized the fit with a stricter convergence criterion.

We report the maximum likelihood values of $\Gamma$ and $E_{\rm cut}$ from the phase-averaged analysis and $TS$ values
in Tables \ref{tbl-psrspec} and \ref{tbl-mspspec} for young and millisecond pulsars, respectively. 
In addition, we report the integrated photon and energy fluxes in the 0.1 to 100 GeV energy band ($F_{100}$ and $G_{100}$, respectively), obtained from the fits as
\begin{eqnarray}
F_{100} = \int_{\rm 100 \, MeV}^{\rm 100 \, GeV} \frac{{\rm d} N}{{\rm d} E} {\rm d} E, \; \; \; \; \; \;
G_{100} = \int_{\rm 100 \, MeV}^{\rm 100 \, GeV} E \frac{{\rm d} N}{{\rm d} E} {\rm d} E.
\end{eqnarray}
\noindent{}Statistical uncertainties on $F_{100}$ and $G_{100}$ are obtained using derivatives with respect to the primary parameters and the covariance matrix obtained from the fitting process.

We tested the validity of modeling the pulsar spectrum as a power law with a simple exponential cutoff shape (PLEC1, Eq.~\ref{expcutoff} with $b\equiv$1) 
by repeating the analysis using a pure power-law shape (PL), and a power law with a more general exponential cutoff shape leaving the $b$ parameter free (PLEC).  
For a number of pulsars, a PLEC1 spectral model is not significantly better than a PL. 
We identified these by computing $TS_{\rm cut} \equiv 2\Delta \log(likelihood)$ (comparable to a $\chi^2$ distribution with one degree of freedom) between the models with and without the cutoff.
We say that the PLEC1 model is not significantly preferred over the PL model for pulsars with $TS_{\rm cut} < 9$, listed in Tables \ref{tbl-psrspec} and \ref{tbl-mspspec}.
Similarly, we calculated $TS_{\rm b\,free} \equiv 2\Delta \log(likelihood)$ between the PLEC1 and PLEC models, also listed in the Tables.

In all cases where PLEC is significantly preferred over the PLEC1 model ($TS_{\rm b\,free} \geq 9$), the maximum likelihood value of $b$ is significantly less than 1, indicating a sub-exponential cutoff.
As noted by \citet{LATVela2} and \citet{CJ11}, a sub-exponential cutoff is a functional form which approximates the superposition of several PLEC1 models with  varying values of $\Gamma$ and $E_{\rm cut}$ as different regions of the pulsar magnetosphere cut across our line of sight.
This is further supported by analysis of PSR J1057$-$5226 for which the $b$ parameter in a PLEC fit is consistent with 1 and the spectral parameters show very little variation with phase \citep{LAT3EGRETpulsars}.
Thus, no physical quantities can be derived from the PLEC best-fit parameters, 
whereas the PLEC1 best-fit parameters can be used if taken as flux-weighted, average measures of $E_{\rm cut}$ and $\Gamma$.
Further, in our PLEC fits, the $E_{\rm cut}$ value was often at the minimum boundary (0.1 GeV) with unrealistically small uncertainties.
Therefore, we do not report the PLEC fit values in the Tables and instead indicate pulsars with high $TS_{\rm b\,free}$ as interesting candidates for phase-resolved spectral analysis, 
a task beyond the scope of this catalog. For pulsars with $TS_{\rm b\,free}\geq 9$, 
the PLEC fit results are included in the spectral plots (see the examples amongst Figures \ref{fig-J0101spec} to \ref{fig-J1954spec} 
in Appendix \ref{App-Samples}) and in the auxiliary files (Appendix \ref{online}).
We encourage comparing predicted spectra from different emission models to the energy sub-band fluxes, which are also included in the auxiliary files.
In some pulsars, particularly the Crab, a preference for $b<$1 may indicate either the presence of a secondary spectral component that dominates above $\gtrsim$10 GeV or 
that the curvature radiation assumption is incorrect \citep[as argued by][for example]{Lyutikov2012}.
Such a determination is difficult using the LAT data alone (see Section \ref{TeVPSRs}). 

\clearpage
\begin{landscape}
\begin{deluxetable}{lrrrrrrrcc}
\tabletypesize{\scriptsize}
\tablecaption{Spectral fitting results for young LAT-detected pulsars
\label{tbl-psrspec}}
\tablewidth{0pt}

\tablehead{
\colhead{PSR\tablenotemark{a}} & \colhead{Photon Flux}                    & \colhead{Energy Flux}                     & \colhead{$\Gamma$} & \colhead{$E_{\rm cut}$} & \colhead{$TS$} & \colhead{$TS_{\rm cut}$} & \colhead{$TS_{\rm b\,free}$} & \colhead{Luminosity}			& \colhead{Efficiency\tablenotemark{b}}\\
\colhead{}        & \colhead{($\rm ph \, cm^{-2} \,s^{-1}$)} & \colhead{($\rm erg \, cm^{-2}\, s^{-1}$)} & \colhead{}         & \colhead{($\rm GeV$)}   & \colhead{}   & \colhead{}		  & \colhead{}  	     & \colhead{($\rm 10^{33}\, erg \,s^{-1}$)} & \colhead{(\%)} \\
\colhead{}        & \colhead{($\times 10^{-8}$)}              & \colhead{($\times  10^{-11}$)}             & \colhead{}         & \colhead{}              & \colhead{}   & \colhead{}		  & \colhead{}  	     & \colhead{}				& \colhead{}
}

\startdata
  J0007+7303 &  $32.9 \pm 0.4 $ & $40.1 \pm 0.4 $ &   $1.4 \pm 0.1 $ & $4.7 \pm 0.2 $ & 43388 & 1884 & 4 &  $94 \pm 1 \pm 40$ &   $21.0 \pm 0.2\pm 8$ \\ 
  J0106+4855 &  $1.7 \pm 0.4 $ & $1.9 \pm 0.2 $ &   $1.2 \pm 0.2 $ & $2.7 \pm 0.6 $ & 544 & 58 & 2 &  $21 \pm 2^{{+20}}_{{-8}} $ &   $71 \pm 7^{{+60}}_{{-30}} $  \\ 
  J0205+6449 &  $10.5 \pm 0.7 $ & $5.4 \pm 0.2 $ &   $1.8 \pm 0.1 $ & $1.6 \pm 0.3 $ & 1019 & 86 & 7 &  $24 \pm 1 \pm 1.0$ &   $0.09 \pm 0.01\pm 0.01$ \\ 
  J0248+6021 &  $9.9 \pm 1.3 $ & $5.2 \pm 0.4 $ &   $1.8 \pm 0.1 $ & $1.6 \pm 0.3 $ & 578 & 61 & 0 &  $25 \pm 2 \pm 5$ &   $12 \pm 1\pm 2$ \\ 
  J0357+3205 &  $9.0 \pm 0.4 $ & $6.4 \pm 0.2 $ &   $1.0 \pm 0.1 $ & $0.8 \pm 0.1 $ & 3468 & 461 & 2 &  \nodata &   \nodata \\ 
  J0534+2200 &  $208 \pm 1 $ & $129.3 \pm 0.8 $ &   $1.9 \pm 0.1 $ & $4.2 \pm 0.2 $ & 102653 & 1461 & 13 &  $619 \pm 4 \pm 300 $ &   $0.14 \pm 0.01\pm 0.1 $  \\ 
  J0622+3749 &  $2.0 \pm 0.3 $ & $1.4 \pm 0.1 $ &   $0.6 \pm 0.4 $ & $0.6 \pm 0.1 $ & 302 & 91 & 0 &  \nodata &   \nodata \\ 
  J0631+1036 &  $6.4 \pm 0.6 $ & $4.7 \pm 0.3 $ &   $1.8 \pm 0.1 $ & $6 \pm 1 $ & 621 & 39 & 1 &  $5.6 \pm 0.3^{{+3}}_{{-2}} $ &   $3.2 \pm 0.2^{{+2}}_{{-1}} $  \\ 
  J0633+0632 &  $9.7 \pm 1.1 $ & $9.4 \pm 0.5 $ &   $1.4 \pm 0.1 $ & $2.7 \pm 0.3 $ & 2448 & 203 & 9 &  \nodata &   \nodata \\ 
  J0633+1746 &  $416 \pm 1 $ & $423.3 \pm 1.2 $ &   $1.2 \pm 0.1 $ & $2.2 \pm 0.1 $ & 906994 & 33861 & 277 &  $31.7 \pm 0.1^{{+90}}_{{-20}} $ &   $97.4 \pm 0.3^{{+300}}_{{-50}} $  \\ 
  J0659+1414 &  $7.1 \pm 0.6 $ & $2.5 \pm 0.2 $ &   $1.7 \pm 0.5 $ & $0.4 \pm 0.2 $ & 419 & 33 & 0 &  $0.24 \pm 0.02 \pm 0.05$ &   $0.62 \pm 0.04\pm 0.1$ \\ 
  J0729$-$1448 \dag&  \nodata & \nodata &  \nodata &  \nodata & 54 & 26 & 0 &  \nodata &   \nodata \\ 
  J0734$-$1559 &  $10.8 \pm 0.7 $ & $5.6 \pm 0.2 $ &   $2.0 \pm 0.1 $ & $3.2 \pm 0.9 $ & 916 & 39 & 9 &  \nodata &   \nodata \\ 
  J0742$-$2822 &  $3.2 \pm 0.6 $ & $1.7 \pm 0.2 $ &   $1.7 \pm 0.3 $ & $1.6 \pm 0.8 $ & 112 & 11 & 2 &  $9 \pm 1 \pm 4$ &   $6.2 \pm 0.7\pm 3$ \\ 
  J0835$-$4510 &  $1088 \pm 2 $ & $906 \pm 2 $ &   $1.5 \pm 0.1 $ & $3.0 \pm 0.1 $ & 1659005 & 43084 & 916 &  $89.3 \pm 0.2 \pm 10$ &   $1.3 \pm 0.1\pm 0.1$ \\ 
  J0908$-$4913 &  $7.9 \pm 1.3 $ & $4.4 \pm 0.4 $ &   $1.0 \pm 0.4 $ & $0.5 \pm 0.2 $ & 315 & 82 & 0 &  $35 \pm 3^{{+30}}_{{-20}} $ &   $7.1 \pm 0.7^{{+6}}_{{-4}} $  \\ 
  J0940$-$5428 \dag&  \nodata & \nodata &  \nodata &  \nodata & 14 & 13 & 8 &  \nodata &   \nodata \\ 
  J1016$-$5857 &  $6.9 \pm 2.4 $ & $5.4 \pm 0.9 $ &   $1.8 \pm 0.2 $ & $6 \pm 3 $ & 290 & 13 & 0 &  $55 \pm 9^{{+30}}_{{-50}} $ &   $2.1 \pm 0.4^{{+1}}_{{-2}} $  \\ 
  J1019$-$5749 \dag&  \nodata & \nodata &  \nodata &  \nodata & 21 & 0 & 0 &  \nodata &   \nodata \\ 
  J1023$-$5746 &  $30 \pm 3 $ & $19.5 \pm 1.2 $ &   $1.7 \pm 0.1 $ & $2.5 \pm 0.4 $ & 2926 & 162 & 20 &  \nodata &   \nodata \\ 
  J1028$-$5819 &  $31 \pm 2 $ & $24.3 \pm 0.8 $ &   $1.7 \pm 0.1 $ & $4.6 \pm 0.5 $ & 5096 & 235 & 28 &  $158 \pm 5 \pm 40$ &   $18.9 \pm 0.6\pm 5$ \\ 
  J1044$-$5737 &  $26 \pm 1 $ & $15.6 \pm 0.5 $ &   $1.8 \pm 0.1 $ & $2.8 \pm 0.3 $ & 3380 & 202 & 19 &  \nodata &   \nodata \\ 
  J1048$-$5832 &  $25 \pm 2 $ & $19.6 \pm 0.6 $ &   $1.6 \pm 0.1 $ & $3.0 \pm 0.3 $ & 5389 & 325 & 30 &  $176 \pm 5 \pm 40$ &   $8.8 \pm 0.3\pm 2$ \\ 
  J1057$-$5226 &  $32 \pm 1 $ & $29.5 \pm 0.3 $ &   $1.0 \pm 0.1 $ & $1.4 \pm 0.1 $ & 27848 & 2377 & 5 &  $4.3 \pm 0.1^{{+5}}_{{-3}} $ &   $14.4 \pm 0.2^\pm 10 $  \\ 
  J1105$-$6107 &  $7.8 \pm 1.6 $ & $4.9 \pm 0.6 $ &   $1.5 \pm 0.3 $ & $1.3 \pm 0.6 $ & 309 & 42 & 8 &  $150 \pm 20 \pm 50$ &   $5.9 \pm 0.7\pm 2$ \\ 
  J1112$-$6103 &  $1.9 \pm 0.9 $ & $2.0 \pm 0.5 $ &   $1.6 \pm 0.3 $ & $6 \pm 3 $ & 58 & 6 & 0 &  $360 \pm 90^{{+600}}_{{-200}} $ &   $8 \pm 2^{{+10}}_{{-4}} $  \\ 
  J1119$-$6127 &  $11 \pm 2 $ & $7.1 \pm 0.5 $ &   $1.8 \pm 0.1 $ & $3.2 \pm 0.8 $ & 661 & 37 & 13 &  $600 \pm 40 \pm 60$ &   $26 \pm 2\pm 2$ \\ 
  J1124$-$5916 &  $10 \pm 1 $ & $6.2 \pm 0.4 $ &   $1.8 \pm 0.1 $ & $2.1 \pm 0.4 $ & 1058 & 79 & 6 &  $170 \pm 10^{{+50}}_{{-70}} $ &   $1.4 \pm 0.1^{{+0.4}}_{{-0.6}} $  \\ 
  J1135$-$6055 &  $7.4 \pm 0.9 $ & $4.8 \pm 0.3 $ &   $1.7 \pm 0.1 $ & $2.4 \pm 0.5 $ & 498 & 61 & 3 &  \nodata &   \nodata \\ 
  J1357$-$6429 &  $7.8 \pm 1.1 $ & $3.4 \pm 0.3 $ &   $1.8 \pm 0.4 $ & $0.9 \pm 0.5 $ & 187 & 20 & 0 &  $25 \pm 2^{{+10}}_{{-8}} $ &   $0.82 \pm 0.08^{{+0.4}}_{{-0.3}} $  \\ 
  J1410$-$6132 \dag&  $3 \pm 3 $ & $3 \pm 1 $ &   \nodata &  \nodata & 40 & 9 & 0 &  $800 \pm 300^{{+900}}_{{-400}} $ &   $8 \pm 3^{{+9}}_{{-4}} $  \\ 
  J1413$-$6205 &  $16 \pm 2 $ & $15.7 \pm 0.6 $ &   $1.5 \pm 0.1 $ & $4.1 \pm 0.5 $ & 1795 & 180 & 1 &  \nodata &   \nodata \\ 
  J1418$-$6058 &  $38 \pm 3 $ & $30.2 \pm 1.4 $ &   $1.8 \pm 0.1 $ & $5.5 \pm 0.5 $ & 3487 & 172 & 1 &  $92 \pm 4^{{+100}}_{{-60}} $ &   $1.9 \pm 0.1^{{+2}}_{{-1}} $  \\ 
  J1420$-$6048 &  $26 \pm 3 $ & $17.0 \pm 1.4 $ &   $1.5 \pm 0.1 $ & $1.6 \pm 0.2 $ & 1220 & 51 & 2 &  $640 \pm 50 \pm 200 $ &   $6.2 \pm 0.5 \pm 2 $  \\ 
  J1429$-$5911 &  $12 \pm 1 $ & $8.0 \pm 0.4 $ &   $1.6 \pm 0.1 $ & $2.2 \pm 0.3 $ & 822 & 124 & 0 &  \nodata &   \nodata \\ 
  J1459$-$6053 &  $24 \pm 2 $ & $12.9 \pm 0.4 $ &   $2.0 \pm 0.1 $ & $2.9 \pm 0.5 $ & 2046 & 103 & 15 &  \nodata &   \nodata \\ 
  J1509$-$5850 &  $20 \pm 2 $ & $12.7 \pm 0.7 $ &   $1.9 \pm 0.1 $ & $4.6 \pm 0.9 $ & 1152 & 67 & 0 &  $105 \pm 6 \pm 40$ &   $20 \pm 1\pm 7$ \\ 
  J1513$-$5908 \dag       &  $10  \pm 2 $    & $3.2 \pm 0.7 $ &   \nodata &  \nodata & 98 & 5 & 1 &  $70  \pm 10 \pm 20$ &   $0.4 \pm 0.1\pm 0.1$ \\ 
  J1531$-$5610 \dag$\star$&  $0.1 \pm 0.05 $ & $0.2 \pm 0.1 $ &   \nodata &  \nodata & 2  & 2 & 1 &  $1.0 \pm 0.6 \pm 0.4 $ &   $0.11 \pm 0.07 \pm 0.04 $  \\ 
  J1620$-$4927 &  $15 \pm 2 $ & $15.7 \pm 0.8 $ &   $1.3 \pm 0.1 $ & $2.5 \pm 0.3 $ & 1407 & 199 & 4 &  \nodata &   \nodata \\ 
  J1648$-$4611 &  $5.3 \pm 1.6 $ & $5.4 \pm 0.8 $ &   $1.6 \pm 0.3 $ & $6 \pm 4 $ & 176 & 17 & 0 &  $160 \pm 20 \pm 40$ &   $80 \pm 10\pm 20$ \\ 
  J1702$-$4128 $\star$&  $4.4 \pm 5.2 $ & $2.8 \pm 2.7 $ &   $1.1 \pm 0.9 $ & $0.8 \pm 0.5 $ & 62 & 7 & 0 &  $80 \pm 70 \pm 20$ &   $20 \pm 20\pm 5$ \\ 
  J1709$-$4429 &  $160 \pm 2 $ & $135 \pm 1 $ &   $1.6 \pm 0.1 $ & $4.2 \pm 0.1 $ & 96893 & 3433 & 132 &  $853 \pm 6 \pm 200$ &   $25.1 \pm 0.2\pm 5$ \\ 
   J1718$-$3825 &  $14 \pm 1 $ & $8.9 \pm 0.5 $ &   $1.5 \pm 0.1 $ & $1.4 \pm 0.1 $ & 462 & 81 & 1 &  $138 \pm 8 \pm 30$ &   $11.0 \pm 0.6\pm 3$ \\ 
   J1730$-$3350 $\star$&  $3.4 \pm 0.1 $ & $2.4 \pm 0.4 $ &   $1.5 \pm 0.3 $ & $1.2 \pm 0.3 $ & 100 & 21 & 1 &  $36 \pm 6 \pm 9$ &   $2.9 \pm 0.5\pm 0.7$ \\ 
   J1732$-$3131 &  $17 \pm 1 $ & $19.4 \pm 0.6 $ &   $1.0 \pm 0.1 $ & $1.9 \pm 0.1 $ & 2821 & 550 & 0 &  $8.6 \pm 0.3 \pm 2$ &   $5.9 \pm 0.2\pm 1$ \\ 
   J1741$-$2054 &  $17 \pm 1 $ & $11.7 \pm 0.4 $ &   $1.1 \pm 0.1 $ & $0.9 \pm 0.1 $ & 3014 & 464 & 0 &  $2.1 \pm 0.1 \pm 0.2$ &   $22 \pm 1 \pm 3$ \\ 
   J1746$-$3239 &  $10 \pm 1 $ & $7.2 \pm 0.5 $ &   $1.4 \pm 0.1 $ & $1.5 \pm 0.2 $ & 654 & 109 & 0 &  \nodata &   \nodata \\ 
  J1747$-$2958 &  $33 \pm 3 $ & $21.1 \pm 1.0 $ &   $1.6 \pm 0.1 $ & $1.9 \pm 0.1 $ & 1689 & 211 & 2 &  $570 \pm 30 \pm 200$ &   $23 \pm 1\pm 7$ \\ 
   J1801$-$2451 $\star$&  $1.4 \pm 1.0 $ & $1.2 \pm 0.5 $ &   $1.5 \pm 0.5 $ & $3 \pm 2 $ & 58 & 10 & 6 &  $40 \pm 10^{{+9}}_{{-7}} $ &   $1.5 \pm 0.6^{{+0.4}}_{{-0.3}} $  \\ 
  J1803$-$2149 &  $11 \pm 2 $ & $9.2 \pm 0.8 $ &   $1.6 \pm 0.1 $ & $3.6 \pm 0.8 $ & 410 & 40 & 1 &  \nodata &   \nodata \\ 
  J1809$-$2332 &  $60 \pm 2 $ & $47.4 \pm 0.8 $ &   $1.6 \pm 0.1 $ & $3.4 \pm 0.2 $ & 15781 & 901 & 41 &  $164 \pm 3^{{+200}}_{{-100}} $ &   $38 \pm 1^{{+60}}_{{-30}} $  \\ 
  J1813$-$1246 &  $45 \pm 2 $ & $25.3 \pm 0.6 $ &   $1.9 \pm 0.1 $ & $2.6 \pm 0.3 $ & 4664 & 272 & 1 &  \nodata &   \nodata \\ 
  J1826$-$1256 &  $54 \pm 3 $ & $38.3 \pm 0.9 $ &   $1.6 \pm 0.1 $ & $2.2 \pm 0.2 $ & 5160 & 533 & 23 &  \nodata &   \nodata \\ 
  J1833$-$1034 &  $7.0 \pm 1.1 $ & $5.9 \pm 0.5 $ &   $0.9 \pm 0.2 $ & $0.9 \pm 0.2 $ & 258 & 78 & 0 &  $160 \pm 10 \pm 30$ &   $0.46 \pm 0.04\pm 0.08$ \\ 
  J1835$-$1106 \dag$\star$&  $0.4 \pm 0.1 $ & $0.6 \pm 0.2 $ &   \nodata &  \nodata & 30 & 18 & 0 &  $6 \pm 2 \pm 2$ &   $3 \pm 1\pm 0.9$ \\ 
  J1836+5925 &  $63 \pm 1 $ & $60.6 \pm 0.4 $ &   $1.2 \pm 0.1 $ & $2.0 \pm 0.1 $ & 142427 & 5747 & 28 &  $20.4 \pm 0.1^{{+30}}_{{-20}} $ &   $180 \pm 1^{{+200}}_{{-100}} $  \\ 
  J1838$-$0537 &  $22 \pm 2 $ & $18.8 \pm 0.9 $ &   $1.6 \pm 0.1 $ & $4.1 \pm 0.4 $ & 1325 & 114 & 1 &  \nodata &   \nodata \\ 
  J1846+0919 &  $1.4 \pm 0.3 $ & $2.4 \pm 0.2 $ &   $0.7 \pm 0.3 $ & $2.2 \pm 0.5 $ & 428 & 79 & 0 &  \nodata &   \nodata \\ 
  J1907+0602 &  $34 \pm 2 $ & $25.4 \pm 0.6 $ &   $1.6 \pm 0.1 $ & $2.9 \pm 0.3 $ & 3773 & 390 & 35 &  $314 \pm 8 \pm 60$ &   $11.1 \pm 0.3\pm 2$ \\ 
  J1952+3252 &  $17 \pm 1 $ & $13.8 \pm 0.3 $ &   $1.5 \pm 0.1 $ & $2.5 \pm 0.2 $ & 4469 & 365 & 11 &  $66 \pm 2^{{+40}}_{{-30}} $ &   $1.8 \pm 0.1^{{+1.0}}_{{-0.8}} $  \\ 
  J1954+2836 &  $13 \pm 1 $ & $10.3 \pm 0.4 $ &   $1.6 \pm 0.1 $ & $3.3 \pm 0.4 $ & 1592 & 168 & 11 &  \nodata &   \nodata \\ 
  J1957+5033 &  $4.0 \pm 0.4 $ & $2.6 \pm 0.1 $ &   $1.3 \pm 0.2 $ & $1.0 \pm 0.2 $ & 846 & 105 & 0 &  \nodata &   \nodata \\ 
  J1958+2846 &  $11 \pm 1 $ & $9.1 \pm 0.4 $ &   $1.4 \pm 0.1 $ & $2.0 \pm 0.3 $ & 1519 & 206 & 25 &  \nodata &   \nodata \\ 
  J2021+3651 &  $68 \pm 2 $ & $49.4 \pm 0.8 $ &   $1.7 \pm 0.1 $ & $3.0 \pm 0.2 $ & 17821 & 998 & 23 &  $5910 \pm 90^{{+3000}}_{{-4000}} $ &   $175 \pm 3^{{+80}}_{{-100}} $  \\ 
  J2021+4026 &  $138 \pm 2 $ & $95.5 \pm 0.9 $ &   $1.6 \pm 0.1 $ & $2.6 \pm 0.1 $ & 53955 & 2343 & 72 &  $257 \pm 2^{{+200}}_{{-100}} $ &   $225 \pm 2^{{+200}}_{{-100}} $  \\ 
  J2028+3332 &  $5.8 \pm 0.9 $ & $5.8 \pm 0.4 $ &   $1.2 \pm 0.2 $ & $1.9 \pm 0.3 $ & 1058 & 161 & 0 &  \nodata &   \nodata \\ 
  J2030+3641 &  $2.3 \pm 0.6 $ & $3.1 \pm 0.3 $ &   $0.7 \pm 0.4 $ & $1.5 \pm 0.4 $ & 313 & 91 & 0 &  $34 \pm 4^{{+30}}_{{-20}} $ &   $110 \pm 10^{{+80}}_{{-60}} $  \\ 
  J2030+4415 &  $9.3 \pm 1.0 $ & $5.8 \pm 0.4 $ &   $1.6 \pm 0.1 $ & $1.7 \pm 0.3 $ & 504 & 74 & 0 &  \nodata &   \nodata \\ 
  J2032+4127 &  $7.4 \pm 1.1 $ & $10.6 \pm 0.6 $ &   $1.1 \pm 0.1 $ & $3.2 \pm 0.5 $ & 1383 & 162 & 0 &  $169 \pm 10 \pm 50$ &   $62 \pm 4\pm 20$ \\ 
  J2043+2740 &  $1.5 \pm 0.4 $ & $1.0 \pm 0.1 $ &   $1.4 \pm 0.4 $ & $1.2 \pm 0.6 $ & 97 & 18 & 2 &  $3.8 \pm 0.6 \pm 1$ &   $7 \pm 1\pm 2$ \\ 
  J2055+2539 &  $6.2 \pm 0.4 $ & $5.4 \pm 0.2 $ &   $1.0 \pm 0.1 $ & $1.1 \pm 0.1 $ & 2751 & 361 & 1 &  \nodata &   \nodata \\ 
  J2111+4606 &  $5.3 \pm 0.7 $ & $4.4 \pm 0.3 $ &   $1.7 \pm 0.1 $ & $5 \pm 1 $ & 731 & 45 & 0 &  \nodata &   \nodata \\ 
  J2139+4716 &  $3.1 \pm 0.5 $ & $2.3 \pm 0.2 $ &   $1.3 \pm 0.2 $ & $1.3 \pm 0.3 $ & 369 & 71 & 2 &  \nodata &   \nodata \\ 
  J2229+6114 &  $38 \pm 1 $ & $25.3 \pm 0.4 $ &   $1.8 \pm 0.1 $ & $4.3 \pm 0.3 $ & 12101 & 424 & 54 &  $19.4 \pm 0.3 \pm 8$ &   $0.09 \pm 0.01\pm 0.04$ \\ 
  J2238+5903 &  $9.3 \pm 0.9 $ & $6.4 \pm 0.3 $ &   $1.6 \pm 0.1 $ & $2.1 \pm 0.3 $ & 1165 & 123 & 10 &  \nodata &   \nodata \\ 
  J2240+5832 &  $1.1 \pm 0.7 $ & $1.1 \pm 0.3 $ &   $1.5 \pm 0.5 $ & $3 \pm 2 $ & 54 & 11 & 0 &  $80 \pm 20 \pm 10$ &   $40 \pm 10\pm 6$ \\

\enddata

\tablenotetext{a}{A dagger (\dag) means the spectral fit is unreliable (see text). A star ($\star$) means that the spectrum was calculated using the on-peak data only.}
\tablenotetext{b}{Overestimated distances or the assumed beaming factor, $f_\Omega = 1$, can result in an efficiency $>$ 100\%.}

\tablecomments{Unbinned maximum likelihood spectral fit results  
for the young LAT gamma-ray pulsars, using the PLEC1 model (Eq. \ref{expcutoff} in Section \ref{spectralSection}). 
Columns 2 and 3 list the phase-averaged integral photon and energy fluxes in the 0.1 to 100 GeV energy band, $F_{\rm 100}$ and $G_{\rm 100}$. 
Columns 4 and 5 list the photon index $\Gamma$ and cutoff energy $E_{\rm cut}$.  
Columns 6, 7, and 8 list the source significance $TS$, significance $TS_{\rm cut}$ of the exponential 
cutoff compared to a simple power law, and significance $TS_{\rm b\,free}$ of the PLEC compared to a PLEC1 shape. A value $TS_{\rm cut} < 9$ indicates that a spectral cutoff is not significantly detected. 
Columns 9 and 10 give the total gamma-ray luminosity $L_\gamma$ in the 0.1 to 100 GeV energy band, and the gamma-ray conversion efficiency $\eta \equiv L_\gamma/\dot{E}$, assuming
$f_\Omega=1$ as described in Section \ref{lumin}. 
The first uncertainty in $L_\gamma$ and $\eta$ comes from the statistical uncertainties in the spectral fit, whereas the second is due to the distance uncertainty.
The strong dependence of these quantities on distance (see Table \ref{tbl-PSRdist}) and beaming factor means that these values
should be considered with care.}

\end{deluxetable}
\normalsize
\noindent

\clearpage
\end{landscape}


\clearpage
\begin{landscape}
\begin{deluxetable}{lrrrrrrrcc}
\tabletypesize{\scriptsize}
\tablecaption{Spectral fitting results for LAT-detected millisecond pulsars
\label{tbl-mspspec}}
\tablewidth{0pt}

\tablehead{
\colhead{PSR\tablenotemark{a}} & \colhead{Photon Flux}                    & \colhead{Energy Flux}                     & \colhead{$\Gamma$} & \colhead{$E_{\rm cut}$} & \colhead{$TS$} & \colhead{$TS_{\rm cut}$} & \colhead{$TS_{\rm b\,free}$} & \colhead{Luminosity}			   & \colhead{Efficiency\tablenotemark{b}}\\
\colhead{}        & \colhead{($\rm ph \, cm^{-2} \,s^{-1}$)} & \colhead{($\rm erg \, cm^{-2}\, s^{-1}$)} & \colhead{}         & \colhead{($\rm GeV$)}   & \colhead{}   & \colhead{}               & \colhead{}  	         & \colhead{($\rm 10^{32}\, erg \,s^{-1}$)}& \colhead{(\%)} \\
\colhead{}        & \colhead{($\times 10^{-8}$)}              & \colhead{($\times  10^{-11}$)}             & \colhead{}         & \colhead{}              & \colhead{}   & \colhead{}               & \colhead{}  	         & \colhead{}				   & \colhead{}
}

\startdata
 J0023+0923 &  $1.2 \pm 0.4 $ & $0.80 \pm 0.12 $ &   $1.4 \pm 0.4 $ & $1.4 \pm 0.6 $ & 131 & 16 & 7 &  $4.6 \pm 0.7^{{+3}}_{{-1}} $ &   $3.0 \pm 0.5^{{+2}}_{{-1}} $  \\ 
 J0030+0451 &  $6.6 \pm 0.3 $ & $6.14 \pm 0.18 $ &   $1.2 \pm 0.1 $ & $1.8 \pm 0.2 $ & 4788 & 316 & 10 &  $5.8 \pm 0.2^{{+5}}_{{-2}} $ &   $16 \pm 1^{{+13}}_{{-6}} $  \\ 
 J0034$-$0534 &  $2.2 \pm 0.3 $ & $1.62 \pm 0.12 $ &   $1.4 \pm 0.2 $ & $1.8 \pm 0.4 $ & 563 & 45 & 0 &  $5.7 \pm 0.4^{{+3}}_{{-2}} $ &   $3.3 \pm 0.2^{{+1.5}}_{{-1.1}} $  \\ 
 J0101$-$6422 &  $0.75 \pm 0.14 $ & $1.05 \pm 0.09 $ &   $0.7 \pm 0.3 $ & $1.5 \pm 0.4 $ & 491 & 59 & 1 &  $3.8 \pm 0.3 \pm 1$ &   $3.8 \pm 0.3^{{+1.3}}_{{-1.0}} $  \\ 
 J0102+4839 &  $1.3 \pm 0.3 $ & $1.32 \pm 0.16 $ &   $1.4 \pm 0.3 $ & $3.2 \pm 1.1 $ & 251 & 29 & 1 &  $90 \pm 10^{{+40}}_{{-30}} $ &   $49 \pm 6^{{+23}}_{{-16}} $  \\ 
 J0218+4232 &  $7.7 \pm 0.7 $ & $4.56 \pm 0.24 $ &   $2.0 \pm 0.1 $ & $4.6 \pm 1.2 $ & 1313 & 38 & 1 &  $380 \pm 20^{{+400}}_{{-200}} $ &   $16 \pm 1^{{+15}}_{{-7}} $  \\ 
 J0340+4130 &  $1.5 \pm 0.2 $ & $2.04 \pm 0.15 $ &   $1.1 \pm 0.2 $ & $2.6 \pm 0.6 $ & 553 & 73 & 4 &  $73 \pm 6 \pm 20$ &   $92.6 \pm 7.0 \pm 29$  \\ 
 J0437$-$4715 &  $2.7 \pm 0.3 $ & $1.67 \pm 0.11 $ &   $1.4 \pm 0.2 $ & $1.1 \pm 0.3 $ & 687 & 67 & 1 &  $0.49 \pm 0.03 \pm 0.01$ &   $1.7 \pm 0.1 \pm 0.1$  \\ 
 J0610$-$2100 &  $0.78 \pm 0.25 $ & $0.66 \pm 0.11 $ &   $1.2 \pm 0.4 $ & $1.6 \pm 0.8 $ & 98 & 14 & 1 &  $100 \pm 20^{{+500}}_{{-50}} $ &   $1180 \pm 190^{{+6400}}_{{-570}} $  \\ 
 J0613$-$0200 &  $2.7 \pm 0.4 $ & $2.99 \pm 0.19 $ &   $1.2 \pm 0.2 $ & $2.5 \pm 0.5 $ & 760 & 88 & 0 &  $29 \pm 2^{{+30}}_{{-10}} $ &   $24.1 \pm 1.5^{{+26}}_{{-9.5}} $  \\ 
 J0614$-$3329 &  $8.5 \pm 0.3 $ & $10.94 \pm 0.27 $ &   $1.3 \pm 0.1 $ & $3.9 \pm 0.3 $ & 9408 & 416 & 11 &  $470 \pm 10 \pm 200 $ &   $215 \pm 5^{{+110}}_{{-72}} $  \\ 
 J0751+1807 &  $1.1 \pm 0.2 $ & $1.33 \pm 0.12 $ &   $1.1 \pm 0.2 $ & $2.6 \pm 0.7 $ & 427 & 46 & 1 &  $2.5 \pm 0.2^{{+3}}_{{-1}} $ &   $3.5 \pm 0.3^{{+4.4}}_{{-1.5}} $  \\ 
 J1024$-$0719 \dag&  $0.2 \pm 0.2 $ & $0.3 \pm 0.1 $ &   \nodata &  \nodata & 46 & 11 & 0 &  $0.6 \pm 0.2 \pm 0.1$ &   $12 \pm 4 \pm 2.3$  \\ 
 J1124$-$3653 &  $0.94 \pm 0.23 $ & $1.21 \pm 0.13 $ &   $1.1 \pm 0.3 $ & $2.5 \pm 0.7 $ & 293 & 39 & 0 &  $43 \pm 5 \pm 20 $ &   $25.0 \pm 2.7^{{+14}}_{{-9.4}} $  \\ 
 J1125$-$5825 &  $1.1 \pm 0.5 $ & $0.89 \pm 0.20 $ &   $1.7 \pm 0.2 $ & $4.8 \pm 2.4 $ & 41 & 4 & 1 &  $70 \pm 20^ \pm 20 $ &   $9.1 \pm 2.0^{{+3.0}}_{{-2.4}} $  \\ 
 J1231$-$1411 &  $9.2 \pm 0.4 $ & $10.28 \pm 0.26 $ &   $1.2 \pm 0.1 $ & $2.7 \pm 0.2 $ & 7931 & 446 & 3 &  $24 \pm 0.6 \pm 5$ &   $45.9 \pm 1.2 \pm 9.9$  \\ 
 J1446$-$4701 &  $0.73 \pm 0.31 $ & $0.74 \pm 0.14 $ &   $1.4 \pm 0.4 $ & $3.0 \pm 1.7 $ & 71 & 9 & 0 &  $19 \pm 4 \pm 5$ &   $5.2 \pm 1.0 \pm 1.4$  \\ 
 J1514$-$4946 &  $4.1 \pm 0.6 $ & $4.50 \pm 0.28 $ &   $1.5 \pm 0.1 $ & $5.3 \pm 1.1 $ & 999 & 66 & 0 &  $48 \pm 3 \pm 10$ &   $29.7 \pm 1.8 \pm 7.1$  \\ 
 J1600$-$3053 $\star$&  $0.22 \pm 0.16 $ & $0.53 \pm 0.28 $ &   $0.40 \pm 0.47 $ & $2.0 \pm 0.7 $ & 147 & 26 & 0 &  $17 \pm 9^{{+7}}_{{-5}} $ &   $23 \pm 12^{{+9.6}}_{{-7.0}} $  \\ 
 J1614$-$2230 &  $2.0 \pm 0.4 $ & $2.44 \pm 0.20 $ &   $0.96 \pm 0.22 $ & $1.9 \pm 0.4 $ & 599 & 81 & 1 &  $12 \pm 1 \pm 2$ &   $32.6 \pm 2.6 \pm 4.8$  \\ 
 J1658$-$5324 &  $5.7 \pm 0.7 $ & $2.89 \pm 0.23 $ &   $1.8 \pm 0.2 $ & $1.4 \pm 0.4 $ & 327 & 34 & 0 &  $30 \pm 2 \pm 8$ &   $9.9 \pm 0.8 \pm 2.6$  \\ 
 J1713+0747 &  $1.3 \pm 0.4 $ & $1.02 \pm 0.14 $ &   $1.6 \pm 0.3 $ & $2.7 \pm 1.2 $ & 126 & 17 & 2 &  $13 \pm 2^{{+2}}_{{-1}} $ &   $39.0 \pm 5.6^{{+4.6}}_{{-3.6}} $  \\ 
 J1741+1351 \dag&  $0.12 \pm 0.04 $ & $0.24 \pm 0.08 $ &   \nodata &  \nodata & 23 & 7 & 0 &  $3 \pm 1 \pm 0.3$ &   $1.5 \pm 0.5 \pm 0.1$  \\ 
 J1744$-$1134 &  $4.6 \pm 0.7 $ & $3.25 \pm 0.25 $ &   $1.3 \pm 0.2 $ & $1.2 \pm 0.3 $ & 394 & 65 & 0 &  $6.8 \pm 0.5 \pm 0.5$ &   $16.5 \pm 1.3 \pm 1.3$  \\ 
 J1747$-$4036 &  $1.5 \pm 0.7 $ & $0.99 \pm 0.23 $ &   $1.9 \pm 0.3 $ & $5.4 \pm 3.3 $ & 43 & 4 & 0 &  $140 \pm 30^{{+70}}_{{-50}} $ &   $11.7 \pm 2.7^{{+6.3}}_{{-4.7}} $  \\ 
 J1810+1744 &  $4.2 \pm 0.5 $ & $2.34 \pm 0.17 $ &   $1.9 \pm 0.2 $ & $3.2 \pm 1.1 $ & 442 & 23 & 3 &  $112 \pm 8^{{+40}}_{{-30}} $ &   $28.2 \pm 2.1^{{+9.4}}_{{-7.3}} $  \\ 
 J1823$-$3021A &  $1.5 \pm 0.4 $ & $1.07 \pm 0.15 $ &   $1.6 \pm 0.2 $ & $2.5 \pm 0.6 $ & 64 & 0 & 5 &  $700 \pm 100 \pm 80$ &   $8.9 \pm 1.3 \pm 0.9$  \\ 
 J1858$-$2216 &  $0.55 \pm 0.28 $ & $0.72 \pm 0.15 $ &   $0.84 \pm 0.74 $ & $1.7 \pm 1.1 $ & 75 & 17 & 0 &  $8 \pm 2^{{+4}}_{{-2}} $ &   $6.8 \pm 1.4^{{+3.2}}_{{-1.7}} $  \\ 
 J1902$-$5105 &  $3.1 \pm 0.4 $ & $2.16 \pm 0.15 $ &   $1.7 \pm 0.2 $ & $3.4 \pm 1.1 $ & 536 & 31 & 2 &  $36 \pm 2 \pm 10 $ &   $5.2 \pm 0.4^{{+2.1}}_{{-1.7}} $  \\ 
 J1939+2134 \dag&  $1.5 \pm 0.8 $ & $0.9 \pm 0.3 $ &   \nodata &  \nodata & 10 & 4 & 1 &  $140 \pm 50 \pm 30$ &   $1.3 \pm 0.5 \pm 0.2$  \\ 
 J1959+2048 &  $2.4 \pm 0.5 $ & $1.70 \pm 0.19 $ &   $1.4 \pm 0.3 $ & $1.4 \pm 0.4 $ & 185 & 43 & 0 &  $130 \pm 10^{{+20}}_{{-40}} $ &   $16.5 \pm 1.8^{{+2.2}}_{{-5.8}} $  \\ 
 J2017+0603 &  $2.0 \pm 0.3 $ & $3.33 \pm 0.21 $ &   $1.0 \pm 0.2 $ & $3.4 \pm 0.6 $ & 1196 & 100 & 1 &  $98 \pm 6 \pm 20$ &   $75.5 \pm 4.8 \pm 14$  \\ 
 J2043+1711 &  $2.7 \pm 0.3 $ & $2.70 \pm 0.16 $ &   $1.4 \pm 0.1 $ & $3.3 \pm 0.7 $ & 918 & 65 & 0 &  $100 \pm 6^{{+20}}_{{-30}} $ &   $79 \pm 5^{{+14}}_{{-26}} $  \\ 
 J2047+1053 &  $0.83 \pm 0.36 $ & $0.62 \pm 0.14 $ &   $1.5 \pm 0.5 $ & $2.0 \pm 1.1 $ & 69 & 11 & 2 &  $31 \pm 7^{{+10}}_{{-8}} $ &   $29.4 \pm 6.6^{{+9.9}}_{{-7.7}} $  \\ 
 J2051$-$0827 $\star$&  $0.24 \pm 0.13 $ & $0.34 \pm 0.10 $ &   $0.50 \pm 0.76 $ & $1.3 \pm 0.7 $ & 68 & 15 & 0 &  $4 \pm 1 \pm 1$ &   $8.1 \pm 2.5 \pm 2.2$  \\ 
 J2124$-$3358 &  $2.7 \pm 0.3 $ & $3.68 \pm 0.16 $ &   $0.78 \pm 0.13 $ & $1.63 \pm 0.19 $ & 1993 & 237 & 3 &  $4.0 \pm 0.2^{{+2}}_{{-1}} $ &   $10.8 \pm 0.5^{{+5.6}}_{{-3.3}} $  \\ 
 J2214+3000 &  $3.0 \pm 0.3 $ & $3.28 \pm 0.15 $ &   $1.2 \pm 0.1 $ & $2.2 \pm 0.3 $ & 1689 & 146 & 2 &  $93 \pm 4 \pm 20$ &   $48.4 \pm 2.2 \pm 11$  \\ 
 J2215+5135 &  $1.0 \pm 0.3 $ & $1.18 \pm 0.14 $ &   $1.3 \pm 0.3 $ & $3.4 \pm 1.0 $ & 183 & 25 & 1 &  $130 \pm 20 \pm 30$ &   $24.6 \pm 3.0 \pm 5.7$  \\ 
 J2241$-$5236 &  $3.0 \pm 0.3 $ & $3.33 \pm 0.16 $ &   $1.3 \pm 0.1 $ & $3.0 \pm 0.5 $ & 2150 & 115 & 1 &  $10.5 \pm 0.5 \pm 3$ &   $4.0 \pm 0.2 \pm 1.1$  \\ 
 J2302+4442 &  $2.6 \pm 0.3 $ & $3.67 \pm 0.17 $ &   $0.94 \pm 0.12 $ & $2.1 \pm 0.3 $ & 1716 & 189 & 2 &  $62 \pm 3^{{+10}}_{{-20}} $ &   $162.6 \pm 7.5^{{+26}}_{{-57}} $  \\ 

\enddata

\tablenotetext{a}{A dagger (\dag) means the spectral fit is unreliable (see text). A star ($\star$) means that the spectrum was calculated using the on-peak data only.}
\tablenotetext{b}{Overestimated distances or the assumed beaming factor, $f_\Omega = 1$, can result in an efficiency $>$ 100\%.}

\tablecomments{Unbinned maximum likelihood spectral fit results  
for the LAT MSPs, using the PLEC1 model (Eq. \ref{expcutoff}, Section \ref{spectralSection}). 
Columns 2 and 3 list the phase-averaged integral photon and energy fluxes in the 0.1 to 100 GeV energy band, $F_{\rm 100}$ and $G_{\rm 100}$. 
Columns 4 and 5 list the photon index $\Gamma$ and cutoff energy $E_{\rm cut}$.  
Columns 6, 7, and 8 list the source significance $TS$, significance $TS_{\rm cut}$ of the exponential 
cutoff compared to a simple power law, and significance $TS_{\rm b\,free}$ of the PLEC compared to a PLEC1 shape. A value $TS_{\rm cut} < 9$ indicates that a spectral cutoff is not significantly detected. 
Column 9 gives the total gamma-ray luminosity $L_\gamma$ in the 0.1 to 100 GeV energy band.
The gamma-ray conversion efficiency $\eta \equiv L_\gamma/\dot{E}^{\rm int}$ in Column 10 assumes a beam correction factor
$f_\Omega=1$ as described in Section \ref{lumin}, and the Shklovskii-corrected $\dot{E}^{\rm int}$ values from Table \ref{tbl-doppler} in Section \ref{doppler}.
The first uncertainty in $L_\gamma$ and $\eta$ comes from the statistical uncertainties in the spectral fit, whereas the second is due to the distance uncertainty.
The strong dependence of these quantities on distance (see Table \ref{tbl-doppler}) and beaming factor means that these values
should be considered with care.
}
\end{deluxetable}
\normalsize
\noindent

\clearpage
\end{landscape}


For some pulsars with low duty cycles, the phase-averaged analysis returned a low $TS$ value and/or could not constrain the spectral parameters well.
We selected the off-peak intervals for these sources and followed the same prescription as for the phase-averaged analyses but without the pulsar in the model.
We then fixed the parameters of all sources $>4^{\circ}$ from the pulsar, left the normalization parameters of the remaining point sources and diffuse models free, added the pulsar back to the model, and performed an on-peak spectral analysis.
These pulsars are indicated by a star ($\star$) in Tables \ref{tbl-psrspec} and \ref{tbl-mspspec}.  The values of $F_{100}$ and $G_{100}$ reported for these pulsars have been corrected to phase-averaged values.
%
Additionally, the spatial residuals for PSR J1702$-$4128 revealed a large deficit near the pulsar attributed to the Galactic diffuse model.
In lieu of a new Galactic diffuse template, we increased the minimum event energy to 300 MeV and performed an on-peak spectral analysis as described previously.
Thus, for PSR J1702$-$4128 the reported values of $F_{100}$ and $G_{100}$ are extrapolations below the energy range of the data, which increases the quoted uncertainties beyond the statistical values.

To estimate systematic uncertainties on the maximum likelihood spectral parameters we selected eight pulsars with representative characteristics (J0218+4232, J0248+6021, J0357+3205, J0614$-$3329, J0631+1036, J1658$-$5324, J1833$-$1034, and J1846+0919) and studied how their spectra changed when perturbing the Galactic diffuse emission and LAT effective area ($A_{\rm eff}$).

The distribution of Galactic diffuse normalization parameters, from all the fits, has a mean of 1.01 with 1$\sigma$ deviation of 4\%.
To estimate possible systematic effects due to an imperfect knowledge of this diffuse component, we repeated the spectral analysis with the normalization of the Galactic diffuse emission fixed to (1$\pm$0.06) times the best-fit value, corresponding to $\pm$1.5$\sigma$ deviations.
The average and largest deviations for $\Gamma$, $E_{\rm cut}$, $F_{100}$, and $G_{100}$ from this test are listed in the first row of Table \ref{tbl-systs}.

\begin{deluxetable}{lrrrrrrrr}
\tabletypesize{\scriptsize}
\tablecaption{Systematic Deviations on Pulsar Spectral Parameters
\label{tbl-systs}}
\tablewidth{0pt}
\tablehead{
\colhead{Systematic} & \colhead{$\langle \Delta\Gamma \rangle$} & \colhead{$\langle \Delta E_{\rm cut} \rangle$}  & \colhead{$\langle \Delta F_{100} \rangle$}  & \colhead{$\langle \Delta G_{100} \rangle$} & \colhead{$\max(\Delta \Gamma)$} & \colhead{$\max(\Delta E_{\rm cut})$}  & \colhead{$\max(\Delta F_{100})$}  & \colhead{$\max(\Delta G_{100})$}\\
\colhead{} & \colhead{(\%)} & \colhead{(\%)} & \colhead{(\%)}  & \colhead{(\%)} & \colhead{(\%)} & \colhead{(\%)} & \colhead{(\%)}  & \colhead{(\%)}
}

\startdata
Galactic Diffuse & 14 & 4 & 16 & 12 & 80 & 27 & 65 & 46\\
Bracketing IRFs & 5 & 4 & 8 & 6 & 21 & 11 & 13 & 8\\
\enddata

\end{deluxetable}
\normalsize


Systematic uncertainties on $A_{\rm eff}$ are estimated to be 10\% for $\log_{10}{E/1\ \rm MeV}\leq2$, 
5\% for $\log_{10}{E/1\ \rm MeV}=2.75$, and 10\% for $\log_{10}{E/1\ \rm MeV}\geq4$ with linear extrapolation in between, 
in log space \citep{P7Paper}.  To estimate the effects of these uncertainties we generated bracketing IRFs in which the usual $A_{\rm eff}$ was replaced by,
\begin{equation}\label{bracket}
A_{\rm B}(E)\ =\ A_{\rm eff}(E) (1 + err(E) B(E)),
\end{equation}
\noindent{}where $err(E)$ represents the $A_{\rm eff}$ uncertainties with 
$B(E)=\pm 1$ for the normalization factor $K$ and $B(E)\ =\ \pm \tanh(log_{10}(E/E_{0})/\kappa)$ for $\Gamma$ and $E_{\rm cut}$. Choosing $\kappa=0.13$ smoothes over twice the energy resolution.  
When using bracketing IRFs, it is important to isolate changes in the source of interest caused by the modified $A_{\rm eff}$ from changes in the diffuse background spectrum 
introduced by this perturbation. Since the diffuse background spectra were derived from flight data, we multiply the spectrum of the Galactic diffuse model by $(1+err(E)B(E))^{-1}$, 
ensuring that the predicted counts spectra from the diffuse background remains unchanged during the bracketing studies.

We integrate the fit results for the different bracketing IRFs to obtain $F_{100}$ and $G_{100}$.
The second row of Table \ref{tbl-systs} lists the average and largest deviations for $\Gamma$, $E_{\rm cut}$, $F_{100}$, and $G_{100}$ due to bracketing.
$G_{100}$ is more robust than $F_{100}$.  
The considerably lower average deviations (columns 2 through 5) than maximum values (columns 6 through 9) shows that most deviations are much smaller than the outlying values.

Finally, to plot the spectra, we divided the 0.1 to 100 GeV energy band into 4 (2) bins per decade for pulsars with $TS$ above (below) 250 and
fit a power law in each bin, curvature within the bins being negligible.  
We fixed the spectral parameters of all sources more than $4^{\circ}$ from the pulsar of interest and of the diffuse components at their full band fit results. 
The spectra of the pulsar of interest and the other point sources within the $4^{\circ}$ region were modeled as power laws with index fixed at 2. 
Their flux levels in each energy band were obtained from a two-step fit. In the first step, sources with $TS\leq0$ were removed, 
from that energy band only, as leaving them can adversely affect the fit uncertainties. The second step is to re-fit with the modified model.
Sample spectra are shown in Figures \ref{fig-J0101spec} to \ref{fig-J1954spec} in Appendix \ref{App-Samples}.  
For pulsars with $TS_{\rm b\,free}\geq 9$, the Figures show both the PLEC1 and PLEC fits.
The spectra obtained for all pulsars with $TS\geq$25 and reliable spectral fits are included in the online material.

\subsection{Spectral Results}
\label{spectralResultsSection}
Table \ref{tbl-psrspec} lists the phase-averaged spectral results for the non-recycled pulsars, and Table \ref{tbl-mspspec} for the MSPs.
For ten pulsars, flagged with a \dag{} symbol, the spectral fits were unreliable, for different reasons.
Five are undetected as point sources, having $TS < 25$. 
For two of these the likelihood analysis fails and we report no spectral parameters, 
as is the case for PSR J0729$-$1448 in spite of its larger $TS$ value. 
For the other three the integrated flux is robust and we report $F_{100}$ and $G_{100}$.
Similarly, we report only the integrated fluxes for four more pulsars, 
either because the maximum likelihood fit solution favored $\Gamma \approx 0$, 
or because the parameter uncertainties were of order 100\%. 
For PSRs J1112$-$6103 and J1410$-$6132, analysis of the off-peak phase intervals (Section \ref{unpulsed}) showed significant extended emission, 
which we added to the phase-averaged source model, improving the spectral results.

\begin{figure}[!ht]
\centering
\includegraphics[width=0.8\textwidth]{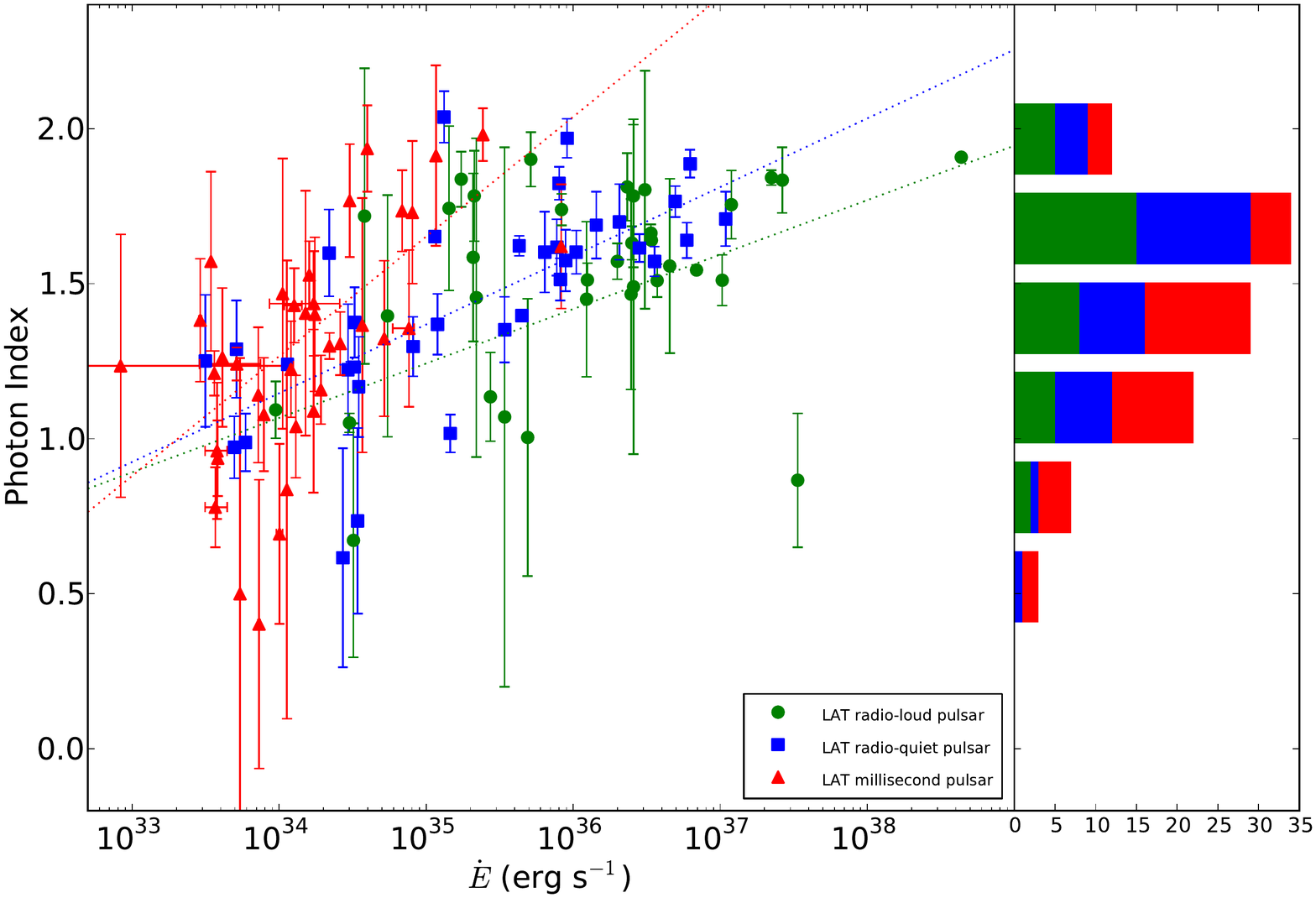}
\caption{
Power-law index $\Gamma$ for the exponentially cutoff gamma-ray spectra versus the Shlovkskii-corrected spindown power, 
for the pulsars bright enough in gamma rays to allow spectral analysis (see text). The straight-line fit results are in Section \ref{spectralResultsSection}.
The markers are the same as in Figure \ref{PPdotplot}.
The uppermost line is for millisecond gamma-ray pulsars. The middle line fits young, radio-quiet gamma-ray pulsars, 
while the lowest line is for young, radio-loud gamma-ray pulsars. 
A histogram of photon index values is projected onto the axis.
\label{Gamma_Edot}}
\end{figure}

Figure \ref{Gamma_Edot} shows correlation between $\Gamma$ and $\dot{E}$. 
The Pearson correlation factor is $0.68$ for the young, radio-quiet pulsars and $0.58$ for the MSPs, 
with probabilities of occurring by chance (two-sided p-values) of $5\times 10^{-6}$ and $1.5\times 10^{-4}$, respectively.
For the young radio-loud pulsars, the correlation is smaller ($0.40$) and marginally significant (two-sided p-value of $0.017$).
For all pulsars together the correlation factor is $0.57$, with high significance ($2\times 10^{-10}$), even allowing for trials.
We fit the measurements with $\Gamma = \mathcal{A} \log(\dot{E}) + \mathcal{B}$ (dashed lines in the figure).
For young pulsars we find similar trends in the radio-loud and radio-quiet populations with $\mathcal{A} \approx 0.2$ and $\mathcal{B} \approx -5$.  
The exact fit values are sensitive to the outliers with small statistical uncertainties, such as the Crab.
The MSPs have more dispersion in $\Gamma$ and a narrower $\dot{E}$ range, with a steeper 
slope than for the young population, $\mathcal{A} \approx 0.4$ and $\mathcal{B} \approx -12$.

\begin{figure}[!ht]
\centering
\includegraphics[width=1.0\textwidth]{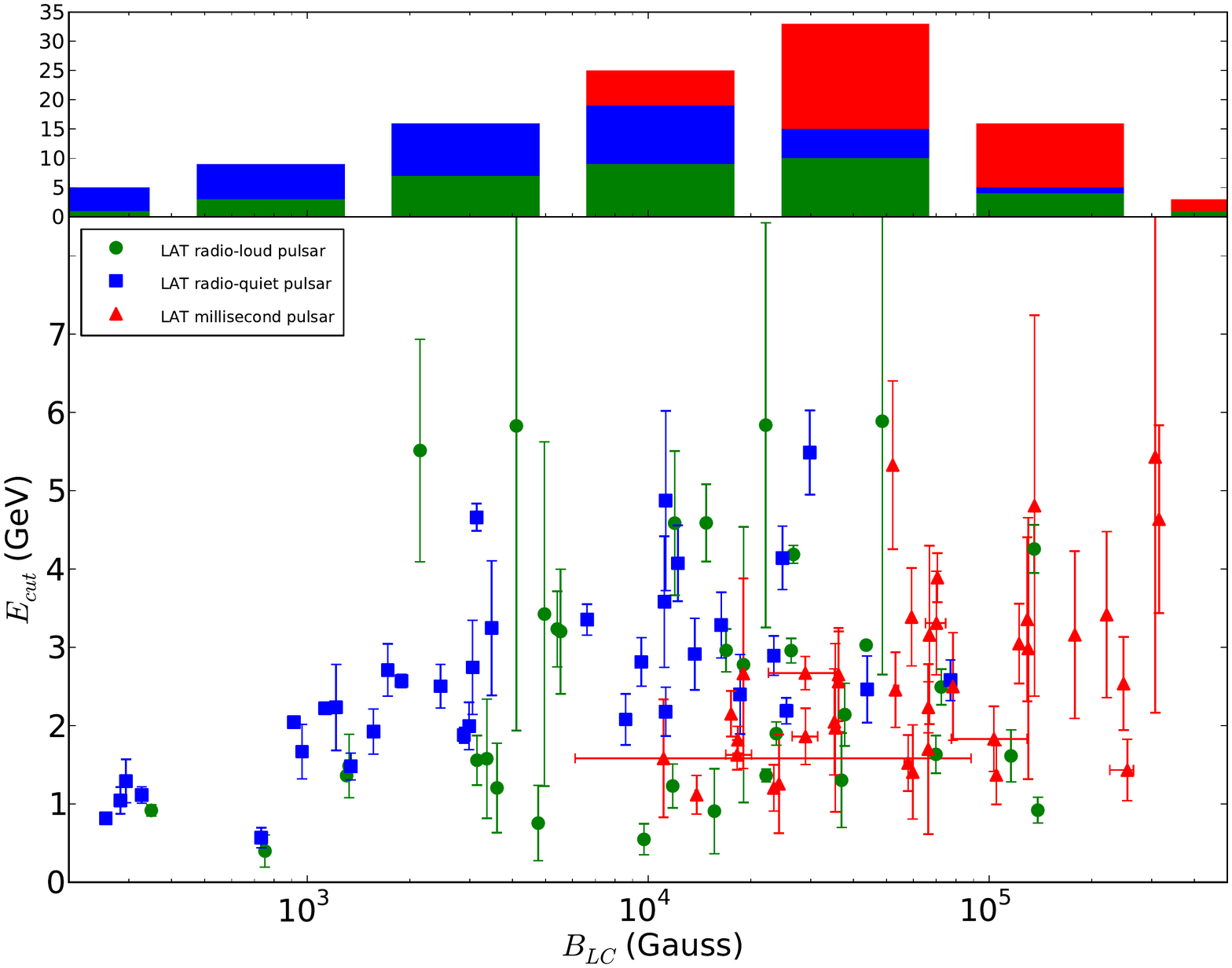}
\caption{The best-fit cutoff energy versus magnetic field at the light cylinder, $B_{\rm LC}$.  The young, radio-quiet gamma-ray pulsars have lower $B_{\rm LC}$ than other gamma-ray pulsars.
The markers are the same as in Figure \ref{PPdotplot}. The histogram highlights the different $B_{LC}$ distributions for the three pulsar classes.
\label{Ecut_BLC}}
\end{figure}

In 1PC we noted a possible correlation between $E_{\rm cut}$ and the magnetic field strength at the light cylinder 
($B_{\rm LC} = 4\pi^{2} (1.5 I_{0} \dot{P})^{1/2}(c^{3} P^{5})^{-1/2}$, assuming an orthogonal rotator).
For the radio-quiet pulsars, Figure \ref{Ecut_BLC} confirms the trend, with a Pearson correlation factor of $0.64$ (p-value $4\times 10^{-5}$). 
The factor is $0.52$ for the MSPs, with p-value $0.0007$. 
Here too, the correlation for the young, radio-loud pulsars is small ($0.24$) and insignificant (p-value $0.17$).

\subsection{Luminosity}
\label{lumin}


Gamma-ray emission models predict different relations between the spindown power $\dot E$ and the gamma-ray luminosity, 
\begin{equation}
 L_\gamma = 4\pi d^2 f_\Omega G_{100},
\label{LumEq}
\end{equation}
making this a discriminating observable when applied to a large sample of gamma-ray pulsars.
Following \citet{AtlasII}, we define the beam correction factor $f_\Omega$ as
\begin{equation}\label{eq-fOmega}
 f_\Omega(\alpha,\zeta_E) = 
{\int F_\gamma(\alpha;\zeta,\phi)\sin \zeta {\rm d}\zeta {\rm d}\phi \over 2\int F_\gamma(\alpha;\zeta_E,\phi){\rm d}\phi}, 
\end{equation}
to extrapolate the observed flux to the full sky for some beam shape model. 
The angle $\alpha$ between the neutron star's magnetic and rotation axes is one model parameter. 
The angle $\zeta_E$ between the rotation axis and the Earth line of sight describes the inclination of the system relative to Earth.
The numerator integrates emission into all space (all inclinations $\zeta$) whereas the
denominator is the observed flux integrated over a neutron star rotation, with pulsar phase $\phi$.
In the past, the gamma-ray beam was conventionally assumed to sweep out a 1 sr solid angle, in which case
$L_\gamma = d^2  G_{100}$. Such a beam is appropriate to near-surface polar cap emission and
corresponds to $f_\Omega = {1 \over 4\pi} = 0.08$. An outer magnetosphere fan-like beam sweeping the entire sky ($4\pi$ steradians) gives $f_\Omega \approx 1$, which is the value we adopt for calculating $L_{\gamma}$.  However, \citet{Pierbattista2012} found a large spread in $f_\Omega$ values for different emission models and for radio-loud versus radio-quiet young pulsars.
Values of $f_\Omega$ exceeding 1 correspond to beams that are narrow in $\phi$, extended in $\zeta$,
and/or have average intensity exceeding the value sampled at $\zeta_E$. 

\begin{figure}[!ht]
\centering
\includegraphics[width=1.0\textwidth]{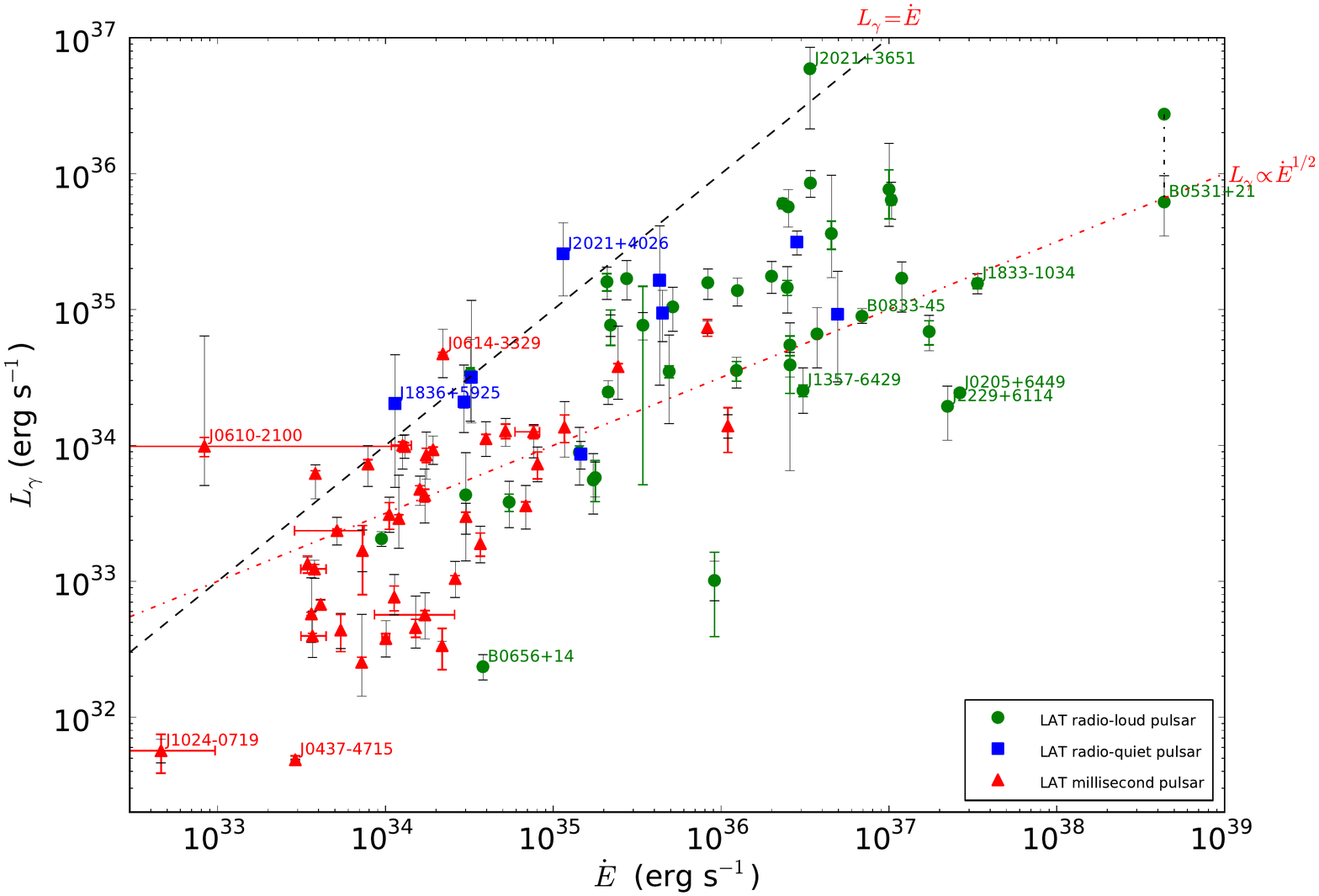}
\caption{
Gamma-ray luminosity $L_\gamma = 4\pi f_\Omega d^2 G_{100}$ in the 0.1 to 100 GeV energy band versus spindown power $\dot E$.
The vertical error bars from the statistical uncertainty on the energy flux $G_{100}$ are colored in the electronic journal version.  
The vertical error bars due to the distance uncertainties are black, and generally larger. 
Doppler corrections (Section~\ref{doppler}) have been applied to MSPs with known proper motions, 
leading to visible horizontal error bars in some cases.
The upper diagonal line indicates 100\% conversion of spindown power into gamma-ray flux: for pulsars above this line, the distance
$d$ may be smaller, and/or the assumed beam correction $f_\Omega \equiv 1$ is wrong. 
The lower diagonal line indicates the heuristic luminosity $L_\gamma^h = \sqrt{10^{33} \dot E}$ erg s$^{-1}$, to guide the eye.
The upper of the two Crab points, at far right, includes the X-ray energy flux (see Section~\ref{Xrays}). 
The markers are the same as in Figure \ref{PPdotplot}. 
\label{EDotLumG}}
\end{figure}

\begin{figure}[!ht]
\centering
\includegraphics[width=1.0\textwidth]{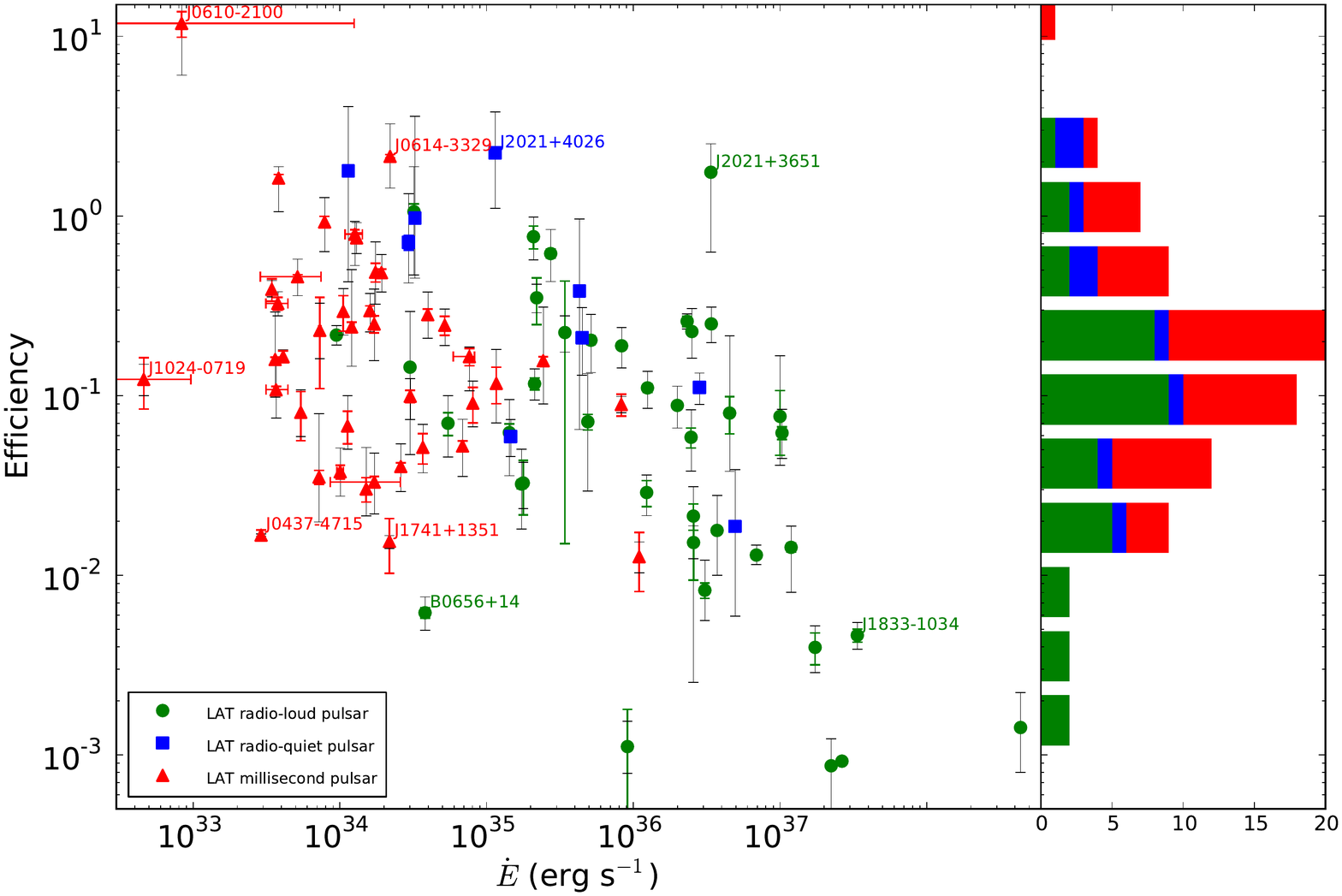}
\caption{
Gamma-ray efficiency $\eta= L_\gamma/\dot E$ versus spindown power $\dot E$. The error bars are as
in Figure \ref{EDotLumG}.
The markers and the side histogram use the same color coding as in Figure \ref{PPdotplot}.
\label{Eff_Edot}}
\end{figure}

Figure \ref{EDotLumG} shows $L_{\gamma}$ versus $\dot{E}$. 
Pulsars with poor spectral fits have been excluded. 
The open field-line voltage is $V \simeq 3.18 \times 10^{-3} \sqrt{\dot E }$ volts.
Above some threshold voltage, gamma-ray emitting electron-positron cascades occur, 
and a linear dependence of $L_\gamma$ on $V$ would give $L_\gamma \propto \sqrt{\dot{E}}$ \citep{Arons96},
as for the lower diagonal line. 
With arbitrary normalization, we call this the ``heuristic'' gamma-ray pulsar luminosity, 
\begin{equation}
 L_\gamma^h = 10^{33} \sqrt{\dot E / 10^{33}} = \sqrt{10^{33} \dot E} \, {\rm erg \, s}^{-1}.
 \label{heurLumEq}
 \end{equation}
At low $\dot E$ values $L_\gamma$ seems to be falling below $L_\gamma^h$. 
The upper diagonal line shows $ L_\gamma = \dot E$, that is, 
a 100\% efficiency $\eta= L_\gamma/\dot E$ for converting spindown power into gamma rays.
A few pulsars appear above this line, likely due to over-estimated distances or $f_{\Omega}$ values. 
Figure \ref{Eff_Edot} plots $\eta$ versus $\dot{E}$. 
Overall, the expected $\eta \propto 1 / \sqrt{\dot{E}}$ trend is roughly respected.
However, the large dispersion due to the distance uncertainties, as well as beaming effects, limits the extent to which the data constrain the theory.

The Doppler correction to $\dot E$ is small ($\left| \xi \right| < 10$\%) for 9 of the 20 pulsars with proper motion corrections (Table \ref{tbl-doppler}).
For the five with $20\% < \xi < 60$\%, the correction refines their positions in e.g. the $L_\gamma$ vs $\dot E^{\rm \, int}$ plane. 
The Doppler correction for PSR J0437$-$4715, with a large $\xi=75$\%, produces a qualitative change: 
observed $\dot E = 12 \times 10^{33}$ erg s$^{-1}$ decreases to 
corrected $\dot E^{\rm \, int}= 3 \times 10^{33}$ erg s$^{-1}$, right at the apparent deathline, 
and the efficiency changes from the lowest outlier amongst MSPs, to a low, but typical, $\eta = 1.7$\%. 

\begin{figure}[!ht]
\centering
\includegraphics[width=1.0\textwidth]{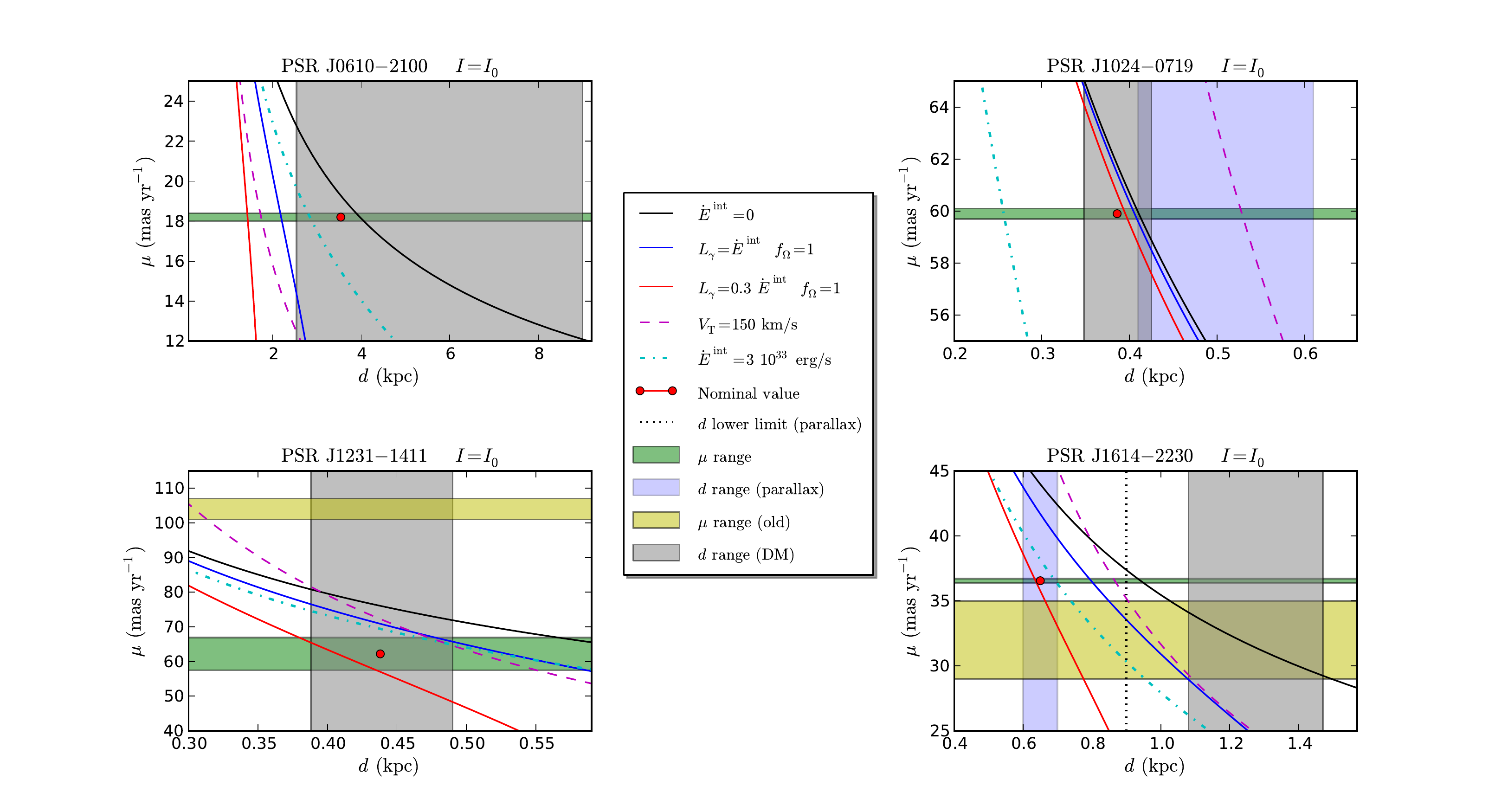}
\caption{Constraints on the proper motion $\mu$ and the distance $d$ for the four MSPs with Doppler corrections that lead to unphysical or out-of-family 
values of the spindown power $\dot E^{\rm \, int}$ (see Section \ref{doppler}). Unphysical or uncommon values are to the right of the curves. 
The uppermost solid curve (black) requires  $\dot E^{\rm \, int} > 0$. The second solid curve (blue) requires $\dot E^{\rm \, int} > L_\gamma$.
The bottom solid curve (red) assumes a large, but common, efficiency  $\eta = 30$ \%, that is, $\eta \dot E^{\rm \, int} > L_\gamma$.
The violet dashed line corresponds to a pulsar transverse space velocity of 150 km s$^{-1}$. Finally, the green dot-dashed line
traces the empirical gamma-ray deathline value of  $\dot E^{\rm \, int} = 3\times 10^{33}$ erg s$^{-1}$.
The red dot is at the adopted values of $\mu$ and the distance $d$.
The distances, with uncertainties, from parallax and from DM with the NE2001 model are represented by light blue and gray zones, respectively.
The green zone indicates the newer proper motion measurement we use, and the yellow zone shows
earlier values. The black dotted line shows the lower limit of parallax distance for J1614$-$2230 from \citet{dpr+10}.
\label{PMvsDist}}
\end{figure}

The remaining four pulsars with $\xi >60 $\% bear special discussion.
Figure \ref{PMvsDist} plots lines of constant $\dot E$, $L_\gamma$, and transverse velocity $v_T$ in $\mu$ vs.~$d$ space for different assumptions: 
$\dot E = 0$, $\dot E = L_\gamma$, and $\eta \dot E = L_\gamma$ with $\eta = 30$ \%, at the high end of the observed range.
The curve for $v_T = \mu d = 150$ km s$^{-1}$ is the $3\sigma$ extremum of the MSP velocity distribution 
of \citet{ParkesSurveyII}. Faster recycled pulsars are possible, but unusual.
Allowed (or favored) regions are to the left of the curves. 
The curve for $\dot E = 3\times 10^{33}$ erg s$^{-1}$ shows how an $\dot E$ value lower than those seen to date would compare with the other constraints. 
The shaded zones correspond to the measurements and their uncertainties adopted in this paper along with previous measurements, for comparison. 
We recall that here, as throughout the paper, we adopt the moment of inertia $I_{0}=10^{45}$ gm cm$^2$, 
corresponding to a neutron star mass of $1.4$ M$_\odot$ and radius of 10 km.

PSR J0610$-$2100 was recently discussed by \citet{egc+12}:
the intrinsic spindown power is well below the empirical deathline, the space velocity is much higher than typical, 
and the gamma-ray efficiency exceeds 100\%. In Figure \ref{PMvsDist} the nominal ($\mu$,  $d$) point for this pulsar is to the right of most of the curves.
These apparent paradoxes are resolved if the pulsar is closer than 2 kpc. Upon inspection of NE2001,
minor changes in the ``Local Super Bubble'' shape and density for this line of sight could yield a distance less than 2~kpc. 
\citet{egc+12} also describe material along the line of sight that is unmodeled in NE2001.

They also discussed PSR J1024$-$0719. We show both the corrected parallax distance from \citet{Verbiest12} as well as the NE2001 distance,
for which the 10\% uncertainty is probably underestimated given the small DM $=6.5$ pc cm$^{-3}$ in that direction.
However, any distance compatible with the parallax yields the lowest intrinsic spindown power $\dot E^{\rm \, int}$ of any gamma-ray pulsar, 
and well below the $3 \times 10^{33}$ erg s$^{-1}$ current minimum. 
 
For PSR J1231$-$1411, discovered in an unassociated LAT source, \citet{rrc+11} measured a large proper motion, $\mu >100$ mas yr$^{-1}$, indicated in Figure \ref{PMvsDist}.
The Doppler correction using that large $\mu$ gave negative $\dot E^{\rm \, int}$ values, which is unphysical. 
Radio observations since the initial measurement allowed us to update the timing model, 
resulting in the smaller proper motion value listed in Table \ref{tbl-doppler}
and plotted in Figure \ref{PMvsDist}. The Doppler corrections now lead to rather typical parameters for this pulsar.

PSR J1614$-$2230 is one of the rare pulsars with a neutron star mass measurement.
\citet{dpr+10} measured the Shapiro delay precisely in this binary system to obtain $M = 1.97 \pm 0.04$ M$_\odot$, 
well above the Chandrasekhar mass of $1.4$ M$_\odot$. The DM distance, 1.27 $\rm kpc$, along with the proper motion measured 
in the same paper yields implausible $\dot E^{\rm \, int}$ and $\eta$ values. A moment of inertia greater than
$I_{0}$ would improve the situation, favoring ``rigid'' neutron star equations-of-state.
However, the parallax distance and proper motion recently measured by \citet{antoine+13} brings this pulsar
back in line with the rest of the population.  

\citet{dpr+10} conclude that $M>1.4$ M$_\odot$ is probably true for many or most MSPs. 
For $I > I_0$, $\dot E^{\rm \, int}$ is larger than the ``standard'' values in Table \ref{tbl-doppler}. 
This would shrink the spread between the $\dot E^{\rm \, int}$ distributions for the young and MSP gamma-ray pulsars.

\section{Unpulsed Magnetospheric Emission}
\label{unpulsed}

Some pulsars have magnetospheric emission over their full rotation phase with similar spectral characteristics to the emission seen through their peaks.
This emission appears in the observed light curves as a low-level,
unpulsed component above the estimated background level (i.e., not
attributable to diffuse emission or nearby point sources) and can be a
powerful discriminator for the emission models.

On the other hand a PWN around the pulsar, or a photon excess due to imprecise knowledge of
diffuse emission around the pulsar, would not be modulated at the rotational period and could be confused with a constant magnetospheric signal. 
Including the discovery of the GeV PWN 3C 58 associated with PSR J0205+6449 described in this section, 
the LAT sees 17 sources potentially associated with PWNe at GeV energies \citep{Rousseau2013}. 
Some are highlighted in Appendix~\ref{App-off_peak_individual_source_discussion}.
This off-peak emission should be properly modeled when searching for pulsar emission at all rotation phases.


We can discriminate between these two possible
signals through spectral and spatial analysis.  If the emission is
magnetospheric, it is more likely to appear as a non-variable point source
with an exponentially cutoff spectrum with a well-known range of cutoff energies.  
On the other hand, PWNe and diffuse excesses have spectra with a power-law shape and either a hard index continuing up to tens of GeV in the PWN case or present only at lower energies with a very soft index in the diffuse case.  
In addition, PWNe
are often spatially resolvable at GeV energies \citep[e.g., Vela-X has been spatially resolved with the LAT and \textit{AGILE}
and HESS J1825$-$137 with the LAT;][respectively]{LAT_collaboration_Vela_X_2010,AGILE_VelaX,LAT_collaboration_HESS_J1825_2011}
so an extended source would argue against a magnetospheric origin
of the emission.  However, given the finite angular resolution of
the LAT (see Section~\ref{obsvSection}) not all PWNe will appear
spatially extended at GeV energies.  The Crab Nebula, for instance,
cannot be resolved by the LAT but can be distinguished from the
gamma-bright Crab pulsar, in the off-peak interval, by its hard
spectrum above $\sim$1 GeV \citep{FermiCrab}.  In addition, GeV
emission from the Crab Nebula was discovered to be time-variable
\citep[e.g.,][]{LAT_Collaboration_Crab_Flare_2011} providing another
possible way to discern the nature of any observed off-peak signal.

Therefore, to identify pulsars with magnetospheric emission across
the entire rotation, we define and search the off-peak intervals of
the pulsars in this catalog for significant emission, except PSR J2215+5135 for
which the rotation ephemeris covers a short time interval and the profile is noisy.
We then evaluate the spectral and spatial characteristics of any off-peak emission
to determine if it is likely magnetospheric, related to the pulsar wind,
or physically unrelated to the pulsar (e.g., unmodeled diffuse emission).

\subsection{Off-peak Phase Selection}
\label{peak_definition}

We first developed a systematic, model-independent, and computationally-efficient method 
to define the off-peak interval of a pulsar light curve.

We begin by deconstructing the light curve 
into simple Bayesian Blocks using the algorithm described in
\citet{Jackson_Bayesian_Blocks_2003} and \citet{Scargle2013}.  
We could not apply the Bayesian Block algorithm to the weighted-counts
light curves because they do not follow Poisson statistics, required
by the algorithm.
We therefore use an unweighted-counts light curve in which the angular radius and minimum energy selection have been varied to maximize the H-test statistic.  To produce Bayesian Blocks on a periodic light curve,
we extend the data over three rotations, by copying and shifting the
observed phases to cover the phase range from $-$1 to 2.  We do, however,
define the final blocks to be between phases 0 and 1.  
To avoid potential contamination from the trailing or leading edges of
the peaks, we reduce the extent of the block by 10\% on either side,
referenced to the center of the block.

There is one free parameter in the Bayesian Block algorithm called
ncp$\rm _{prior}$ which modifies the probability that the
algorithm will divide a block into smaller intervals.
We found that, in most cases, setting ncp$\rm _{prior}=8$ protects against
the Bayesian Block decomposition containing unphysically small blocks.
For a few marginally-detected pulsars, the algorithm failed 
to find more than one block and we had to decrease ncp$\rm _{prior}$ until the
algorithm found a variable light curve. Finally, for a few pulsars the 
Bayesian-block decomposition of the light curves failed to model 
weak peaks found by the light-curve fitting method
presented in Section~\ref{lightCurveFitting} or extended
too far into the other peaks. For these pulsars,
we conservatively shrink the off-peak region.

For some pulsars, the observed light curve has two well-separated peaks
with no significant bridge emission, which leads to two well-defined
off-peak intervals.  We account for this possibility by finding the second-lowest 
Bayesian block and accepting it as a second off-peak interval if
the emission is consistent with that in the lowest block (at the 99\%
confidence level) and if the extent of the second block is at least half
that of the first block.

\begin{figure}
  \ifdefined\bwfigures
  \plotone{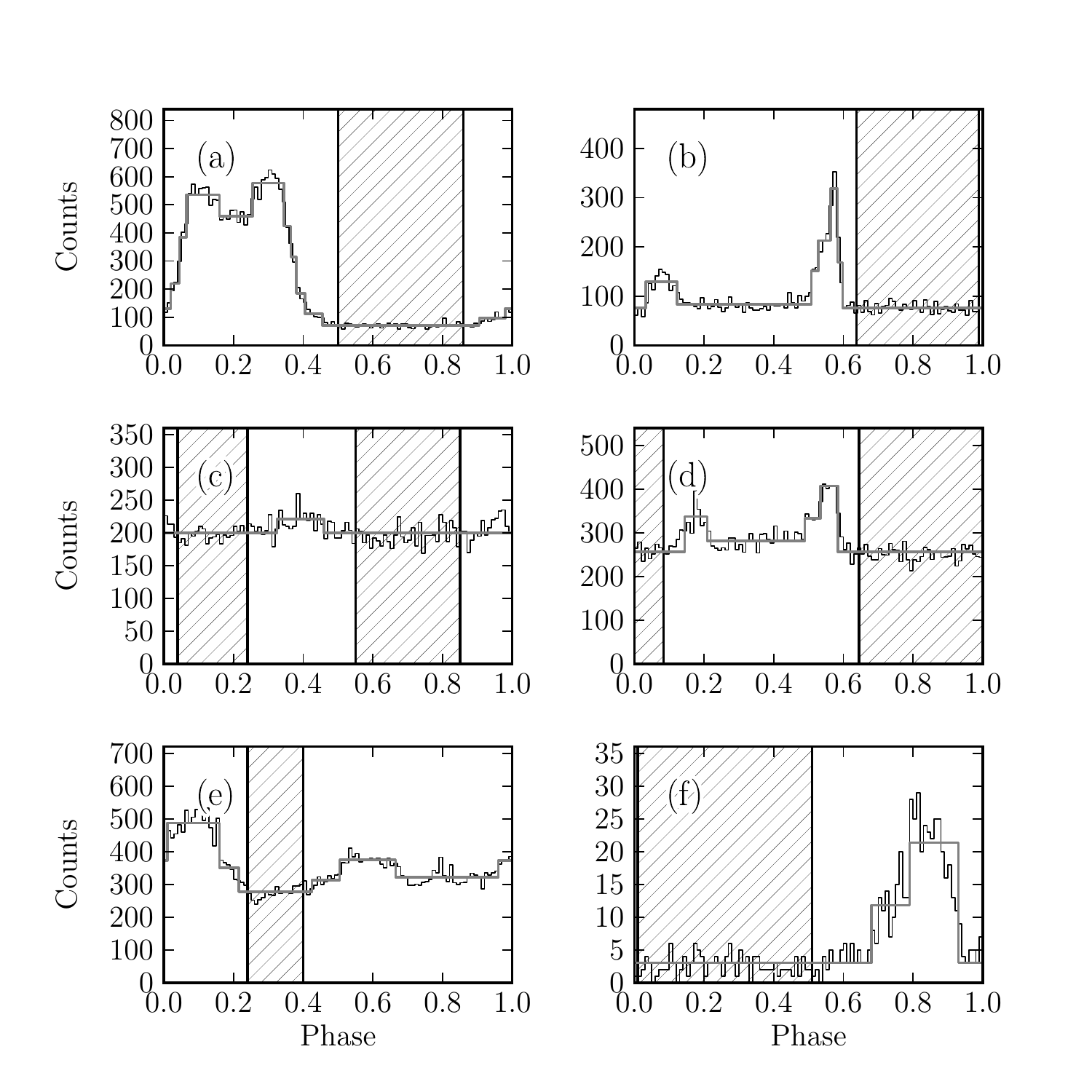}
  \else
  \plotone{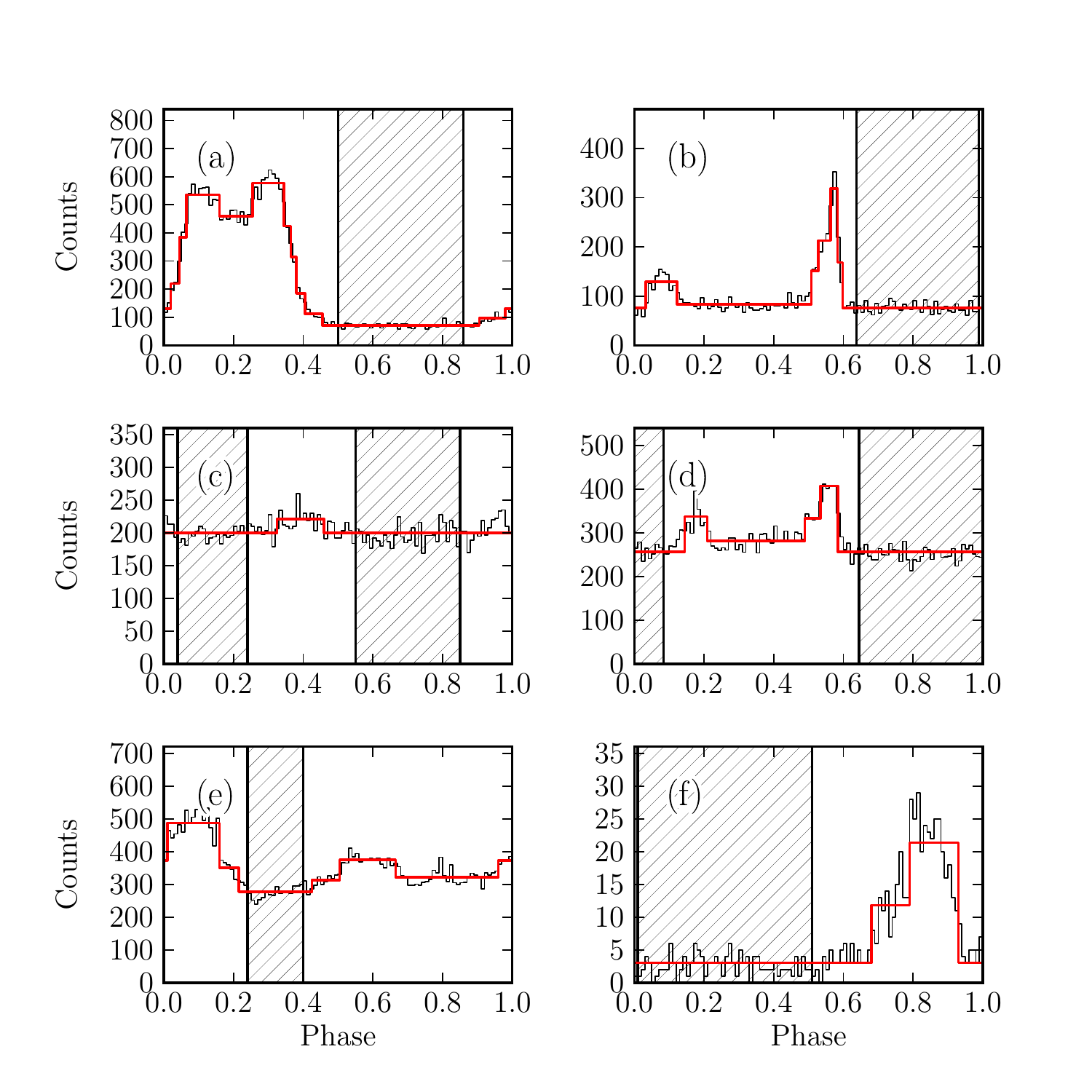}
  \fi
  \caption{The energy-and-radius optimized light curve, Bayesian block decomposition of the        
  light curve, and off-peak interval for
  (a) PSR J0007+7303, (b) PSR J0205+6449, (c) PSR J1410$-$6132,
  (d) PSR J1747$-$2958, (e) PSR J2021+4026, and (f) PSR J2124$-$3358.
  The black histograms represent the light curves,
  the gray lines (colored red in the electronic version)
  represent the Bayesian block decompositions of the pulsar light curves, and
  the hatched areas represent the off-peak intervals selected by this method.}
  \label{off_peak_select}
\end{figure}

Figure~\ref{off_peak_select} shows the energy-and-radius optimized
light curves, the Bayesian block decompositions, 
and the off-peak intervals for six pulsars.  
Figures \ref{J0034-0534lc} to \ref{J2240+5832lc} overlay off-peak intervals
over the weighted light curves of several pulsars.
The off-peak
intervals for all pulsars in this catalog are given in Tables \ref{tbl-pulsePSR} and \ref{tbl-pulseMSP}.

\subsection{Off-peak Analysis Method}
\label{off_peak_analysis}

Characterizing both the spatial and spectral
characteristics of any off-peak emission helps discern its origin.
We employ a somewhat different analysis procedure here than
for the phase-averaged analysis described in Section
\ref{spectralMethodSection}.  To evaluate the spatial characteristics of
any off-peak emission we use the likelihood fitting package \pointlike
\citep[detailed in][]{LAT_collaboration_extended_search_2012}, and
to fit the spectrum we use \gtlike in binned mode via {\it pyLikelihood} as was done for the
phase-averaged analysis.

For each pulsar we start from
the same temporal and spatial event selections described in Section
\ref{obsvSection} but we increase the maximum energy to 400 GeV (the
highest event energy for any ROI under this selection is $\sim$316 GeV).
For the \pointlike analysis we further select a 10$\degree$ radius ROI
and for \gtlike a $14\degree\times14\degree$ square ROI, both centered
on the pulsar postition.  Finally, we only consider photons with 
pulse phases within the corresponding off-peak interval.

We search for off-peak emission
assuming a point source and (except for the Crab Nebula and Vela-X, described below) 
a power-law spectral model.  We fit the position of this
putative off-peak source using \pointlike as described by \citet{2FGL}
and then use the best-fit position in a spectral analysis with \gtlike.
From the spectral analysis we require $TS\geq25$ (just over $4\sigma$)
to claim a detection.  If $TS<25$, we compute upper limits on the flux 
in the energy range from 100 \mev to 316 \gev
assuming
a power law with photon index fixed to 2.0 and a PLEC1 model with
$\Gamma=1.7$ and $E_{\rm cut}=3$ GeV.

The spectrum of the Crab Nebula (associated with PSR J0534+2200) is
uniquely challenging because the GeV spectrum contains both a falling
synchrotron and a rising inverse Compton component \citep{FermiCrab}.
For this particular source we used the best-fit two-component spectral
model from \cite{LAT_collaboration_crab_2012} and fit only the overall
normalization of the source. 
In addition, for Vela-X (associated with PSR J0835-4510) we took the
best-fit spectral model from \cite{FermiVelaX2nd} and fit only the overall
normalization of this source. This spectrum has a smoothly broken power
law spectral model and was fit assuming Vela-X to have an elliptical
disk spatial model.

If the off-peak source is significant, we test whether the spectrum shows
evidence for a cutoff, as described in Section \ref{spectralMethodSection}
and by \citet{LAT_collaboration_PWNCAT_2011}, assuming the source is at
the pulsar position.  We say that the off-peak emission shows evidence for
a cutoff if $TS_{\rm cut}\geq9$, corresponding to a $3\sigma$ detection.

For a significant off-peak point source,
we use \pointlike to test if the emission is significantly
extended.  We assume a radially-symmetric Gaussian source
and fit the position and extension parameter ($\sigma$) as described
in \citet{LAT_collaboration_extended_search_2012}.  The best-fit
extended source parameters are then given to \gtlike, which is used
to fit the spectral parameters and the significance of the extension
over a point source, $TS_{\rm ext}$, evaluated as described in
\citet{LAT_collaboration_extended_search_2012}.  
That paper established that $TS_{\rm ext}\geq16$ means highly probable source extension.
In the present work we aim only to flag possible extension, and use  $TS_{\rm ext}\geq9$.

To test for variability, even without significant emission over the 3-year time
range, we divide the dataset into 36 intervals and fit the point-source
flux independently in each interval, 
computing $TS_{\rm var}$ as in 2FGL.  
For sources with potential magnetospheric off-peak emission and for
regions with no detection, we performed the test at the pulsar's position.
Otherwise, we test at the best-fit position.
The off-peak emission is said
to show evidence for variability if $TS_{\rm var}\geq91.7$, corresponding
to a $4\sigma$ significance.  
As noted in Section \ref{obsvSection}, our timing solutions for PSRs J0205+6449 and J1838$-$0537 are not  
coherent across all three years.  
For these two pulsars, we excluded the time ranges without ephemerides  
and only tested for variability during months that were completely covered.  
For J1838$-$0537 only one month is lost, whereas for J0205+6449 the 7\% data loss is spread across
three separate months.
As a result, $TS_{\rm var}$ for these pulsars is a conservative estimate of variability significance.

The procedure described above, especially the extension analysis,
is particularly sensitive to sources not included in 2FGL that are near the pulsar of interest,
for two reasons.
First, we are using an additional year of data and second,
when ``turning off'' a bright pulsar nearby, faint sources become more
important to the global fit.  Therefore, in many situations we had to run
the analysis several times, iteratively improving the model by including
new sources, until we removed all $TS>25$ residuals. The final
\gtlike-formatted XML source model for each off-peak 
region is included in the auxiliary material.

There are still, however, pulsars for which we were unable to obtain
an unbiased fit of the off-peak emission,
most likely due to inaccuracies in the model of the Galactic diffuse emission
and incorrectly modeled nearby sources.  The most common symptom
of a biased fit is an unphysically large extension.  In these cases,
the extended source attempts to account for multiple point sources or
incorrectly-modeled diffuse emission,
not just the putative off-peak emission.  Systematics associated
with modeling extended sources are discussed more thoroughly in
\citet{LAT_collaboration_extended_search_2012}.  For the purposes of this
catalog, we have flagged the pulsars where off-peak analysis suffered from
these issues and do not attempt a complete understanding of the emission.

\subsection{Off-peak Results}
\subseclabel{off_peak_results}

\begin{figure}
  \ifdefined\bwfigures
  \plotone{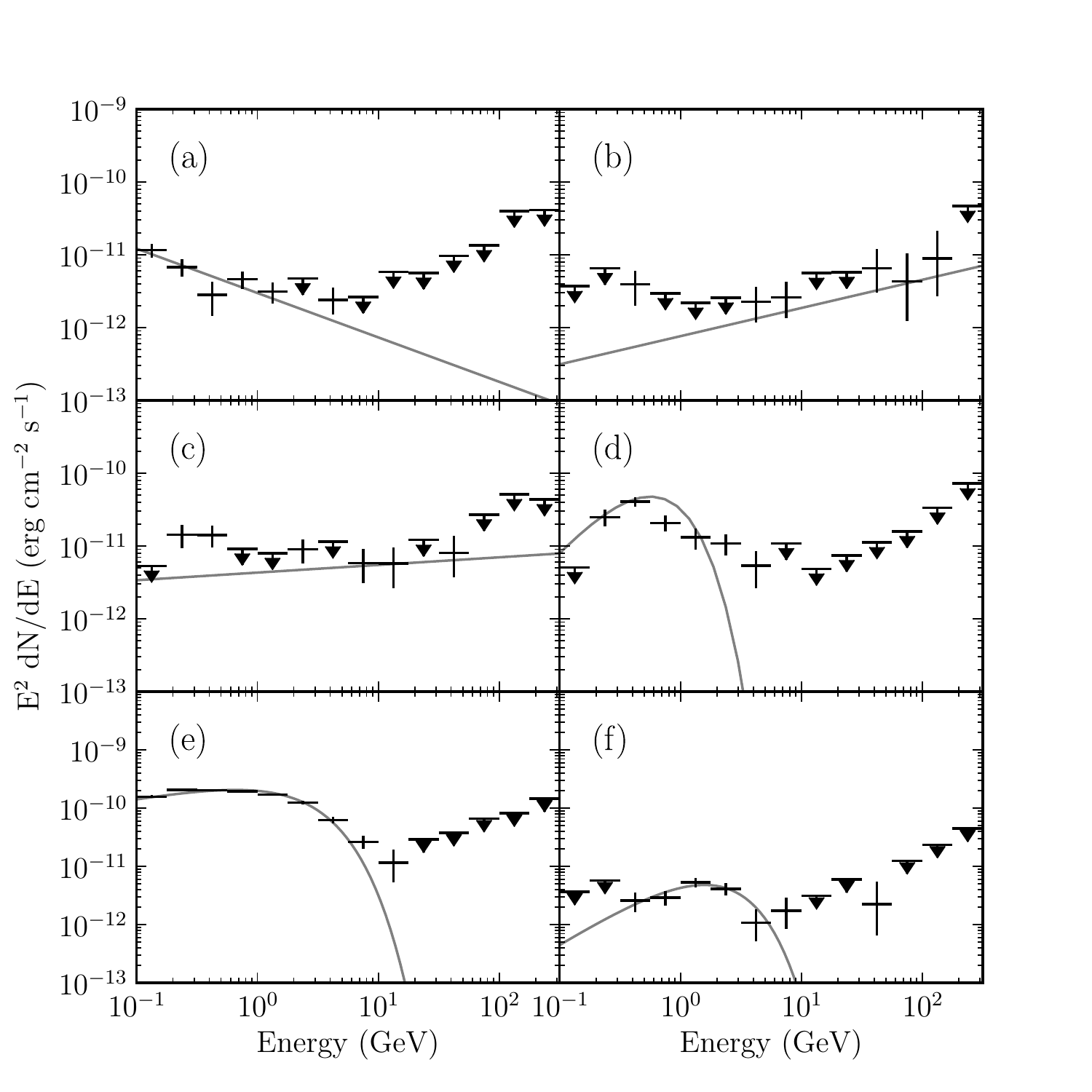}
  \else
  \plotone{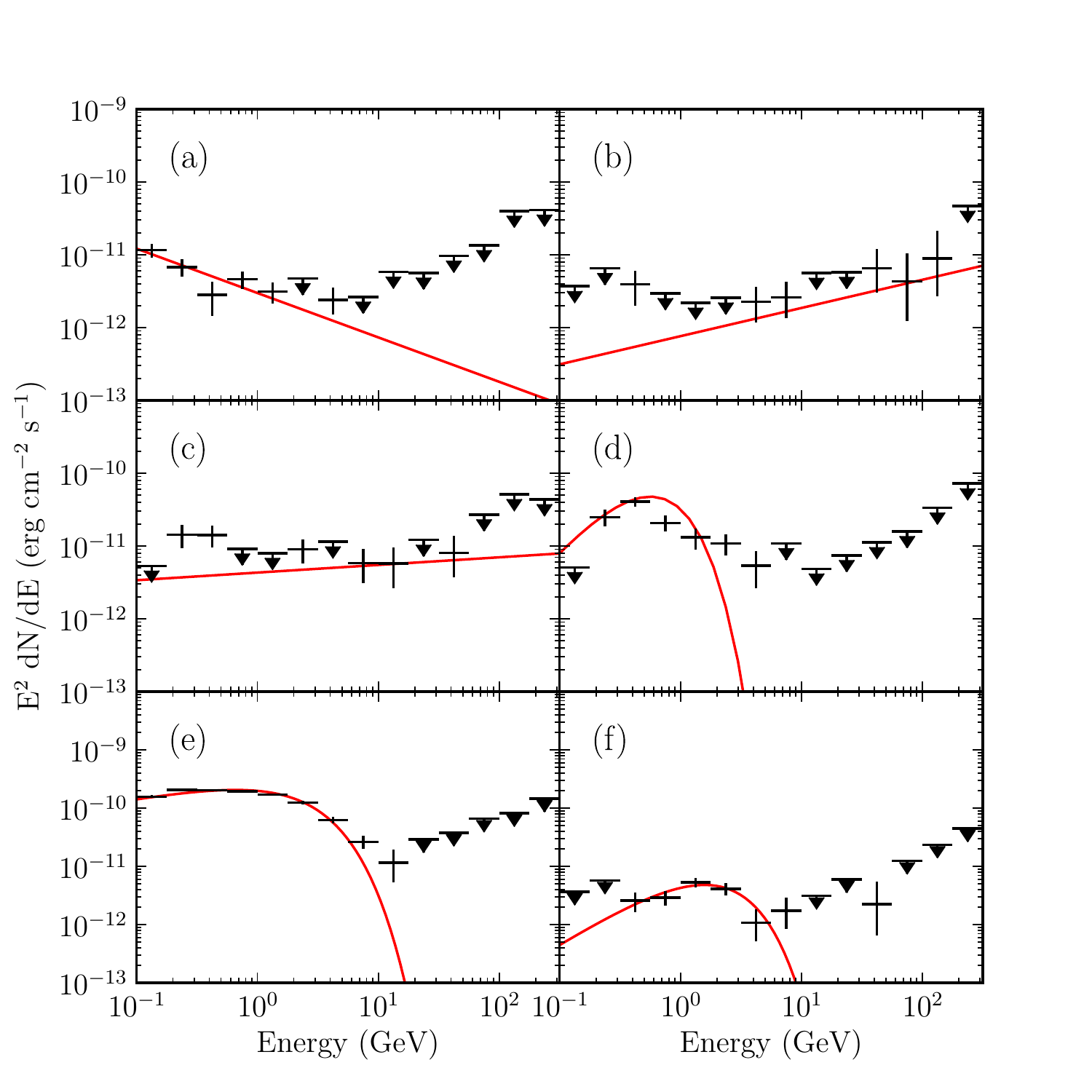}
  \fi
  \caption{Spectral energy distributions for the off-peak
  phase intervals around
  (a) PSR J0007+7303 (b) PSR J0205+6449, (c) PSR J1410$-$6132,
  (d) PSR J1747$-$2958, (e) PSR J2021+4026, and (f) PSR J2124$-$3358.
  We plot a detection in those energy bands in which the source is found with $TS\geq4$ (a $2\sigma$ detection) and report a Bayesian 95\% confidence-level upper limit otherwise.  The best-fit spectral model, using the full energy range, is also shown for comparison.
  }
  \label{off_peak_seds}
\end{figure}

The off-peak intervals of 54 LAT-detected pulsars have
been evaluated by \citet{LAT_collaboration_PWNCAT_2011} using 16 months of sky survey observations.  
This led to the discovery of PWN-like emission
in the off-peak interval of PSR J1023$-$5746, coincident with HESS
J1023$-$575, and identification of 5 pulsars that appear to have near 100\%
duty cycles.  Our results, summarized in Table \ref{tbl-off_peak_table},
extend the analysis to 116 pulsars over 3 years of data.  
Sample off-peak spectra are shown in Figure \ref{off_peak_seds}.
Using the procedures outlined in Sections \ref{peak_definition}
and \ref{off_peak_analysis}, we have identified 34 pulsars that
have significant emission ($TS\geq25$) in their off-peak intervals.
We classify the likely nature of the emission as follows.

If the spectrum cuts off ($TS_{\rm cut}\geq9$), the emission could be magnetospheric (`M').
An indication of spatial extension ($TS_{\rm ext}\geq9$) flags sources 
where the emission may instead be an artifact of defects in the Galactic diffuse
emission model and we list such sources as type `U', for `unidentified'. 
Similarly, emission from sources without evidence for a spectral cutoff 
could originate in the pulsar wind, type `W'. 
Spatial extension alone is not a sufficient indicator, since the LAT PSF is larger than many PWNe. 
A hard spectral index also suggests a PWN contribution.
The table lists the four solid PWN detections.
PSR J0205+6449 in 3C58 is a new detection at GeV energies.
Only one of the four, the previously identified Vela-X PWN \citep{LAT_collaboration_Vela_X_2010}, 
is spatially extended for the LAT. 

We identify 19 type `U' regions, and 11 type `M' sources,
significantly expanding the number of pulsars that perhaps have detectable magnetospheric emission across all rotational phases. 
One caution is that many of these `M' pulsars, especially the young objects, are in regions of particularly bright diffuse gamma-ray
emission, where small fractional uncertainties in the level of diffuse emission can account for much of the apparent unpulsed emission. 
However, if established as true magnetospheric components, these will be important test cases for
pulsar emission models. 
For type `M' and `U' sources, we present the best-fit spectral parameters using
a point source at the pulsar's position with a PLEC1 spectral model in
Table \ref{tbl-off_peak_table}.  For all other
sources (except the Crab Nebula described in Section~\ref{off_peak_analysis}), 
we present the spectral results using a power-law spectral
model and the best-fit spatial representation.

For a few sources, the spectral analysis
performed here disagrees with that in
\citet{LAT_collaboration_PWNCAT_2011}. For soft and faint sources
(including J1044$-$5737 and J1809$-$2332), the spectral discrepancy is
mainly caused by our use of a newer Galactic diffuse model. At lower
energies, small changes in the diffuse model can have a significant
impact on the analysis of a region.  For bright magnetospheric sources
(including J0633+1746 and J2021+4026), the spectral discrepancy is mainly
due to using different phase ranges (see Section~\ref{peak_definition}).

\figref{off_peak_luminosity_vs_edot} shows that only
a small fraction of the spindown power goes into 
the gamma-ray emission from LAT-detected PWNe.
Similarly, 
Figure~\ref{P0_sqEDOTD2} shows that the LAT only detects
PWNe from the youngest pulsars with the highest spindown power.
GeV emission from the Crab Nebula is highly time variable (Section \ref{off_peak_analysis}).  
Indeed, we find
$TS_{\rm var}=373$ for the Crab Nebula; however
no other source demonstrated flux variability (all have $16 <TS_{\rm var} <65)$.  
Other GeV PWNe may be variable, but the combination of lower fluxes
and less-extreme variations limits our ability to identify them as such.

The off-peak results for several interesting sources are
presented in Appendix~\ref{App-off_peak_individual_source_discussion}.
The complete off-peak search results can be found in the 
auxiliary information described in Appendix \ref{online}.
For regions where we find $TS<25$, the auxiliary information contains
upper limits computed for both a power-law spectral model and a
PLEC1 model with $E_{\rm cut}=3$ GeV and $\Gamma=1.7$.  The auxiliary
information also contains $TS_{\rm var}$ for each off-peak interval.

\begin{deluxetable}{ll*{8}c}
\tablecolumns{8}
\tablewidth{0pt}
\tabletypesize{\scriptsize}
\tablecaption{Off-Peak Spatial and Spectral Results
\label{tbl-off_peak_table}
}
\tablehead{
\colhead{PSR} & \colhead{Type} & \colhead{$\tspoint$} & \colhead{$\tsext$} & \colhead{$\tscutoff$}& \colhead{Energy Flux} & \colhead{$\Gamma$} & \colhead{$\Ecutoff$}\\ 
\colhead{ }   & \colhead{ }    & \colhead{ }          & \colhead{ }        & \colhead{ }          & \colhead{($10^{-11}\,\efluxunits$)} & \colhead{ } & \colhead{(GeV)}}
\startdata
\multicolumn{8}{c}{Young Pulsars} \\
\hline
J0007+7303   & U & 71.4 & 10.8 & 0.0   & $1.98 \pm 0.26$ & $2.61 \pm 0.14$ &  \\
J0205+6449   & W & 33.7 & 0.5  & 0.0   & $1.75 \pm 0.68$ & $1.61 \pm 0.21$ &  \\
J0534+2200   & W & 5247.& 0.0  & 0.3   & $67.2 \pm 1.6$ & \tablenotemark{a} &  \\
J0631+1036   & U & 33.1 & 0.0  & 5.4   & $1.70 \pm 0.33$ & $2.38 \pm 0.14$ &  \\
J0633+1746   & M & 3666.& 2.3  & 239.  & $41.4 \pm 1.1$ & $1.37 \pm 0.09$ & $0.93 \pm 0.10$ \\
J0734$-$1559 & U & 28.3 & 12.4 & 30.8  & $1.61 \pm 0.24$ & $0.01 \pm 0.08$ & $0.17 \pm 0.03$ \\
J0835$-$4510 & W & 473. & 283. & 22.8  & $30.3 \pm 1.2$ & \tablenotemark{b} &  \\
J0908$-$4913 & U & 65.1 & 41.4 & 60.4  & $3.04 \pm 1.07$ & $0.15 \pm 0.59$ & $0.30 \pm 0.01$ \\
J1023$-$5746 & U & 59.7 & 30.0 & 10.9  & $5.35 \pm 1.17$ & $0.57 \pm 0.80$ & $0.49 \pm 0.21$ \\
J1044$-$5737 & U & 42.0 & 98.1 & 22.4  & $3.12 \pm 0.75$ & $0.80 \pm 0.93$ & $0.40 \pm 0.18$ \\
J1105$-$6107 & U & 33.3 & 37.5 & 21.7  & $3.81 \pm 0.77$ & $0.92 \pm 0.56$ & $0.48 \pm 0.22$ \\
J1112$-$6103 & U & 65.0 & 71.1 & 0.9   & $5.10 \pm 0.74$ & $2.17 \pm 0.09$ &  \\
J1119$-$6127 & U & 61.3 & 1.0  & 0.9   & $4.11 \pm 0.63$ & $2.22 \pm 0.09$ &  \\
J1124$-$5916 & M & 95.9 & 0.0  & 18.2  & $2.87 \pm 0.71$ & $1.31 \pm 0.91$ & $1.43 \pm 1.42$ \\
J1410$-$6132 & U & 27.5 & 71.2 & 0.4   & $4.29 \pm 1.05$ & $1.90 \pm 0.15$ &  \\
J1513$-$5908 & W & 102. & 3.5  & 0.0   & $4.95 \pm 0.83$ & $1.78 \pm 0.12$ &  \\
J1620$-$4927 & M & 28.9 & 0.5  & 35.2  & $5.25 \pm 0.96$ & $0.35 \pm 0.94$ & $0.57 \pm 0.29$ \\
J1746$-$3239 & U & 53.3 & 34.3 & 34.2  & $3.65 \pm 0.59$ & $0.94 \pm 0.31$ & $0.60 \pm 0.10$ \\
J1747$-$2958 & M & 45.5 & 5.4  & 49.8  & $8.41 \pm 2.84$ & $0.02 \pm 0.32$ & $0.28 \pm 0.01$ \\
J1809$-$2332 & U & 32.5 & 13.6 & 21.9  & $4.10 \pm 0.80$ & $0.24 \pm 0.83$ & $0.31 \pm 0.11$ \\
J1813$-$1246 & M & 62.8 & 0.0  & 9.0   & $6.31 \pm 1.40$ & $1.60 \pm 0.73$ & $0.99 \pm 0.95$ \\
J1836+5925   & M &10407.& 0.0  & 365.  & $36.9 \pm 0.7$ & $1.47 \pm 0.03$ & $1.98 \pm 0.09$ \\
J1838$-$0537 & U & 51.3 & 32.9 & 21.9  & $8.35 \pm 1.31$ & $1.39 \pm 0.54$ & $2.55 \pm 2.48$ \\
J2021+4026   & M & 1717.& 8.7  & 244.  & $64.0 \pm 1.4$ & $1.64 \pm 0.02$ & $1.82 \pm 0.04$ \\
J2032+4127   & U & 53.6 & 76.1 & 1.5   & $4.36 \pm 0.77$ & $2.07 \pm 0.12$ &  \\
J2055+2539   & M & 123. & 0.0  & 30.0  & $1.63 \pm 0.19$ & $1.05 \pm 0.28$ & $0.64 \pm 0.12$ \\
\cutinhead{Millisecond Pulsars}
J0034$-$0534 & U & 41.0 & 0.0 & 6.0  	& $0.82 \pm 0.16$ & $2.40 \pm 0.19$ &  \\
J0102+4839   & U & 49.7 & 0.0 & 7.4  	& $1.29 \pm 0.20$ & $2.51 \pm 0.14$ &  \\
J0218+4232   & U & 50.1 & 0.0 & 6.8  	& $2.13 \pm 0.33$ & $2.72 \pm 0.26$ &  \\
J0340+4130   & M & 26.9 & 0.1 & 16.3 	& $0.53 \pm 0.11$ & $0.02 \pm 0.22$ & $0.94 \pm 0.28$ \\
J1658$-$5324 & U & 42.3 & 0.0 & 1.9  	& $1.69 \pm 0.29$ & $2.52 \pm 0.76$ &  \\
J2043+1711   & U & 52.5 & 0.0 & 8.8  	& $1.46 \pm 0.27$ & $2.29 \pm 0.14$ &  \\
J2124$-$3358 & M & 129. & 0.0 & 19.8 	& $1.08 \pm 0.15$ & $0.70 \pm 0.51$ & $1.21 \pm 0.49$ \\
J2302+4442   & M & 114. & 0.0 & 9.8  	& $1.45 \pm 0.20$ & $1.54 \pm 0.40$ & $1.61 \pm 0.82$ \\
\enddata

\tablenotetext{a}{The spectral shape of the Crab Nebula was taken from \citet{LAT_collaboration_crab_2012}.}
\tablenotetext{b}{The spectral shape of Vela-X was taken from \cite{FermiVelaX2nd}.}

\tablecomments{Off-peak regions with a significant detection of emission.
The source classification is `M' for likely magnetospheric, 
`W' for likely pulsar wind, and `U' for unidentified.
The table includes the significance of the source ($TS$),  
of the source extension ($TS_{\rm ext}$), and of a spectral cutoff ($TS_{\rm cut}$).
The best-fit energy flux and photon index are computed in the energy range from 100 \mev to 316 \gev.
Exponential cutoff energies are listed for sources with large $TS_{\rm cut}$. 
The quoted errors are statistical only. A few sources are discussed in Appendix~\ref{App-off_peak_individual_source_discussion}.
}

\end{deluxetable}

\begin{figure}
  \ifdefined\bwfigures
  \includegraphics{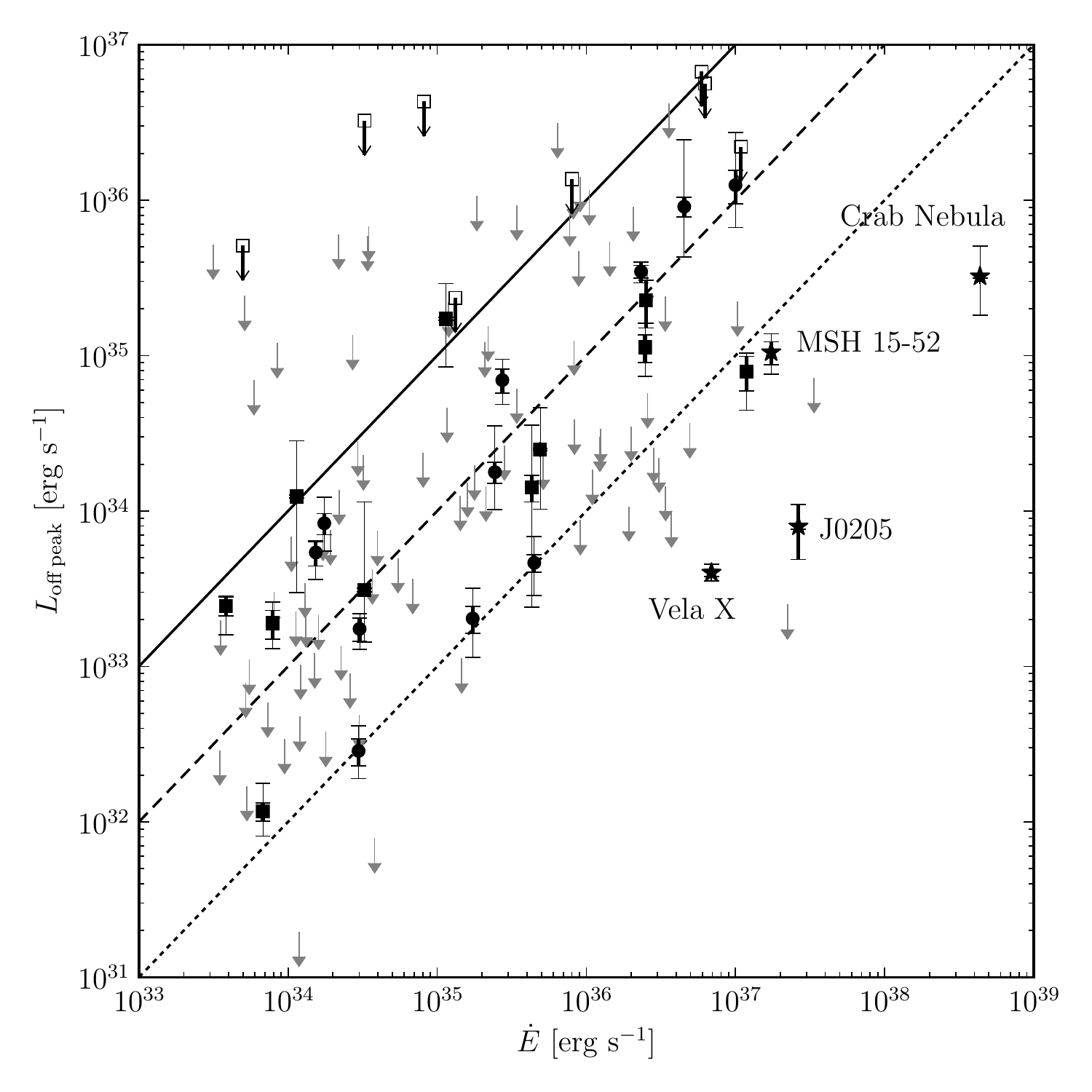}
  \else
  \includegraphics{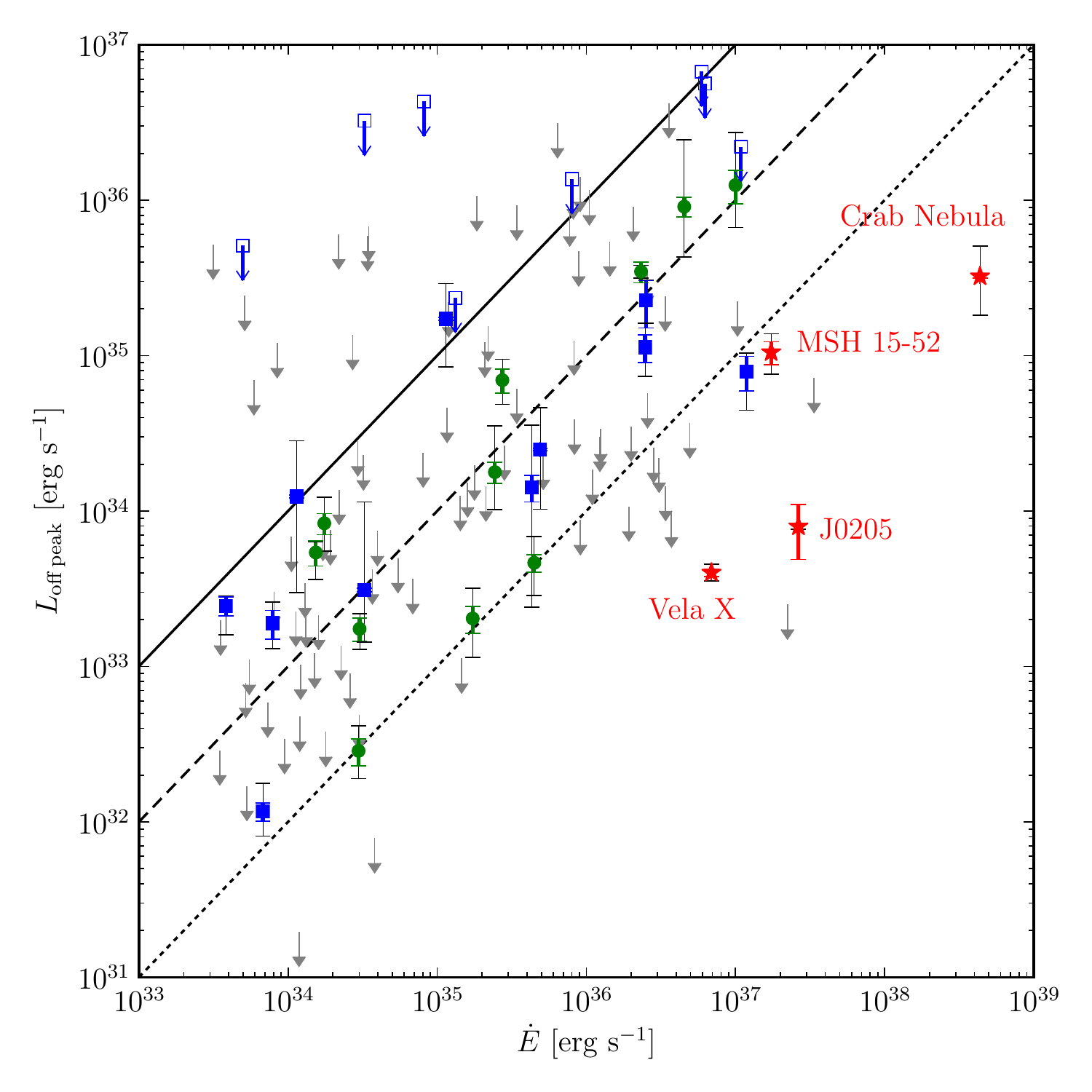}
  \fi
  \caption{The off-peak luminosity compared to the observed pulsar spindown power. 
  The luminosity is computed and plotted with the same convention as
  Figure~\ref{EDotLumG}. A luminosity upper limit is plotted
  when there is no significant off-peak emission or when there
  is only a distance upper limit.
  The
  star-shaped markers (colored red in the online version) represent
  type `W' sources, the square-shaped markers (colored blue) represent type `M' sources, 
  circular markers (colored green) represent type `U' sources,
  and the gray arrows represent non-detections.
  The filled blue square-shaped markers represent `M' sources
  with a detected luminosity and the unfilled markers represent luminosity
  upper limits where there is only a distance upper limit.
  The solid, dashed, and dotted diagonals show 100\%, 10\%, and 1\% efficiency
  (respectively).
  }
  \figlabel{off_peak_luminosity_vs_edot}
\end{figure}

\section{The Pulsars Not Seen}
\label{not_seen}
This catalog is a milestone in the progress toward the long-term goal of acquiring the most uniform sample of neutron stars possible, so that comparisons
with model predictions \citep[e.g.,][]{Gonthier07,2011ApJ...727..123W,Pierbattista2012} will allow improved tests of emission models and of 
their links with their parent population of massive stars or with diffuse Galactic emission.
Selection biases can be subtle and the advantage of pulsar searches in the coming years is not so much to increase
the absolute numbers, but to be sure to have explored the dark corners of parameter space. 
Continued support from pulsar radio astronomers is crucial to maintain sensitivity to the more unusual gamma-ray pulsars in the coming years of the \textit{Fermi} mission.
Here, we consider pulsars that might have been expected to be seen with the LAT, but were not, to highlight ``gamma-quiet'' or ``sub-luminous'' pulsars.

%

\subsection{High Spindown Power Pulsars}	
\label{undetect}
\begin{figure}[!ht]
\centering
\includegraphics[width=0.8\textwidth]{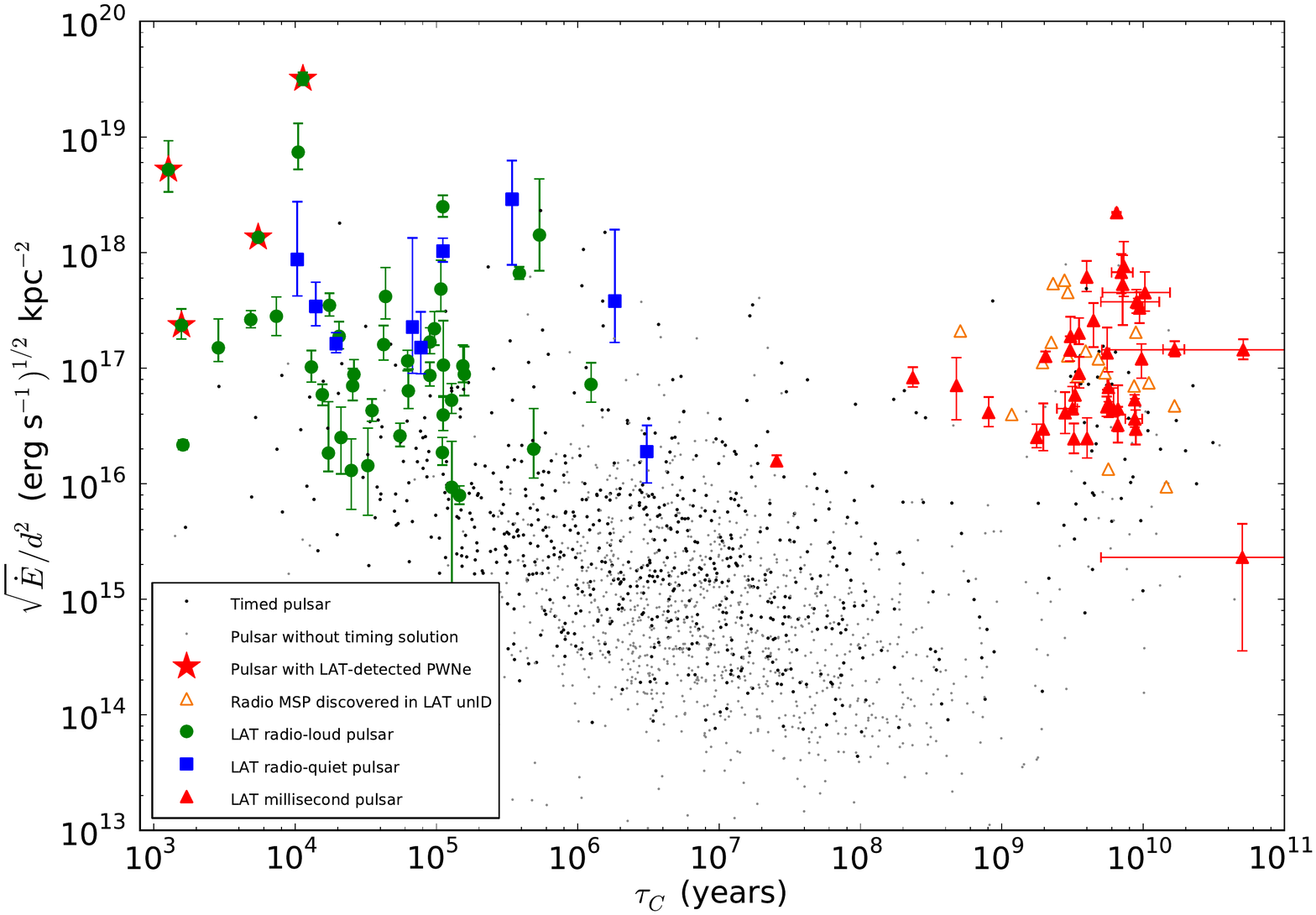}
\caption{Figure-of-merit for the gamma-ray flux from a given pulsar, $\sqrt{\dot E} / d^2$, versus the pulsar characteristic age $\tau_c$.
For comparison with the integral energy flux $G_{100}$, a scale factor $2.65\times10^{-28}$ can be applied to the y-axis 
(see Section \ref{undetect}).
The markers are the same as in Figure \ref{PPdotplot}, with red stars added for the four pulsars associated with GeV PWNe (see Section \ref{unpulsed}). 
Black and grey dots include all $\dot E$ values, even though only high $\dot E$ pulsars have been seen in gamma rays to date.
\label{P0_sqEDOTD2}}
\end{figure}

Of the 64 known RPPs with $\dot E > 10^{36}$ erg s$^{-1}$, Table \ref{Edot36Tab} lists the 28 for which we did \textit{not} see gamma-ray pulsations
when the data set for this paper was frozen.
When no steady LAT point source lies within $0\fdg2$ we provide a 95\% confidence level upper limit (UL) on $G_{100}$ (Section \ref{sensitivity}). 
The Galactic latitude $b$ roughly indicates the diffuse background level.
The last column compares the UL with the ``heuristic'' energy flux, $\sqrt{\dot E} /(4\pi d^2)$.
Figure \ref{P0_sqEDOTD2} plots $\sqrt{\dot E} /d^2$ for all pulsars. Note the absence of the factor $4\pi$ in the latter case.
To convert the plot's scale, in units of erg$^{1/2}$ s$^{-1/2}$ kpc$^{-2}$, to the units used for $G_{100}$ and the flux ULs,
using $L_\gamma^h$, (Eq. \ref{heurLumEq}), we multiply the scale of Figure \ref{P0_sqEDOTD2} by
\begin{equation}
 {\sqrt{10^{33}\, {\rm erg \,s}^{-1}} \over 4\pi (3.08\times10^{21} \, {\rm cm \, kpc}^{-1})^{2} } 
 = 2.65\times10^{-28}\, {\rm erg}^{1/2} \,{\rm s}^{-1/2}\, {\rm kpc}^{2} \, {\rm cm}^{-2}.
\end{equation}

For several of the pulsars, the predicted flux is less than twice the UL. These pulsars seem to be near our current sensitivity limit,
sensitive to distance, beaming, or local background uncertainties. 
We highlight a few pulsars from Table \ref{Edot36Tab}, in order of decreasing spindown power.
The large distance (54 kpc) to the Large Magellanic Cloud (LMC) is the simplest explanation for the non-detections of PSRs J0537$-$6910 and J0540$-$6919.
Using highly accurate rotation ephemerides based on RXTE X-ray observations and varied data
selection cuts, we confirm the non-detection reported by \cite{LAT_LMC}, with over three times as much data. 

\clearpage
\tabletypesize{\scriptsize}
\begin{deluxetable}{lccrcc}
\tablewidth{0pt}
\tablecaption{Undetected RPPs with spindown power $\dot E > 10^{36}$ erg s$^{-1}$
\label{Edot36Tab}}

\tablehead{
\colhead{PSR} & \colhead{$\dot E$}  & \colhead{Distance}& \colhead{$b$}        & \colhead{Flux UL\tablenotemark{a}}                & \colhead{$\sqrt{10^{33}\dot E}/(4\pi d^2)$} \\
              & \colhead{($10^{36}$ erg s$^{-1}$)}            & \colhead{(kpc)}   & \colhead{($^\circ$)} & \colhead{($10^{-12}$ erg cm$^{-2}$ s$^{-1}$)} & \colhead{($10^{-12}$ erg cm$^{-2}$ s$^{-1}$)} } 

\startdata

 J0537$-$6910 & 488  &  53.7   & $-$31.7 & LMC          & 2 \\ 
 J0540$-$6919 & 146  &  53.7   & $-$31.5 & LMC          & 1 \\ 
 J1813$-$1749 &  56  &  $>5$\tablenotemark{b}   & $-$0.02 & $<$26    &$<80$ \\ 
 J1400$-$6325 &  51  &  11.3   & $-$1.6  & $<6$       & 15 \\ 
 J1747$-$2809 &  44  &  13.3   & 0.08  & $<15$       & 10 \\ 
 J2022+3842   &  30  &  10.0   & 0.96  & SNR G$76.9+1.0$ & 14 \\ 
 J1617$-$5055 &  16  &   6.8   & $-$0.28 & PWN\tablenotemark{c}      & 23 \\ 
 J1930+1852   &  12  &   7.0   & 0.27  & $<16$      & 18 \\ 
 J1849$-$0001 & 9.8  &  7.0   &  0.53  & $<15$                & 17   \\ 
 J1846$-$0258 & 8.1  &  5.8   & $-$0.24  & $<26$                 &  22 \\ 
 J1811$-$1925 & 6.5  &  5\tablenotemark{b}  & $-$0.35  &  $<24$                & 27  \\ 
 J1838$-$0655 & 5.6  &\nodata & $-$0.20  & UnId\tablenotemark{c}  &  \nodata  \\ 
 J1856$+$0245 & 4.6  &  9.0   & 0.06   & $<27$                &   7 \\ 
 J1935$+$2025 & 4.6  &  6.2   & $-$0.20  & $<18$                 &  15  \\ 
 J1524$-$5625 & 3.2  &  2.8   & 0.35   & $<6$                  &  61 \\ 
 J1913$+$1011 & 2.9  &  4.8   & $-$0.17  & $<23$                 &  20 \\ 
 J1826$-$1334 & 2.8  &  3.9   & $-$0.69  & PWN\tablenotemark{c}  &  29 \\ 
 J1803$-$2137 & 2.2  &  4.4   & 0.15   & J1803.3$-$2148 (110)       &  20 \\ 
 J1837$-$0604 & 2.0  &  6.4   & 0.27   & $<31$                 &  9 \\ 
 J1809$-$1917 & 1.8  &  3.5   & 0.08   &  $<26$                &  28 \\ 
 J1301$-$6305 & 1.7  &  6.7   & $-$0.24  &  $<14$                &   8 \\ 
 J1614$-$5048 & 1.6  &  7.9   & 0.17   &   $<31$               &   5 \\ 
 J1828$-$1101 & 1.6  &  6.6   & 0.04   & $<23$                 &   8 \\ 
 J1928$+$1746 & 1.6  &  5.8   & 0.11   & J1928.8+1740 (33)    &  10 \\ 
 J1341$-$6220 & 1.4  &  11.1  & $-$0.04  & $<18$                &   3 \\ 
 J1437$-$5959 & 1.4  &  8.1   & 0.23   & $<19$                &   5 \\ 
 J0855$-$4644 & 1.1  &  $<0.9$\tablenotemark{b}    & $-$1.0   &  $<10$        &  $>$340 \\ 
 J1831$-$0952 & 1.1  &  4.0   & $-$0.13  & $<17$                 &  17 \\ 

\enddata

\tablenotetext{a}{An energy flux upper limit is given if the pulsar is not within $0\fdg2$ of a LAT 2FGL source; otherwise the source name is given, with the integral energy flux above 100 MeV in parentheses. LMC is the Large Magellanic Cloud \citep{LAT_LMC}.}
\tablenotetext{b}{Distance constraint references:
PSR J0855$-$4644 from \citet{Acero2013}; 
PSR J1811$-$1925 by association with SNR G11.2$-$0.3 \citep{TamRoberts2003}; 
PSR J1813$-$1749 from \citet{Halpern2012}.
}
\tablenotetext{c}{The LAT detects a (possible) PWN extending to the pulsar position: 
HESS J1616$-$508, centered $\approx 0\fdg2$ away \citep{LAT_collaboration_extended_search_2012};
HESS J1825$-$137, centered $\approx 0\fdg5$ away \citep{LAT_collaboration_HESS_J1825_2011};
HESS J1837$-$069, centered $\approx 0\fdg4$ away, 
may be a PWN powered by this pulsar \citep{LAT_collaboration_extended_search_2012}.}

%
%
%

\tablecomments{Of the 64 known RPPs with spindown power $\dot E > 10^{36}$ erg s$^{-1}$,
the above 28 were unpulsed in GeV gamma rays as this catalog was being prepared. 
Column 3 is distance (the ATNF database DIST1 parameter unless noted otherwise) and Column 4 is Galactic latitude. 
The upper limit in Column 5 is calculated using the all-sky model as described in Section \ref{sensitivity}. 
The heuristic spindown luminosity in Column 6 corresponds to the lower diagonal in Figure \ref{EDotLumG}.
}

\end{deluxetable}

\normalsize

PSR J2022+3842 was thought to have the second-highest spindown power, after the Crab, of any known pulsar in the Milky Way \citep{J2022p3842discovery}. 
However, recent XMM data revealed that the period is 48 ms, twice the period previously seen with RXTE and GBT, 
with a $4\times$ smaller $\dot E$ value\citep{J2022p3842update}.
Its DM distance is 10 kpc but, situated in the heart of the Cygnus region, it (or PSR J2021+3651) may be as close as 2 kpc. 
For $d = 10$ kpc, $\sqrt{\dot E} / d^2$ is $\approx 5\times 10^{16}$ erg $^{1/2}$ s$^{-1/2}$ kpc$^{-2}$. The LAT detects a source $0\fdg06$ away, with a
spectrum adequately modeled by a power law with or without an exponential cutoff. The pulsar is difficult to detect in radio and suffers large
timing noise; consequently, phase-folding the LAT data is difficult and blind period searches of the LAT data are hampered by the intense background.
If unpulsed emission is confirmed, this could be an example of magnetospheric emission with low modulation and 100\% duty cycle.
The second-highest spindown power now belongs to the radio-quiet, X-ray PSR J1813$-$1749. It has a highly uncertain distance, probably greater than 5 kpc \citep{Halpern2012},
and is undetected with the LAT.

PSRs J1400$-$6325 and J1747$-$2809 are also distant, with slightly smaller $\sqrt{\dot E} / d^2$. 
The latter lies toward the Galactic center where the diffuse background level is intense and source density is high.  The former is undetected in radio and was timed in
X-rays; however, phase-connected ephemerides cover the \Fermi mission epoch only partially. The nearest source with $TS\geq25$ in an internal 3-year source list, constructed in a similar fashion to 2FGL, is more than 1$\degr$ away. Curiously, Geminga and PSR J0007+7303 are the only X-ray loud, radio-quiet pulsars detected by the LAT. 

Both J1617$-$5055 and J1930+1852 are roughly in the same $\sqrt{\dot E} / d^2 \approx 10^{17}$  erg $^{1/2}$ s$^{-1/2}$ kpc$^{-2}$ range as the previous pulsars.
J1617$-$5055 has particularly strong timing noise and the weekly to monthly Parkes observations did not allow a phase-connected timing solution covering the mission epoch. 
The rotation ephemeris was finally obtained in 2011 but gamma-ray phase-folding reveals no hint of pulsations\footnote{R. Shannon, private communication}.
The nearest LAT source ($0\fdg2$ away) is the PWN HESS J1616$-$508 \citep {LAT_collaboration_extended_search_2012}.
PSR J1930+1852 has an accurate radio ephemeris but no hint of gamma-ray pulsations.
The nearest 2FGL source with $TS\geq25$ is $0\fdg55$ away, nine times the 95\% error radius for that source.
Of the 14 pulsars for which \citet{Ng2008} fit the nebular torus seen in X-rays, PSR J1930+1852, in PWN G$54.1+03$ \citep{Rousseau2013}, 
has the smallest angle between the spin axis and the line of sight. 
The radio beam intersects the Earth but the equatorial gamma-ray emission may not.  

Skipping down the list, PSR J1928+1746 is within twice the 95\% error radius of a source in an internal 3-year source list,
but phase-folding provides no significant pulsations.
Similar explorations for $10^{35} < \dot E < 10^{36}$ erg s$^{-1}$ led to the discovery of PSR J1913+0904 as a gamma-ray pulsar,
listed in Table \ref{Beyond117}.
Three pulsars for which the nearby high-$TS$ LAT source is a gamma-ray pulsar are PSRs J1524$-$5625, J1803$-$2137, and J1831$-$0952. 
Searching the off-peak phase interval of the nearby pulsar reduces background but did not allow us to detect pulsations.
PSR J1524$-$5625 bears special mention because of the large spread between the upper limit and the heuristic energy flux.
If the NE2001 DM distance is accurate this could be a gamma-quiet pulsar candidate. The same is also true for PSR J0855$-$4644, 
again with a distance caveat.

In 2FGL, 83 sources have a `PSR' identification and are in this catalog. An additional 27 sources have a
`psr' association, meaning that the pulsar lies within the error ellipse of the source but $5\sigma$ pulsations were not seen when 2FGL was
written. Since then, pulsations for 12 of the 27 have allowed firm identification, included in this catalog. Of the remaining 15, 12 are radio
MSPs discovered at LAT source positions (Section \ref{psc}). 
The remaining three spatial associations of young RPPs with 2FGL sources are PSRs J1632$-$4818, J1717$-$5800, and J1928+1746.
The last was discussed above.
PSR J1632$-$4818 is a typical gamma-ray candidate ($\dot E = 4.8\times 10^{34}$ erg s$^{-1}$, $d\approx 8$ kpc, $b = -0\fdg21$) 
except that it has one of the highest surface B-fields of any RPP. It shows no hint of pulsations and the 2FGL association could be a chance spatial coincidence,
or PSR J1632$-$4818 may be a candidate for unpulsed magnetospheric emission.
PSR J1717$-$5800 is almost certainly a chance spatial coincidence: it is well below the empirical deathline ($\dot E = 2.3\times 10^{32}$ erg s$^{-1}$) and not nearby ($d \approx 3.5$ kpc),
although with low background so far from the Galactic plane ($b = -11\fdg5$). Folding with an archival ephemeris showed no hint of pulsations. 

Figure \ref{P0_sqEDOTD2} illustrates the utility of the commonly used detectability
metric $\sqrt{\dot E}/d^2$, showing an approximate LAT threshold of $\sim 10^{16} {\rm \, erg^{1/2}s^{-1/2}kpc^{-2}}$. While the difficulty of establishing
reliable distances for many pulsars (Section~\ref{Distances}) makes quantitative comparison
challenging, it is clear that this is a good predictor of pulsar detectability.
However, as emphasized by \citet{subluminous} 
there are a number of pulsars with a high detectability metric not seen
by the LAT (Section \ref{not_seen}), indicated by black dots above the LAT threshold in Figure \ref{P0_sqEDOTD2}. 
In some cases, underestimated distances may explain the non-detections; however, in other 
cases the pulsars have accurate parallax measurements and the gamma-ray beam 
must either miss Earth (e.g., a pulsar viewed at small $\zeta$ for outer-magnetosphere models) 
or have light curves with very small modulation amplitude (e.g., a large, unpulsed component for SG or aligned 
polar cap models). We note that in many cases no LAT flux is detected 
in the pulsar direction, so beaming presents the most likely explanation.

In summary: many non-detections are due to large distances and/or background. In a few rare cases, our rotation ephemeris allows
inadequate phase-folding. However, we are also accumulating a sample of ``gamma-quiet'' pulsars as well as a sample of possibly unpulsed gamma-ray pulsars, that is, for which the emission from the magnetosphere is unmodulated.



\subsection{Flux Upper Limits and Sensitivity}
\label{sensitivity}
\begin{figure}[!ht]
\centering
\includegraphics[width=1.0\textwidth]{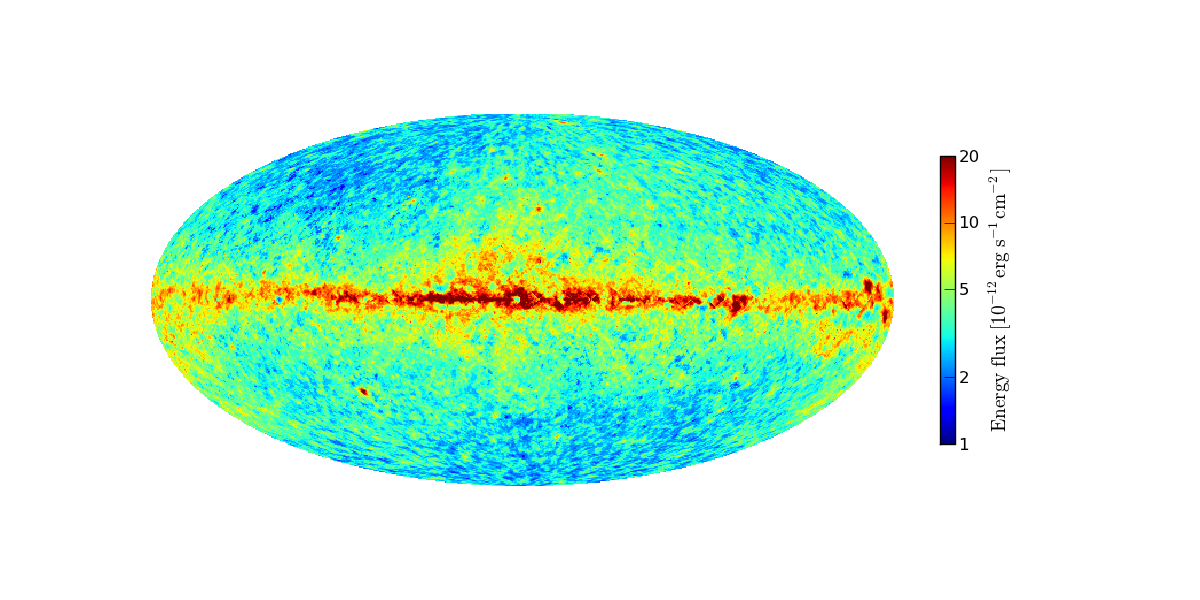}
\caption{Hammer-Aitoff projection of the LAT 3-year sky-survey energy flux sensitivity above 100 MeV,
assuming a pulsar-like exponentially cutoff power law energy spectrum.
\label{SeattleSkySensitivity}}
\end{figure}

Figure \ref{SeattleSkySensitivity} maps the LAT sensitivity on the sky for the phase-averaged detection of a point source with a pulsar-like spectrum,
for the 3-year data set.
To build the map, we started from the all-sky source model. 
For each point on a $0\fdg15$ grid we added an additional pulsar-like point source with fixed 
parameters of $\Gamma = 1.8$ and $E_{\rm cut} = 2$ GeV for the PLEC1 spectral shape (see Section \ref{spectralSection}) and re-fit the data.
We then use the corresponding likelihood, as a function of the flux, to determine 95\% confidence-level upper limits. 
This underestimates the actual sensitivity for two reasons.
First, for weak sources near the threshold, leaving $\Gamma$ and $E_{\rm cut}$ as free parameters during the data fitting increases the likelihood function peak width by a factor of two.
Second, we claim a point-source detection only if the signal can be localized. 
To account for these effects, we increased the derived flux limit by a factor of two.
The result is consistent with the measured fluxes of detected sources.
The apparent fluctuations along the Galactic plane are a consequence of the limitations of the interstellar emission model used to represent the diffuse background. 
Discontinuities result from different optimized normalizations for the regions of interest selected for the all-sky analysis that were used to calculate the sensitivity limits. 
The upper limits in Table \ref{Edot36Tab} correspond to the values nearest the positions of those pulsars.

\begin{figure}[!ht]
\centering
\includegraphics[width=1.0\textwidth]{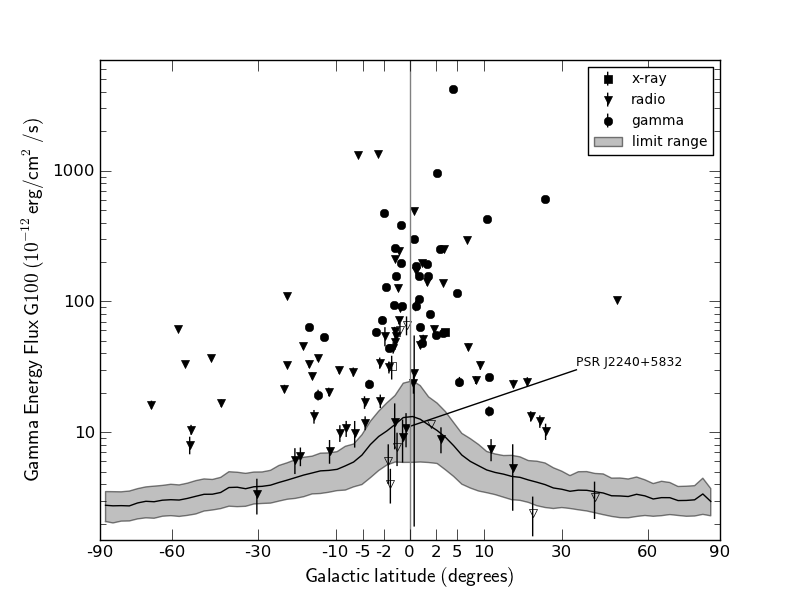}
\caption{Integral energy flux from 0.1 to 100 GeV, $G_{100}$, versus Galactic latitude $b$ (scaled as $b^{0.65}$ for clarity). 
Circles indicate gamma-selected pulsars discovered in blind period searches, 
while triangles and squares indicate previously known pulsars discovered in gamma rays by phase-folding with rotation ephemerides
obtained from radio or X-ray data. Open symbols indicate $TS<25$. 
The solid curve shows the 3-year mean sensitivity for point-source detection averaged over longitude, and the gray bands show the 10\% to 90\% percentile sensitivity range.
\label{F100b}}
\end{figure}

Figure \ref{F100b} shows the LAT pulsars' integral energy flux from 0.1 to 100 GeV ($G_{100}$, see Section \ref{spectralSection}), versus their Galactic latitude $b$.
Also shown is the latitude dependence of the sensitivity from Figure \ref{SeattleSkySensitivity},
averaged over longitude, with the 10\% and 90\% percentile limits.
The minimum detectable flux increases with the background level, 
causing a selection bias against low-latitude pulsars seen against the bright Galactic background. 
Below $20 \times 10^{-12}$ erg cm$^{-2}$ s$^{-1}$ all but one pulsar was discovered in gamma rays by phase-folding using rotation ephemerides
obtained from radio or X-ray data. The gamma-selected pulsars, those discovered in a blind period search
of LAT data, are brighter in gamma rays. \citet{Dormody_2011} found that the blind-search sensitivity is a factor of 2.5 worse than
for searches using ephemerides. 
The nine pulsars directly on the plane, and/or with very low fluxes, having phase-averaged significances below the formal detection threshold ($TS<25$) were all found
with ephemerides. The observed minimum energy fluxes for LAT pulsars are below $5 \times 10^{-12}$ erg cm$^{-2}$ s$^{-1}$ far from the Galactic plane.
PSR J2240+5832 ($l,b = 106\fdg6, -0\fdg11 $) is an example of a pulsar with measured flux right at the sensitivity threshold.
Figure \ref{J2240+5832lc} shows that it has particularly narrow peaks, 
facilitating its pulsed detection without improving its phase-averaged significance.
Pulsars with spectral parameters far from the average values assumed for the sensitivity sky map also lead to outlying points.

\subsection{SNRs and PWNe without Detected Gamma-ray Pulsars}
\label{assoc}
%
Table~\ref{tab:snrs} compiles pulsars explored in \textit{Fermi}-LAT studies of 19 SNRs.
Eleven of the SNRs are spatially extended at GeV energies. Of these, seven have known CCOs: IC 443, Puppis A, RX J1713.7$-$3946, S147, W30, W44, and W28.
The remaining four (Cygnus Loop, HB21, RX J0852.0$-$4622 and W51C) have associated CCOs, or candidate PWNe indicating the likely presence of a CCO.
Their gamma-ray emission is more consistent with a single extended source than a composite system of a pulsar and a remnant.
Three of the remaining GeV SNRs (Cassiopeia A, Tycho and W49B) show no evidence for extended emission at GeV or TeV energies.
The final five SNRs (3C 58, Crab, Vela, MSH 15-52, and MSH 11-62) contain gamma-ray pulsars.
MSH 11-62 has no off-peak detection, meaning the gamma-ray emission is completely due to nearby PSR J1105$-$6107 with no detected SNR contribution.

\tabletypesize{\scriptsize}
\begin{landscape}
\begin{deluxetable}{lllcllc}
\tablecaption{GeV-detected SNRs with dedicated studies\label{tab:snrs}}
\tablewidth{0pt}

\tablehead{
\colhead{SNR name}      & 
\colhead{Common name(s)}   & 
\colhead{Objects within Region of Interest}      &
\colhead{Gamma-ray}	&
\colhead{Emission}      & 
\colhead{Method}      & 
\colhead{Ref.} \\
\colhead{}      & 
\colhead{}      & 
\colhead{}      & 
\colhead{ pulsar?}      & 
\colhead{source}      & 
\colhead{}      & 
\colhead{}}
\startdata
SNR 120.1+01.4    &  Tycho, 3C 10, SN 1572  & None & \nodata & remnant only & spectral & 15 \\
SNR 130.7+03.1    &  3C 58, SN 1181         & PSR J0205+6449 $a$  & Y & composite & off-pulse & 1 \\
SNR 184.6$-$05.8  &  Crab                   & PSR J0534+2200 $a$  & Y & composite & off-pulse & 3 \\
SNR 180.0$-$01.7  &  S147                   & PSR J0538+2817  & N & remnant only & spatial & 18 \\
SNR 189.1+03.0    &  IC443                  & PWN CXOU J061705.3+222127  & \nodata & remnant only & spectral & 5 \\
SNR 260.4$-$03.4  &  Puppis A               & PSR J0821$-$4300 $a$ & \nodata &  remnant only &  spatial  &  20 \\
SNR 263.9$-$03.3  &  Vela                   & PSR J0831$-$4406, PSR J0835$-$4510 $a$, PWN Vela-X  & Y & composite & off-pulse & 6 \\
SNR 266.2$-$01.2  &  Vela Jr.               & PSR J0855$-$4644, PSR J0855$-$4658 & N  & remnant only & spectral & 12 \\
SNR 291.0$-$00.1  &  MSH 11-62              & PSR J1105$-$6107  & Y & pulsar only & off-pulse & 17 \\
SNR 320.4$-$01.2  &  MSH 15-52              & PSR J1513$-$5908 $a$ & Y & composite & off-pulse & 7 \\
SNR 347.3$-$00.5  &  RX J1713.7$-$3946      & PSR J1712$-$391, J1712$-$392, J1713$-$3949  & \nodata & remnant only & spectral & 11 \\
SNR 006.4$-$00.1  &  W28                    & PSR J1759$-$2307, PSR J1800$-$2343  & N & remnant only & spatial & 8 \\
SNR 008.7$-$00.1  &  W30                    & PSR J1803$-$2137 $a$, PSR J1806$-$2125  & N & remnant only & spatial & 14 \\
SNR 034.7$-$00.4  &  W44                    & PSR J1856+0113 $a$  & N & remnant only & spatial & 10, 16 \\
SNR 043.3$-$00.2  &  W49B                   & None & N & remnant only & spectral & 9 \\
SNR 049.2$-$00.7  &  W51c                   & PWN CXOU J192318.5+140335  & N & remnant only & spatial & 2 \\
SNR 074.0$-$08.5  &  Cygnus Loop            & PWN 2XMM J204920.2+290106  & \nodata & remnant only & spatial & 13, 19 \\
SNR 089.0+04.7    &  HB21                   &  None & \nodata & remnant only & spatial & 21 \\
SNR 111.7$-$02.1  &  Cassiopeia A           &  CCO  & N & remnant only & spectral & 4 
\enddata

\tablecomments{Nineteen supernova remnants with dedicated investigations of their GeV emission mechanisms. 
An ``$a$'' after a pulsar name indicates physical association with the SNR.
Four of the five gamma-ray pulsars (`Y' in Column 4) are associated with SNRs, while PSR J1105$-$6107 merely overlaps.}

\tablerefs{
(1) \citet{LATPSR0205};
(2) \citet{LATW51C};
(3) \citet{FermiCrab};
(4) \citet{LAT_CasA};
(5) \citet{LATIC443};
(6) \citet{LAT_collaboration_Vela_X_2010};
(7) \citet{LAT_PSR_B1509-58};
(8) \citet{LATW28};
(9) \citet{LATW49B};
(10) \citet{LATW44_Science};
(11) \citet{LATRXJ1713};
(12) \citet{LATRXJ0852};
(13) \citet{LATCygnusLoop};
(14) \citet{LATSNRG8.7-0.1};
(15) \citet{LATTycho};
(16) \citet{LATW44_ApJ2012};
(17) \citet{LATMSH11-62};
(18) \citet{LATS147};            
(19) \citet{XrayPWN_CygnusLoop};
(20) \citet{LATPuppisA};              
(21) \citet{GeVHB21}; }

\end{deluxetable}
\end{landscape}
\normalsize



\subsection{Towards TeV Energies}\label{TeVPSRs}
Both the MAGIC and VERITAS atmospheric Cherenkov telescopes (ACTs) detected pulsations from the Crab,
with an integral photon flux above 100 GeV of $\sim 6 \times 10^{-12}$ cm$^{-2}$ s$^{-1}$ \citep{magic-crab,veritas-crab,magic-phase-resolved-crab}.
Extrapolating the LAT Crab pulsar spectrum (Table \ref{tbl-psrspec}) predicts an integral flux above 100 GeV of 
less than $10^{-19}$ cm$^{-2}$ s$^{-1}$ for a power law with a pure exponential cutoff ($b = 1$ in Eq. \ref{expcutoff}).
Fitting the cutoff shape yields $b = 0.44$ and the extrapolation again yields a value far below their measurement.
In consequence these authors fit the joint LAT-ACT data with a broken power law, bridging any dip that may exist
between the LAT and ACT energy ranges.
We have examined whether the high-energy tails of the LAT data point to other pulsars that may be detectable
by ground-based instruments, or that may help distinguish between the various emission models.

For our brightest and hardest pulsars
(some LAT events aligned in position and phase with Vela's second gamma-ray peak exceed 50 GeV)
we fit the LAT data to a broken power law. 
We also simulated 3-year data sets using the standard LAT tool \texttt{gtobssim} and fit the simulated data in the same way. 
We find that the extrapolations to ACT energies are unreliable. The main problem is that the few gamma rays with the highest energies greatly influence 
the spectral parameters. The fits are also sensitive to the low-energy bound of the data set.
Varying the functional form further broadens the range of extrapolated fluxes, predicting anything from quick to
extremely difficult detections by ground-based instruments. We chose to make no such predictions here and advise caution
in extrapolating the GeV data. The LAT collaboration is currently preparing a catalog of sources detected above 10 GeV \citep{1FHL}.
Of the 27 sources associated with known pulsars, 20 (11) have significant pulsations in the range $>10$ GeV ($>25$ GeV).

%

\section{Multiwavelength Counterparts}
\label{multiwavelength}

%

\subsection{X-ray Properties}
\label{XraySect}
\label{Xrays}
Gamma-ray pulsars are usually observed to release most of their pulsed energy in the GeV range, but they are inherently multiwavelength objects. X-ray emission associated with individual pulsars is often detected with a significance higher than 5$\sigma$. Extensive work on X-ray pulsars was enabled by \textit{ROSAT} and \textit{ASCA} \citep{BT97}.
The X-ray flux can be pulsed non-thermal emission from the magnetosphere; blackbody thermal emission from the neutron star surface, either pulsed or unpulsed; or extended emission from a PWN energized by particles accelerated by the pulsar.
Information from X-ray observations should be included in any audit of the rotational energy loss of a given pulsar.

To characterize the X-ray spectra of LAT-detected pulsars, we use only photons
with energies from 0.3-10 keV collected by any of the major contemporary observatories operating
in the soft X-ray band: {\it Chandra/ACIS} \citep{Garmire2003}, 
{\it XMM-Newton} \citep{Struder2001,Turner2001}, 
{\it SWIFT/XRT} \citep{Burrows2005}, and {\it Suzaku} \citep{Mitsuda2007}.
Unlike in the gamma-ray band, the X-ray coverage of LAT
pulsars is uneven since the majority of the newly discovered 
pulsars (which account for half of the entries in the present catalog) have never been the targets 
of deep X-ray observations, while for
other well-known gamma-ray pulsars, such as Crab, Vela, and Geminga extensive obeservations have been carried out. 
However, all LAT pulsars do have some degree of X-ray coverage, ranging from few-ksec shallow snapshots with {\it SWIFT/XRT} to orbit-long, deep observations by {\it Chandra}, {\it XMM-Newton} or {\it Suzaku}.

Tables \ref{tbl-Xrays-PSR} and \ref{tbl-Xrays-MSP} compile X-ray spectral results for all pulsars in this catalog.
Given the complex phenomenology of pulsar X-ray emission, we have attempted to categorize the fluxes in a manner that will 
support comparisons and statistical studies focused on the system energy audit.

The status and quality for X-ray detections are indicated as follows:
`0' indicates no confirmed X-ray counterpart (or a purely thermal emission without a
non-thermal spectral component), 
`1' indicates that a counterpart has been identified but with too few counts for further characterization, and 
`2' indicates sufficient information for spectral characterization \citep[e.g.,][]{Ray2011,Fermi_CTA1_2012}.  
An ad hoc analysis was performed for some pulsars for which the standard
analysis could not be applied (e.g., owing to the very intense thermal component
of the spectrum of Vela or to an active galaxy near PSR J1418$-$6058); we designate them as
type `2*' pulsars.

We consider an X-ray counterpart to be established if (i) X-ray pulsations have been detected, (ii) X-ray and radio pulsar positions coincide, or (iii) LAT timing \citep{Ray2011} yields a position good enough
to claim a high confidence identification with an X-ray source.
The probability of finding a serendipitous source located inside a typical {\it Chandra} error circle is less 
than 0.0005 \citep{Ebisawa2005,Novara2006}; however, the probability increases by a factor of $\sim$50 for {\it Suzaku} 
observations owing to a more-limited spatial resolution.
Thus, we label all the objects found by {\it Suzaku} as `1*' to indicate that there is a non-negligible possibility
of a chance coincidence.
All the pulsar and nebular spectra have been modeled as absorbed power laws. Blackbody components have been added to the spectra when statistically needed.
Absorption along the
line of sight has been evaluated through the fitting procedure. However, for pulsars
with very low statistics we used values derived from observations taken in
different bands, when available.
We note that six MSPs can be fitted only with a thermal model: thus, owing to the lack of any non-thermal component,
we designate them as `type 0'. 
According to our classification scheme, we have 50 type 0, 11 type 1 and 56
type 2 pulsars. In total, 67 gamma-ray pulsars (30 radio-loud, 19 radio-quiet, and 18 millisecond) have an X-ray counterpart
with a non-thermal spectral component.

For each type 2 pulsar, we checked for a possible PWN contribution.
We analyzed all the data to search for extended emission through a radial brilliance study.
When a PWN was found, its contribution was evaluated as follows:
we extracted photons from an inner circular region containing $\sim$95\% of the point-like source counts,
following the prescriptions suggested for each telescope.
Such a region contains both the PSR and the PWN so that its X-ray flux must be fitted with two absorbed (PWN and PSR) power laws
(plus a blackbody, if needed).
We also selected an ad hoc outer region containing the brightest part of the nebula and fitted it
with a single (PWN) power law, forcing the $N_{\rm H}$ and the PWN photon index values to be identical for the two (inner and outer) spectra.
The PWN fluxes listed in Tables \ref{tbl-Xrays-PSR} and \ref{tbl-Xrays-MSP} are the spatially integrated 
fluxes for the two regions.
Details on data analysis and fitting procedures for each telescope can be found in
\citet{Marelli2011} and \citet{Marelli2012}.

For pulsars with a confirmed counterpart but too few photons to 
distinguish the spectral shape (type `1') the unabsorbed flux is 
estimated assuming a single power-law spectrum with photon 
index 2 to characterize all components, an absorbing column 
obtained by rescaling the Galactic column density estimated
from the Leiden-Argentine-Bonn Survey of 
Galactic H~{\sc I} \citep{Kalberla2005}\footnote{\url{http://heasrc.gsfc.nasa.gov/cgi-bin/Tools/w3nh/w3nh.pl}}, and 
the distances from Tables \ref{tbl-PSRdist} and \ref{tbl-doppler}.  
For type 1 pulsars we assume that the combined PWN and PSR thermal contributions account for 30\% of the total
source flux, a value comparable to the mean value obtained for type 2 objects.
For pulsars without a confirmed counterpart (type `0') 
an upper limit is shown, again 
derived assuming a photon index of 2 and a 
signal-to-noise ratio of 3.

\begin{figure}[bp]
\centering
\includegraphics[width=0.7\textwidth]{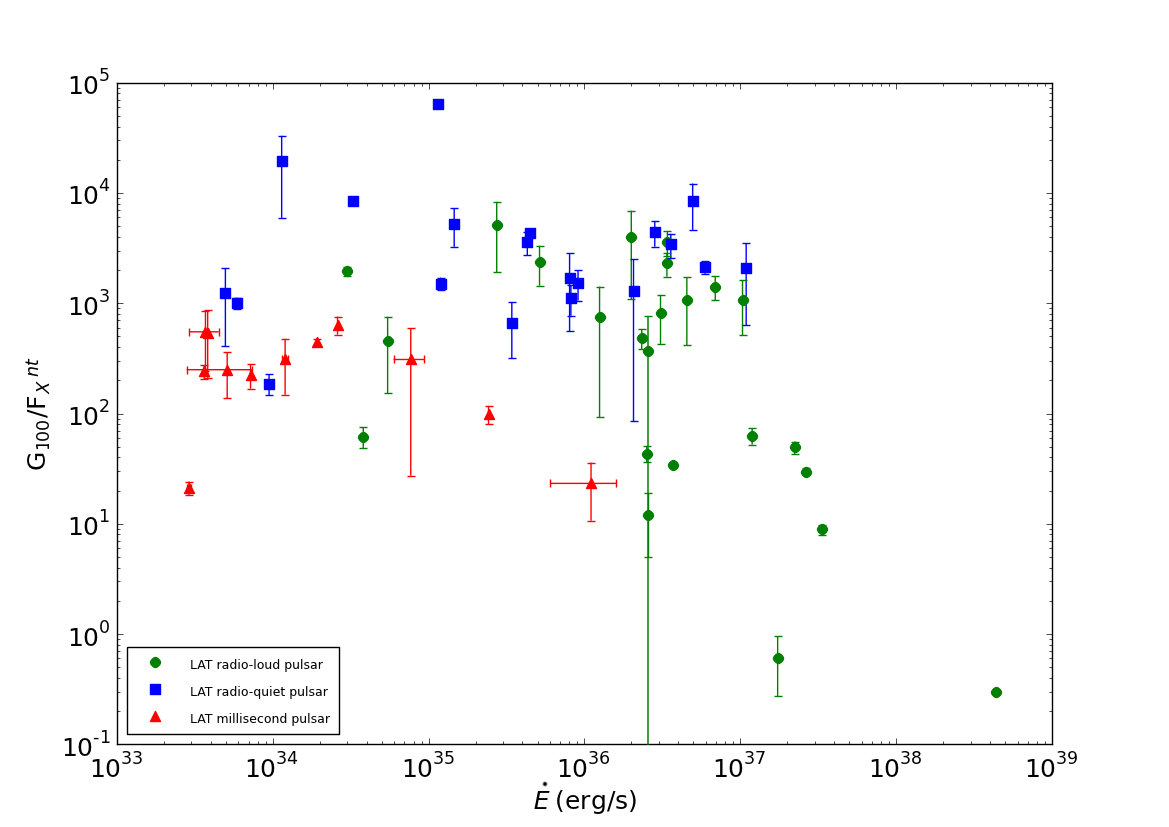}
\caption{The ratio of $G_{\rm 100}$ to the unabsorbed non-thermal X-ray flux versus $\dot{E}$ for pulsars
with good X-ray spectral measurements (`type 2'). 
All $G_{\rm 100}$ values are included, even when flagged with a \dag in Tables \ref{tbl-psrspec} and \ref{tbl-mspspec}.
Correlation with $\dot{E}$, if any, is weak.
The markers are the same as in Figure \ref{PPdotplot}.
The young radio-loud pulsars have $\langle\log(G_{100}/F_X)\rangle\,=2.37$
(standard deviation of 1.10), the radio-quiet population has
$\langle\log(G_{100}/F_X)\rangle\,=3.48$ (s.d.~of 0.49)
while the MSPs have
$\langle\log(G_{100}/F_X)\rangle\,=2.31$ (s.d.~of 0.48).
\label{G100FXnt}}
\end{figure}

A plot $G_{100}/F_X$ as a function of $\dot{E}$ (see Figure \ref{G100FXnt}) for type 2 pulsars 
shows a three decade spread in the $G_{100}/F_X$ values for a given value of $\dot{E}$.
This lack of correlation between gamma-ray and non-thermal
X-ray fluxes may point to important
(yet poorly-understood) differences in the geometry and/or height
of the X- and gamma-ray emitting regions within the magnetosphere.

In general, radio-quiet pulsars are 
characterized by fainter X-ray counterparts than radio-loud pulsars. Indeed, the X-ray fluxes of LAT-discovered radio-quiet
population have less scatter than the radio-loud pulsars.
MSPs have the lowest gamma-to-X flux ratio with less apparent scatter than that observed in young pulsars.
These results confirm and expand those obtained by \citet{Marelli2011} and \citet{Marelli2012} for 
smaller samples of LAT pulsars.

\clearpage
\begin{deluxetable}{lcccccc}
\tabletypesize{\scriptsize}
\tablewidth{0pt}
\tablecaption{X-ray spectral parameters of young LAT-detected pulsars and their nebulae
\label{tbl-Xrays-PSR}}

\tablehead{
\colhead{PSR\tablenotemark{a}}&\colhead{Inst\tablenotemark{b}}& \colhead{$N_{\rm H}$}              &\colhead{F$_X^{nt}$}                      & \colhead{Sp.~Type,} & \colhead{$G_{100}$/F$_{X}^{nt}$} & \colhead{F$_{X}^{pwn}$}\\
\colhead{}                    &\colhead{}                     & \colhead{10$^{20}$ cm$^{-2}$}&\colhead{10$^{-13}$ erg cm$^{-2}$s$^{-1}$}& \colhead{pulsed\tablenotemark{d}}   & \colhead{} & \colhead{10$^{-13}$ erg cm$^{-2}$s$^{-1}$}
}
\startdata
J0007+7303$^{2}$ & X+C & 16.6$_{-7.6}^{+8.9}$ & 0.98$\pm$0.01 & BB+Pow, P & 4320$\pm$70 & 21.40$_{-0.17}^{+0.14}$\\
J0106+4855$^{0}$ & Su & 10\tablenotemark{c} & $<$0.84 & \nodata & $>$229 & \nodata \\
J0205+6449$^{2}$ & C & 45.0$_{-1.1}^{+1.3}$ & 19.70$\pm$0.70 & BB+Pow, P & 29.7$\pm$2.1 & 24.0$\pm$0.5\\
J0248+6021$^{0}$ & S & 80\tablenotemark{c} & $<$9.00 & \nodata & $>$57.4 & \nodata \\
J0357+3205$^{2}$ & C & 8.0$\pm$4.0 & 0.64$_{-0.06}^{+0.09}$ & Pow, P & 1000$_{-100}^{+150}$ & 3.72$_{-1.36}^{+0.62}$\\
J0534+2200$^{2}$ & L & 34.5$\pm$0.2 & 44300$\pm$1000 & Pow, P & 0.296$\pm$0.007 & 396000$\pm$1000\\
J0622+3749$^{0}$ & C & 10\tablenotemark{c} & $<$2.58 & \nodata & $>$56.1 & \nodata \\
J0631+1036$^{0}$ & X & 20\tablenotemark{c} & $<$0.23 & \nodata & $>$2070 & \nodata \\
J0633+0632$^{2}$ & C & 6.08$_{-6.08}^{+21.91}$ & 0.63$\pm$0.05 & BB+Pow & 1510$\pm$170 & 2.92$_{-0.81}^{+0.79}$\\
J0633+1746$^{2}$ & L & 1.07\tablenotemark{c} & 4.97$_{-0.27}^{+0.09}$ & BB+Pow, P & 8520$_{-460}^{+160}$ & 0.172$\pm$0.001\\
J0659+1414$^{2}$ & L & 4.3$\pm$0.2 & 4.06$_{-0.59}^{+0.03}$ & BB+Pow, P & 61.8$_{-10.9}^{+6.3}$ & N\\
J0729$-$1448$^{0}$ & C & 50\tablenotemark{c} & $<$0.37 & \nodata & $>$318 & \nodata \\
J0734$-$1559$^{0}$ & S & 20\tablenotemark{c} & $<$2.36 & \nodata & $>$236 & \nodata \\
J0742$-$2822$^{0}$ & X & 20\tablenotemark{c} & $<$0.23 & \nodata & $>$771 & \nodata \\
J0835$-$4510$^{2*}$ & L & 2.2$\pm$0.5 & 65.1$\pm$15.7 & BB+Pow, P & 1410$\pm$340 & 128$\pm$1\\
J0908$-$4913$^{0}$ & C+X & 80\tablenotemark{c} & $<$0.39 & \nodata & $>$1130 & \nodata \\
J0940$-$5428$^{0}$ & C & 50\tablenotemark{c} & $<$0.13 & \nodata & $>$314 & \nodata \\
J1016$-$5857$^{2}$ & C & 57.5$_{-19.5}^{+23.5}$ & 1.47$_{-1.31}^{+0.40}$ & Pow & 370$_{-343}^{+137}$ & 3.53$_{-2.77}^{+0.26}$\\
J1019$-$5749$^{0}$ & S & 150\tablenotemark{c} & $<$1.50 & \nodata & $>$51.4 & \nodata \\
J1023$-$5746$^{2*}$ & C & 117$_{-33}^{+37}$ & 0.94$_{-0.60}^{+0.19}$ & Pow & 2070$_{-1320}^{+460}$ & 0.853$_{-0.593}^{+0.193}$\\
J1028$-$5819$^{1}$ & C+Su & $<$15.0 & 0.45$\pm$0.13 & Pow & 5390$\pm$1660 & \nodata \\
J1044$-$5737$^{1*}$ & Su & $<$12.9 & 0.92$_{-0.59}^{+0.26}$ & Pow & 1700$_{-1090}^{+490}$ & \nodata \\
J1048$-$5832$^{2*}$ & C+X & 46.0$\pm$2.3 & 0.49$_{-0.34}^{+0.18}$ & Pow & 4000$_{-2800}^{+1490}$ & 0.608$_{-0.426}^{+0.224}$\\
J1057$-$5226$^{2}$ & C+X & 2.7$\pm$0.2 & 1.51$_{-0.13}^{+0.02}$ & BB+Pow, P & 1950$_{-170}^{+40}$ & N\\
J1105$-$6107$^{0}$ & C & 50 & $<$0.08 & \nodata & $>$6130 & \nodata \\
J1112$-$6103$^{2}$ & C & 121$_{-50}^{+76}$ & 0.57$\pm$0.28 & Pow & 1070$\pm$560 & 0.63$\pm$0.24\\
J1119$-$6127$^{2}$ & C+X & 185$_{-38}^{+42}$ & 1.48$\pm$0.21 & BB+Pow, P & 483$\pm$84 & 0.601$\pm$0.194\\
J1124$-$5916$^{2}$ & C & 30.0$_{-4.8}^{+2.8}$ & 9.78$_{-1.08}^{+1.18}$ & BB+Pow, P & 63.1$_{-9.0}^{+9.5}$ & 5.17$_{-0.30}^{+0.24}$\\
J1135$-$6055$^{2}$ & C & 41.9$_{-15.2}^{+18.9}$ & 0.37$_{-0.32}^{+0.15}$ & Pow & 1290$_{-1130}^{+520}$ & 1.87$_{-1.05}^{+0.39}$\\
J1357$-$6429$^{2}$ & X & 18.9$_{-4.5}^{+4.8}$ & 0.42$\pm$0.16 & BB+Pow, P & 809$\pm$324 & 3.81$_{-0.52}^{+0.38}$\\
J1410$-$6132$^{0}$ & S & 200\tablenotemark{c} & $<$1.80 & \nodata & $>$366 & \nodata \\
J1413$-$6205$^{1*}$ & C+Su & 40\tablenotemark{c} & 1.41$\pm$0.38 & Pow & 1120$\pm$310 & \nodata \\
J1418$-$6058$^{2}$ & C+X & 225$_{-45}^{+52}$ & 0.36$\pm$0.14 & Pow & 8400$\pm$3420 & Y\\
J1420$-$6048$^{2*}$ & X & 202$_{-106}^{+161}$ & 1.60$\pm$0.70 & Pow & 1060$\pm$480 & \nodata \\
J1429$-$5911$^{0}$ & Su & 80\tablenotemark{c} & $<$0.73 & \nodata & $>$1100 & \nodata \\
J1459$-$6053$^{2}$ & X & 42.0$_{-18.0}^{+47.0}$ & 0.85$\pm$0.23 & Pow & 1520$\pm$420 & N\\
J1509$-$5850$^{2}$ & C+X & 79.5$_{-16.5}^{+22.1}$ & 0.53$_{-0.18}^{+0.20}$ & Pow & 2380$_{-830}^{+900}$ & 2.47$_{-0.54}^{+0.32}$\\
J1513$-$5908$^{2*}$ & C+L$^{f}$ & 91.8$\pm$0.2 & 52$\pm$18 & Pow & 0.612$\pm$0.284 & 1459.5$_{-12.7}^{+12.9}$\\
J1531$-$5610$^{1}$ & C & 40\tablenotemark{c} & 1.62$\pm$0.53 & Pow & \nodata & \nodata \\
J1620$-$4927$^{0}$ & X & 40\tablenotemark{c} & $<$0.67 & \nodata & $>$2330 & \nodata \\
J1648$-$4611$^{0}$ & C & 100\tablenotemark{c} & $<$0.22 & \nodata & $>$2520 & \nodata \\
J1702$-$4128$^{1}$ & C+X & 100\tablenotemark{c} & 0.23$\pm$0.07 & Pow & 3150$_{-3150}^{+4500}$ & \nodata \\
J1709$-$4429$^{2}$ & C+X & 45.6$_{-2.9}^{+4.4}$ & 3.78$_{-0.94}^{+0.37}$ & BB+Pow, P & 3560$_{-890}^{+350}$ & 8.36$_{-0.67}^{+0.52}$\\
J1718$-$3825$^{2}$ & X & 40.7$_{-15.5}^{+14.6}$ & 1.18$_{-0.97}^{+0.58}$ & Pow & 753$_{-622}^{+375}$ & 1.33$_{-0.95}^{+0.55}$\\
J1730$-$3350$^{0}$ & C+X & 100\tablenotemark{c} & $<$0.26 & \nodata & $>$3280 & \nodata \\
J1732$-$3131$^{2}$ & C & 9.39$_{-9.39}^{+28.58}$ & 0.37$\pm$0.13 & Pow & 5260$\pm$1870 & \nodata \\
J1741$-$2054$^{2}$ & C & 15.3$_{-3.6}^{+5.1}$ & 6.24$_{-1.14}^{+0.34}$ & BB+Pow & 187$_{-35}^{+13}$ & 1.68$_{-0.34}^{+0.28}$\\
J1746$-$3239$^{0}$ & S & 10\tablenotemark{c} & $<$1.74 & \nodata & $>$416 & \nodata \\
J1747$-$2958$^{2*}$ & C+X & 256$_{-6}^{+9}$ & 48.70$_{-6.00}^{+21.30}$ & Pow & 43.3$_{-6.1}^{+19.2}$ & 84.5$_{-4.0}^{+10.2}$\\
J1801$-$2451$^{2}$ & C+X & 374$_{-108}^{+120}$ & 9.97$\pm$2.02 & Pow & 75.3$\pm$45.9 & 3.27$\pm$1.24\\
J1803$-$2149$^{0}$ & Su & 50\tablenotemark{c} & $<$0.46 & \nodata & $>$2030 & \nodata \\
J1809$-$2332$^{2}$ & C+X & 49.2$_{-5.5}^{+6.8}$ & 1.32$\pm$0.30 & BB+Pow & 3590$\pm$820 & 14.4$_{-2.2}^{+1.6}$\\
J1813$-$1246$^{1}$ & Su & 153$_{-50}^{+61}$ & 1.37$_{-0.45}^{+0.24}$ & Pow & 1840$_{-610}^{+330}$ & Y\\
J1826$-$1256$^{2}$ & C & 126$_{-46}^{+53}$ & 1.12$\pm$0.25 & Pow & 3420$\pm$770 & 1.52$\pm$0.33\\
J1833$-$1034$^{2}$ & X+C & 210$\pm$1 & 66.30$\pm$1.50 & Pow & 8.89$\pm$1.15 & 721$\pm$5\\
J1835$-$1106$^{0}$ & C & 90\tablenotemark{c} & $<$0.28 & \nodata & \nodata & \nodata \\
J1836+5925$^{2}$ & X+C & 0.7$_{-0.7}^{+10.6}$ & 0.31$_{-0.21}^{+0.04}$ & BB+Pow & 19500$_{-13400}^{+2300}$ & N\\
J1838$-$0537$^{1*}$ & Su & 100\tablenotemark{c} & 0.88$\pm$0.07 & Pow & 2130$\pm$230 & \nodata \\
J1846+0919$^{0}$ & S & 20\tablenotemark{c} & $<$2.92 & \nodata & $>$83.3 & \nodata \\
J1907+0602$^{2}$ & X+C+Su & 41.1$_{-3.0}^{+3.5}$ & 0.58$\pm$0.14 & Pow & 4410$\pm$1050 & N\\
J1952+3252$^{2}$ & X+C & 33.3$\pm$0.9 & 40.70$\pm$1.50 & BB+Pow & 33.9$\pm$1.8 & 77.7$\pm$1.5\\
J1954+2836$^{0}$ & Su & 50\tablenotemark{c} & $<$0.75 & \nodata & $>$1370 & \nodata \\
J1957+5033$^{0}$ & Su & 10\tablenotemark{c} & $<$0.33 & \nodata & $>$810 & \nodata \\
J1958+2841$^{2}$ & C+Su & 122$_{-54}^{+71}$ & 1.37$\pm$0.66 & Pow & 667$\pm$325 & N\\
J2021+3651$^{2}$ & C+X & 63.8$_{-0.39}^{+0.50}$ & 2.15$_{-0.49}^{+0.24}$ & BB+Pow & 2300$_{-530}^{+260}$ & 10.4$\pm$0.6\\
J2021+4026$^{2}$ & C & 65.2$_{-37.3}^{+30.5}$ & 0.15$\pm$0.01 & BB+Pow & 64600$\pm$4000 & N\\
J2028+3332$^{0}$ & S & 10\tablenotemark{c} & $<$1.57 & \nodata & $>$370 & \nodata \\
J2030+3641$^{0}$ & S & 80\tablenotemark{c} & $<$4.52 & \nodata & $>$69.5 & \nodata \\
J2030+4415$^{0}$ & S & 40\tablenotemark{c} & $<$2.53 & \nodata & $>$228 & \nodata \\
J2032+4127$^{2}$ & C+X & 47.8$_{-14.9}^{+13.1}$ & 0.27$_{-0.16}^{+0.14}$ & Pow & 5110$_{-2950}^{+2630}$ & \nodata \\
J2043+2740$^{2}$ & X & $<$3.62 & 0.22$_{-0.11}^{+0.03}$ & Pow & 453$_{-255}^{+117}$ & \nodata \\
J2055+2539$^{2}$ & X & 15.1$_{-11.7}^{+14.2}$ & 0.43$_{-0.28}^{+0.12}$ & Pow & 1240$_{-800}^{+350}$ & 2.61$\pm$0.84\\
J2111+4606$^{0}$ & S & 30\tablenotemark{c} & $<$2.25 & \nodata & $>$196 & \nodata \\
J2139+4716$^{0}$ & S & 10\tablenotemark{c} & $<$3.20 & \nodata & $>$73.1 & \nodata \\
J2229+6114$^{2}$ & C+X & 30$_{-4}^{+9}$ & 51.30$_{-5.80}^{+9.30}$ & Pow, P & 49.4$_{-5.7}^{+9.0}$ & 11.4$_{-1.0}^{+0.8}$\\
J2238+5903$^{0}$ & S & 70\tablenotemark{c} & $<$4.49 & \nodata & $>$143 & \nodata \\
J2240+5832$^{0}$ & S & 70\tablenotemark{c} & $<$4.60 & \nodata & $>$23.5 & \nodata \\

\enddata

\tablenotetext{a}{Superscripts: 
0: no X-ray detection, or no non-thermal component to the X-ray spectrum; 
1: X-ray spectrum is poorly constrained; 
2: Non-thermal X-ray source (see Section \ref{Xrays}). 
An asterisk means an ad hoc analysis was necessary. The 1* indicates a Suzaku detection positionally consistent with a \textit{Fermi} source, see \citet{Marelli2011}.}
\tablenotetext{b}{C = {\it Chandra}/ACIS,  X = {\it XMM-Newton}/PN+MOS,  S = {\it SWIFT}/XRT,  Su = {\it Suzaku}/XIS. 
For L, the results were taken from \citet{Kar2008} for the Crab, from \citet{DeLuca2005} for Geminga and J0659+1414, and from \citet{Mori2004} for Vela.}
\tablenotetext{c}{The column density $N_{\rm H}$ was set to the Galactic value for the pulsar direction obtained with Webtools (\url{http://heasarc.gsfc.nasa.gov/docs/tools.html}), 
scaled for the distance.}
\tablenotetext{d}{``P'' indicates observation of pulsed X-rays.}

\tablecomments{X-ray characteristics of young LAT pulsars.
The listed fluxes are unabsorbed and non-thermal in the 0.3-10 keV energy band.
The model used is an absorbed power law, plus a blackbody (BB) when statistically necessary. 
The exceptions are PSRs J0633+1746 and J0659+1414 (double BB plus power law).
The errors are at the 90\% confidence level.
For type 1 and 1* pulsars we assumed that the PWN and pulsar thermal contributions are 30\% of the entire source flux.
X-ray nebulae have been detected (or excluded) through brilliance profile analyses; when spectral analysis was possible the flux is in the last column, otherwise the confirmed (or not) presence of a PWN is noted by `Y' (`N').}
\end{deluxetable}
\normalsize

\clearpage
\begin{deluxetable}{lcccccc}
\tabletypesize{\scriptsize}
\tablewidth{0pt}
\tablecaption{X-ray spectral parameters of LAT-detected MSPs 
\label{tbl-Xrays-MSP}}

\tablehead{
\colhead{PSR\tablenotemark{a}}& \colhead{Inst\tablenotemark{b}}& \colhead{$N_{\rm H}$}              & \colhead{F$_X^{nt}$}                      & \colhead{Sp.Type,} & \colhead{$G_{100}$/F$_{X}^{nt}$} & \colhead{F$_{X}^{pwn}$}\\
\colhead{}                    & \colhead{}                     & \colhead{10$^{20}$ cm$^{-2}$}& \colhead{10$^{-13}$ erg cm$^{-2}$s$^{-1}$}& \colhead{pulsed\tablenotemark{d}} & \colhead{} & \colhead{10$^{-13}$ erg cm$^{-2}$s$^{-1}$}
}
\startdata
J0023+0923$^{1}$ & C & 5\tablenotemark{c} & 0.21$_{-0.17}^{+0.20}$ & Pow & 381$_{-321}^{+374}$ & \nodata \\
J0030+0451$^{2}$ & X & 6.4$_{-2.4}^{+3.4}$ & 2.55$\pm$0.29 & BB+Pow, P & 241$\pm$29 & N\\
J0034$-$0534$^{0}$ & X & $<$56.3 & $<$0.06 & BB & $>$2800 & N\\
J0101$-$6422$^{0}$ & S & 1\tablenotemark{c} & $<$2.31 & \nodata & $>$45.3 & \nodata \\
J0102+4839$^{0}$ & Su & 5\tablenotemark{c} & $<$0.17 & \nodata & $>$777 & \nodata \\
J0218+4232$^{2}$ & X & 2.70$_{-2.70}^{+3.76}$ & 4.62$_{-0.63}^{+0.43}$ & Pow, P & 98.7$_{-15.5}^{+12.0}$ & N\\
J0340+4130$^{0}$ & X & 5\tablenotemark{c} & $<$0.20 & \nodata & $>$1020 & \nodata \\
J0437$-$4715$^{2}$ & X+C & 1.58$_{-1.09}^{+0.93}$ & 7.91$_{-0.60}^{+0.50}$ & BB+Pow, P & 21.1$_{-2.6}^{+2.5}$ & N\\
J0610$-$2100$^{0}$ & S & 8\tablenotemark{c} & $<$1.15 & \nodata & $>$57.1 & \nodata \\
J0613$-$0200$^{2*}$ & X & $<$3.30 & 0.96$_{-0.44}^{+0.69}$ & Pow & 312$_{-147}^{+225}$ & N\\
J0614$-$3329$^{1}$ & X+Su & 6.44$_{-2.01}^{+6.32}$ & 1.41$_{-0.58}^{+0.48}$ & Pow & 776$_{-320}^{+266}$ & Y\\
J0751+1807$^{2}$ & X & 8.74$_{-8.74}^{+2.10}$ & 0.59$\pm$0.09 & BB+Pow & 224$\pm$47 & N\\
J1024$-$0719$^{0}$ & X & $<$3.58 & $<$0.11 & BB & $>$286 & \nodata \\
J1124$-$3653$^{1}$ & S & 5\tablenotemark{c} & 0.45$\pm$0.25 & Pow & 269$\pm$156 & \nodata \\
J1125$-$5825$^{0}$ & \nodata & \nodata & \nodata & \nodata & \nodata & \nodata \\
J1231$-$1411$^{2}$ & X & 11.3$\pm$5.1 & 4.12$_{-1.74}^{+0.88}$ & BB+Pow & 250$_{-106}^{+54}$ & N\\
J1446$-$4701$^{0}$ & S & 10\tablenotemark{c} & $<$1.50 & \nodata & $>$49.5 & \nodata \\
J1514$-$4946$^{0}$ & C & 50\tablenotemark{c} & $<$0.16 & \nodata & $>$2760 & \nodata \\
J1600$-$3053$^{0}$ & X & 10\tablenotemark{c} & $<$0.07 & BB & $>$2500 & \nodata \\
J1614$-$2230$^{0}$ & C+X & 2.9$_{-2.9}^{+4.3}$ & $<$0.29 & BB & $>$852 & N\\
J1658$-$5324$^{1}$ & C & $<$14.9 & 1.24$_{-0.80}^{+0.39}$ & Pow & 233$_{-153}^{+78}$ & \nodata \\
J1713+0747$^{0}$ & S & 5\tablenotemark{c} & $<$1.80 & \nodata & $>$56.5 & \nodata \\
J1741+1351$^{0}$ & S & 5\tablenotemark{c} & $<$2.30 & Pow & $>$10.5 & \nodata \\
J1744$-$1134$^{0}$ & C & 9.40$_{-9.40}^{+11.50}$ & $<$0.26 & BB & $>$1270 & N\\
J1747$-$4036$^{0}$ & S & 20\tablenotemark{c} & $<$1.80 & \nodata & $>$55.1 & \nodata \\
J1810+1744$^{1}$ & C & 6$_{-6}^{+20}$ & 0.18$\pm$0.07 & Pow & 1290$\pm$520 & \nodata \\
J1823$-$3021A$^{0}$ & X+C & \nodata & \nodata & \nodata & \nodata & \nodata \\
J1858$-$2216$^{0}$ & S & 10\tablenotemark{c} & $<$1.95 & \nodata & $>$37.1 & \nodata \\
J1902$-$5105$^{0}$ & Su & 3\tablenotemark{c} & $<$0.56 & \nodata & $>$382 & \nodata \\
J1939+2134$^{2}$ & C & 109$_{-44}^{+63}$ & 3.95$\pm$0.71 & Pow, P & 23.2$\pm$13.3 & N\\
J1959+2048$^{2}$ & X+C & 3.72$_{-3.72}^{+3.79}$ & 0.55$_{-0.44}^{+0.10}$ & BB+Pow & 309$_{-253}^{+76}$ & 0.168$_{-0.071}^{+0.061}$\\
J2017+0603$^{1}$ & C & 10\tablenotemark{c} & 0.11$\pm$0.02 & Pow & 3030$\pm$620 & \nodata \\
J2043+1711$^{0}$ & S & 6\tablenotemark{c} & $<$0.98 & \nodata & $>$276 & \nodata \\
J2047+1053$^{0}$ & S & 7\tablenotemark{c} & $<$1.90 & \nodata & $>$32.5 & \nodata \\
J2051$-$0827$^{0}$ & C+X & $<$17.1 & $<$0.04 & BB & $>$1530 & \nodata \\
J2124$-$3358$^{2}$ & X & 2.76$_{-2.76}^{+4.87}$ & 0.67$_{-0.34}^{+0.15}$ & BB+Pow, P & 550$_{-286}^{+129}$ & 0.140$_{-0.069}^{+0.094}$\\
J2214+3000$^{2}$ & C & $<$21.3 & 0.74$\pm$0.03 & Pow & 441$\pm$34 & \nodata \\
J2215+5135$^{1}$ & C & 10\tablenotemark{c} & 0.77$\pm$0.35 & Pow & 153$\pm$75 & \nodata \\
J2241$-$5236$^{2}$ & C & $<$24.8 & 0.52$\pm$0.07 & Pow & 638$\pm$99 & \nodata \\
J2302+4442$^{2}$ & X & 13.0$_{-5.2}^{+9.1}$ & 0.68$_{-0.38}^{+0.14}$ & Pow & 539$_{-305}^{+117}$ & N\\

\enddata

\tablenotetext{a}{Superscripts: 
0: no X-ray detection, or no non-thermal component to the X-ray spectrum; 
1: X-ray spectrum is poorly constrained; 
2: Non-thermal X-ray source (see Section \ref{Xrays}).  
An asterisk means an ad hoc analysis was necessary.}
\tablenotetext{b}{C = {\it Chandra}/ACIS,  X = {\it XMM-Newton}/PN+MOS,  S = {\it SWIFT}/XRT,  Su = {\it Suzaku}/XIS.}
\tablenotetext{c}{The column density $N_{\rm H}$ was set to the Galactic value for the pulsar direction obtained with Webtools,
scaled for the distance.}
\tablenotetext{d}{``P'' indicates observation of pulsed X-rays.}

\tablecomments{X-ray characteristics of LAT MSPs.
The listed fluxes are unabsorbed and non-thermal in the 0.3-10 keV energy band.
The model used is an absorbed power law, plus a blackbody (BB) component when statistically necessary. 
The exceptions is PSR J0437$-$4715 (double BB plus power law).
The errors are at the 90\% confidence level.
For type 1 and 1* pulsars we assumed that the PWN and PSR thermal contributions are 30\% of the entire source flux.
X-ray nebulae have been detected (or excluded) through brilliance profile analyses; when spectral analysis was possible the flux is in the last column of the table, otherwise the confirmed (or not) presence of a PWN is noted by `Y' (`N').}

\end{deluxetable}
\normalsize

\subsection{Optical Properties}
\label{OptSect}
Only seven of the \Fermi pulsars are firmly identified at ultraviolet (UV), optical and/or infrared (IR) wavelengths.  
Six are solitary young--to--middle-aged pulsars  (Crab, Vela, PSR B1509$-$58, PSR B0656+14, Geminga, PSR B1055$-$52), 
all detected in the optical and some also in the IR and UV. One is an MSP in a binary system, only detected in the near-UV (PSR J0437$-$4715). 
All were identified prior to the launch of \textit{Fermi}, mostly in the 1990s \citep{Mignani2011a}. 
In the last decade possible counterparts were found for PSR B1951+32 \citep{Butler2002} and the solitary MSP PSR J1024$-$0719 \citep{Sutaria2003}, 
prior to their detection as gamma-ray pulsars.  Furthermore, PSR J1124$-$5916 has been associated with a bright, optical PWN, 
although no point source has been identified as a potential counterpart \citep{Zharikov2008}. 
While companion stars have been identified for eight of the binary Fermi pulsars, with four more obtained recently \citep{Breton2013}, 
our discussion is focused on the optical emission properties of the pulsars only, and not of their binary companions.
Table \ref{OptTab} summarizes these results, and includes 49 \Fermi pulsars, both solitary and binary.  

Since the launch of \Fermi there have been no deep systematic optical observations of gamma-ray pulsars. 
A quick-look survey carried out with 2--4 m class telescopes did not discover any potential counterparts \citep{Collins2011}.  
Dedicated follow-up observations with either the \textit{HST} or 8 m-class telescopes have been made in a few cases, 
e.g. for PSR B1055$-$52 \citep{Mignani2010a,Mignani2010b}, PSR J1357$-$6429 \citep{Mignani2011b,Danilenko2012}, PSR J1048$-$5832 \citep{Mignani2011b,Razzano2013}, 
PSR J1028$-$5819 \citep{Mignani2012a}, PSR J0205+6449 \citep{Shearer2013}, and PSR J0007+7303 \citep{Mignani2013}. 
Apart from PSR B1055$-$52 for which an optical counterpart had been previously identified, new counterparts were detected only for PSR J1357$-$6429 and PSR J0205+6449. 
A bright, near-IR PWN was also found coincident with PSR J1833$-$1034 \citep{Zajczyk2012}. 

We derived optical upper limits for gamma-ray pulsars in two ways: by compiling information from previous publications and by searching public optical archives for unpublished or
serendipitous observations of \Fermi pulsars. We did not include observations from optical/IR sky surveys, as the limiting fluxes of these surveys are usually too shallow to derive
constraining upper limits on the pulsar optical/IR emission. We considered both solitary pulsars and pulsars in binary systems since, in the latter case, the upper limit applies to
both the pulsar and its companion. 
For pulsars associated with PWNe or with a detected binary companion, we assumed the nebula or the companion flux as a very conservative upper limit on the pulsar emission.
In all, upper limits exist for 38 \Fermi pulsars, though with different sensitivity limits. 

Pulsars are located at different distances and are affected by different amounts of interstellar extinction. 
To investigate their emission properties at optical energies, we first computed the extinction-corrected energy fluxes in the different pass bands. 
To avoid bias in comparing with X-ray energy flux densities, we used the hydrogen column density $N_{\rm H}$ derived from the X-ray spectral fits \citep{Marelli2011} for this calculation. 
We derived the interstellar reddening $E(B-V)$ from $N_{\rm H}$, using the relation of \cite{Predehl1995} with $R_V \equiv A_V/E(B-V)=3.1$, 
and computed the extinction in the different pass bands according to the differential extinction coefficients of \cite{Fitzpatrick1999}. 
We note that the uncertainties of the extinction values derived in this way are dominated by the accuracy on the $N_{\rm H}$ determination from the X-ray spectral fits 
and the uncertainty on the $N_{\rm H}/E(B-V)$ ratio that depends on the dust--to--gas ratio along the line of sight and the grain properties. 
Ideally, to mitigate these uncertainties, one should directly measure the $E(B-V)$ using color-magnitude diagram comparison techniques, as done by \cite{Mignani2012b}, for example.  
However, this requires suitable multiband optical/IR data over a sufficiently large angular scale for all pulsars listed in Table \ref{OptTab} 
and photometric analysis for each field, which is beyond the goal of this work. 
For consistency, we use V-band measurements where available. 
For the other cases, we extrapolated the measured flux or upper limit in the pass band closest to the V band assuming either the measured spectrum for the identified pulsars or, 
as a first approximation, a power law with spectral index $\alpha_{\rm O}=0$ for the unidentified ones. 
While the true spectral index of a given pulsar may differ from that we have assumed, we note that in most cases the uncertainty of the extinction-corrected flux upper limits due to this assumption is negligible compared to the uncertainty of the interstellar extinction correction. 

\begin{figure}[!ht]
\centering
\includegraphics[width=0.7\textwidth]{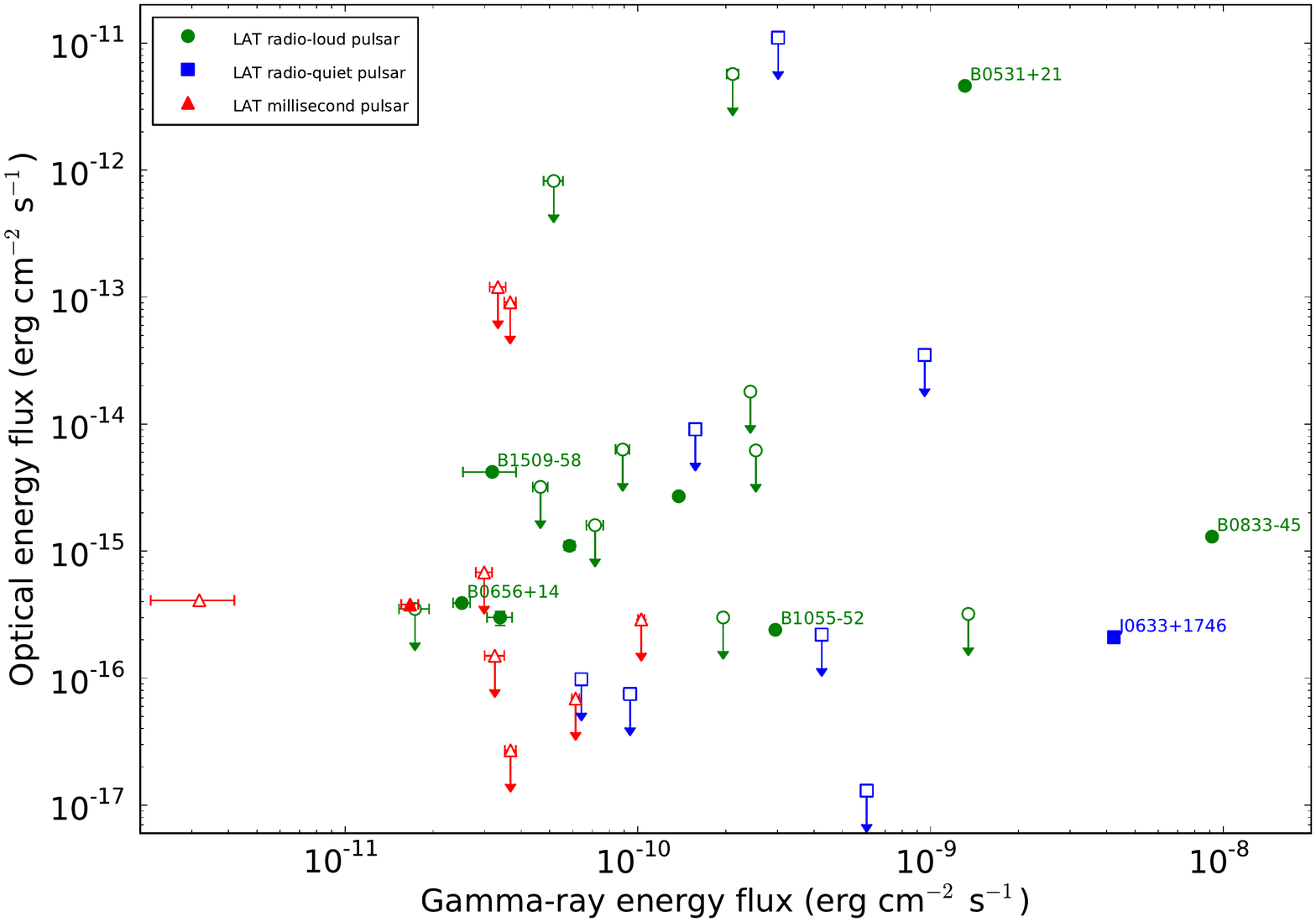}
\caption{Optical energy fluxes and upper limits when available (see Table \ref{OptTab}) versus the 0.1 to 100 GeV gamma-ray energy fluxes. 
All $G_{\rm 100}$ values are included, even when flagged with a \dag \,in Tables \ref{tbl-psrspec} and \ref{tbl-mspspec}.
The markers are the same as in Figure \ref{PPdotplot}.
\label{G100Fopt}}
\end{figure}

Figure \ref{G100Fopt} shows the extinction-corrected optical energy flux plotted against the gamma-ray energy flux for 38 pulsars from Table \ref{OptTab} 
(binary pulsars where the companion star is detected are excluded).
No correlation is apparent but it is clear that the gamma-ray energy flux is much greater than the optical energy flux for all of these pulsars. 
This is due, in part, to the fact that only seven pulsars are
firmly detected in the optical. In addition, the different sensitivities of the observations produce a rather inhomogeneous set of upper limits. If one
excludes the Crab and considers only the faintest detected pulsars (J0437$-$4715, J1057$-$5226, and Geminga) the optical energy flux appears to be
independent of the gamma-flux  across about three orders of magnitude. This is surprising as the optical luminosity \citep[$L_{opt,IR} \propto
\dot{E}^{1.70\pm0.03}$,][]{Mignani2012c} and the gamma-ray luminosity ($L_{\gamma} \propto \dot{E}^{1/2}$, see Figure \ref{EDotLumG}) both scale as a power of the rotational energy
loss $\dot{E}$, albeit with different slopes. Because of this mutual dependence, one would expect the optical luminosity, and hence the unabsorbed optical
energy flux, to scale with $L_{\gamma}$. However, the luminosity-$\dot{E}$ relation in the optical is computed from a very limited sample of objects and is sensitive to possible outliers. 
Expanding the sample of gamma-ray pulsars detected in the optical is therefore crucial to establishing possible correlations between the emission in the two energy bands.

\clearpage


\tabletypesize{\scriptsize}
\begin{deluxetable}{lcclccc} 

\tablecaption{Optical observations of LAT pulsars}
\tablewidth{0pt}

\tablehead{
\colhead{PSR} & \colhead{Magnitude} & \colhead{Filter} & \colhead{Extinction} & \colhead{Corrected Flux} & \colhead{Type\tablenotemark{a}} & \colhead{ References} \\
              &                     &                  &                      & \colhead{($10^{-15}$ erg cm$^{-2}$ s$^{-1}$)}  &       &   } 
\startdata
J0007+7303 & $>$27.6 & R & $0.92^{+0.49}_{-0.42}$ & $<0.076$ & U & \citet{Mignani2013} \\
J0023+0923 &  25          & g & 0.37                               & $<$0.86        & BC   & \citep{Breton2013} \\ 
J0030+0451 & $>$27.0 & V & $0.36^{+0.19}_{-0.13}$ & $<0.08$ & U  & \citet{Koptsevich2003} \\
J0034$-$0534 & 24.80 & I & $3.13$ & $<0.66$ & BC & \citet{Lundgren1996} \\
J0205+6449 & $>$25.50 & R & $2.50^{+0.07}_{-0.06}$ & $<1.10$ & U  & Collins et al. (in prep.)\\
J0218+4232 & 24.20 & V & $0.15^{+0.21}_{-0.15}$ & $<0.75$ & BC & \citet{Bassa2003} \\
J0248+6021 & $>$20.9 & U & $4.44 $ & $<3400$ & U  & \citet{Theureau2011} \\
J0357+3205 & $>$27.3 & V & $0.44\pm 0.22$ & $<0.07$ & U & \citet{DeLuca2011} \\
J0437$-$4715 & 24.80 & V & $0.01 $ & $0.38\pm 0.001$ & BP & \citet{Kargaltsev2004} \\
J0534+2200 & 16.50 & V & $1.92\pm 0.01$ & $4600 \pm 47$ & P*  & \citet{Cocke1969} \\
J0610$-$2100 & 26.70 & V & $0.44 $ & $<0.098$ & BU & \citet{Pallanca2012} \\
J0613$-$0200 & $>$26.0 & B & $0.18 $ & $<0.28$ & BU  &  Collins et al. (in prep.)\\
J0631+1036 & $>$24.2 & V & $1.11 $ & $<1.90$ & U  &  Collins et al. (in prep.) \\
J0633+0632 & $>$27.4 & R & $0.34^{+1.22}_{-0.34}$ & $<0.10$ & U  &  Collins et al. (in prep.) \\
J0633+1746 & 25.50 & V & $0.06 $ & $0.21$ & P*  & \citet{Bignami1993} \\
J0659+1414 & 25.00 & V & $0.24\pm 0.01$ & $0.39 \pm 0.004$ & P*  & \citet{Caraveo1994} \\
J0742$-$2822 & $>$26.0 & V & $1.11 $ & $<0.35$ & U & \nodata \\
J0751+1807 & 25.08 & R & $0.49^{+0.12}_{-0.49}$ & $<0.41$ & BC  & \citet{Bassa2006} \\
J0835$-$4510 & 23.60 & V & $0.12\pm 0.03$ & $1.3 \pm 0.03$ & P*  & \citet{Lasker1976} \\
J1024$-$0719 & 24.90 & V & $0.20 $ & $0.41$ & P+ & \citet{Sutaria2003} \\
J1028$-$5819 & $>$25.4 & B & $2.78 $ & $<12$ & U  & \citet{Mignani2012a} \\
J1048$-$5832 & $>$27.6 & V & $2.56\pm 0.13$ & $<0.34$ & U  & \citet{Mignani2011b} \\
J1057$-$5226 & 25.43 & V & $0.15\pm 0.01$ & $0.24 \pm 0.003$ & P &  \citet{Mignani2010a} \\
J1119$-$6127 & $>$24.0 & J & $10.3 \pm 2.2$ & $<2.90$ & U & \citet{Mignani2007} \\
J1124$-$5916 & 24.93 & V & $1.67^{+0.16}_{-0.27}$ & $1.5$ & N  & \citet{Zharikov2008} \\
J1231$-$1411 & $>$26.3 & V & $0.63\pm 0.28$ & $<0.23$ & BU &  Collins et al. (in prep.) \\
J1357$-$6429 & 24.60 & I & $1.05^{+0.27}_{-0.25}$ & $0.3$ & P+  & \citet{Mignani2011b} \\
J1413$-$6205 & $>$23.0 & R & $2.22 $ & $<9.1$ & U  & \nodata \\
J1418$-$6058 & $>$23.0 & R & $12.5^{+2.9}_{-2.5}$ & $<80000$ & U  & \nodata  \\
J1513$-$5908 & 26.00 & R & $5.10\pm 0.01$ & $4.2 \pm 0.032$ & P  & \citet{Wagner2000} \\
J1614$-$2230 & 24.30 & R & $0.16^{+0.24}_{-0.16}$ & $<0.67$ & BC  & \citet{Bhalerao2011} \\
J1709$-$4429 & $>$27.5 & V & $2.53^{+0.24}_{-0.16}$ & $<0.40$ & U & \citet{Mignani1999} \\
J1713+0747 & 26.00 & V & $0.00 $ & $<0.12$ & BC & \citet{Lundgren1996} \\
J1718$-$3825 & $>$24.0 & V & $2.3^{+0.8}_{-0.9}$ & $<13.3$ & U  & \nodata  \\
J1744$-$1134 & $>$26.3 & V & $0.52^{+0.64}_{-0.52}$ & $<0.27$ & U   & \citet{Sutaria2003} \\
J1747$-$2958 & $>$25.0 & R & $14.22^{+0.33}_{-0.50}$ & $<7200$ & U & \nodata  \\
J1833$-$1034 & 15.86 & K & $11.67\pm 0.06$ & $64$ & N  & \citet{Zajczyk2012} \\
J1810+1744    & 20.20 	& g  &	 0.43 	                &$<  76$  &	 BC & \cite{Breton2013} \\
J1836+5925 & $>$28.5 & V & $0.04^{+0.59}_{-0.04}$ & $<0.02$ & U  & \citet{Halpern2002} \\
J1952+3252 & 24.50 & V & $1.85\pm 0.05$ & $2.70$ & P+  & \citet{Butler2002} \\
J1959+2048 & 20.00 & V & $0.21\pm 0.21$ & $<38$ & BC & \citet{Kulkarni1988} \\
J2017+0603 & $>$19.1 & V & $0.56 $ & $<120$ & BU   & \citet{Cognard2011} \\
J2021+4026 & $>$25.2 & R & $3.62^{+1.69}_{-2.07}$ & $<10.20$ & U  & \citet{Weisskopf2011} \\
J2051$-$0827 & 22.3 & R & $0.95 $ & $<7.2$ & BC & \citet{Stappers1996} \\
J2124$-$3358 & $>$27.8 & V & $0.15^{+0.27}_{-0.15}$ & $<0.03$ & U  & \citet{Mignani2004} \\
J2215+5135  &18.70 	& g 	 & 1.15 	                      &	 $<600$ 	& BC & \cite{Breton2013} \\
J2229+6114 & $>$23.0 & R & $1.67^{+0.50}_{-0.22}$ & $<8.80$ & U & \citet{Halpern2001a} \\
J2256-1024 & 26.80 	&  g 	& 0.14 	                          & $<0.13$ 	 & BC & \cite{Breton2013} \\
J2302+4442 & $>$19.6 & V & $0.72^{+0.51}_{-0.29}$ & $<145$ & BU  & \citet{Cognard2011} \\ 
\enddata
\label{OptTab}
\tablenotetext{a}{P indicates that the optically detected object is the pulsar, and P* means the detection is pulsed.
P+ means that the pulsar may have been detected (candidate). 
B indicates a binary system, and C means that the optical detection is of the companion star. 
N is for an optical detection of the host nebula.
U means that no optical detection was achieved, and magnitude lower limits are provided.
Upper limits for binary systems apply to both the pulsar and the companion.
For undetected pulsars, a measured PWN or companion flux is taken as a conservative upper limit on the pulsar flux.}
\tablecomments{
~LAT pulsars with optical or infrared detections or upper limits. 
Sensitivities vary between magnitude 22 and 27, depending on the telescope used, the pass band, the instrument, the exposure time, and the observing conditions. 
The Table reports the observed magnitudes (or limits) in a given band, either for the pulsar or its binary companion, the computed interstellar reddening based upon $N_{\rm H}$, 
and the unabsorbed optical flux of the pulsar (or upper limit) in the V-band (peak wavelength $\lambda=5500$\AA; bandwidth $\Delta \lambda = 890$ \AA). }
\end{deluxetable}
\normalsize
\clearpage

\section{Discussion}
\label{discuss}

This catalog expands on the results of 1PC, with the uniformly analyzed
pulsar sample growing from 46 (six months) to 117 (36 months). Nearly half these
pulsars were unknown before {\it Fermi}, having either been discovered in
blind gamma-ray searches or in LAT-directed radio searches. These pulsars
fall nearly equally into three main classes: young radio-loud, young radio-quiet,
and millisecond. 
Compared to 1PC, the larger sample and detailed, uniform analysis of the 117 gamma-ray pulsars in this catalog
enables more extensive population studies and evaluations of pulsar models. Although the present work is not an
exhaustive review of such work, the following sections outline some of the implications of the catalog. A
striking change in the pulsar sample is the dramatic increase in the MSP fraction, 34\% (40/117). This doubles
the 1PC fraction and is a testimony to the remarkable success of radio follow-up studies \citep{Ray2012psc}.


\subsection{Radio and Gamma-ray Detectability}
\label{Detectability}
In our sample, 53\% (41/77) of the young pulsars are radio loud (see Table \ref{tab:tallies}), close to the 55\% (21/38) fraction in 1PC. Because the sensitivity for blind searches
is lower than for simple folding (see Section \ref{sensitivity}), the parent population must
contain substantially more radio-quiet than radio-loud pulsars (to a given gamma-ray flux
limit). This again is quite similar to the inferrence 1PC. 
By contrast, all known gamma-ray MSPs are radio loud \citep[MSP J1311$-$3430 was detected in a blind search of LAT data, but radio pulsations were subsequently detected,][]{Pletsch_2012c,RayJ1311}.
Although this circumstance might seem to be just a result of the difficulty of blind searches for MSPs, an analysis first done by \citet{Romani2012} showed that the lack of gamma-ray MSPs without detectable radio emission is not an artifact.
Among the 250 brightest 2FGL sources, for which the counterpart 
identifications are nearly complete and all sources have sensitive blind searches, only four remain unassociated with objects seen at longer wavelengths.
Within this sample, only 41\% (17/41) of the young pulsars are radio loud, while all 12 of the MSPs are radio loud.
Even if all four of the remaining unassociated sources in this sample are gamma-ray MSPs with radio beams that do not cross our line of sight (a highly unlikely scenario) the radio-loud MSP fraction can be no smaller than 75\%, a much larger fraction than seen for young pulsars.

\begin{figure}[h]
\centering
\includegraphics[width=0.7\textwidth]{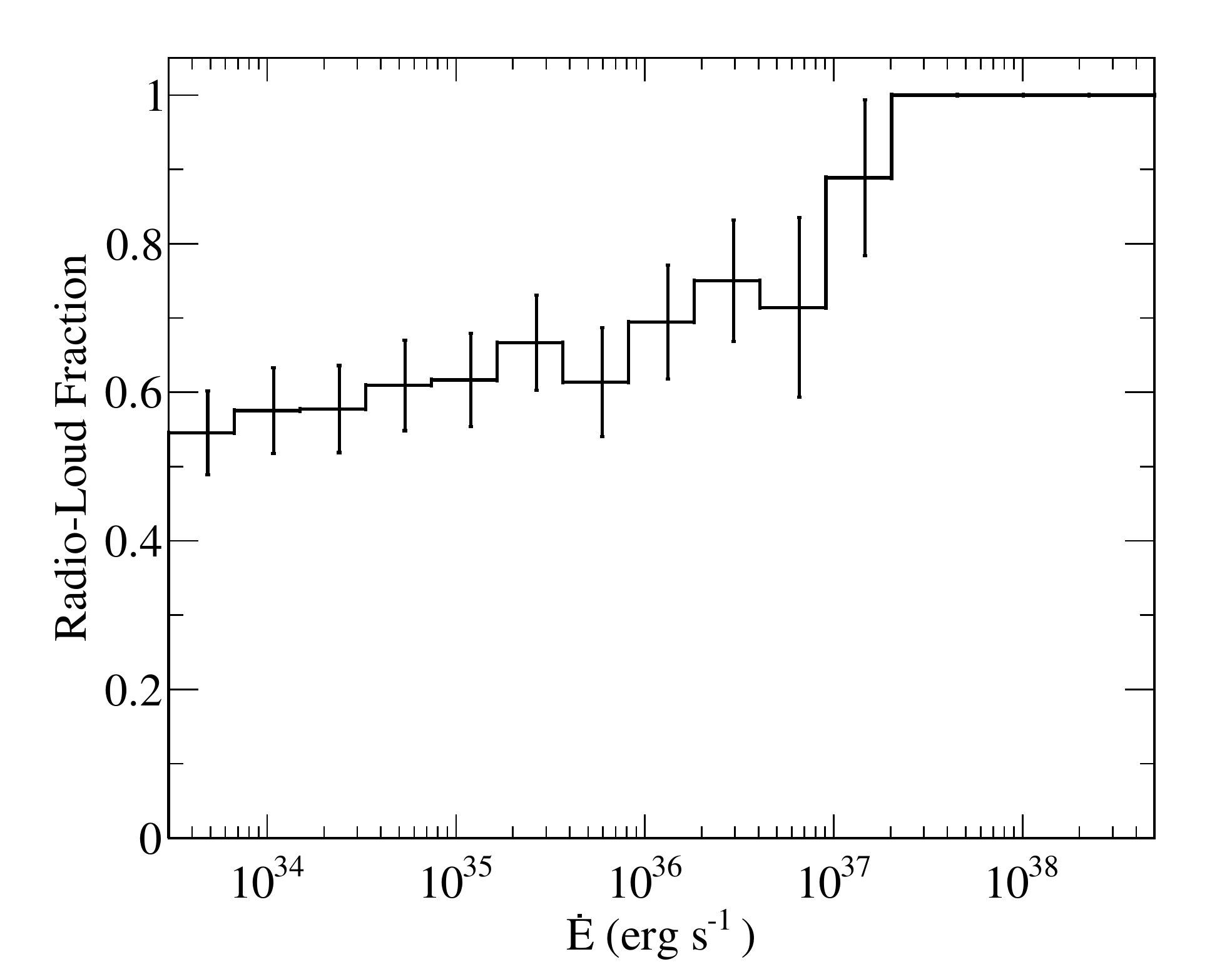}
\caption{Fraction of young gamma-ray pulsars above a given $\dot{E}$ which are radio-loud, ranging from $\sim$0.5 for the least energetic to 1.0 for the most energetic.
\label{fig-Ravi}}
\end{figure}

The detected gamma-ray pulsars are clearly highly energetic, with no young pulsars and only
a few MSPs detected below ${\dot E} \approx 3 \times 10^{33} {\rm erg\,s^{-1}}$ 
(Figure \ref{PPdotplot}). There are apparent differences in detectability between
the pulsar classes, which likely reflect differences in the radio and gamma-ray beaming. 
Figure \ref{fig-Ravi} shows that the fraction of radio-loud young gamma-ray pulsars increases with $\dot{E}$ \citep[a feature first noted by][]{Ravi2010}.  For $\dot{E} > 1 \times 10^{37}$ erg s$^{-1}$, only one of the nine pulsars is radio-quiet.
This may be an effect of the size of the magnetosphere rather than the spindown power, supported by the observation that there is 
only one radio-quiet pulsar with $P<70$\,ms. Thus, for gamma-ray pulsars with light-cylinder radius
$R_{\rm LC} < 200 R_{\rm NS}$, one nearly always detects the radio beam.

\citet{2011ApJ...727..123W} 
showed that these trends can indeed be explained by beaming evolution. They found that
if radio beams move to high altitudes for young, short period pulsars, as argued by
\citet{KaraJohnston2007}, then nearly all such gamma-ray pulsars are also radio detectable. 
Also, the increased fraction of low-${\dot E}$ radio-quiet pulsars was shown 
to be a natural consequence of outer-magnetosphere models with emission dominated 
by regions above the null charge surface, especially if the pulsars evolve toward 
alignment (decreasing $\alpha$) on Myr timescales \citep{Young2010PsrAlignEvolve}. 
This is because
the radio beam, increasingly aligned to the spin pole, is seldom visible
when one views the gamma-ray beam, which is concentrated to the spin equator.
\citet{Pierbattista2012} also discuss the lack of high-${\dot E}$ radio-quiet pulsars
and the increase in the radio-quiet fraction at low-${\dot E}$, plausibly
addressed by an evolution toward spin alignment.
However, these authors note that such alignment on Myr timescales cannot address 
the discrepancy seen at high-${\dot E}$ because of the young age of those objects.
Also, while \citet{2011ApJ...727..123W} find 
that the outer gap \citep[OG, see][]{Cheng86,Romani95} model is preferred over the lower-altitude 
two-pole caustic \citep[TPC, see][]{Dyks2003} model with very large statistical significance,
\citet{Pierbattista2012} argue that a slot gap \citep[SG, see][]{Muslimov04} 
model extending to the light cylinder can provide acceptable numbers of pulsar detections and radio-loud/radio-quiet 
ratios.
They also stress that, unless the radio beam is broader than currently modeled, both the SG and OG geometries fail to reproduce the 100\% radio-loud fraction found at high ${\dot E}$.
Further work is needed to determine whether any of these models can reproduce the
detailed ratios and their evolution with spin period and spindown power.

Compared with the young, radio-loud pulsars, we see in Figures \ref{Aitoff} and \ref{galaxie} that MSPs are often detected at smaller distances because of their lower luminosities and at larger Galactic latitude $|b|$ because of the much greater age of this population.
The radio-quiet young pulsars are intermediate in this respect, reflecting the increased tendency
noted above for gamma-only detection as the pulsars spin down. In general MSPs
lie within 2\,kpc, although the detection of the MSP J1823$-$3021A in the 
globular cluster NGC 6624 \citep{J1823m3021A} illustrates that very energetic (young) 
MSPs in low background regions can be detectable by the LAT across much of the Galaxy.


In Figure \ref{P0_sqEDOTD2}, it is apparent that a large fraction of the MSPs
above the LAT sensitivity threshold are detected as gamma-ray pulsars. 
Thus, the absence of gamma-ray only MSPs and the paucity of above-threshold 
radio MSPs undetected by the LAT suggests that MSP radio and gamma-ray beams 
cover a comparably-sized, and nearly-coincident, fraction of the sky. 

\subsection{Light Curve Trends}
\label{lcTrends}
Additional clues to the pulsar beaming and detectability can be extracted
from patterns in the radio and gamma-ray light curves. For young pulsars, the
trends visible in 1PC are strengthened and extended in this catalog. Most (58/77) 
of these pulsars show two strong, caustic peaks significantly separated, often with significant bridge emission.
The most prominent light curve trend is the anti-correlation (Figure \ref{deltaDelta}) between 
$\Delta$ and $\delta$, shown by \citet{Romani1995} to
be a general property of outer-magnetosphere models with caustic pulses. However, 
the sample in this catalog makes it clear that the trend is not universally followed. Certainly, MSPs show 
less correlation, and a significant number of young pulsars have only one strong gamma-ray peak.
\citet{2011ApJ...727..123W} argued that many single-peak, young
pulsars fall in the $\delta \approx 0.3-0.6$ range, which can be attributed to 
either missing the first peak (P1) or a blended combination of P1 and the second peak (P2). However several objects, 
especially MSPs, depart strongly from this pattern. 
The LAT MSPs show, on average, larger radio lags than young pulsars, as expected since 
MSPs have smaller magnetospheres and hence stronger aberration of the radio pulses
at typical radio emission heights.  The distribution in $\Delta$ for MSPs,
shown in the side histogram of Figure \ref{deltaDelta}, also shows a larger fraction 
with $\Delta>0.5$ than for young pulsars.

One feature noted in 1PC that persists in the current sample is that for pulsars with
two strong caustic peaks, the P2/P1 ratio increases with energy. This often helps
us identify the harder P2 component. However, when $\Delta \sim 0.5$ the
spectral evolution of the peak ratio is often weak, making peak assignment more difficult 
(e.g., PSR J0908$-$4913). For a few single-peak pulsars with sharp trailing edges, we can see 
the peak strength grow with photon energy -- for these objects we suspect that the observed peak is a P2
component and P1 may be detectable only below the LAT energy band.

The uniformity of the light curve fitting in this catalog (Section \ref{lightCurveFitting}) should greatly facilitate use of
the $\delta-\Delta$ distribution to test magnetospheric models.  For example, in such
a treatment of the 1PC sample, \citet{2011ApJ...727..123W} found
that the data gave a strong statistical preference for an OG geometry to that of the TPC model. However, accurate $\delta$ values 
need the true phase of the magnetic axis, likely requiring careful modeling of the 
radio light curve and polarization. In addition, magnetospheric currents
\citep[see][]{Kalapotharakos2012a} can provide systematic $\delta$ shifts with
respect to the vacuum approximation. Even when radio emission is not detected
the $\Delta$ distribution can provide a statistical test of the beaming model.
For example, the preponderance of $\Delta\approx 0.5$ is very natural in
SG or TPC models.  Figure \ref{DeltaEdot} does not show any strong correlation of
$\Delta$ with $\dot{E}$, although some increase in the incidence of smaller $\Delta$
is expected in OG models for lower $\dot{E}$.

\begin{figure}[h]
\centering
\includegraphics[height=0.5\textheight]{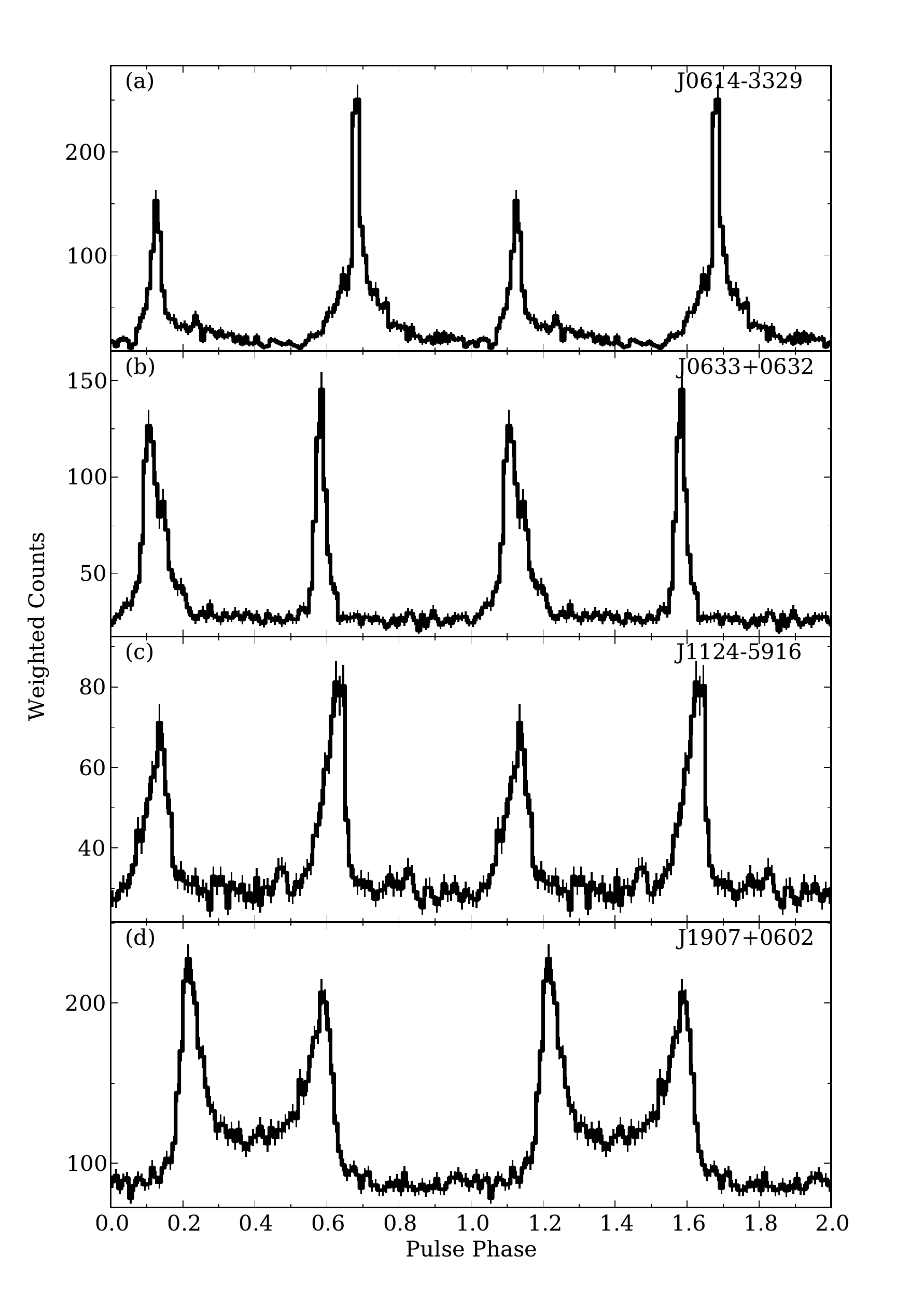}
\caption{
Examples of four types of two-peak gamma-ray pulse profiles:
(a) PSR J0614$-$3329: sharpest peak edge between the two peaks,
(b) PSR J0633+0632: sharp leading edge for both peaks,
(c) PSR J1124$-$5916:  sharp trailing edge for both peaks, and
(d) PSR J1907+0602: ``outside'' peak edges are sharpest.
\label{fig-LCtypes}}
\end{figure}

The majority of pulsars have two peaks: three-quarters for the young pulsars, and 60\% of the MSPs.
Figure \ref{fig-LCtypes} shows a sample of four recurring profile shapes,
which we classify using the ratio $R_{rf}^i$ of the half-widths of the rising and falling peak edges, included in the auxiliary files.
Here, the index $i=1$, 2 indicates the first and second gamma-ray peaks following phase $\phi=0$.
In panel (a) of the figure, the sharpest edges are the trailing edge of the first peak, $R_{rf}^1>1$,
and the leading edge of the second peak, $R_{rf}^2 <1$. 
In panel (d), it is the opposite: the ``outer'' peak edges are the most abrupt, $R_{rf}^1<1$ and $R_{rf}^2 >1$.
In panel (b), the leading edges are sharpest, $R_{rf}^{1,2}<1$ while
for panel (c), the trailing edges are steep, $R_{rf}^{1,2}>1$.
About half of the profiles are classifiable with good statistical significance,
and nearly all of these have sharp outer edges as in panel d. 
This is the expected pattern for caustics from a hollow cone. 
A few pulsars depart significantly from this pattern,
most prominently those for which both peaks fall sharply (panel c).  
These tend to have $\Delta \approx 0.5$, suggesting emission from both poles. 
Many MSPs do not fit this simple scheme and cannot be classified in this way with any confidence.  
More detailed analysis of the pulse width statistics may uncover other trends.

For two of our highest-statistics young pulsars, Vela and J1057$-$5226,
the bridge emission shows persistent structure that can be identified as a third
peak. In addition, two MSPs (J1231$-$1411 and J0751+1807) show significant 
third peaks. For Vela we have sufficient statistics to see that the phase of the
third peak shifts with energy \citep{LATVela2}; this is not expected for a simple caustic-induced peak.
To explain such structure we must go beyond simple geometrical approximation and
employ models with a full treatment of the radiation spectrum and its variation 
through the magnetosphere \citep[e.g.,][]{AnnularGap_Vela, 2layerOG}. As LAT statistics improve, we can expect further such `P3' 
detections and better constraints on the origin of such pulse components.

Table \ref{tbl-off_peak_table} lists 11 pulsars with significant `M'-type off-peak emission, suggesting the possibility of nearly constant magnetospheric emission.
However, several of these pulsars are faint or reside in regions of high and/or complex diffuse background, which complicates the spectral analysis.  Therefore, we identify only four young pulsars (PSRs J0633+1746 `Geminga', J1836+5925, J2021+4026, and J2055+2539) and two MSPs (PSRs J2124$-$3358 and J2302+4442) that have strong evidence for exponentially cutoff off-peak emission well in excess of the inferred diffuse background (including estimated systematic uncertainty).
For the young pulsars this is a serious challenge to outer-magnetosphere models radiating only above
the null-charge surface; such weakly pulsed emission should be rare, being 
expected only for nearly aligned pulsars with $\zeta \approx \pi/2$. In contrast, lower
altitude emission \citep[such as from SG or extended polar cap models,][]{Dyks2004,Venter2009}
provides a natural explanation for off-peak, non-caustic emission.
The remaining five pulsars with `M'-type off-peak emission 
(J0340+4130, J1124$-$5916, J1620$-$4927, J1747$-$2958, and J1813$-$1246)
should be treated with caution as the uncertainty in the flux from systematic error in distinguishing the diffuse background from magnetospheric emission for these sources makes it difficult to probe faint off-peak emission.  However, lower-altitude emission may also be present in a number of these cases.
For MSPs, and especially for radio-quiet young pulsars, we expect that the magnetic impact
angle $|\beta|\equiv|\alpha-\zeta|$ is relatively large, i.e.
the Earth line of sight passes far from the radio pole.  Thus, off-peak emission for these objects may be associated with large $|\beta|$.
The separatrix layer in the wind zone just outside the magnetosphere has
also been suggested \citep{BaiSpitko2010} as a site for such gamma-ray flux.

The majority of MSPs (27) have profiles that are very similar to those 
of young pulsars, with a variety of double and single-peaked profiles with radio lags following 
the $\delta-\Delta$ trend seen in the normal radio-loud pulsars.  A standard SG or OG geometry with narrow gaps 
can fit these \citep{Venter2009}, requiring surprisingly high-multiplicity pair cascades in these old pulsars. 
The rest have very different profile types.  In particular, a relatively large fraction (6) have phase-aligned
radio and gamma-ray pulses, four have $\delta\approx0.99$ while two (PSRs J0030+0451 and J1810+1744) have $\delta\approx0.85$ due to the particular method of defining the fiducial radio phase.  The peak alignment suggests that the radio as well as the gamma-ray pulses result from caustic emission at altitudes that are a large fraction of the light cylinder in 
the relatively small MSP magnetospheres \citep{Venter2012}. The Crab pulsar, with
sharp (caustic) main and inter-pulse radio components composed 
of `giant pulse' emission and aligned with the two gamma-ray pulses,
likely represents a similar case.  If the radio emission is caustic, a very low level of 
polarization may result from depolarization of emission over a large range of altitudes \citep{Dyks2004}.  
In fact, at least four of the six MSPs that have aligned profiles do show little or no linear polarization, as well as 
unusually steep radio spectra and high $B_{\rm LC}$ \citep{Venter2012,egc+12}.  There are also MSP 
profiles (7) where the gamma-ray peaks lead the main radio peak by $\sim$0.2--0.3 in phase, a profile class that is so far not seen in 
the young gamma-ray pulsar population.  A `pair-starved polar cap'  (PSPC) model provides good 
fits for this class \citep{Venter2009}, implying emission from nearly the full open zone at all altitudes
and the same pole that dominates the radio pulse.  
Three of these MSPs, PSRs J1124$-$3653, J1744$-$1134 and J2214+3000, do not have the expected value of $\delta$ in Table \ref{tbl-pulseMSP}; 
this is because we chose to use the opposite hemisphere when defining the fiducial radio phase.  The best-fit magnetic inclination angles for 
all of these profile classes show a very wide range, from nearly aligned to nearly orthogonal (Johnson et al., in prep.).  
The $|\beta|$ values show a large range since MSPs have relatively larger radio beams than young, shorter-period pulsars, allowing the radio pulse to be visible at larger $\zeta$.

The gamma-ray light curves in this catalog strongly suggest that the gamma-ray emission is distributed in a narrow gap bordering the closed field line boundary.  Although the light curves show a large degree of regularity, none of the currently proposed emission models alone is able to account for all of the observed properties.  While many light curves follow the OG pattern, some, particularly MSPs and some radio-quiet pulsars, do not fit this model and may require additional emission from other zones, such as below the null-charge surface.

\subsection{Luminosity and Spectral Trends}
\label{lumANDspec}
As in 1PC, the dependence of gamma-ray luminosity on spindown power is
one of our most important results. Figures \ref{EDotLumG} and \ref{Eff_Edot} confirm the general
trend of 1PC: young pulsars show increased efficiency for producing gamma rays
as the spindown luminosity decreases toward ${\dot E} \approx 5\times10^{35} {\rm erg\,s^{-1}}$, an observation which is in conflict with the two-layer OG model of \citet{Takata10} that predicts a much flatter trend at high $\dot{E}$. 
Above $\approx 5\times10^{35} {\rm erg\,s^{-1}}$ MSPs show a similar trend but there is a wide range of efficiencies below this $\dot{E}$.
Figure \ref{Eff_Edot} shows this more directly, and emphasizes the point that for most gamma-ray pulsars the apparent efficiency exceeds 10\% and for the lowest ${\dot E}$ efficiencies approach unity. This tension may be
slightly mitigated if MSP moments of inertia, and thus spindown luminosities, are
a few times larger than the standard $I_{0}=10^{45}{\rm g\,cm^2}$ assumed here \citep{dpr+10}.
Nevertheless, the high efficiencies are remarkable,
meaning that gamma rays trace the bulk energetics of the pulsar machine
and implying that studies correlating this output with other observables
provide excellent prospects for understanding pulsar magnetosphere physics.

Two factors complicate such studies. The first is the difficulty
of obtaining accurate pulsar distances, which we have discussed in Section \ref{Distances}.
Improvements to the distance determinations \citep[e.g. from additional VLBI parallaxes
for the radio-bright pulsars,][]{Deller2011}
provides the best hope for eliminating this large uncertainty. An additional
challenge is the uncertainty in the geometry-dependent beaming correction $f_\Omega$
(Eq.~\ref{eq-fOmega}). While for most models and viewing geometries this is
within a factor of a few of unity, for some situations \citep[e.g., OG geometries at
${\dot E} \approx 10^{34} {\rm erg\,s^{-1}}$ where $f_\Omega \sim 0.1$,][]{Pierbattista2012} the correction
can be quite substantial. Further, physical effects such as $\alpha$-dependent
variation of the gap width, changes in the gamma-ray emissivity along the gap,
or differences in the detected pulse spectrum can increase the variation of
$f_\Omega$ beyond simple geometrical factors.

The large scatter in Figures \ref{EDotLumG} and \ref{Eff_Edot} may well 
be due to these two factors.
Attempts to quantitatively constrain pulsar physics by fitting the luminosity 
evolution should certainly marginalize over these uncertainties. However, the current
sample does provide a greatly improved testbed for comparing predictions
of the heuristic $L_\gamma \propto {\dot E}^{1/2}$ law with more detailed
predictions \citep[e.g.,][]{Takata10}. 

We can also check the ${\dot E}$ dependence of the spectral fit
parameters $\Gamma$ (Figure \ref{Gamma_Edot}) and $E_{\rm cut}$.
As for 1PC we see that the lowest measured $\Gamma$ are near the limit of $\Gamma = 2/3$ 
for single-particle curvature or synchrotron radiation. In addition, there is
a trend toward a softer spectrum (larger $\Gamma$) at high spindown luminosity
with $\Gamma \propto {\dot E}^{0.2}$. One explanation for this trend is increased 
pair formation activity in high-$\dot{E}$ pulsars, leading to pair cascades
and steep radiating particle spectra.
With this catalog we now have the statistics to separately probe the trend in MSPs, which 
appears steeper with $\Gamma \propto {\dot E}^{0.4}$. 

There is no apparent trend of $E_{\rm cut}$ with $\dot{E}$.  If this cutoff follows the radiation-reaction limited energy, and 
$L_\gamma \propto \sqrt{\dot E}$, then we expect $E_{\rm cut} \propto P^{-1/4} {\dot E}^{1/8}$ so the dependence on spindown
power should indeed be weak.  One might also expect that MSPs with small $P$ would have larger $E_{\rm cut}$. However, the small radius of curvature in their compact magnetospheres tends to ensure that the primaries reach lower maximum energy, if radiation-reaction limited.


We note two important caveats to those wishing to use these phase-average spectral
fits to constrain models. First, for the fainter pulsars (especially the MSPs)
the covariance between $\Gamma$ and $E_{\rm cut}$ is substantial and the apparent
trends may be affected. Second, it is clear from our phase-resolved studies
of the brighter pulsars \citep{FermiCrab,LATVela2,Geminga1} that phase-dependent
variations in the spectral parameters can be nearly as large as the variations
in the full population. Additional evidence for varying $E_{\rm cut}$ appears when we allow
the exponent $b$ to vary in the spectral fits, as discussed in Section \ref{spectralMethodSection}.
Nevertheless, non-exponential cutoffs may be present in some cases as discussed in Section \ref{TeVPSRs}.
%
	
\subsection{Pulsar Population and the Millisecond Pulsar Revolution}
\label{popANDrev}
As the LAT pulsar sample grows, our ability to make statistically
powerful statements about the Galactic neutron stars and their evolution increases.
Compared to radio surveys, the LAT provides a new and differently biased sample of the
energetic pulsars, so many conclusions drawn from the classic radio samples
need to be revisited.

Perhaps the most dramatic progress presented in this catalog
is the major increase in the MSP sample from 8 objects to 40, with discoveries
continuing (Table \ref{Beyond117}). In fact, the LAT has proved such an excellent signpost
to nearby energetic MSPs that the LAT-guided discoveries represent a large
and increasing fraction of the known energetic Galactic MSPs. For example, 
70 Galactic (non-globular cluster) MSPs were known before {\it Fermi}; there are now 120
such MSPs, 39 of which are in the present catalog.  This dominance
is especially obvious for $P<3$\,ms; as noted by \citet{Ray2012psc} the LAT-detected
MSPs are a shorter-period, more-energetic population than radio-selected MSPs.  

Also, \citet{Roberts2012} has noted that {\it Fermi}-guided pulsar searches have resulted
in a dramatic ten-fold increase in the number of Galactic MSPs in 
tight binaries with pulsar-driven companion winds, the so-called `black widows' 
and `redbacks'. This is because the gamma-ray signal penetrates the companion wind,
flagging the source as a possible MSP and guiding repeated radio searches for
the intermittently visible radio pulsations.  

In addition to the Galactic MSPs, there is a population of some 120 pulsars in the globular
cluster system, discovered through radio searches. As these clusters are relatively
distant it is not surprising that, to date, the LAT has detected pulsed signals from
only two of the youngest, most energetic cluster MSPs \citep{J1823m3021A,LAT_M28}. However, the
detection of gamma-ray sources coincident with a dozen globular clusters \citep{GlobClustPop,Kong_2010,Tam_2011}
indicates a substantial MSP population \citep[first predicted by][]{Chen91}. Indeed, the cluster gamma-ray flux appears to correlate with the expected MSP formation rate \citep{GlobClustPop}. It had been noticed that a large
fraction of the radio MSPs in globular clusters are tight, often evaporating, black-widow-type
binaries and it was suggested that this was a true difference from the Galactic field
population \citep{KDB03}. However, it now seems that this was largely an artifact of the very long radio
dwell times used for cluster searches that allowed discovery of radio-intermittent MSPs generally 
undetectable to field surveys. With the LAT unassociated sources providing a signpost for
deep targeted searches, a similar field black widow population has now been discovered.
Thus, the LAT has uncovered a new sample of MSPs with less (or at least different)
bias than the classical radio population, demanding a re-assessment of the
MSP population and its evolution. 

For young pulsars, there is good hope that the LAT can similarly provide a much more
complete census of massive star remnants in the nearby Galaxy.
For example, \citet{2011ApJ...727..123W} found that the 1PC sample
implied an energetic young pulsar birthrate of 1.69 $\pm$ 0.24 pulsars/100 yr, a substantial
fraction of the 2.4/100 yr OB star birthrate and the 2.30 $\pm$ 0.48 SNe/100 yr rate
\citep{Li2010SNrates}. This catalog will allow further refinement of this comparison.

Finally, we conclude by noting at least one expected pulsar population
remains missing from the present catalog. As in 1PC, no pulsed detection
of a young spin-powered object in a massive binary has yet been made.
Several percent of the pulsar population is expected to pass through this channel
(the progenitor of the double neutron star binaries). A few such pulsars are
known from the radio, but like PSR B1259$-$63 are detected only at large 
orbital separation. For short period systems, plasma from the companion wind presumably prevents radio
pulsar detection. However, gamma rays provide an excellent signpost to
such wind-absorbed pulsars as witnessed by recent success in detecting short-period
black widows and redbacks in LAT sources. It is probable that some of the
gamma-ray detected massive binaries host spin-powered pulsars, a possibility discussed by \citet{Dubus06}. With improved
search techniques, such objects may appear in future LAT pulsar catalogs.


\clearpage




\textit{Acknowledgments:}
The {\it Fermi}-LAT Collaboration acknowledges generous ongoing support from a number of agencies and institutes that have supported both the development and the operation of the LAT as well as scientific data analysis. These include the National Aeronautics and Space Administration and the Department of Energy in the United States, the Commissariat \`a l'Energie Atomique and the Centre National de la Recherche Scientifique / Institut National de Physique Nucl\'eaire et de Physique des Particules in France, the Agenzia Spaziale Italiana and the Istituto Nazionale di Fisica Nucleare in Italy, the Ministry of Education, Culture, Sports, Science and Technology (MEXT), High Energy Accelerator Research Organization (KEK) and Japan Aerospace Exploration Agency (JAXA) in Japan, and the K.~A.~Wallenberg Foundation, the Swedish Research Council and the Swedish National Space Board in Sweden.

Additional support for science analysis during the operations phase is gratefully acknowledged from the Istituto Nazionale di Astrofisica in Italy and the Centre National d'\'Etudes Spatiales in France.

The Parkes radio telescope is part of the Australia Telescope which is funded by the Commonwealth Government for operation as a National Facility managed by CSIRO.  The Green Bank Telescope is operated by the National Radio Astronomy Observatory, a facility of the National Science Foundation operated under cooperative agreement by Associated Universities, Inc. The Arecibo Observatory is part of the National Astronomy and Ionosphere Center (NAIC), a national research center operated by Cornell University under a cooperative agreement with the National Science Foundation. The Nan\c cay Radio Observatory is operated by the Paris Observatory, associated with the French Centre National de la Recherche Scientifique (CNRS). The Lovell Telescope is owned and operated by the University of Manchester as part of the Jodrell Bank Centre for Astrophysics with support from the Science and Technology Facilities Council of the United Kingdom. The Westerbork Synthesis Radio Telescope is operated by Netherlands Foundation for Radio Astronomy, ASTRON.
This work made extensive use of the ATNF pulsar  catalog \citep{ATNFcatalog}.

\bibliographystyle{apj}
\bibliography{2ndPulsarCatalog}

\appendix

\section{Appendix: Sample light curves and spectra.}\label{App-Samples}
\renewcommand{\thefigure}{A-\arabic{figure}}
\setcounter{figure}{0}

\makeatletter
\@addtoreset{figure}{section}
\makeatother

\begin{figure}[!ht]
\centering
\includegraphics[width=0.7\textwidth]{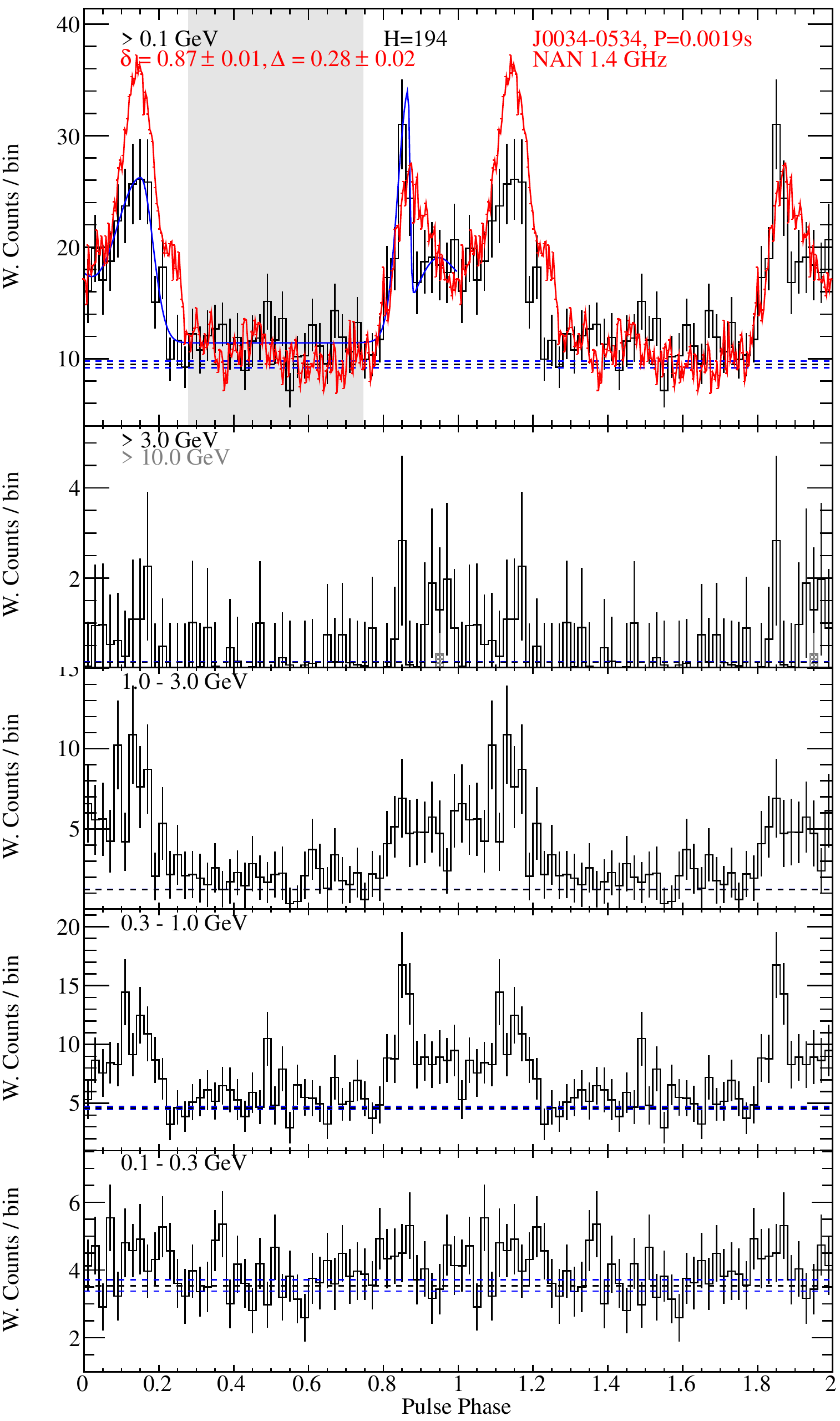}
\caption{Light curves of PSR J0034$-$0534. Appendix A text  describes the different frames. This MSP has radio and gamma-ray peaks occuring at the same phase and is a strong case for near-constant magnetospheric emission.
Phase 0 (the ``fiducial point'', see Section \ref{profiles}) for this pulsar was set to the median point between the
two radio peaks.
\label{J0034-0534lc}}
\end{figure}

\begin{figure}[!ht]
\centering
\includegraphics[width=0.7\textwidth]{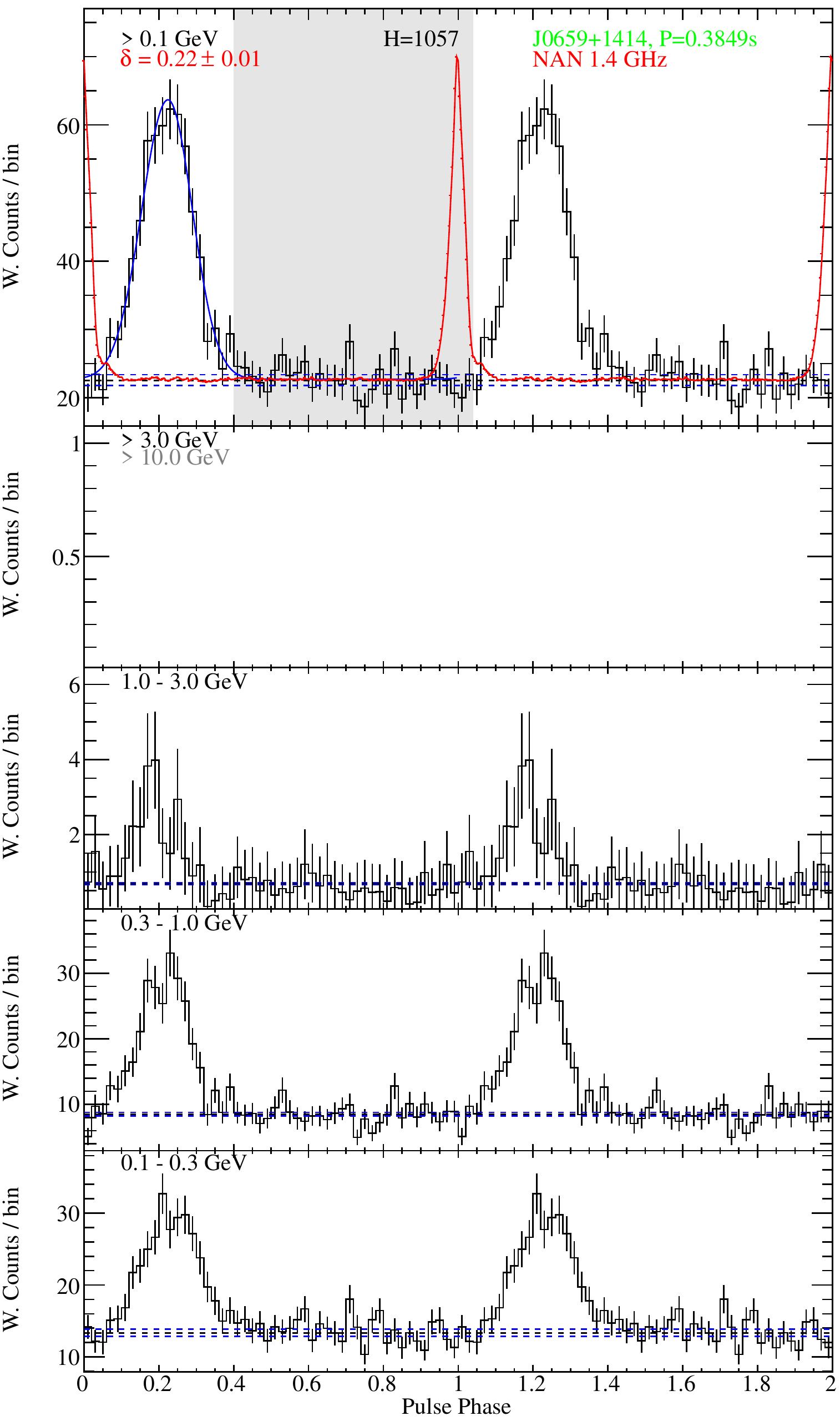}
\caption{Light curves of PSR J0659+1414. Appendix A text  describes the different frames.  This is an example of a single-peaked gamma-ray light curve.
\label{J0659+1414lc}}
\end{figure}

\begin{figure}[!ht]
\centering
\includegraphics[width=0.7\textwidth]{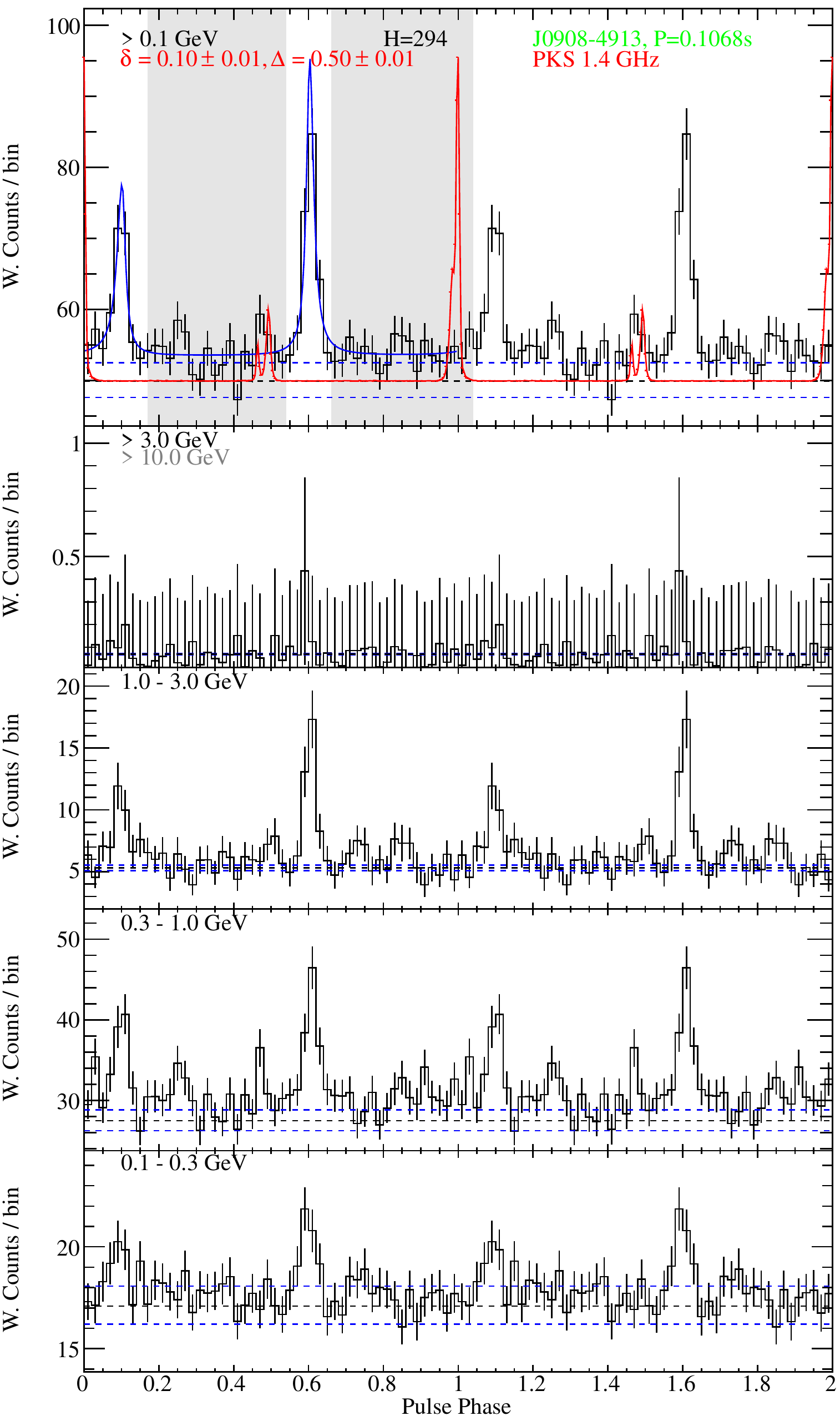}
\caption{Light curves of PSR J0908$-$4913. Appendix A text  describes the different frames.  This is an example of a two-peaked gamma-ray light curve with no apparent emission between the peaks, leading to a disconnected off-peak region as evident from the shaded regions in the figure.
\label{J0908-4913lc}}
\end{figure}

\begin{figure}[!ht]
\centering
\includegraphics[width=0.7\textwidth]{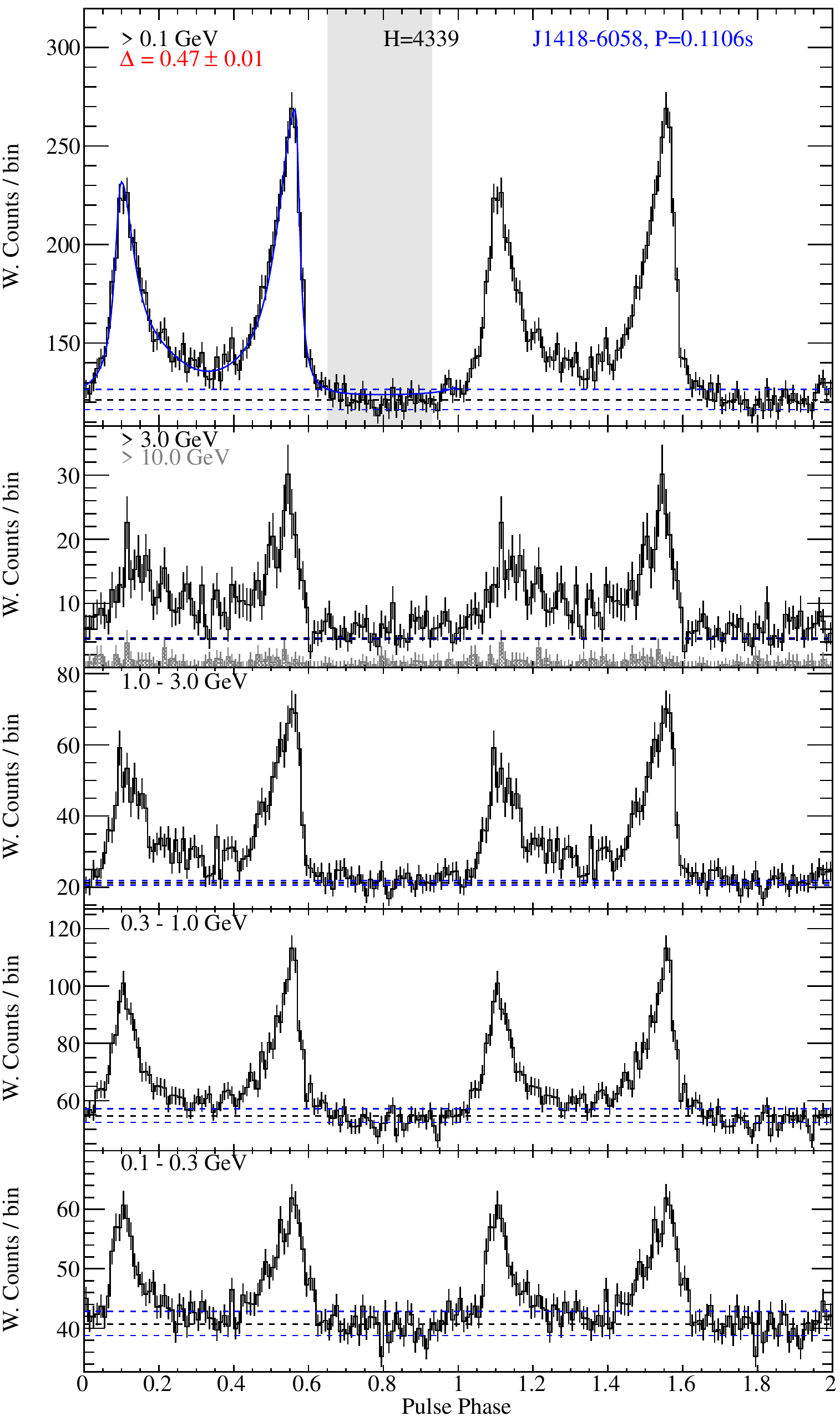}
\caption{Light curves of PSR J1418$-$6058. Appendix A text  describes the different frames.  This is an example of a radio-quiet pulsar with no radio detection and demonstrates the common two-peaks with sharp, asymmetric structure.
\label{J1418-6058lc}}
\end{figure}

\begin{figure}[!ht]
\centering
\includegraphics[width=0.7\textwidth]{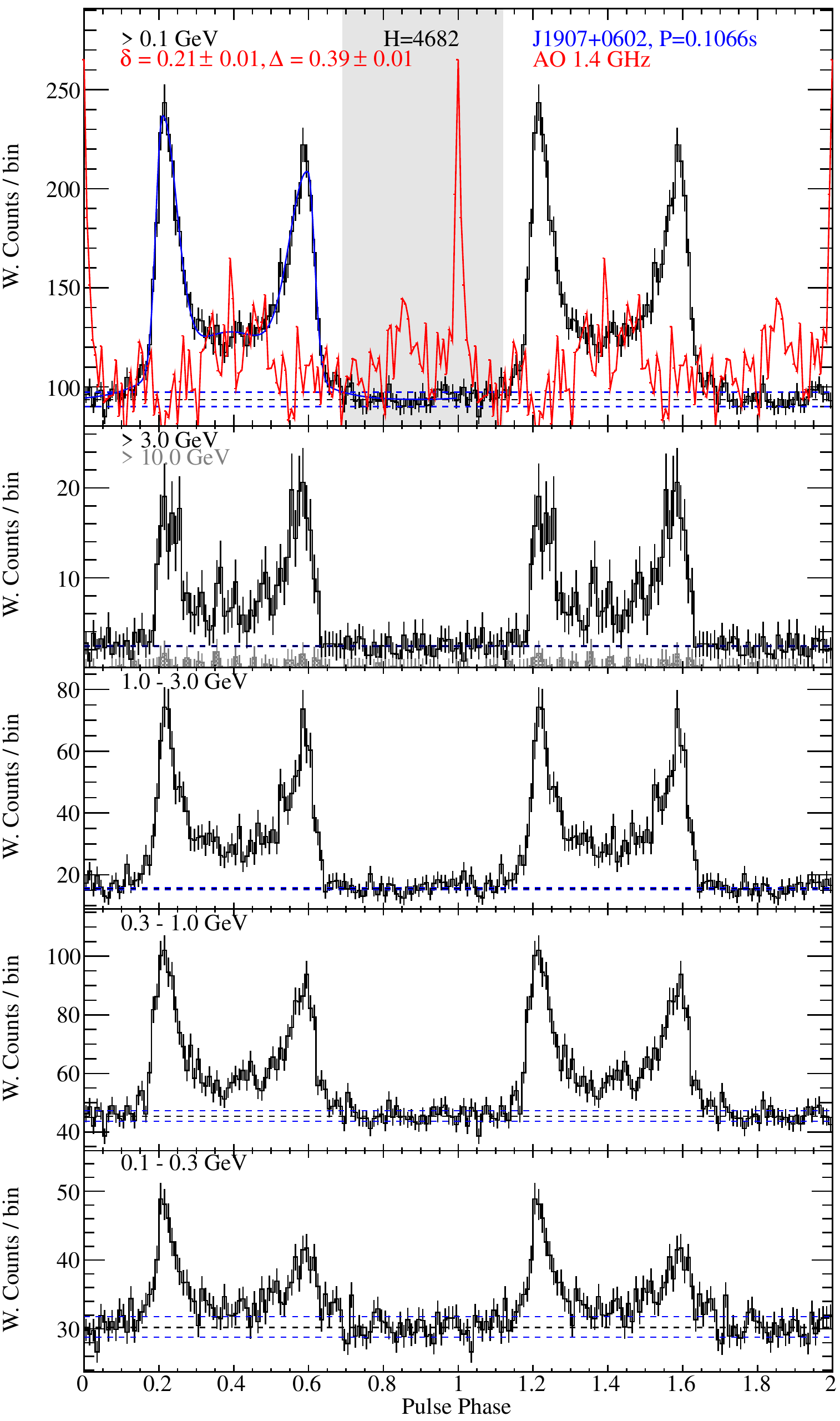}
\caption{Light curves of PSR J1907+0602s. Appendix A text describes the different frames.  This gamma-ray light curve illustrates the sharp, asymmetric peak structure described in Section \ref{lcTrends}.
\label{J1907+0602lc}}
\end{figure}

\begin{figure}[!ht]
\centering
\includegraphics[width=0.7\textwidth]{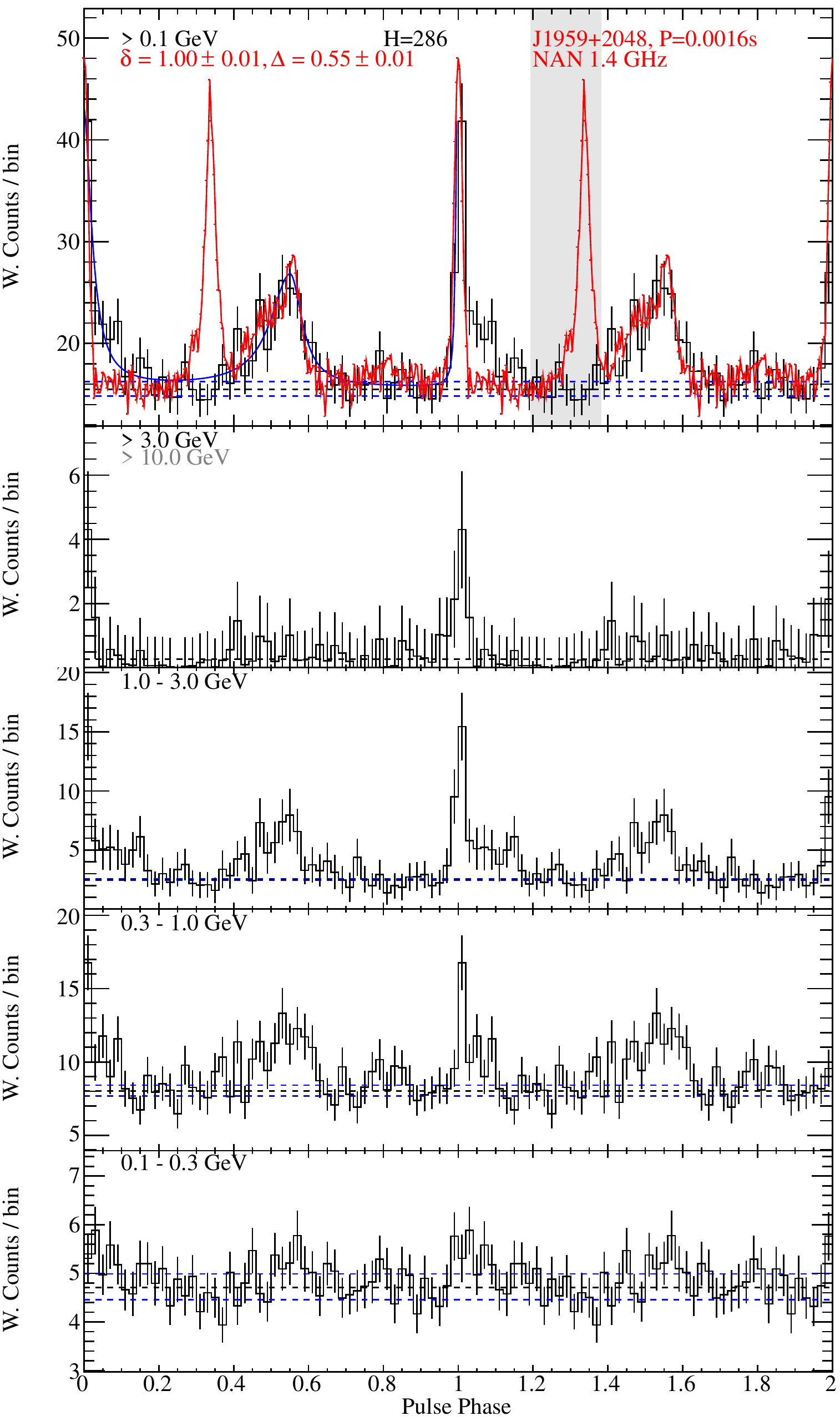}
\caption{Light curves of PSR J1959+2048. Appendix A text  describes the different frames.  This MSP has gamma-ray and radio peaks occuring at the same phase but with one unmatched radio peak that is not present at lower radio frequencies.
\label{J1959+2048lc}}
\end{figure}

\begin{figure}[!ht]
\centering
\includegraphics[width=0.7\textwidth]{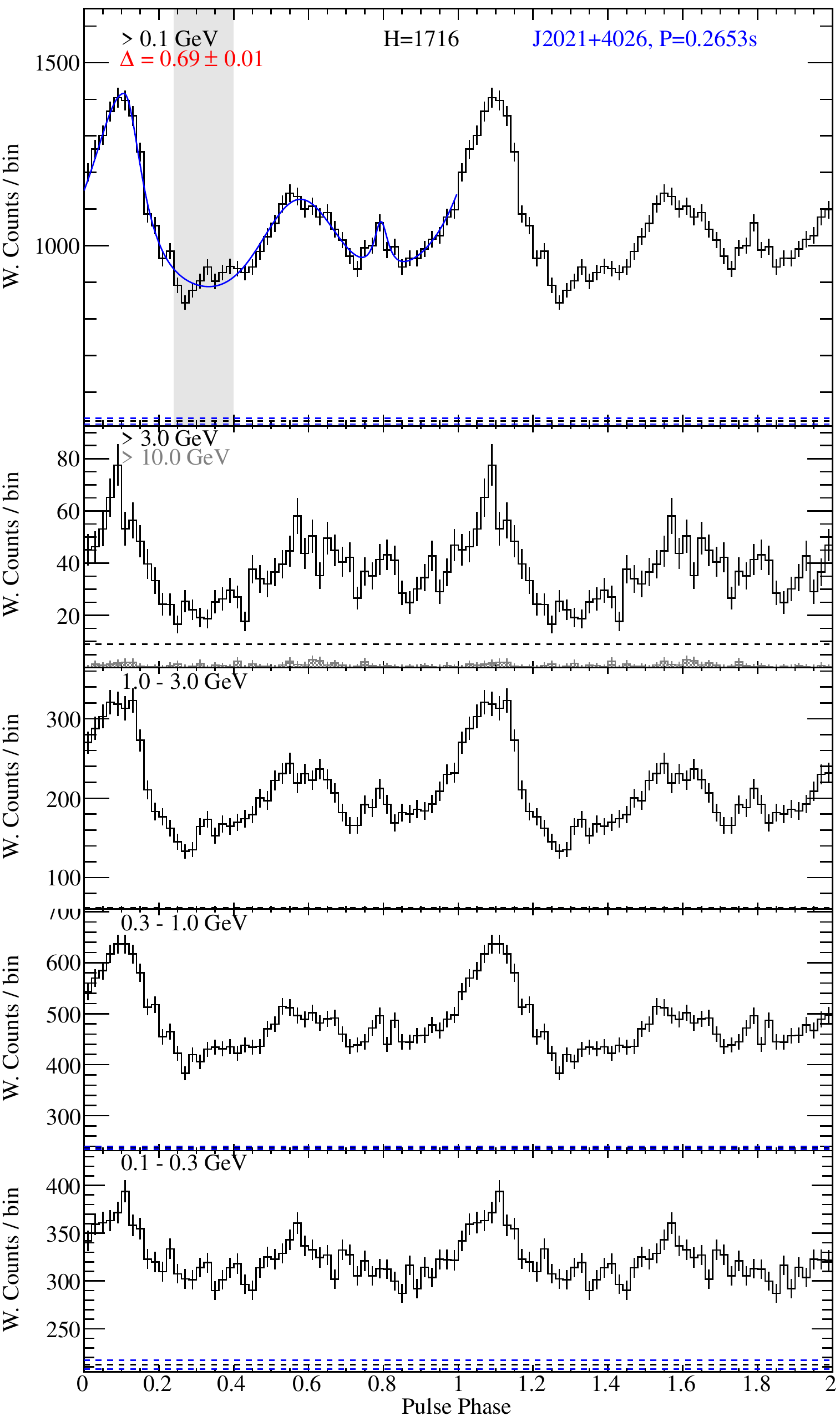}
\caption{Light curves of PSR J2021+4026. Appendix A text  describes the different frames.  This is an example of a radio-quiet pulsar with no radio detection and also a strong case for magnetospheric emission across the entire pulse.
\label{J2021+4026lc}}
\end{figure}

\begin{figure}[!ht]
\centering
\includegraphics[width=0.7\textwidth]{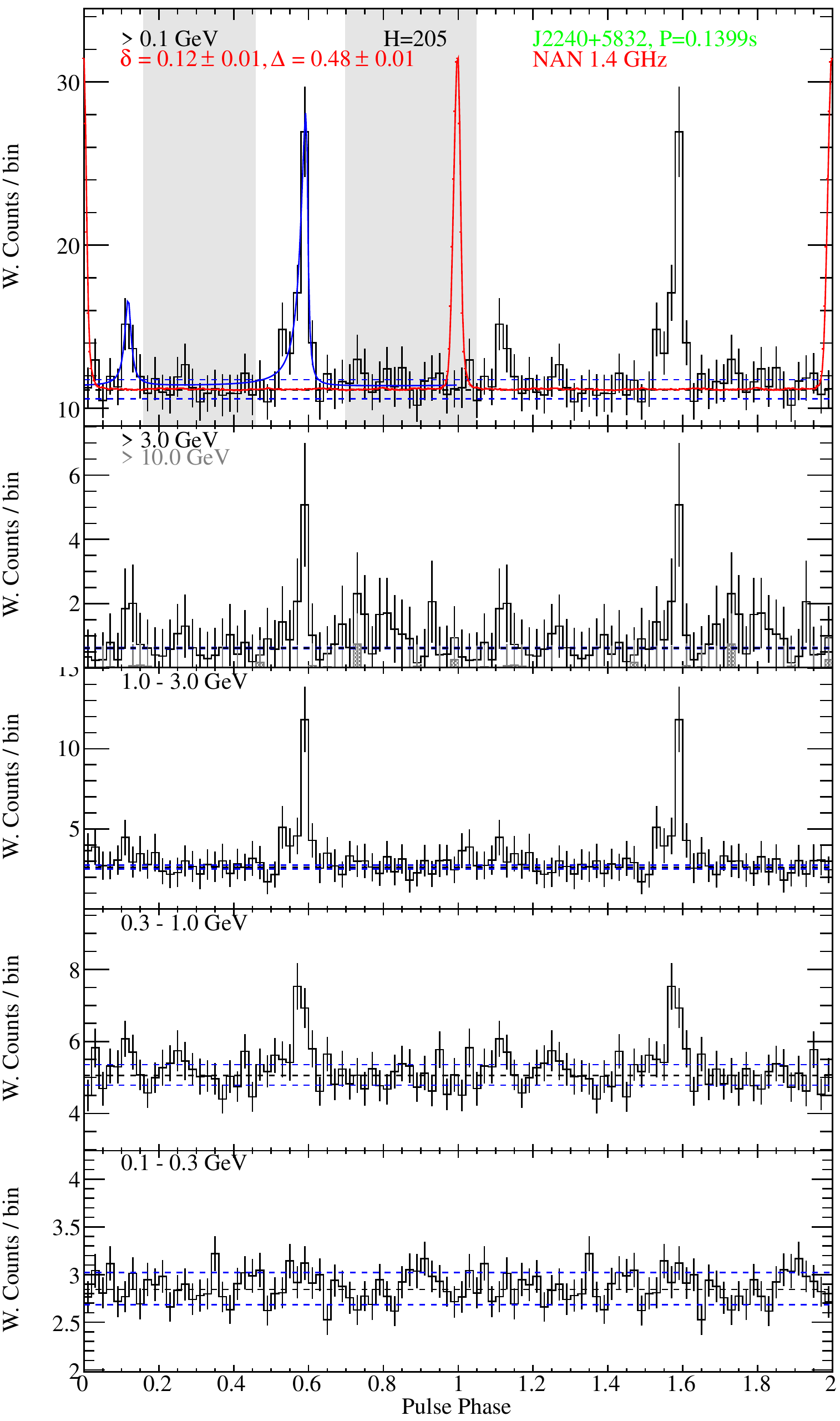}
\caption{Light curves of PSR J2240+5832. Appendix A text  describes the different frames.  This is an example of a two-peaked gamma-ray profile with clear evolution of the P2/P1 ratio with energy.
\label{J2240+5832lc}}
\end{figure}

\begin{figure}[!ht]
\centering
\includegraphics[width=0.7\textwidth]{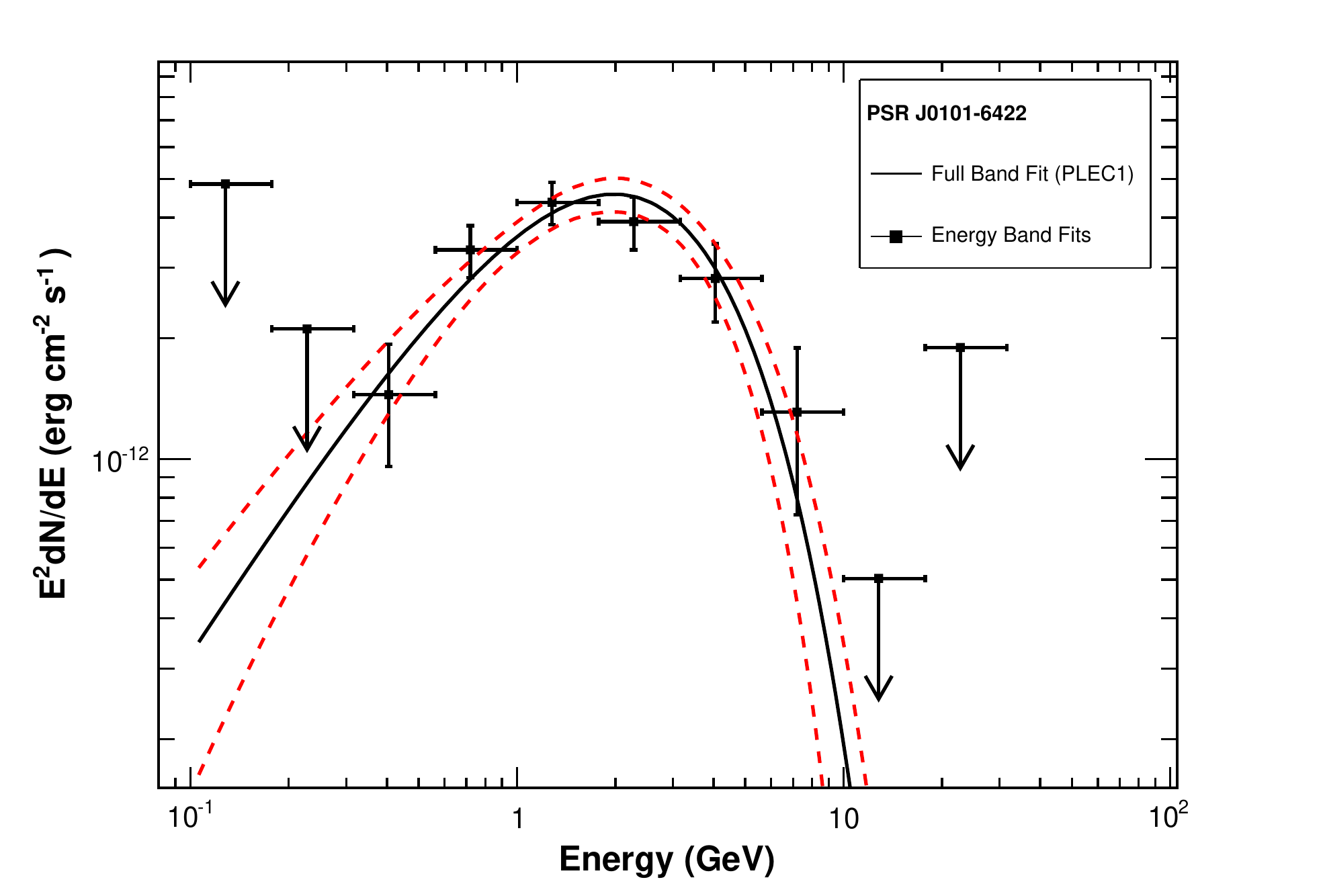}
\caption{Gamma-ray spectrum of PSR J0101$-$6422.  Appendix A text describes the figure components.
This spectrum demonstrates that pulsars are typically most significant near 1 GeV and often cannot be detected as point sources near 0.1 GeV.
\label{fig-J0101spec}}
\end{figure}

\begin{figure}[!ht]
\centering
\includegraphics[width=0.7\textwidth]{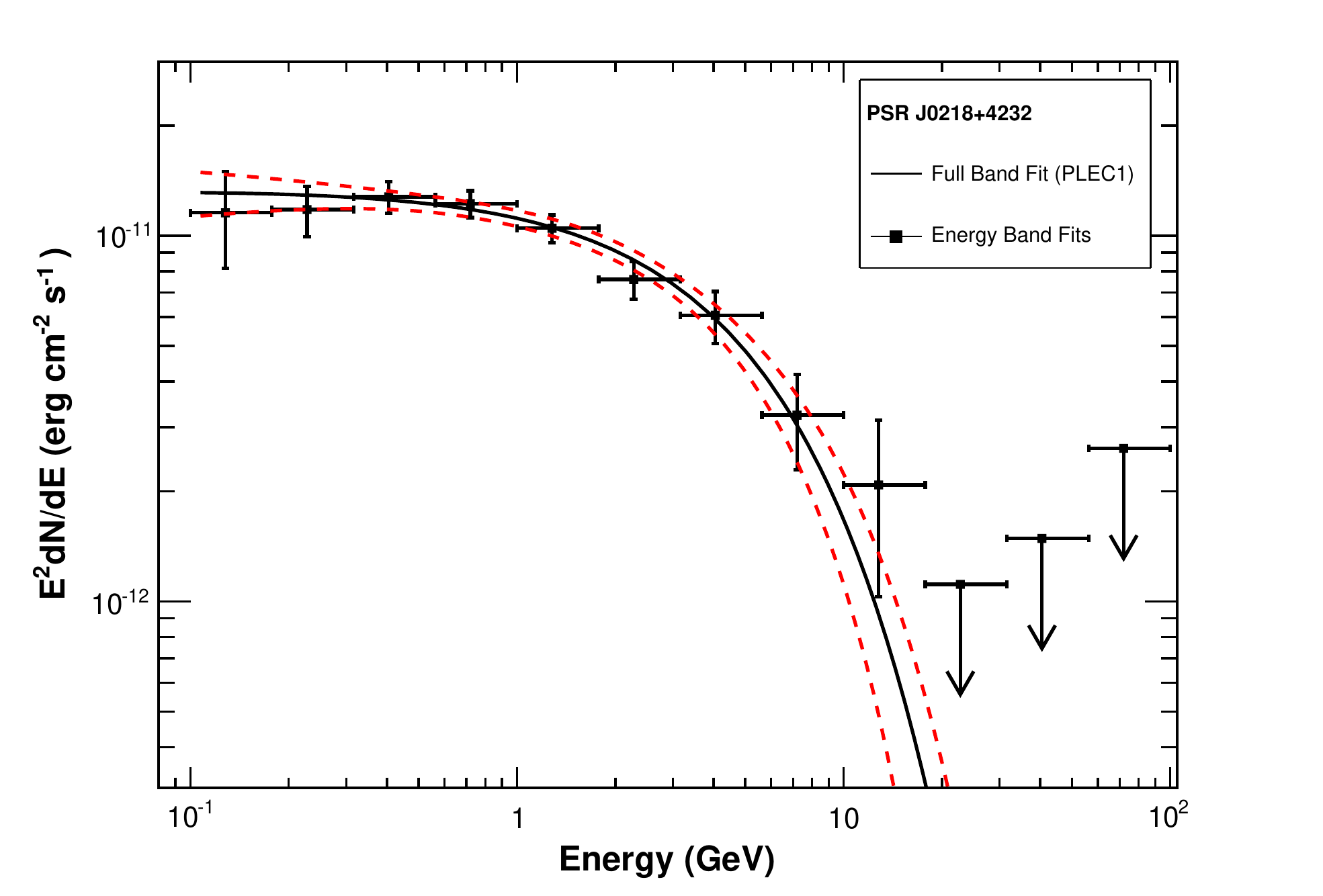}
\caption{Gamma-ray spectrum of PSR J0218+4232.  Appendix A text describes the figure components. 
This spectrum has a cutoff on the high-energy tail of the $E_{\rm cut}$ distribution.
\label{fig-J0218spec}}
\end{figure}

\begin{figure}[!ht]
\centering
\includegraphics[width=0.7\textwidth]{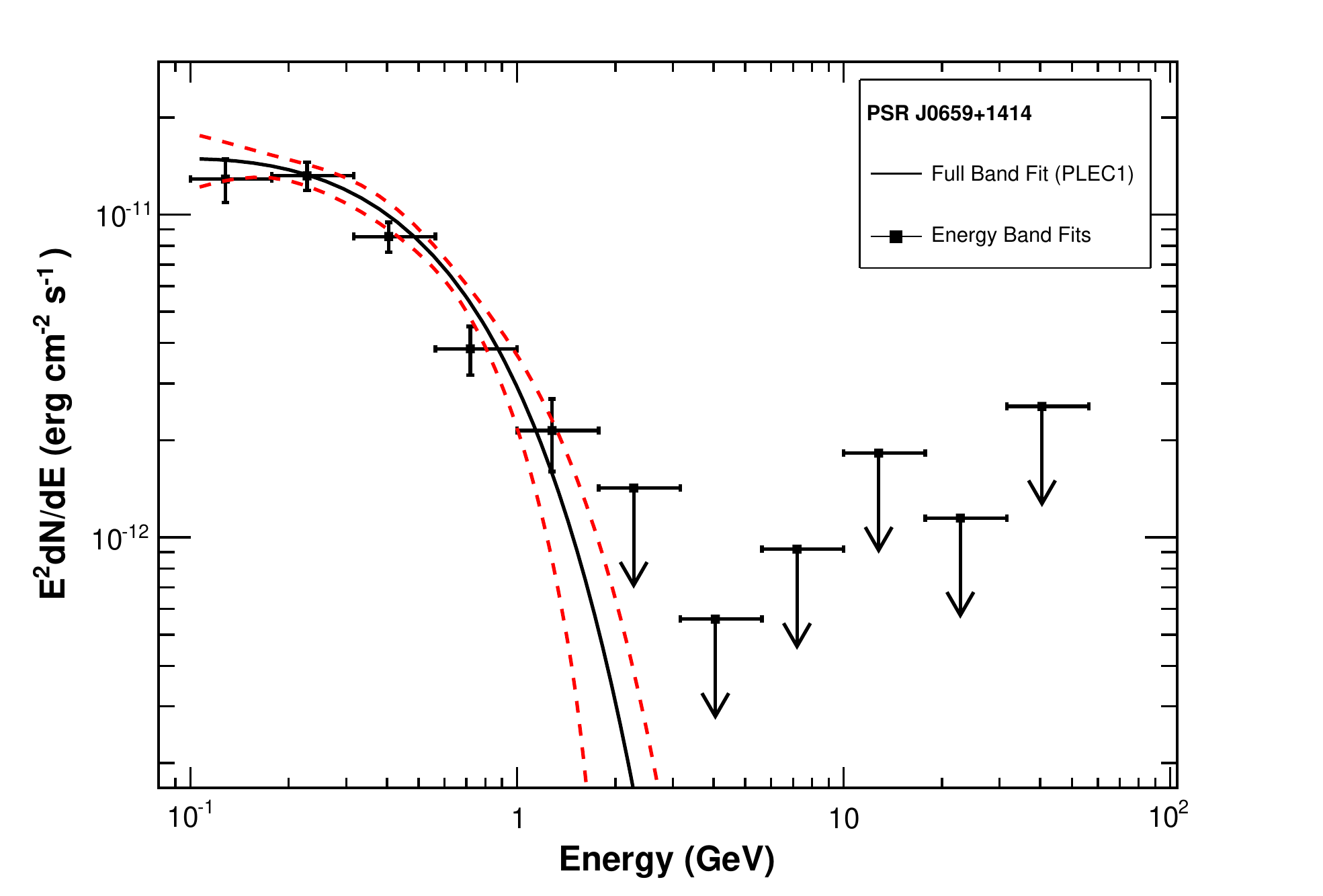}
\caption{Gamma-ray spectrum of PSR J0659+1414.  Appendix A text describes the figure components. 
This spectrum has a cutoff on the low-energy tail of the $E_{\rm cut}$ distribution.
\label{fig-J0659spec}}
\end{figure}

\begin{figure}[!ht]
\centering
\includegraphics[width=0.7\textwidth]{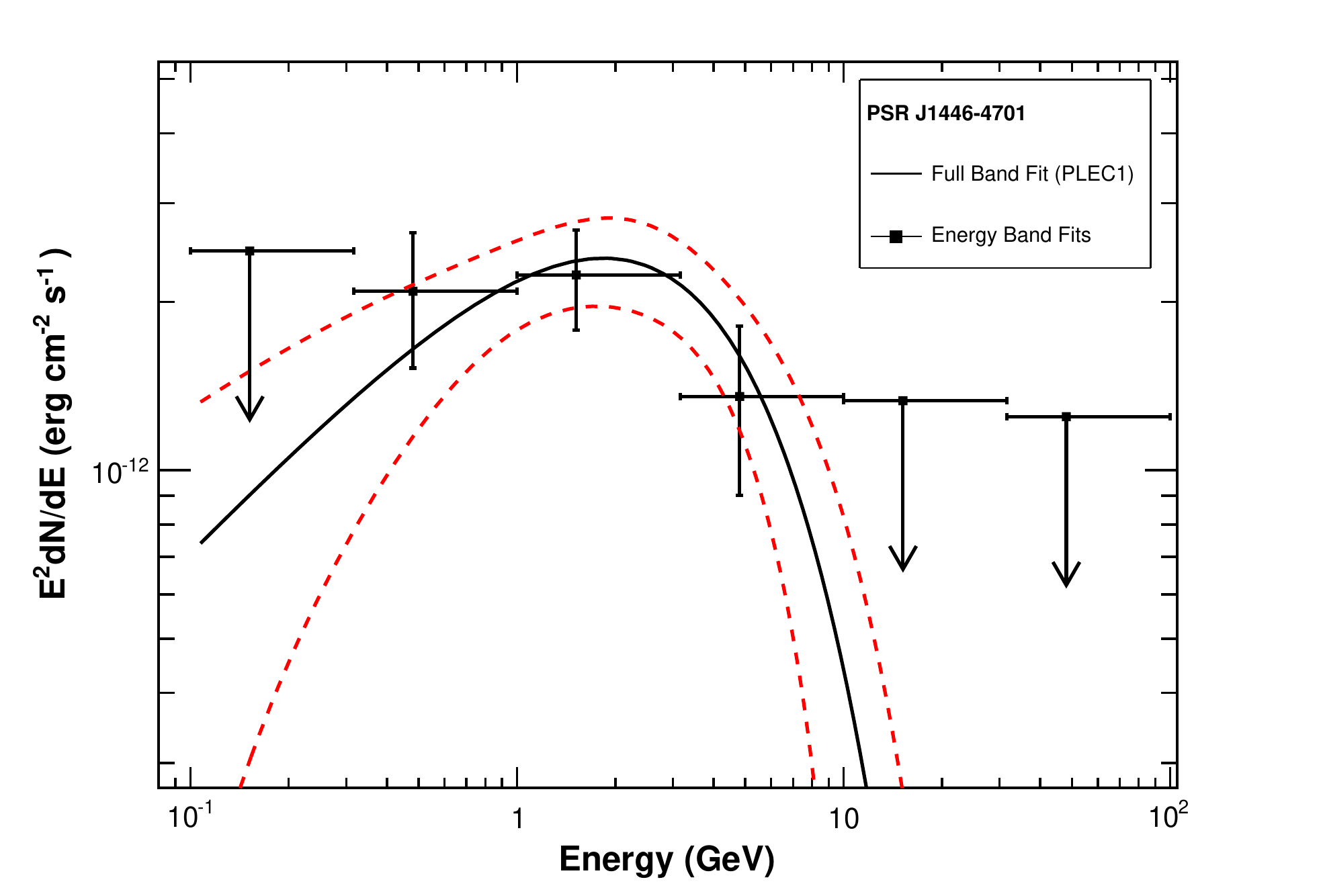}
\caption{Gamma-ray spectrum of PSR J1446$-$4701.  Appendix A text describes the figure components. 
This is a faint pulsar with only a few points in the spectrum.
\label{fig-J1446spec}}
\end{figure}

\begin{figure}[!ht]
\centering
\includegraphics[width=0.7\textwidth]{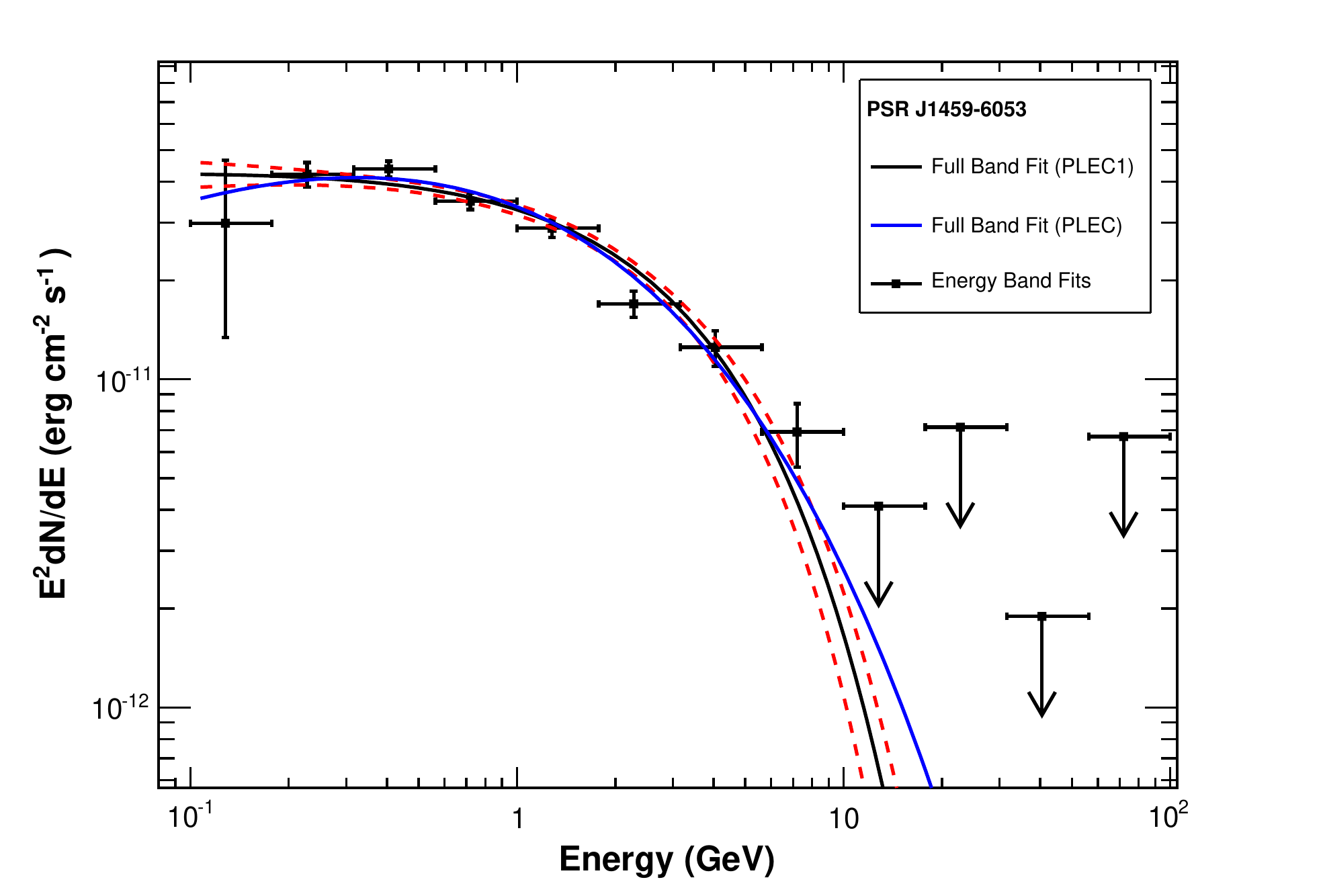}
\caption{Gamma-ray spectrum of PSR J1459$-$6053.  Appendix A text describes the figure components. 
For this pulsar, $TS_{\rm b\,free}\geq 9$ and the PLEC fit (exponential cutoff parameter not fixed to $b=1$) is shown 
(blue in the online version).
The full energy range maximum likelihood fits are, typically, a better measure of the spectrum than fits to the points.
\label{fig-J1459spec}}
\end{figure}

\begin{figure}[!ht]
\centering
\includegraphics[width=0.7\textwidth]{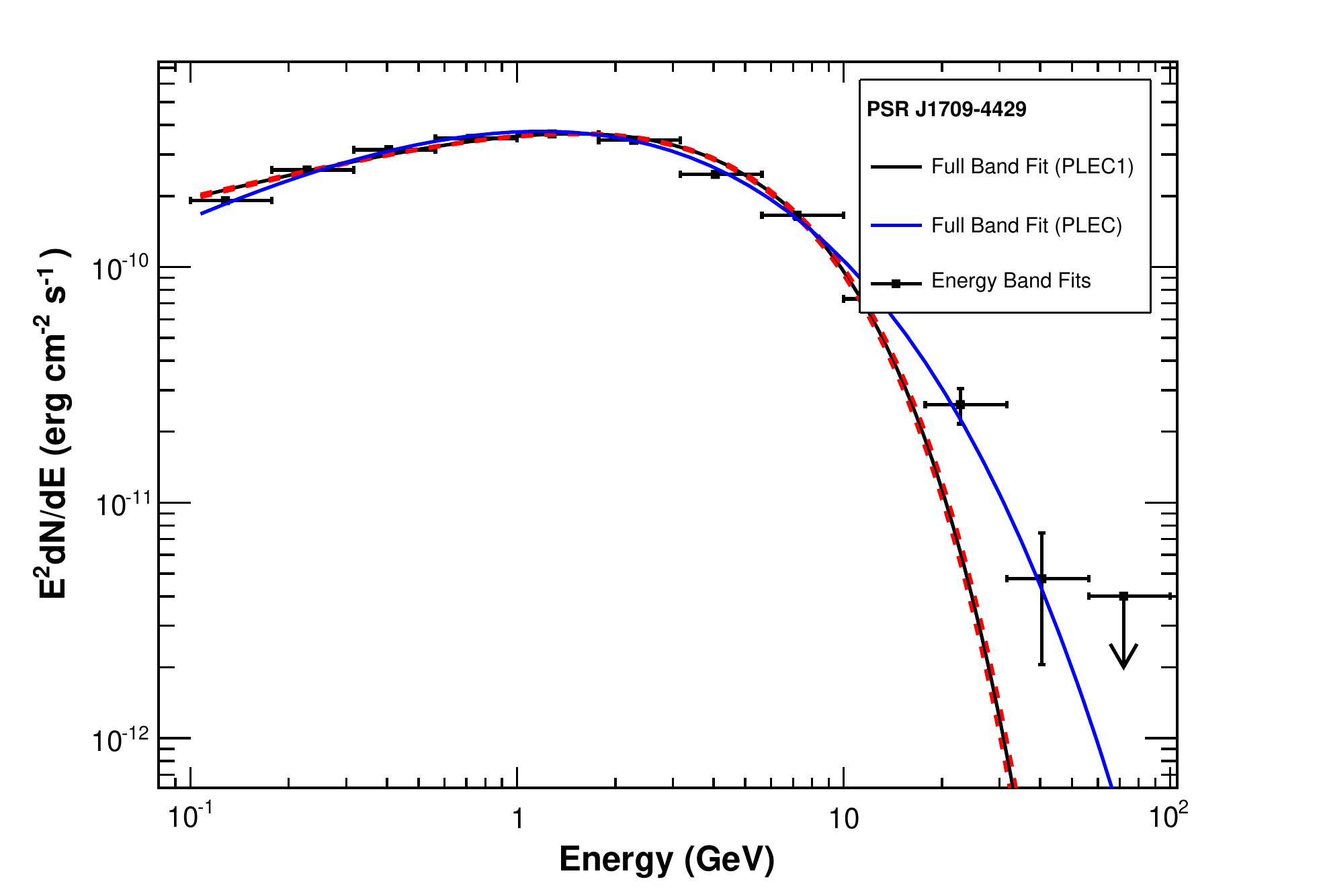}
\caption{Gamma-ray spectrum of PSR J1709$-$4429.  Appendix A text describes the figure components. 
This spectrum illustrates deviations from the PLEC1 model above 1 GeV (PLEC fit shown in blue online).
\label{fig-J1709spec}}
\end{figure}

\begin{figure}[!ht]
\centering
\includegraphics[width=0.7\textwidth]{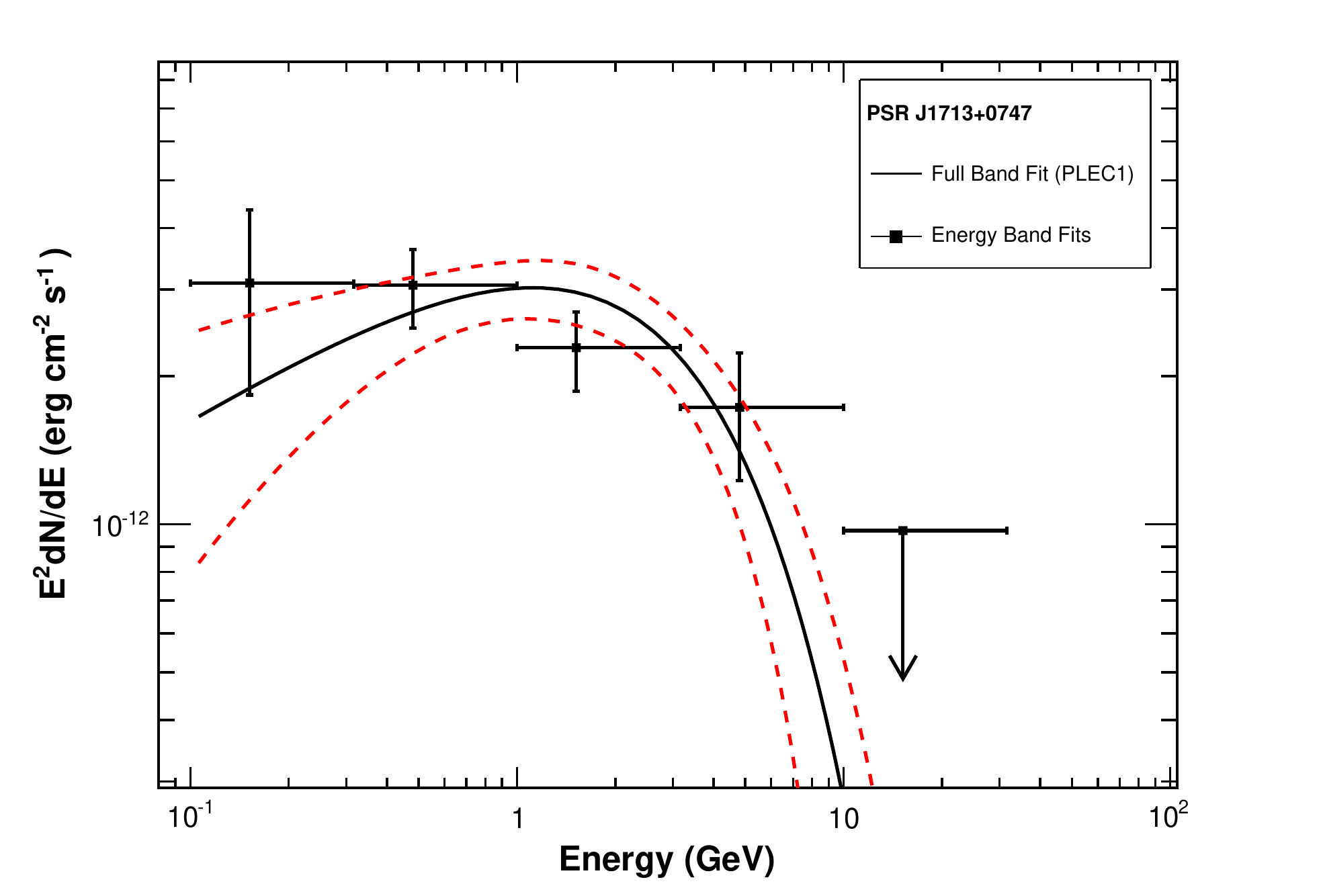}
\caption{Gamma-ray spectrum of PSR J1713+0747, Appendix A text describes the figure components. 
This is a faint pulsar with a reliable spectral fit and significant emission out to at least a few GeV.
\label{fig-J1713spec}}
\end{figure}

\begin{figure}[!ht]
\centering
\includegraphics[width=0.7\textwidth]{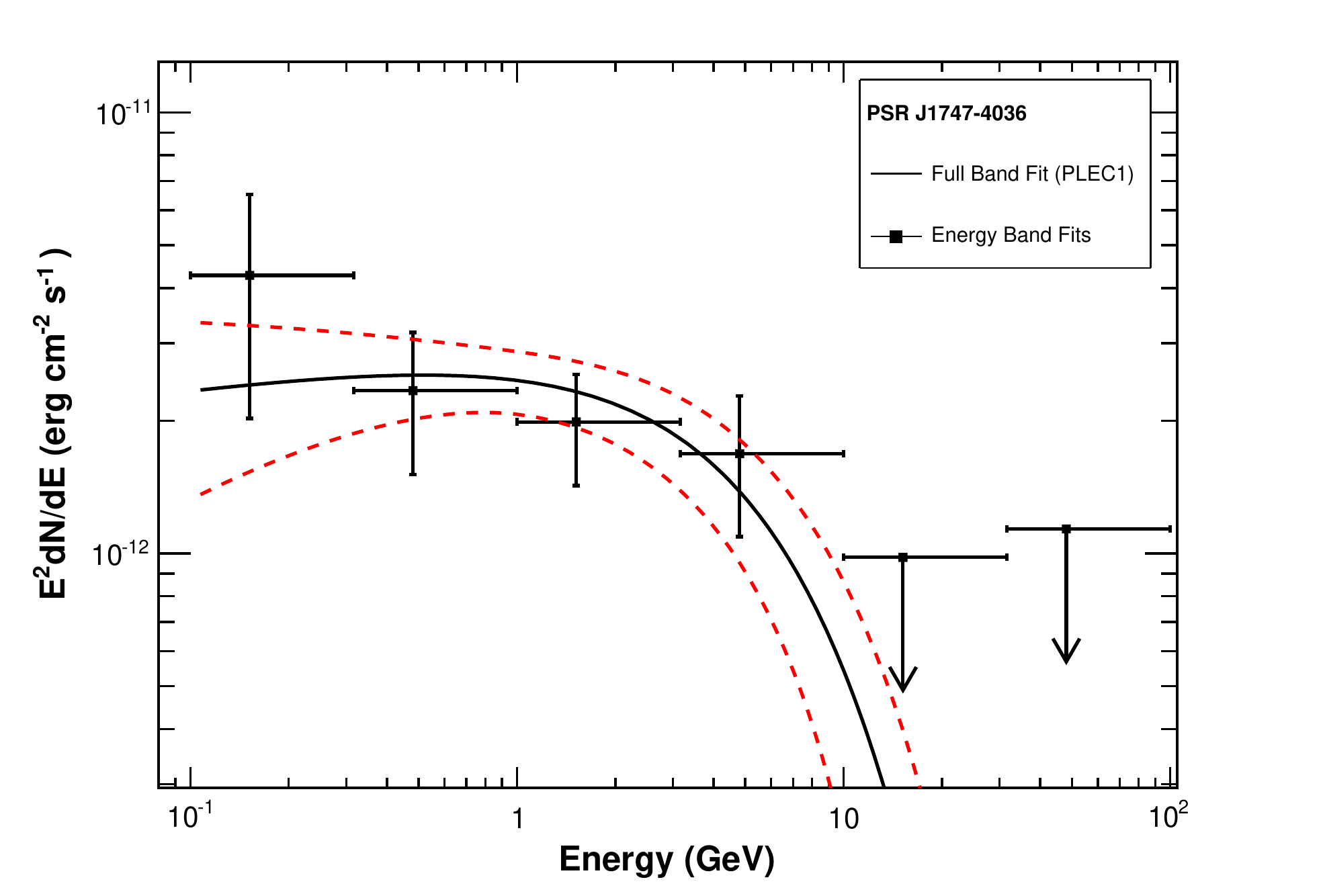}
\caption{Gamma-ray spectrum of PSR J1747$-$4036.  Appendix A text describes the figure components. 
This spectrum is an example of a faint pulsar for which significant spectral curvature cannot be seen with 3 years of data.
\label{fig-J1747spec}}
\end{figure}

\begin{figure}[!ht]
\centering
\includegraphics[width=0.7\textwidth]{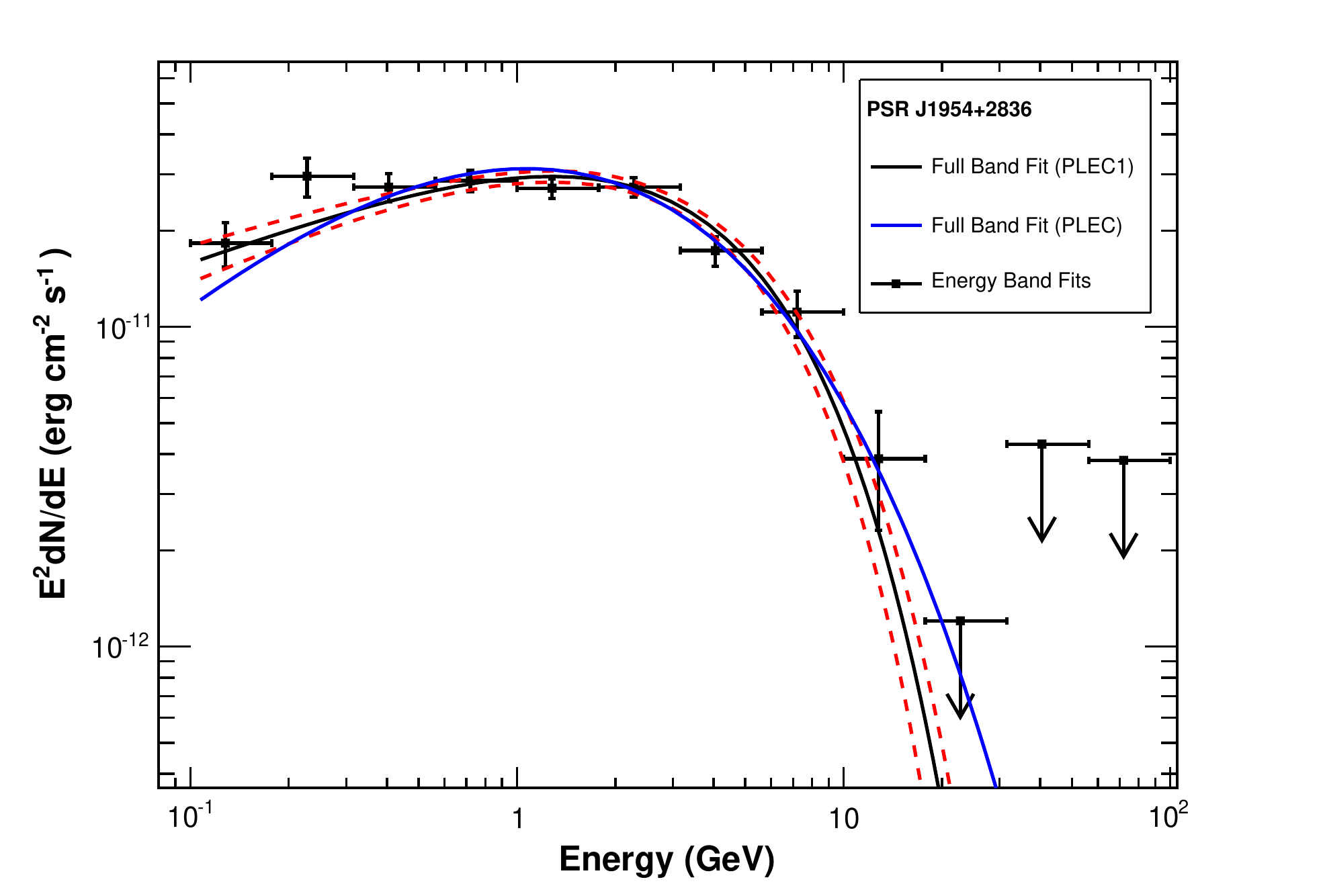}
\caption{Gamma-ray spectrum of PSR J1954+2835.  Appendix A text describes the figure components. 
This spectrum has a photon index on the soft (larger) end of the $\Gamma$ distribution.
For this pulsar, $TS_{\rm b\,free}\geq 9$ and the PLEC fit (exponential cutoff parameter not fixed to $b=1$) is shown 
(blue in the online version).
\label{fig-J1954spec}}
\end{figure}

Figures \ref{J0034-0534lc} to \ref{J2240+5832lc} show sample multiband light curves.  Figures \ref{fig-J0101spec} to \ref{fig-J1954spec} show sample gamma-ray spectra.  
Plots for all pulsars in this catalog are available in the on-line auxiliary material, detailed in Appendix \ref{online}.

For each pulsar, the top frame of the light curve figure  
shows the 0.1 to 100 GeV gamma-ray light curve, with the same data repeated over two rotations, to clarify structures near phases 0 and 1.  
The pulsar name follows the same color-code as the markers in the plots in the body of the text: red for MSPs ; blue for young radio-quiet, and green for young loud.
The letter P gives the rotation period. The letter H gives the H-test value for the gamma-ray light curve in the top frame. 
The letters d and D give the lag $\delta$ of the first gamma peak relative to the radio fiducial phase, and the separation $\Delta$ of the outermost gamma peaks, respectively.
The dark (blue online) curve over the phase range $\phi\in[0.0,1.0)$ in the top frame is the 0.1 to 100 GeV profile fit described in Section \ref{profiles}.  
The radio profile, when it exists, is drawn (red curve online) in the top frame, also repeated over 2 rotations, with the observing frequency as indicated on the figure. 
The radio telescopes that provided the profiles shown are indicated by NAN (Nan\c cay), PKS (Parkes), JBO (Jodrell Bank), AO (Arecibo), GBT (Green Bank), and WSRT (Westerbork).
The shaded gray region in the top frame represents the off-peak interval as defined in Section \ref{peak_definition}.  
The lower frames show the gamma-ray light curves in the indicated energy bands.  
The horizontal, dashed lines indicate the estimated background levels, for the gamma-ray light curves, and the associated uncertainties as described in Section \ref{profiles}.

For each pulsar, the gamma-ray spectral points are from individual energy-band fits in which the pulsar spectrum is approximated as a pure power law.  
For energy bands in which the pulsar is detected with $TS<9$ we report 95\% confidence level upper limits.  
The solid black line corresponds to the best-fit PLEC1 model from the full energy range fit, while the red (online) dashed lines represent the 1$\sigma$ confidence region on the best-fit model. 
For pulsars where a pure exponential cutoff ($b=1$) is disfavored ($TS_{\rm b\,free}\geq 9$) we also show the PLEC fit (blue online). 

\clearpage

\clearpage

\clearpage

\clearpage

\clearpage

\clearpage

\clearpage

%

\clearpage

\clearpage

\clearpage

\section{Appendix: Description of the online catalog files
\label{online}}

The complete results of the analyses described in this paper are reported in the online supplemental material. 
This supplement is an electronic archive, provided as a tarred and gzipped file (2PC\_auxiliary\_files\_v\#\#.tgz). 
Inside the archive is a directory structure containing FITS files with tables reporting the analysis results, images of the light curves and spectral results for each pulsar, 
a text file containing the rotation ephemeris used in the analysis of each pulsar, and individual FITS files for each pulsar with the light curves and spectra in numerical form.
This structure is described in Table \ref{tbl-files}. 
The online material is available at \url{http://fermi.gsfc.nasa.gov/ssc/data/access/lat/2nd_PSR_catalog/}.
%

Additional information, such as for example the half-widths at half-maximum of the leading and trailing edges of the peak fits described in Section \ref{profiles}, 
and their uncertainties, is also provided.
Another example is that in addition to the exponentially cut-off power law spectral parameters listed in Tables \ref{tbl-psrspec} and \ref{tbl-mspspec}, 
we provide the results of the pure power-law fits, and of the fits with the b parameter free for pulsars with $TS_{b\,free}\geq9$ (see Section \ref{spectralSection}).
Detailed column descriptions for the main FITS tables are in Section \ref{fits_main}, 
and detailed column descriptions for the individual pulsar FITS tables are in Section \ref{fits_individual}.


\subsection{Detailed column descriptions of Main catalog FITS tables
\label{fits_main}}


The main catalog file, 2PC\_catalog\_v\#\#.fits contains summary results for all 117 pulsars. The file contains four FITS table extensions: PULSAR\_CATALOG, SPECTRAL, OFF\_PEAK, and REFERENCES. 
The primary extension is empty.  Table \ref{main_pc_hdu} details the contents of the PULSAR\_CATALOG extension.
This file is also duplicated in ascii format (2PC\_catalog\_v\#\#.asc) with the four tables appended in the order listed above.

The PULSAR\_CATALOG extension contains one row for each pulsar with most of the information provided in the Tables in this paper, 
as well as some additional quantities that can be computed from these results. 
Exceptions are the complete spectral analysis (reported in the SPECTRAL extension, Table \ref{main_spec_hdu}), 
and the results of the off-peak analysis (reported in the OFF\_PEAK extension, Table \ref{main_offpeak_hdu}).

\begin{deluxetable}{p{1in} p{2in} p{2.5in}}
\tablewidth{0pt}
\tabletypesize{\scriptsize}
\tablecaption{LAT $2^{nd}$ Pulsar Catalog Auxiliary Files Description
\label{tbl-files}
}
\tablewidth{6in}
\tablehead{\colhead{Description} & \colhead{Filename(s) and path} & \colhead{Description of file contents}}
\startdata
 Main catalog FITS table  &   2PC\_catalog\_v\#\#.fits   &    This file contains summary results for all 117 pulsars. The file 
 										      contains four FITS table extensions: PULSAR\_CATALOG, SPECTRAL, 
										      OFF\_PEAK, and REFERENCES. The primary extension is empty. 
										      Section~\ref{fits_main} provides descriptions of these extensions.  \\
 Main catalog ascii file  &   2PC\_catalog\_v\#\#.asc   &    This file contains the same summary results for all 117 pulsars as 
 											in the previous file, but in ascii format. The four tables are appended
										   	in the ascii file in the order listed above.   \\
 Light curve  images  &   /images/lightcurves/PNG 
                                     \&  \newline /images/lightcurves/PDF   &    These directories contain plots in both~.pdf and~.png format for each pulsar 
 											of the 100 MeV - 100 GeV and best-fit gamma-ray 
											light curves, light curves in the 100 - 300 MeV, 300 MeV - 1 GeV, 
											1 - 3 GeV, greater than 3 GeV, and greater than 10 GeV bands, 
											as well as the radio profile for the pulsar, when one exists. \\
SED images  &   /images/SED/PNG 
                         \& \newline  /images/SED/PDF   &    These directories contain plots in both~.pdf and~.png format for each pulsar of the 
 								band-by-band energy fits plus the best-fit spectrum using the \fitsformat{\PLSuperExpCutoff} 
								model (a power law with exponential cutoff model where the exponential index is fixed at a value of 1).    \\
 Pulsar timing files  &  /par\_files   &  This directory contains text files with the ephemerides and other timing parameters
 						        used in this work. The files are in \textsc{Tempo2} format.\\
 Individual pulsar data FITS files  &   /PSR\_data/FITS   &    This directory contains FITS format tables for each pulsar with the spectral,
 											  light curve, and model data reported in this work. Each data file
 										           contains a number of FITS table extensions: PULSAR\_SED, MODEL\_SED, 
										           GAMMA\_LC, BEST\_FIT\_LC, and RADIO\_PROFILE 
										           (where a radio profile exists). The primary extension is empty.  Detailed
										           descriptions of these extensions are provided in Section~\ref{fits_individual}  \\
 Individual pulsar data ascii files  &   /PSR\_data/ascii   &    This directory contains the same results for each pulsar as in the previous
 											    directory, but in ascii format. The relevant extensions are appended
										   	    in the ascii file in the order listed above.   \\
\enddata
\end{deluxetable}
\clearpage

\begin{deluxetable}{p{1.5in} p{1in} p{3.5in}}
\tablewidth{0pt}
\tabletypesize{\scriptsize}
\tablecaption{LAT $2^{nd}$ Pulsar Catalog FITS format:  PULSAR\_CATALOG Extension
\label{main_pc_hdu}
}
\tablewidth{6in}
\tablehead{\colhead{Name} & \colhead{Units} & \colhead{Description}}
\startdata
 PSR\_Name   &  \nodata  &   Pulsar name   \\
 RAJ2000, 
 DEJ2000   &   deg   &   The pulsar position in celestial coordinates  (J2000).  \\
 GLON, 
 GLAT   &   deg   &   The pulsar position in galactic coordinates.   \\
 Period   &   ms   &   The pulsar rotation period.  \\
 P\_Dot   &  s s$^{-1}$   &   The period first derivative, uncorrected for Shklovskii 
                                                 effect or Galactic acceleration, see Section \ref{doppler}.  \\
 E\_Dot   &   erg s$^{-1}$   &   The pulsar spin-down luminosity, uncorrected for 
                                                 Shklovskii effect or Galactic acceleration, see Section \ref{doppler}.  \\
 F100, 
  Unc\_F100   &   \fluxunits   &   The best-fit photon flux and statistical error, integrated from 100 \mev to 100 \gev.
                                                        NULL values indicate unreliable spectral fits.\\
 G100, 
 Unc\_G100   &   \efluxunits  &   The best-fit energy flux and statistical error, integrated from 100 \mev to 100 \gev. NULL values indicate unreliable spectral fits.  \\
 TS\_DC   &  \nodata  &   The test statistic obtained at the position of the pulsar, assuming a 
                                             \fitsformat{\PLSuperExpCutoff} spectral model with the exponential index 
                                             fixed to 1. The fit uses data from 100 \mev to 100 \gev, and 
                                             includes all pulse phases except as noted in Tables \ref{tbl-psrspec} and \ref{tbl-mspspec}.\\
 TS\_Cutoff   &  \nodata  &   The significance of the spectral cutoff, obtained from the 
                                                 improvement in log(Likelihood) from the \fitsformat{\PLSuperExpCutoff} 
                                                 spectral model fit over the \fitsformat{PowerLaw} spectral fit. \\
 TS\_bfree   &  \nodata  &   The improvement in the test statistic when the exponential index is left free 
                                                 in the \fitsformat{\PLSuperExpCutoff} spectral fit.  If there is no improvement, 
                                                 or the fit is worse, this value is zero.  \\
 Photon\_Index, 
 Unc\_Photon\_Index   &  \nodata  &   The best-fit photon index and statistical error 
                                                                   for the \fitsformat{\PLSuperExpCutoff} spectral model. NULL values indicate unreliable spectral fits. \\
 Cutoff,
 Unc\_Cutoff   &   \mev   &  The best-fit cutoff energy and the statistical error
                                               for the \fitsformat{\PLSuperExpCutoff} spectral model. NULL values indicate unreliable spectral fits.   \\
 Distance, \newline
 Neg\_Unc\_Distance, 
 Pos\_Unc\_Distance   &   pc   &   The pulsar's distance measurement and its uncertainties. NULL values indicate that only an upper limit has been determined.   \\
 UL\_Distance         &   pc   &   Upper limit on pulsar distance when no measured value has been determined, NULL values indicate that a distance measurement has been determined.\\  
 Distance\_Method,   &  \nodata  &   The method used to determine the pulsar's distance. Methods are: 
                                                        \fitsformat{K} for the kinematic model, 
                                                        \fitsformat{DM} for the dispersion measure using the NE2001 model of \citet{Cordes2002}, 
                                                        \fitsformat{O} for optical measurements, and \fitsformat{X} for X-ray 
                                                        measurements. \fitsformat{DMM} means that the distance to the Galaxy's 
                                                        edge, as determined by the maximum DM value provided by the NE2001
                                                        model for that line of sight, is taken as an upper limit. \\
 Distance\_Ref   &  \nodata  &   Numerical reference for the distance measurement. The full reference is 
                                                        in the REFERENCES extension of this FITS file.\\
 Prop\_Motion, 
 Unc\_Prop\_Motion   &   mas yr$^{-1}$   &   The proper motion and errors for the pulsar when available.   \\
 Prop\_Motion\_Ref   &  \nodata  &   Numerical reference for the proper motion measurement. The full reference is 
                                                                 in the REFERENCES extension of this FITS file.  \\
 P\_Dot\_Int, 
 Neg\_Unc\_P\_Dot\_Int , 
 Pos\_Unc\_P\_Dot\_Int   &   s s$^{-1}$   &   The intrinsic P-dot and associated errors, after contributions from 
                                                                               the Shklovskii effect and Galactic acceleration have been removed, see Section \ref{doppler}.   \\
 E\_Dot\_Int, 
 Neg\_Unc\_E\_Dot\_Int, 
 Pos\_Unc\_E\_Dot\_Int   &   erg s$^{-1}$   &   The intrinsic spin down power and associated errors.   \\
 \tablebreak

 Luminosity, 
 Unc\_Luminosity, 
 Neg\_Sys\_Luminosity, 
 Pos\_Sys\_Luminosity   &    erg s$^{-1}$    &   The pulsar luminosity, statistical error, and systematic 
                                                                                    errors. Systematic errors are derived from the distance uncertainty.  
										    Values are NULL when only an upper limit exists.   \\
UL\_Luminosity       &      erg s$^{-1}$    &    Upper limit on the luminosity when no value has been determined.  Entries are NULL when a value has been determined.
                                               NULL values in all Luminosity columns indicate unreliable spectral fits.\\
 Efficiency, \newline
 Unc\_Efficiency, 
 Neg\_Sys\_Efficiency, 
 Pos\_Sys\_Efficiency   &   percent   &   The pulsar efficiency, statistical error, and systematic errors
                                                                      from the distance measurement.  Values are NULL when only an upper limit has been determined.   \\
UL\_Efficiency          &   percent   &   Upper limit on the pulsar efficiency.  Entries are NULL when a value has been determined. NULL values in all Efficiency columns indicate unreliable spectral fits.\\
 S1400   &   mJy   &   Radio flux density at 1400 MHz. In some cases, documented in Section \ref{S1400}, this value is extrapolated from measurements at other frequencies.  Entries are NULL when only an upper limit has been reported.   \\
 UL\_S1400   &  \nodata  &   Upper limit on the radio flux density at 1400 MHz when no measurement has been reported.  Entries are NULL when a value has been reported.  \\
 S1400\_Ref   &  \nodata  &   Numerical reference for the radio flux density measurement.  The full reference is 
                                                    in the REFERENCES extension of this FITS file.  \\
 Num\_Peaks   &  \nodata  &   Number of peaks in the gamma-ray profile.   \\
 Shift\_Method   &  \nodata  &   Method used to choose the radio fiducial phase. Methods are:  
                                                        \fitsformat{p} for the peak radio intensity,  \fitsformat{h} for an opposite 
                                                        hemisphere shift (0.5 phase shift from the peak intensity),   \fitsformat{s} for
                                                        the point of symmetry in the radio profile, and  \fitsformat{o} for some other 
                                                        method as described in the text (used only for PSR~J0534+2200).  \\
 Radio\_Lag, \newline
 Unc\_Radio\_Lag   &  \nodata  &   Phase separation ($\delta$) between the first gamma-ray peak and the radio 
                                                              peak, and the associated error on that separation.   \\
 Peak\_Sep, \newline
 Unc\_Peak\_Sep   &  \nodata  &   Phase separation ($\Delta$) between the first and last gamma-ray peaks, 
                                                             and the associated error on that separation. This value is \fitsformat{NULL} 
                                                             for pulsars with only a single gamma-ray peak.  \\
 HWHM\_P1\_L,
 Unc\_HWHM\_P1\_L   &  \nodata  &   Half-width half-max and corresponding uncertainty of the leading (left) 
                                                                    first peak edge, as fitted. The best-fit light curve is in the 
                                                                    BEST\_FIT\_LC extension in the individual FITS file for each pulsar.   \\
 HWHM\_P1\_R, 
 Unc\_HWHM\_P1\_R   &  \nodata  &   Half-width half-max and corresponding uncertainty of the trailing (right) 
                                                                    first peak edge, as fitted. The best-fit light curve is in the 
                                                                    BEST\_FIT\_LC extension in the individual FITS file for each pulsar.   \\
 HWHM\_P2\_L,
 Unc\_HWHM\_P2\_L   &  \nodata  &   Half-width half-max and corresponding uncertainty of the leading (left) 
                                                                    second peak edge, as fitted. The best-fit light curve is in the 
                                                                    BEST\_FIT\_LC extension in the individual FITS file for each pulsar.   \\
 HWHM\_P2\_R, 
 Unc\_HWHM\_P2\_R   &  \nodata  &   Half-width half-max and corresponding uncertainty of the trailing (right) 
                                                                    second peak edge, as fitted. The best-fit light curve is in the 
                                                                    BEST\_FIT\_LC extension in the individual FITS file for each pulsar.   \\
 \tablebreak
 H\_ColDensity, 
 Neg\_Unc\_H\_ColDensity, 
 Pos\_Unc\_H\_ColDensity   &    cm$^{-2}$   &   Hydrogen column density and associated systematic errors
                                                                                  from the distance measurement.  The values are \fitsformat{NULL} 
                                                                                  when only an upper limit for the hydrogen column density exists.   \\
UL\_H\_ColDensity          &    cm$^{-2}$   &   Upper limit on the hydrogen column density. Entries are NULL when a value has been reported.\\
 XFlux\_NonTherm,
 Unc\_XFlux\_NonTherm   &    \efluxunits   &   Non-thermal unabsorbed X-ray energy flux and 90\% CL statistical errors, in the 0.3-10 keV energy band. Spectrum is an absorbed 
                                                                                power law, plus black body model when needed.  Exceptions are PSRs J0633+1746 
                                                                                and J0659+1414 where a double black body plus power law model was 
                                                                                used.  Entries are NULL when only an upper limit has been reported. \\
UL\_XFlux\_NonTherm     &   \efluxunits    &   Non-thermal X-ray energy flux upper limit.  Entries are NULL when a value has been reported.\\
 XFlux\_PWN,
 Unc\_XFlux\_PWN   &   \efluxunits   &   Estimated non-thermal X-ray flux and 90\% CL statistical errors, from the brightest part of the associated plerion, in the 
                                                                    0.3-10 keV energy band.   \\
 X\_Qual    &  \nodata &   Quality of X-ray detections: `0' indicates no confirmed counterpart, `1' indicates that a counterpart has been identified but with too few counts for further characterization, and `2' indicates that a counterpart has been identified with sufficient counts for spectral characterization.\\
 Opt\_Mag   &  \nodata  &   Optical magnitude of the optical counterpart for the pulsar or pulsar system, where a counterpart is detected. NULL if no observation available.   \\
 LL\_Opt\_Mag   &  \nodata  &   Y indicates that Opt\_Mag is a lower limit on the optical magnitude.\\
 Opt\_Band   &  \nodata  &   The filter used for the optical observation.   \\
 Opt\_Object  &  \nodata  &   Object to which the mesured optical flux pertains.  The codes are B for binary system; U for upper limit; P = neutron star detected; P* = pulsed optical detection; 
 P+ = pulsar candidate (possible unpulsed pulsar detection); C = companion detected ; N = nebula (PWN) detected.\\
 Extinction,
 Neg\_Unc\_Extinction,
 Pos\_Unc\_Extinction   &  \nodata  &   Optical extinction and associated errors derived from the hydrogen column density and using the relation of \citet{Fitzpatrick1999}.  Entries are NULL when only an upper limit is reported.   \\
 UL\_Extinction         &  \nodata  &   Optical extinction upper limit when no value has been reported.  Entries are NULL when a value has been reported.\\
 Corr\_OptFlux,
 Neg\_Unc\_OptFlux,
 Pos\_Unc\_OptFlux   &   \efluxunits   &   Corrected (unabsorbed) optical energy flux in the V-band, and associated 
                                                                    errors. The optical flux has been corrected for interstellar reddening, and 
                                                                    scaled to the V-band (peak wavelength $\lambda = 5500$\AA, 
                                                                    bandwidth $\Delta\lambda = 890$\AA) where necessary.  Entries are NULL when only an upper limit has been reported.    \\
 UL\_OptFlux   &  \nodata  &    Upper limit on the corrected (unabsorbed) optical energy flux in the V-band when no measurement has been reported.  Entries are NULL when a value has been reported.   \\
 Type   &  \nodata  &   Indicates whether the pulsar is a young radio loud (\fitsformat{YRL}), young radio 
                                       quiet (\fitsformat{YRQ}, S1400 $> 30\, \mu$Jy), or millisecond (\fitsformat{MSP}) pulsar.   \\
 Binary   &  \nodata  &   \fitsformat{Y} indicates the pulsar is in a binary system.   \\
 History   &  \nodata  &   Indicates whether the pulsar was discovered in radio (\fitsformat{Radio}), 
 				     X-rays (\fitsformat{X-rays}), or gamma rays (\fitsformat{Gamma}).  \\
\enddata
\end{deluxetable}
\clearpage

The SPECTRAL extension (Table \ref{main_spec_hdu}) contains the results of the spectral analysis, one row for each pulsar. 
Models used in the spectral analysis are PLEC1, PLEC, and PL. 
The spectral analysis is described in Section~\ref{spectralMethodSection}.

The \fitsformat{Prefactor}, \fitsformat{Scale}, \fitsformat{Photon\_Index}, and \fitsformat{Cutoff} values for each pulsar using the PLEC1 model are provided.  
Results from the fit using the PLEC spectral model are only reported when $TS_{b\,free} \geq 9$. 
The differential spectrum of the PLEC spectral model (\fitsformat{\PLSuperExpCutoff}) is defined as:
\begin{equation}\label{PLEC_definition}
  \dnde = \fitsformat{Prefactor} \left(\frac{E}{\fitsformat{Scale}}\right)^{-\fitsformat{Photon\_Index}} \exp \left\{-\left(\frac{E}{\fitsformat{Cutoff}}\right)^{\fitsformat{Exponential\_Index}}\right\}.
\end{equation}
while the differential spectrum of the PLEC1 model is Eq.~\ref{PLEC_definition} with \fitsformat{Exponential\_Index} fixed to 1. When the PLEC fit is reported in the SPECTRAL extension, it includes the value for the \fitsformat{Exponential\_Index}

The differential spectrum of the PL (\fitsformat{PowerLaw}) spectral model is defined as:
\begin{equation}\label{PL_definition}
  \dnde = \fitsformat{Prefactor} \left(\frac{E}{\fitsformat{Scale}}\right)^{-\fitsformat{Photon\_Index}}
\end{equation}
with the \fitsformat{Prefactor}, \fitsformat{Scale}, and \fitsformat{Photon\_Index} for each pulsar using the PL model given in the SPECTRAL extension.

\begin{deluxetable}{p{1.5in} p{1in} p{3.5in}}
\tablewidth{0pt}
\tabletypesize{\scriptsize}
\tablecaption{LAT $2^{nd}$ Pulsar Catalog FITS format:  SPECTRAL Extension
\label{main_spec_hdu}
}
\tablewidth{6in}
\tablehead{\colhead{Name} & \colhead{Units} & \colhead{Description}}
\startdata
 PSR\_Name   &  \nodata  &   Pulsar name   \\
 On\_Peak   &  \nodata  &   \fitsformat{Y} indicates the spectral fit used only on-peak events.   \\
 TS\_DC   &  \nodata  &   The test statistic obtained at the position of the pulsar, assuming a 
                                             \fitsformat{\PLSuperExpCutoff} spectral model with the exponential index 
                                             fixed to 1. The fit uses data from from 100 \mev to 100 \gev, and 
                                             includes all pulse phases.\\
 TS\_Cutoff   &  \nodata  &   The significance of the spectral cutoff, obtained from the 
                                                 improvement in log(Likelihood) from the \fitsformat{\PLSuperExpCutoff} 
                                                 spectral model fit over the \fitsformat{PowerLaw} spectral fit. \\
 TS\_bfree   &  \nodata  &   The improvement in the test statistic when the photon index is left free 
                                                 in the \fitsformat{\PLSuperExpCutoff} spectral fit.  If there is no improvement, 
                                                 or the fit is worse, this value is zero.  \\
 PLEC1\_Prefactor,
 Unc\_ECPL1\_Prefactor   &   \prefunits   &   The best-fit prefactor and associated error for the 
 								  spectral fit using a power law with exponential cutoff model 
								  where the exponential index is fixed at a value of 1.    \\
 PLEC1\_Photon\_Index,
 Unc\_ECPL1\_Photon\_Index   &  \nodata  &   The best-fit photon index and associated error for the 
 								      spectral fit using a \fitsformat{\PLSuperExpCutoff} model 
								      where the exponential index is fixed at a value of 1.    \\
 PLEC1\_Scale   &   MeV   &   The scaling energy for the spectral fit using a \fitsformat{\PLSuperExpCutoff} model 
						           where the exponential index is fixed at a value of 1.    \\
 PLEC1\_Cutoff, 
 Unc\_PLEC1\_Cutoff   &   MeV   &   The best-fit cutoff energy and associated error for the 
 							spectral fit using a \fitsformat{\PLSuperExpCutoff} model 
							where the exponential index is fixed at a value of 1.    \\
 PLEC1\_Flux,
 Unc\_PLEC1\_Flux   &   \fluxunits   &   The photon flux integrated from 100 MeV to 100 GeV and associated error for the 
 							  spectral fit using a \fitsformat{\PLSuperExpCutoff} model 
							  where the exponential index is fixed at a value of 1.    \\
 PLEC1\_EFlux, 
 Unc\_PLEC1\_EFlux   &   \efluxunits   &   The energy flux  integrated from 100 MeV to 100 GeV and associated error for the 
 							       spectral fit using a \fitsformat{\PLSuperExpCutoff} model 
							       where the exponential index is fixed at a value of 1.    \\
 PLEC\_Prefactor, 
 Unc\_PLEC\_Prefactor   &   \prefunits   &   The best-fit prefactor and associated error for the 
 							   	spectral fit using a \fitsformat{\PLSuperExpCutoff} 
							   	model where the exponential index is left free.    \\ 
 PLEC\_Photon\_Index, 
 Unc\_PLEC\_Photon\_Index   &  \nodata  &   The best-fit photon index and associated error for the 
 								      spectral fit using a \fitsformat{\PLSuperExpCutoff} model 
								      where the exponential index is left free.    \\ 
 PLEC\_Scale   &   MeV   &   The scaling energy for the spectral fit using a \fitsformat{\PLSuperExpCutoff} 
                                                   model where the exponential index is left free.    \\ 
 PLEC\_Cutoff , 
 Unc\_PLEC\_Cutoff   &   MeV   &  The best-fit cutoff energy and associated error for the 
 								      spectral fit using a \fitsformat{\PLSuperExpCutoff} model 
								      where the exponential index is left free.    \\ 
 PLEC\_Exponential\_Index, 
 Unc\_PLEC\_Exponential\_Index   &  \nodata  &   The best-fit value and associated error for the spectral fit using a \fitsformat{\PLSuperExpCutoff} model where the exponential index is left free.    \\
 PLEC\_Flux, 
 Unc\_PLEC\_Flux   &   \fluxunits  &   The photon flux integrated from 100 MeV to 100 GeV and associated error for the spectral fit using a \fitsformat{\PLSuperExpCutoff} model where the exponential index is left free.    \\
 PLEC\_EFlux, 
 Unc\_PLEC\_EFlux   &   \efluxunits   &   The energy flux integrated from 100 MeV to 100 GeV and associated error for the spectral fit using a \fitsformat{\PLSuperExpCutoff} model where the exponential index is left free.    \\
 PL\_Prefactor, 
 Unc\_PL\_Prefactor   &   \prefunits   &   The best-fit prefactor and associated error for the 
 							   spectral fit using a \fitsformat{PowerLaw} model.    \\ 
 PL\_Photon\_Index, 
 Unc\_PL\_Photon\_Index   &  \nodata  &   The best-fit photon index and associated error for the 
 							       spectral fit using a \fitsformat{PowerLaw} model.    \\ 
 PL\_Scale   &   MeV   &   The scaling energy for the spectral fit using a \fitsformat{PowerLaw} 
                                                   model.    \\ 
 PL\_Flux, 
 Unc\_PL\_Flux   &   \fluxunits  &   The photon flux  integrated from 100 MeV to 100 GeV and associated error for the 
 						   spectral fit using a \fitsformat{PowerLaw} model.    \\
 PL\_EFlux, 
 Unc\_PL\_EFlux   &   \efluxunits  &   The energy flux integrated from 100 MeV to 100 GeV and associated error for the 
 						   	spectral fit using a \fitsformat{PowerLaw} model.    \\
\enddata
\end{deluxetable}
\clearpage


The OFF\_PEAK extension (Table \ref{main_offpeak_hdu}) contains the spatial and spectral results of the search for off-peak emission. 
The table contains one row for each pulsar. Details of this analysis are given in Section~\ref{off_peak_analysis}. 

For sources reported with a PL spectral model, the differential spectrum is defined as:
\begin{equation}\eqnlabel{powerlaw_definition}
  \dnde = \fitsformat{Prefactor\_OP} \left(\frac{E}{\fitsformat{Scale\_OP}}\right)^{-\fitsformat{Index\_OP}}.
\end{equation}
The \fitsformat{Prefactor\_OP}, \fitsformat{Index\_OP}, and \fitsformat{Scale\_OP} are given in the OFF\_PEAK extension.

For sources reported with a PLEC1 spectral model (\fitsformat{\PLSuperExpCutoff}), the differential spectrum is defined as:
\begin{equation}\eqnlabel{cutoff_powerlaw_definition}
  \dnde = \fitsformat{Prefactor\_OP} \left(\frac{E}{\fitsformat{Scale\_OP}}\right)^{-\fitsformat{Index\_OP}} \exp\left(-\frac{E}{\fitsformat{Energy\_Cutoff\_OP}}\right)
\end{equation}
and the \fitsformat{Prefactor\_OP}, \fitsformat{Index\_OP}, \fitsformat{Scale\_OP}, and \fitsformat{Energy\_Cutoff\_OP} are given in the OFF\_PEAK extension as described below.

For the Crab Nebula and Vela-X, we took the spectral shape and initial normalization from \citet{LAT_collaboration_crab_2012} and
\cite{FermiVelaX2nd}, respectively, and fit only a multiplicative offset (see Section~\ref{off_peak_analysis}). For these two
sources, the differential spectrum was defined as:

\begin{equation}\eqnlabel{filefunction_definition}
  \dnde = \fitsformat{Normalization\_OP} \left.\frac{dN}{dE}\right|_\mathrm{file}
\end{equation}
and \fitsformat{Normalization\_OP} is provided in the OFF\_PEAK extension of the main pulsar catalog FITS file.

References used for pulsar distances and radio flux values have been assigned a number, and the number is reported in the PULSAR\_CATALOG extension. 
The REFERENCES extension (Table \ref{main_ref_hdu}) provides the full information for each reference.

\begin{deluxetable}{p{1.5in} p{1in} p{3.5in}}
\tablewidth{0pt}
\tabletypesize{\scriptsize}
\tablecaption{LAT $2^{nd}$ Pulsar Catalog FITS format:  OFF\_PEAK Extension
\label{main_offpeak_hdu}
}
\tablewidth{6in}
\tablehead{\colhead{Name} & \colhead{Units} & \colhead{Description}}
\startdata
 PSR\_Name   &  \nodata  &   Pulsar name   \\
 Classification\_OP   &  \nodata  &  Off-peak emission class: \fitsformat{M} for magnetospheric (``pulsar-like''), 
                                                               \fitsformat{W} for possible PWN emission, and \fitsformat{U} for Unidentified.
                                                               \fitsformat{L} is for sources with no significant off-peak emission.   \\
 Min\_Phase\_OP, 
 Max\_Phase\_OP   &  \nodata  &  The minimum and maximum phase that defines the off-peak interval.\\
 Min\_2\_Phase\_OP, 
 Max\_2\_Phase\_OP   &  \nodata  &   For pulsars with two off-peak phase ranges, the 
                                                               minimum and maximum phase for the second off-peak interval.   \\
 TS\_point\_OP   &  \nodata  &   The test statistic obtained at the best-fit position of 
                                                               the assumed point-like source.
                                                               $\tspoint$ is computed at the best-fit position assuming a 
                                                               power-law spectral model (except for PSR~J0534+2200
                                                               as is described in Section \ref{off_peak_analysis}).   \\
 TS\_ext\_OP   &  \nodata  &   The significance of any possible extension 
                                                               for the source.   \\
 TS\_cutoff\_OP   &  \nodata  &   The significance of any spectral cutoff for a source detected in the off-peak
                                                              region. (Computed at the pulsar's position)   \\
 TS\_var\_OP   &  \nodata  &   The significance  of variability in the off-pulse emission.  \\
 Spectral\_Model\_OP   &  \nodata  &   For regions with a significant detection, this is the best spectral model selected by 
                                                               our analysis procedure described in Section \ref{off_peak_analysis}.  The possible
                                                               spectral models are \fitsformat{PowerLaw}, \fitsformat{\PLSuperExpCutoff}, 
                                                               and \fitsformat{FileFunction} and are consistent with
                                                               naming convention in \gtlike.  \\
 Flux\_OP, \newline
 Unc\_Flux\_OP   &   \fluxunits   &   The best-fit photon flux and estimated statistical error. 
                                                               The flux is integrated from 100 \mev to 316 \gev.  \\
 EFlux\_OP, \newline
 Unc\_EFlux\_OP   &   \efluxunits   &   The best-fit energy flux and estimated statistical error. 
                                                               The flux is integrated from 100 \mev to 316 \gev.   \\
 Prefactor\_OP, 
 Unc\_Prefactor\_OP   &   \prefunits   &   The best-fit prefactor and estimated statistical error 
                                                               for the \fitsformat{PowerLaw} and \fitsformat{\PLSuperExpCutoff} spectral models.
                                                               The prefactor is defined in \eqnref{powerlaw_definition} and 
                                                               \eqnref{cutoff_powerlaw_definition} for the two spectral models.   \\
 Normalization\_OP, 
 Unc\_Normalization\_OP   &  \nodata  &   The best-fit normalization and estimated statistical error for \fitsformat{FileFunction} spectral models. 
                                                               The normalization is defined in \eqnref{filefunction_definition}. This spectral model was only
                                                               used for the Crab Nebula and Vela-X.  \\
 Scale\_OP   &   \mev   &   The scaling energy for the \fitsformat{PowerLaw} and \fitsformat{\PLSuperExpCutoff} spectral models.
                                                               The scale is defined in \eqnref{powerlaw_definition} and 
                                                               \eqnref{cutoff_powerlaw_definition} for the two spectral 
                                                               models.  \\
 Index\_OP, \newline Unc\_Index\_OP   &  \nodata  &   The best-fit photon index and estimated statistical error 
                                                               for the \fitsformat{PowerLaw} and \fitsformat{PLSuperExpCutoff} spectral models.
                                                               The photon index is defined in \eqnref{powerlaw_definition} and 
                                                               \eqnref{cutoff_powerlaw_definition} for the two 
                                                               spectral models.  \\
 Energy\_Cutoff\_OP, 
 Unc\_Energy\_Cutoff\_OP   &   \mev   &   The best-fit cutoff energy and the estimated statistical error
                                                               for the \fitsformat{\PLSuperExpCutoff} spectral model. 
                                                               It is defined in \eqnref{cutoff_powerlaw_definition}.  \\
 Spatial\_Model\_OP   &  \nodata  &   For off-peak regions with a significant detection, the spatial model selected by
                                                               our analysis procedure described in Section \ref{off_peak_analysis}.
                                                               The choices are \fitsformat{At\_Pulsar}, \fitsformat{Point}, and \fitsformat{Extended}.  \\
 RAJ2000\_OP,
 DEJ2000\_OP   &   deg   &   The position of the source in celestial coordinates. 
                                                               For upper limits and sources with a best-fit spatial 
                                                               model at the pulsar position, this is the pulsar's position.
                                                               For sources where the localized position is the selected
                                                               spatial model, this is the best-fit position.
                                                               For spatially-extended sources, this is the center
                                                               of the best-fit extended source spatial model.    \\
 GLON\_OP, 
 GLAT\_OP   &   deg   &   This is the same as \fitsformat{RA\_J2000} and
                                                               \fitsformat{DEC\_J2000}, but in Galactic coordinates.   \\
 Unc\_Position\_OP   &   deg   &   For sources with a \fitsformat{Point} spatial model,
                                                               the estimated statistical error on the localization of the source. For sources
                                                               with an \fitsformat{Extended} spatial model, the 
                                                               estimated statistical error on the center of the extended source.   \\
 Extension\_OP,  
 Unc\_Extension\_OP   &   deg   &   For sources with an \fitsformat{Extended} spatial model, the best fit extension
                                                               and estimated statistical error.  \\
 PowerLaw\_Flux\_UL\_OP   &   \fluxunits   &   For regions with no significant detection, this is the 95\% confidence-level photon flux upper limit
                                                               computed assuming a \fitsformat{PowerLaw} spectral model with $\fitsformat{Index}=2$
                                                               and integrated from 100 \mev to 316 \gev.   \\
\tablebreak
 PowerLaw\_EFlux\_UL\_OP   &   \efluxunits   &   The same as \fitsformat{PowerLaw\_Flux\_UL}, but instead the energy flux
                                                               integrated from 100 \mev to 316 \gev.  \\
 Cutoff\_Flux\_UL\_OP   &   \fluxunits  &   For regions with no significant detection, the 95\% confidence-level photon flux upper limit
                                                               assuming a \fitsformat{\PLSuperExpCutoff} spectral model with a canonical pulsar
                                                               spectrum of $\fitsformat{Index}=1.7$ and $\fitsformat{Energy\_Cutoff}=3$.
                                                               This is the flux upper limit integrated from 100 \mev to 316 \gev.  \\
 Cutoff\_EFlux\_UL\_OP   &   \efluxunits   &   The same as \fitsformat{Cutoff\_Flux\_UL}, but instead the energy flux 
                                                                integrated from 100 \mev to 316 \gev.  \\
 SED\_Lower\_Energy\_OP, 
 SED\_Upper\_Energy\_OP, 
 SED\_Center\_Energy\_OP   &    \mev    &   For each region, we computed a Spectral Energy Distribution (SED)
                                                               for the source in 14 energy bins spaced uniformly from 100 \mev 
                                                               to 316 \gev (4 bins per energy decade). Therefore, each 
                                                               \fitsformat{SED\_*} column corresponds to a vector of 14 values,
                                                               one for each energy bin.
                                                               \fitsformat{SED\_Lower\_Energy}, \fitsformat{SED\_Upper\_Energy}, and
                                                               \fitsformat{SED\_Middle\_Energy} are the lower energy, upper energy,
                                                               and energy in the geometric mean of the energy bin for each SED point.    \\
 SED\_TS\_OP   &  \nodata  &   The test statistic obtained for each SED point.   \\
 SED\_Prefactor\_OP, 
 Neg\_Unc\_SED\_Prefactor\_OP, 
 Pos\_Unc\_SED\_Prefactor\_OP, 
 SED\_Prefactor\_UL\_OP   &  \prefunits   &   The best-fit prefactor, asymmetric lower and upper error, and
                                                               95\% confidence-level upper limit computed
                                                               for the source in each energy bin.
                                                               When $\tspoint\geq25$, a detection is quoted 
                                                               when $\fitsformat{SED\_TS}>4$
                                                               and an upper limit is quoted otherwise. When $\tspoint<25$,
                                                               all SED points are quoted as upper limits.   \\
\enddata
\end{deluxetable}

\begin{deluxetable}{p{1in} p{1in} p{3.5in}}
\tablewidth{0pt}
\tabletypesize{\scriptsize}
\tablecaption{LAT $2^{nd}$ Pulsar Catalog FITS format:  REFERENCES Extension
\label{main_ref_hdu}
}
\tablewidth{6in}
\tablehead{\colhead{Name} & \colhead{Units} & \colhead{Description}}
\startdata
 Ref\_Number   &  \nodata  &   Numerical value of the reference from Distance\_Ref and S1400\_Ref columns.   \\
 Citation   &  \nodata  &    Citation for each reference.  \\
 ADS\_URL   &  \nodata  &   URL for the reference at the Astrophysical Data Service (ADS). This webpage 
 					   provides links to the original publishing journal of the referenced paper, article, 
					   or catalog.   \\
 Title   &  \nodata  &   Title of the reference.   \\
\enddata
\end{deluxetable}
\clearpage

\subsection{Individual pulsar FITS files
\label{fits_individual}}

In addition to the summary information for each pulsar contained in the main catalog file, detailed results of the analyses are provided in the individual pulsar FITS files. 
Each file contains a variable number of FITS table extensions: PULSAR\_SED, MODEL\_SED, GAMMA\_LC, BEST\_FIT\_LC, and RADIO\_PROFILE (for radio detected pulsars). 
Again, the primary extension is empty.  These files are also provided in ascii format with the content of the FITS tables appended in the order listed above.

The PULSAR\_SED extension (Table \ref{individ_fitsed_hdu}) contains the photon and energy fluxes for each pulsar in either six or twelve energy bins, fitting the pulsar with a power-law spectral form. These points correspond to the black data points in the pulsar SED image files. The number of energy bins used in the SED varies with the significance of the pulsar. In a few cases, the pulsar is too faint to construct an SED or there were problems with the spectral fit, and this extension is not included.

The MODEL\_SED extension (Table \ref{individ_modelsed_hdu}) contains the model photon flux and bowtie uncertainty using the PLEC1 spectral form fitted over the full energy range. 
A description of the spectral fitting method is provided in Section~\ref{spectralMethodSection}. These points correspond to the red curves in the pulsar SED image files. 
In cases where the pulsar is too faint to construct an SED, this extension is not reported.

The GAMMA\_LC extension (Table \ref{individ_gammalc_hdu}) contains weighted counts and the corresponding uncertainties for light curves in six different energy ranges. The number of points in each light curve varies with the significance of the pulsar.  These points correspond to the light curves shown in black in the pulsar light curve image files.  The values for the background shown in those images are provided as keywords in the header of this FITS extension. 

The BEST\_FIT\_LC extension (Table \ref{individ_bestfitlc_hdu}) reports the fitted light curve that best represents the data, as described in Section~\ref{profiles}. These points correspond to the blue curves shown in the pulsar light curve image files.

The RADIO\_PROFILE extension (Table \ref{individ_radio_hdu}) reports the radio profile for the radio loud pulsars. 
These points correspond to the red curves shown in the pulsar light curve image files. This extension is not included for pulsars undetected in radio.

\begin{deluxetable}{p{1in} p{1in} p{3.5in}}
\tablewidth{0pt}
\tabletypesize{\scriptsize}
\tablecaption{LAT $2^{nd}$ Individual Pulsar FITS file format:  PULSAR\_SED Extension
\label{individ_fitsed_hdu}
}
\tablewidth{6in}
\tablehead{\colhead{Name} & \colhead{Units} & \colhead{Description}}
\startdata
 Energy\_Min, \newline
 Energy\_Max   &   GeV   &  Lower and upper bounds for each SED bin.   \\
 Center\_Energy   &   GeV   &   Central energy for each SED bin.   \\ 
 PhotonFlux   &   \fluxunits   &   Photon flux in bin   \\
 Unc\_PhotonFlux   &   \fluxunits   &   Best-fit value and associated error for the photon flux 
 							in each SED bin using a power law 
 							spectral model. The error is set to zero when the 
							given photon flux is an upper limit.   \\
 EnergyFlux, 
 Unc\_EnergyFlux   &   \efluxunits   &   Best-fit value and associated error for the energy flux 
 							in each SED bin using a power law spectral model. The 
							error is set to zero when the given energy flux is an upper limit.   \\
\enddata
\end{deluxetable}


\begin{deluxetable}{p{1in} p{1in} p{3.5in}}
\tablewidth{0pt}
\tabletypesize{\scriptsize}
\tablecaption{LAT $2^{nd}$ Individual Pulsar FITS file format:  MODEL\_SED Extension
\label{individ_modelsed_hdu}
}
\tablewidth{6in}
\tablehead{\colhead{Name} & \colhead{Units} & \colhead{Description}}
\startdata
 Energy\_Min, \newline
 Energy\_Max   &   GeV   &   Lower and upper bounds for each SED bin.   \\
 Center\_Energy   &   GeV   &   Central energy for each SED bin.   \\ 
 Model\_PhotonFlux   &   \fluxunits   &   Integrated photon flux in each bin calculated from the 
 							   \fitsformat{\PLSuperExpCutoff} model with the exponential index fixed at a value of 1 that
							   has been fitted over the full energy range (from 100 \mev to 100 \gev).    \\
 Bowtie\_Flux   &   \fluxunits   &   One-sigma uncertainty on the Model\_PhotonFlux used to construct the bowtie on the spectral plots. \\
\enddata
\end{deluxetable}


\begin{deluxetable}{p{1.5in} p{1in} p{3.5in}}
\tablewidth{0pt}
\tabletypesize{\scriptsize}
\tablecaption{LAT $2^{nd}$ Individual Pulsar FITS file format:  GAMMA\_LC Extension
\label{individ_gammalc_hdu}
}
\tablewidth{6in}
\tablehead{\colhead{Name} & \colhead{Units} & \colhead{Description}}
\startdata
 Phase\_Min,\newline
 Phase\_Max   &  \nodata  &   Lower and upper bounds for each bin in the gamma-ray light curve.   \\
 GT100\_WtCounts,
 Unc\_GT100\_WtCounts   &  \nodata  &   Weighted counts and associated error in each phase bin 
 								for the gamma-ray light curve using an energy range of 
								100 MeV to 100 GeV.  \\
 GT3000\_WtCounts, 
 Unc\_GT3000\_WtCounts   &  \nodata  &   Weighted counts and associated error in each phase bin 
 								for the gamma-ray light curve using an energy range of 
								3 GeV to 100 GeV.  \\
 GT10000\_WtCounts, 
 Unc\_GT10000\_WtCounts   &  \nodata  &   Weighted counts and associated error in each phase bin 
 								for the gamma-ray light curve using an energy range of 
								10 GeV to 100 GeV.  \\
 100\_300\_WtCounts, 
 Unc\_100\_300\_WtCounts   &  \nodata  &   Weighted counts and associated error in each phase bin 
 								for the gamma-ray light curve using an energy range of 
								100 MeV to 300 MeV.  \\
 300\_1000\_WtCounts, 
 Unc\_300\_1000\_WtCounts   &  \nodata  &   Weighted counts and associated error in each phase bin 
 								for the gamma-ray light curve using an energy range of 
								300 MeV to 1 GeV.  \\
 1000\_3000\_WtCounts, 
 Unc\_1000\_3000\_WtCounts   &  \nodata  &   Weighted counts and associated error in each phase bin 
 								for the gamma-ray light curve using an energy range of 
								1 to 3 GeV.  \\
\enddata
\end{deluxetable}


\begin{deluxetable}{p{1in} p{1in} p{3.5in}}
\tablewidth{0pt}
\tabletypesize{\scriptsize}
\tablecaption{LAT $2^{nd}$ Individual Pulsar FITS file format:  BEST\_FIT\_LC Extension
\label{individ_bestfitlc_hdu}
}
\tablewidth{6in}
\tablehead{\colhead{Name} & \colhead{Units} & \colhead{Description}}
\startdata
 Phase\_Min,
 Phase\_Max   &  \nodata  &    Lower and upper bounds for each bin in the best fit gamma-ray light curve.   \\
 Norm\_Intensity   &  \nodata  &   Normalized gamma-ray intensity for each bin in the best fit gamma-ray 
 						light curve. The intensity is normalized so that the integral of the profile is 
						$\sim1$ (i.e. normalized as a density function).\\
\enddata
\end{deluxetable}


\begin{deluxetable}{p{1in} p{1in} p{3.5in}}
\tablewidth{0pt}
\tabletypesize{\scriptsize}
\tablecaption{LAT $2^{nd}$ Individual Pulsar FITS file format:  RADIO\_PROFILE Extension
\label{individ_radio_hdu}
}
\tablewidth{6in}
\tablehead{\colhead{Name} & \colhead{Units} & \colhead{Description}}
\startdata
 Phase\_Min,
 Phase\_Max   &  \nodata  &   Lower and upper bounds for each bin in the radio light curve.\\
 Norm\_Intensity   &  \nodata  &   Normalized radio flux for each bin in the radio light curve. 
 						The flux is normalized so that the peak flux equals 1.   \\
\enddata
\end{deluxetable}

\clearpage

\section{Appendix: Off-Peak Individual Source Discussion}
\label{App-off_peak_individual_source_discussion}

Here we discuss several interesting sources found in the off-peak
analysis presented in Section~\ref{unpulsed}.

The off-peak emission from PSR J0007+7303 in the SNR CTA1 was previously studied by \cite{Fermi_CTA1_2012}.
They found a soft and not-significantly cut off source in the off-peak region that
is marginally extended.
We find a similar spectrum and extension significance ($TS_{\rm ext}=10.8$), and therefore classify this
source as type `U'.

The new type `W' source is associated with PSR J0205+6449 \citep{LATPSR0205}.
The off-peak spectrum for this source is shown in panel b of Figure
\ref{off_peak_seds}.  The emission is best fit as a point source at
$(l,b)=(130\fdg73,3\fdg11)$ with a 95\% confidence-level radius of $0\fdg03$.
The source has a hard spectrum (power law with $\Gamma=1.61\pm0.21$)
and is therefore consistent with a PWN hypothesis.
This nebula has been observed at infrared
\citep{3C58_IR_PWN} and X-ray \citep{3C58_Xray_PWN} energies. This
suggests that we could be observing the inverse Compton emission from
the same electrons powering synchrotron emission at lower energies.
The PWN hypothesis is supported by the associated pulsar's
very high $\dot E=2.6\times10^{36}$ erg s$^{-1}$ and relatively young
characteristic age, $\tau_c = 5400$ yr. This is consistent with the properties
of other pulsars with LAT-detected PWN, and we favor a PWN
interpretation.
We note that the discrepancy between our spectrum and the upper limit
quoted in \citet{LAT_collaboration_PWNCAT_2011} is mainly caused by
our expanded energy range and because the flux upper limit was computed
assuming a different spectral index.

However, we note that PSR J0205+6449 is associated to the SNR 3C58
(G130.7+3.1).  Given the 2 kpc distance estimate from Section \ref{Distances}
and the density of thermal material estimated by \cite{3C58_Xray_PWN}, we can estimate
the energetics required for the LAT emission to originate in the SNR.
Following the prescription in \cite{Gamma_Visibility_SNRs_Drury_1994},
we assume the LAT emission to be hadronic and estimate a cosmic-ray
efficiency for the SNR of $\sim10$\%, which is energetically allowed.
We therefore cannot rule out the SNR hypothesis.

No TeV detection of this source has been reported, but given the hard photon
index at GeV energies this is a good candidate for observations by
an atmospheric Cherenkov telescope. Improved spectral and spatial observations 
at TeV energies might help to uniquely classify the emission.

We obtain a flux for Vela-X which is $\sim10\%$ larger than the flux
obtained in \cite{FermiVelaX2nd}. This discrepancy is most-likey due to
assuming a different spatial model for the emission (radially-symmetric
Gaussian compared to elliptical Gaussian).

PSR J1023$-$5746 is associated with the TeV PWN HESS J1023$-$575
\citep{HESS_J1023-575_HESS_Collaboration_2007}.  LAT emission from
this PWN was first reported in \citet{LAT_collaboration_PWNCAT_2011}.
Because of the dominant low-energy magnetospheric emission, we classify
this as type `M' and not as a PWN.
A phase-averaged analysis of this source for energies above 10 GeV is
reported in \citet{Rousseau2013}.

PSR J1119$-$6127 \citep{FermiMagnetars} is associated with the TeV source HESS
J1119$-$614\footnote{The discovery of HESS J1119$-$614 was presented at the
``Supernova Remnants and Pulsar Wind Nebulae in the Chandra Era'' in
2009. See \url{http://cxc.harvard.edu/cdo/snr09/pres/DjannatiAtai\_Arache\_
v2.pdf}.}. Our off-peak analysis 
classifies this source as `U' because its spectrum is soft and not significantly
cut off. However, the SED appears to represent a cutoff spectrum at low
energy and a hard rising spectrum at high energy.  
\citet{Rousseau2013} significantly detect this PWN using the analysis procedure as described for J1023$-$575.
We are likely detecting a composite of magnetospheric emission at low energy and pulsar-wind emission
at high energy.

PSR J1357$-$6429 \citep{LAT_J1357} has an associated PWN HESS J1356$-$645 detected at
TeV energies \citep{HESS_HESS_J1356-645_2011}.  Our analysis of the off-peak regions surrounding PSR
J1357$-$6429 shows a source positionally and spectrally consistent with
HESS J1356$-$645, but with significance just below detection threshold ($TS=21.0$).  
\citet{Rousseau2013} present significant emission from this source.

The off-peak region of PSR J1410$-$6132 \citep{OBrien_2008} shows a relatively hard
spectral index of $1.90\pm0.15$, and the spectrum is not significantly
cut off.  There is no associated TeV PWN
and enough low-energy GeV emission is present to caution against a clear
PWN interpretation.  We classify this source as `U', but
further observations could reveal interesting emission.

PSR J2021+4026 is spatially coincident with the
LAT-detected and spatially extended Gamma Cygni SNR
\citep{LAT_collaboration_extended_search_2012}.  The off-peak emission
from this pulsar is consistent with an exponentially-cutoff spectrum and
is therefore classified as type `M'.  The source's marginal extension
($TS_{\rm ext}=8.7$) is likely due to some contamination from the SNR.



\end{document}